\documentclass[a4paper]{book}
\usepackage[utf8x]{inputenc}
\usepackage[left=3cm,right=3cm,top=3cm,bottom=3cm]{geometry}
\usepackage[linktoc=all,hidelinks]{hyperref}
\usepackage{bookmark}
\usepackage{appendix}
\usepackage[nottoc,notlot,notlof,numbib]{tocbibind}
\usepackage{emptypage}
\usepackage[shortlabels]{enumitem}
\usepackage{cite}
\usepackage{booktabs}
\usepackage{multirow}
\usepackage[labelfont=bf]{caption}

\usepackage{amsmath}
\usepackage{amsfonts}
\usepackage{amssymb}

\usepackage[smalltableaux,centertableaux,centerboxes]{ytableau}
\usepackage{tensor}
\usepackage{braket}

\usepackage{tikz}

\newcommand{\Z}{\mathbb{Z}}
\newcommand{\R}{\mathbb{R}}
\newcommand{\C}{\mathbb{C}}

\renewcommand{\a}{\alpha}

\newcommand{\gL}{\Lambda}

\newcommand{\Th}{\Theta}

\newcommand{\cA}{\mathcal{A}}
\newcommand{\cB}{\mathcal{B}}
\newcommand{\cE}{\mathcal{E}}
\newcommand{\cF}{\mathcal{F}}
\newcommand{\cG}{\mathcal{G}}
\newcommand{\cH}{\mathcal{H}}
\newcommand{\cI}{\mathcal{I}}
\newcommand{\cK}{\mathcal{K}}
\newcommand{\cL}{\mathcal{L}}
\newcommand{\cM}{\mathcal{M}}
\newcommand{\cN}{\mathcal{N}}
\newcommand{\cR}{\mathcal{R}}
\newcommand{\cS}{\mathcal{S}}
\newcommand{\cV}{\mathcal{V}}

\newcommand{\tta}{\mathtt{a}}
\newcommand{\ttb}{\mathtt{b}}
\newcommand{\ttc}{\mathtt{c}}
\newcommand{\ttd}{\mathtt{d}}
\newcommand{\tte}{\mathtt{e}}

\newcommand{\ReDeclareMathOperator}[2]{\let #1 \relax \DeclareMathOperator{#1}{#2}}
\newcommand{\nn}{\nonumber}
\newcommand{\up}[1]{^{(#1 )}}
\newcommand{\dwn}[1]{_{(#1 )}}
\newcommand{\wt}{\widetilde}
\usepackage{dsfont}
\newcommand{\1}{\mathds{1}}
\newcommand{\onetab}[1]{\begin{tabular}{c} #1 \end{tabular}}
\newcommand{\pd}{\partial}
\renewcommand{\d}{\partial}
\DeclareMathOperator{\diag}{diag}

\ReDeclareMathOperator{\Im}{Im}
\ReDeclareMathOperator{\Re}{Re}
\DeclareMathOperator{\ad}{ad}
\DeclareMathOperator{\tr}{tr}
\DeclareMathOperator{\Tr}{Tr}
\newcommand{\lap}{\,\Delta}
\newcommand{\SL}{SL(2,\R)}
\newcommand{\lb}{\left[}
\newcommand{\rb}{\right]}
\newcommand{\be}{\begin{equation}}
\newcommand{\ee}{\end{equation}}
\newcommand{\dx}[1][D]{\! d^{#1}\! x}
\newcommand{\dtdx}[1][d]{\! dt \, d^{#1}\! x}
\newcommand{\yng}[1]{\ydiagram{#1}}
\newcommand{\tyng}[1]{\text{\tiny\ydiagram{#1}}}
\newcommand{\col}[2]{\begin{pmatrix} #1 \vspace{2pt} \\ #2 \end{pmatrix}}
\newcommand{\bbar}[1]{\bar{\bar{ #1 }}}
\DeclareMathOperator{\curl}{curl}
\newcommand{\hc}{\text{h.c.}}
\newcommand{\otoprule}{\midrule[\heavyrulewidth]}
\newcommand{\et}{{\widetilde{\epsilon}}}

\newcommand{\half}{\frac{1}{2}}

\newcommand{\ddl}[2]{\frac{\d #1}{\d #2}}
\newcommand{\ddll}[2]{\frac{\d^L #1}{\d #2}}

\newcommand{\vddr}[2]{\frac{\delta^R #1}{\delta #2}}
\newcommand{\vddl}[2]{\frac{\delta #1}{\delta #2}}

\usepackage{fancyhdr}

\newcommand{\nnchapter}[1]{
\chapter*{#1}
\addcontentsline{toc}{chapter}{#1}
\markboth{\MakeUppercase{#1}}{\MakeUppercase{#1}}
}

\newcommand{\figuresandtables}{
\listoffigures
\begingroup
\let\clearpage\relax
\listoftables
\endgroup
}

\makeatletter
\renewcommand\part{%
  \if@openright
    \cleardoublepage
  \else
    \clearpage
  \fi
  \thispagestyle{empty} 
  \if@twocolumn
    \onecolumn
    \@tempswatrue
  \else
    \@tempswafalse
  \fi
  \null\vfil
  \secdef\@part\@spart}
\makeatother

\begin{document}

\frontmatter
\thispagestyle{empty}

\begin{center}

\textbf{\Large Université Libre de Bruxelles} \\ \medskip
\large{Faculté des Sciences \\
Département de Physique \\
Service de Physique Mathématique des Interactions Fondamentales}

\vspace{ \stretch{0.5} }

\hrule
\bigskip
{\huge Aspects of electric-magnetic dualities\\
\medskip
in maximal supergravity}
\bigskip
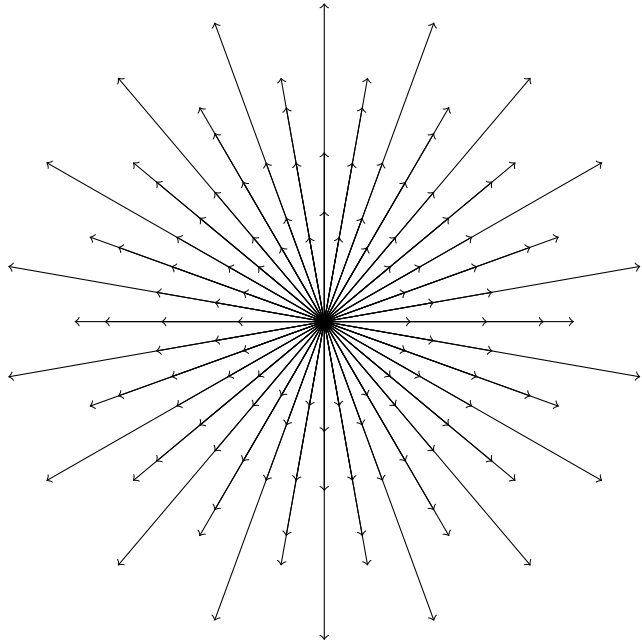
\hrule

\vspace{ \stretch{0.25} }

\begin{tikzpicture}[scale=2.5]
\draw[->] (0,0) -- (0.428525,-0.742227); 
\draw[->] (0,0) -- (0.428525,-0.15597); 
\draw[->] (0,0) -- (0.777862,-0.449099); 
\draw[->] (0,0) -- (1.00588,-0.84403); 
\draw[->] (0,0) -- (1.08506,-1.29313); 
\draw[->] (0,0) -- (0.228013,0.394931); 
\draw[->] (0,0) -- (0.57735,0.101802); 
\draw[->] (0,0) -- (0.805364,-0.293128); 
\draw[->] (0,0) -- (0.884552,-0.742227); 
\draw[->] (0,0) -- (0.57735,0.688059); 
\draw[->] (0,0) -- (0.805364,0.293128); 
\draw[->] (0,0) -- (0.884552,-0.15597); 
\draw[->] (0,0) -- (1.1547,0.0); 
\draw[->] (0,0) -- (1.23389,-0.449099); 
\draw[->] (0,0) -- (1.4619,-0.84403); 
\draw[->] (0,0) -- (-0.148825,0.84403); 
\draw[->] (0,0) -- (0.200512,0.550901); 
\draw[->] (0,0) -- (0.428525,0.15597); 
\draw[->] (0,0) -- (0.507713,-0.293128); 
\draw[->] (0,0) -- (0.200512,1.13716); 
\draw[->] (0,0) -- (0.428525,0.742227); 
\draw[->] (0,0) -- (0.507713,0.293128); 
\draw[->] (0,0) -- (0.777862,0.449099); 
\draw[->] (0,0) -- (0.85705,0.0); 
\draw[->] (0,0) -- (1.08506,-0.394931); 
\draw[->] (0,0) -- (0.0,1.68806); 
\draw[->] (0,0) -- (0.228013,1.29313); 
\draw[->] (0,0) -- (0.307202,0.84403); 
\draw[->] (0,0) -- (0.57735,1.0); 
\draw[->] (0,0) -- (0.656539,0.550901); 
\draw[->] (0,0) -- (0.884552,0.15597); 
\draw[->] (0,0) -- (0.57735,1.58626); 
\draw[->] (0,0) -- (0.656539,1.13716); 
\draw[->] (0,0) -- (0.884552,0.742227); 
\draw[->] (0,0) -- (1.23389,0.449099); 
\draw[->] (0,0) -- (-0.376839,0.449099); 
\draw[->] (0,0) -- (-0.57735,1.0); 
\draw[->] (0,0) -- (-0.57735,1.58626); 
\draw[->] (0,0) -- (-0.228013,1.29313); 
\draw[->] (0,0) -- (0.0,0.898198); 
\draw[->] (0,0) -- (0.0791882,0.449099); 
\draw[->] (0,0) -- (-0.200512,0.550901); 
\draw[->] (0,0) -- (-0.200512,1.13716); 
\draw[->] (0,0) -- (0.148825,0.84403); 
\draw[->] (0,0) -- (0.376839,0.449099); 
\draw[->] (0,0) -- (0.456027,0.0); 
\draw[->] (0,0) -- (0.0,0.586257); 
\draw[->] (0,0) -- (0.349337,0.293128); 
\draw[->] (0,0) -- (0.57735,-0.101802); 
\draw[->] (0,0) -- (0.656539,-0.550901); 
\draw[->] (0,0) -- (0.349337,-0.293128); 
\draw[->] (0,0) -- (0.57735,-0.688059); 
\draw[->] (0,0) -- (0.656539,-1.13716); 
\draw[->] (0,0) -- (0.228013,-0.394931); 
\draw[->] (0,0) -- (0.307202,-0.84403); 
\draw[->] (0,0) -- (0.0791882,-0.449099); 
\draw[->] (0,0) -- (-1.08506,-1.29313); 
\draw[->] (0,0) -- (-1.4619,-0.84403); 
\draw[->] (0,0) -- (-1.66241,-0.293128); 
\draw[->] (0,0) -- (-1.66241,0.293128); 
\draw[->] (0,0) -- (-1.31308,0.0); 
\draw[->] (0,0) -- (-1.08506,-0.394931); 
\draw[->] (0,0) -- (-1.00588,-0.84403); 
\draw[->] (0,0) -- (-0.428525,0.742227); 
\draw[->] (0,0) -- (-0.428525,0.15597); 
\draw[->] (0,0) -- (-0.777862,0.449099); 
\draw[->] (0,0) -- (-1.00588,0.84403); 
\draw[->] (0,0) -- (-1.08506,1.29313); 
\draw[->] (0,0) -- (-0.228013,-0.394931); 
\draw[->] (0,0) -- (-0.57735,-0.101802); 
\draw[->] (0,0) -- (-0.805364,0.293128); 
\draw[->] (0,0) -- (-0.884552,0.742227); 
\draw[->] (0,0) -- (-0.57735,-0.688059); 
\draw[->] (0,0) -- (-0.805364,-0.293128); 
\draw[->] (0,0) -- (-0.884552,0.15597); 
\draw[->] (0,0) -- (-1.1547,0.0); 
\draw[->] (0,0) -- (-1.23389,0.449099); 
\draw[->] (0,0) -- (-1.4619,0.84403); 
\draw[->] (0,0) -- (0.148825,-0.84403); 
\draw[->] (0,0) -- (-0.200512,-0.550901); 
\draw[->] (0,0) -- (-0.428525,-0.15597); 
\draw[->] (0,0) -- (-0.507713,0.293128); 
\draw[->] (0,0) -- (-0.200512,-1.13716); 
\draw[->] (0,0) -- (-0.428525,-0.742227); 
\draw[->] (0,0) -- (-0.507713,-0.293128); 
\draw[->] (0,0) -- (-0.777862,-0.449099); 
\draw[->] (0,0) -- (-0.85705,0.0); 
\draw[->] (0,0) -- (-1.08506,0.394931); 
\draw[->] (0,0) -- (0.0,-1.68806); 
\draw[->] (0,0) -- (-0.228013,-1.29313); 
\draw[->] (0,0) -- (-0.307202,-0.84403); 
\draw[->] (0,0) -- (-0.57735,-1.0); 
\draw[->] (0,0) -- (-0.656539,-0.550901); 
\draw[->] (0,0) -- (-0.884552,-0.15597); 
\draw[->] (0,0) -- (-0.57735,-1.58626); 
\draw[->] (0,0) -- (-0.656539,-1.13716); 
\draw[->] (0,0) -- (-0.884552,-0.742227); 
\draw[->] (0,0) -- (-1.23389,-0.449099); 
\draw[->] (0,0) -- (0.376839,-0.449099); 
\draw[->] (0,0) -- (0.57735,-1.0); 
\draw[->] (0,0) -- (0.57735,-1.58626); 
\draw[->] (0,0) -- (0.228013,-1.29313); 
\draw[->] (0,0) -- (0.0,-0.898198); 
\draw[->] (0,0) -- (-0.0791882,-0.449099); 
\draw[->] (0,0) -- (0.200512,-0.550901); 
\draw[->] (0,0) -- (0.200512,-1.13716); 
\draw[->] (0,0) -- (-0.148825,-0.84403); 
\draw[->] (0,0) -- (-0.376839,-0.449099); 
\draw[->] (0,0) -- (-0.456027,0.0); 
\draw[->] (0,0) -- (0.0,-0.586257); 
\draw[->] (0,0) -- (-0.349337,-0.293128); 
\draw[->] (0,0) -- (-0.57735,0.101802); 
\draw[->] (0,0) -- (-0.656539,0.550901); 
\draw[->] (0,0) -- (-0.349337,0.293128); 
\draw[->] (0,0) -- (-0.57735,0.688059); 
\draw[->] (0,0) -- (-0.656539,1.13716); 
\draw[->] (0,0) -- (-0.228013,0.394931); 
\draw[->] (0,0) -- (-0.307202,0.84403); 
\draw[->] (0,0) -- (-0.0791882,0.449099); 
\draw[->] (0,0) -- (1.08506,1.29313); 
\draw[->] (0,0) -- (1.4619,0.84403); 
\draw[->] (0,0) -- (1.66241,0.293128); 
\draw[->] (0,0) -- (1.66241,-0.293128); 
\draw[->] (0,0) -- (1.31308,0.0); 
\draw[->] (0,0) -- (1.08506,0.394931); 
\draw[->] (0,0) -- (1.00588,0.84403);
\end{tikzpicture}

\vspace{ \stretch{0.25} }

{\Large Victor Lekeu}

\vspace{ \stretch{0.25} }

{\large
\begin{tabular}{r l}
Promoteur: & Marc Henneaux\\Co-promoteur: & Geoffrey Compère
\end{tabular}}

\end{center}

\vspace{ \stretch{0.25} }

{\large \noindent Thèse présentée en vue de l'obtention \hfill Année académique 2017-2018\\du titre de Docteur en Sciences}

\newpage
\thispagestyle{empty}

\vspace*{\fill}

\noindent
This thesis was presented behind closed doors on the 4th of June 2018, and publicly on the 18th of June 2018, at Université Libre de Bruxelles (U.L.B.) in front of the following jury:
\begin{itemize}
\item Riccardo Argurio (President, U.L.B.)
\item Stéphane Detournay (Secretary, U.L.B.)
\item Marc Henneaux (Advisor, U.L.B.)
\item Geoffrey Compère (Co-advisor, U.L.B.)
\item Andrès Collinucci (U.L.B.)
\item Chris Hull (Imperial College, London)
\item Axel Kleinschmidt (Albert Einstein Institute, Potsdam).
\end{itemize}

\vspace*{\fill}

\noindent
The author is a Research Fellow (\textit{Aspirant}) at the Belgian F.R.S.-FNRS.\\

\noindent
Work during the academic year 2014-2015 was partially supported by the ERC Advanced Grant ``SyDuGraM", by FNRS-Belgium (convention FRFC PDR T.1025.14 and convention IISN 4.4514.08) and by the ``Communaut\'e Fran\c{c}aise de Belgique" through the ARC program. \\

\noindent
Cover: root system of $E_7$, projected on the Coxeter plane.

\tableofcontents
\nnchapter{Acknowledgements}

First of all, I feel very lucky for having had Marc Henneaux as an advisor. Many thanks for your continued support, kind availability, and deep yet crystal clear explanations of many subjects throughout the past seven years.\\

I would also like to thank the members of my jury for accepting to be on my committee and taking the time to read this thesis. Thanks for your many useful comments and interesting discussions on this work.\\

I have learned a lot from my co-advisor Geoffrey Compère and my other collaborators Glenn Barnich, Nicolas Boulanger, Bernard Julia, Axel Kleinschmidt, Amaury Leonard, Javier Matulich, Stefan Prohazka and Arash Ranjbar. Thank you all for this nice teamwork, and for showing me that physics is very much a social activity.\\

I am very grateful to all the members of the mathematical physics department for the enjoyable and stimulating work environment they have created, and to Céline Zwikel in particular; I could not have hoped for a better office partnership.\\

Finally, thanks to all my friends and family for their support all along my studies. Among those, I am particularly grateful to Agnès for her careful re-reading of this thesis, and for all the rest.


\nnchapter{Credits}

The original results of this thesis were presented in the published papers
\begin{enumerate}
\item[\cite{Henneaux:2015opa}] M.~Henneaux, A.~Kleinschmidt, and V.~Lekeu, ``{Enhancement of hidden symmetries
  and Chern-Simons couplings},'' {\em Rom. J. Phys.} {\bfseries 61} no.~1-2,
  (2016) 167,
\href{http://arxiv.org/abs/1505.07355}{{\ttfamily arXiv:1505.07355 [hep-th]}}
\item[\cite{Compere:2015roa}] G.~Compère and V.~Lekeu, ``{$E_{7(7)}$ invariant non-extremal entropy},''
  \href{http://dx.doi.org/10.1007/JHEP01(2016)095}{{\em JHEP} {\bfseries 01}
  (2016) 095},
\href{http://arxiv.org/abs/1510.03582}{{\ttfamily arXiv:1510.03582 [hep-th]}}
\item[\cite{Henneaux:2016opm}] M.~Henneaux, V.~Lekeu, and A.~Leonard, ``{Chiral Tensors of Mixed Young
  Symmetry},'' \href{http://dx.doi.org/10.1103/PhysRevD.95.084040}{{\em Phys.
  Rev.} {\bfseries D95} no.~8, (2017) 084040},
\href{http://arxiv.org/abs/1612.02772}{{\ttfamily arXiv:1612.02772 [hep-th]}}
\item[\cite{Henneaux:2017kbx}] M.~Henneaux, B.~Julia, V.~Lekeu, and A.~Ranjbar, ``{A note on ‘gaugings’ in
  four spacetime dimensions and electric-magnetic duality},''
  \href{http://dx.doi.org/10.1088/1361-6382/aa9fd5}{{\em Class. Quant. Grav.}
  {\bfseries 35} no.~3, (2018) 037001},
\href{http://arxiv.org/abs/1709.06014}{{\ttfamily arXiv:1709.06014 [hep-th]}}
\item[\cite{Henneaux:2017xsb}] M.~Henneaux, V.~Lekeu, and A.~Leonard, ``{The action of the (free) (4,
  0)-theory},'' \href{http://dx.doi.org/10.1007/JHEP01(2018)114}{{\em JHEP}
  {\bfseries 01} (2018) 114},
\href{http://arxiv.org/abs/1711.07448}{{\ttfamily arXiv:1711.07448 [hep-th]}}
\item[\cite{Barnich:2017nty}] G.~Barnich, N.~Boulanger, M.~Henneaux, B.~Julia, V.~Lekeu, and A.~Ranjbar,
  ``{Deformations of vector-scalar models},''
  \href{http://dx.doi.org/10.1007/JHEP02(2018)064}{{\em JHEP} {\bfseries 02}
  (2018) 064},
\href{http://arxiv.org/abs/1712.08126}{{\ttfamily arXiv:1712.08126 [hep-th]}}
\end{enumerate}
and in the preprints
\begin{enumerate}
\item[\cite{Lekeu:2018kul}] V.~Lekeu and A.~Leonard, ``{Prepotentials for linearized supergravity},''
\href{http://arxiv.org/abs/1804.06729}{{\ttfamily arXiv:1804.06729 [hep-th]}}
\item[\cite{Henneaux:2018rub}]M.~Henneaux, V.~Lekeu, J.~Matulich, and S.~Prohazka, ``{The Action of the
  (Free) $\mathcal{N} = (3,1)$ Theory in Six Spacetime Dimensions},''
\href{http://arxiv.org/abs/1804.10125}{{\ttfamily arXiv:1804.10125 [hep-th]}}
\end{enumerate}
which have been submitted for publication.
Another paper was published during the course of this Ph.D., whose results fall outside the scope of this thesis and are not presented here:
\begin{enumerate}
\item[\cite{Henneaux:2015gya}] M.~Henneaux and V.~Lekeu, ``{Kac-Moody and Borcherds Symmetries of
  Six-Dimensional Chiral Supergravity},''
  \href{http://dx.doi.org/10.1007/JHEP03(2015)056}{{\em JHEP} {\bfseries 03}
  (2015) 056},
\href{http://arxiv.org/abs/1502.00518}{{\ttfamily arXiv:1502.00518 [hep-th]}}.
\end{enumerate}

\chapter*{Abstract}
\addcontentsline{toc}{chapter}{Abstract - Résumé}
\markboth{\MakeUppercase{Abstract - Résumé}}{}

This thesis is devoted to various aspects of electric-magnetic duality and its gravitational generalization, with an emphasis on the case of maximal supergravity. It is divided into three parts.\\

In the first part, we review the symmetries of maximal supergravity in various dimensions, with a particular focus on the exceptional ``hidden" symmetries that appear upon toroidal dimensional reduction of eleven-dimensional supergravity. Two new results are obtained. First, we prove in detail that these hidden symmetries appear if and only if the Chern-Simons coupling of the eleven-dimensional theory takes the value predicted by supersymmetry. Second, we obtain a manifestly $E_{7(7)}$-invariant formula for the entropy of non-extremal black holes in four-dimensional $\cN = 8$ supergravity.\\

The second part of the thesis concerns the gaugings of extended supergravities in four dimensions. We first show that the embedding tensor formalism does not allow for deformations that cannot be reached by working with the usual Lagrangian in the duality frame picked by the embedding tensor. We then examine through BRST methods the deformations of a large class of non-minimally coupled scalars and abelian vector fields in four dimensions, of which ungauged supergravities offer a prime example. We prove that all local deformations of these models which modify the gauge transformations are of the usual Yang-Mills type, i.e., correspond to the gauging of some rigid symmetries of the undeformed theory. Combined with the first result, this shows that the embedding tensor formalism correctly captures the most general local deformations of these theories.\\

In the third part, we construct self-contained action principles for several types of free fields in six dimensions, whose field strengths satisfy a self-duality condition. These fields are motivated by two considerations. First, their existence allows for a remarkable geometric interpretation of the electric-magnetic duality symmetries of vector fields and linearized gravity in four dimensions. Second, they appear in the spectrum of the chiral $\cN = (4,0)$ and $\cN = (3,1)$ ``exotic supergravities" in place of the usual metric. The free action and supersymmetry transformations for those theories are explicitly constructed. We also check that they reduce to linearized maximal supergravity in five dimensions, thus completing the picture of higher-dimensional parents of $\cN = 8$ supergravity, at least at the free level. Along the way, we also generalize previous works on linearized supergravity by other authors, in which the graviton and its dual appear on the same footing at the level of the action.\\

We conclude with some open questions, perspectives for future work, and a series of technical appendices.\\

\chapter*{Résumé}

Cette thèse est consacrée à divers aspects de la dualité électrique/magnétique et de sa généralisation gravitationnelle, avec un accent sur le cas de la supergravité maximale. Elle est divisée en trois parties.\\

Dans la première partie, nous commençons par passer en revue les symétries de la supergravité maximale en diverses dimensions, en portant une attention particulière aux symétries ``cachées" exceptionnelles apparaissant lors de la réduction dimensionnelle toroïdale de la théorie de supergravité en onze dimensions. Deux résultats nouveaux sont obtenus. D'une part, nous démontrons en détail que ces symétries cachées apparaissent si et seulement si le coefficient de Chern-Simons de la théorie à onze dimensions prend sa valeur particulière déterminée par la supersymétrie. D'autre part, nous construisons une formule manifestement invariante sous le groupe $E_{7(7)}$ pour l'entropie des trous noirs non-extrêmes en supergravité $\cN = 8$ en dimension quatre.\\

La deuxième partie de la thèse concerne les jaugeages de supergravités étendues en dimension quatre. Tout d'abord, nous montrons que le formalisme du tenseur de plongement n'autorise pas de déformations qui ne puissent être obtenues avec un Lagrangien conventionnel (dans le repère de dualité déterminé par le tenseur de plongement). Nous examinons ensuite par des méthodes BRST les déformations d'une large classe de modèles contenant des champs scalaires et vectoriels abéliens avec couplage non-minimal, dont les supergravités étendues sont l'exemple le plus important. Nous démontrons que toutes les déformations locales de ces modèles sont du type Yang-Mills habituel, i.e., correspondent au jaugeage de symétries rigides de l'action de départ. Combiné avec le résultat précédent, ceci montre que le formalisme du tenseur de plongement contient en effet les déformations locales les plus générales de ces théories.\\

Enfin, dans la troisième partie, nous construisons des principes variationnels pour plusieurs types de champs libres en dimension six, dont la courbure satisfait à une condition d'auto-dualité. Ces champs sont motivés par deux considérations. D'une part, leur existence permet une interprétation géométrique remarquable des symétries de dualité pour les champs vectoriels et la gravité linéarisée en dimension quatre. D'autre part, ils apparaissent dans le spectre des ``supergravités exotiques" avec supersymétrie chirale $\cN = (4,0)$ et $\cN = (3,1)$ à la place de la métrique usuelle. L'action libre et les transformations de supersymétrie sont construites explicitement pour ces deux théories. Ceci complète l'arbre des parents de la supergravité $\cN = 8$ en plus grande dimension, en tout cas au niveau libre. Ce faisant, nous généralisons aussi les travaux d'autres auteurs sur la supergravité linéarisée, où le graviton et son dual apparaissent sur un pied d'égalité au niveau de l'action.\\

La thèse conclut enfin par quelques questions ouvertes et perspectives futures, ainsi que par une série d'appendices techniques.\\

\mainmatter
\nnchapter{Introduction}

One of the main challenges of modern theoretical physics is the construction of a consistent theory of \emph{quantum gravity}, to account for gravitational phenomena at microscopic scales and/or very high energies where Einstein's theory breaks down. In that context, the most promising candidate is string theory, in which elementary particles are replaced by excitations of fundamental strings. In the low energy limit, string theory is decribed by theories of \emph{supergravity}, whose rich structure has been the subject of intense research since the seventies.\\

This thesis is devoted to various aspects of electric-magnetic duality and its gravitational generalization in various dimensions, with a particular focus on maximal supergravity. Broadly speaking, it is inscribed in a collective effort towards a better understanding of the symmetry structures of supergravity and string theory. To be more precise, let us first review the two recurrent ideas that appear throughout the thesis: extra dimensions and electric-magnetic dualities.

\subsection*{Higher dimensions and hidden symmetries}

Many physical theories are most simply formulated in dimensions other than four; for example, string theory can only be quantised consistently in ten or twenty-six dimensions. To make contact with four-dimensional physics, one possibility is to consider these higher-dimensional models on manifolds of the form
\[ \cM = \cM\dwn{4} \times \cK \, .\]
The four-dimensional manifold $\cM\dwn{4}$ is our usual space-time, while $\cK$ is compact and very small. Therefore, $\cK$ is unobservable from the four-dimensional point of view: it takes a very high energy (inversely proportional to the size of $\cK$) to excite fields along the compact directions. Still, the effective theory in four dimensions is sensitive to the shape and properties of the compact manifold $\cK$.\\

This is how the maximally supersymmetric $\cN = 8$ supergravity in four dimensions was constructed \cite{Cremmer:1978ds,Cremmer:1979up}, by reducing the (unique) eleven-dimensional theory \cite{Cremmer:1978km} on a flat torus $T^7$. In this reduction, it was noticed that the four-dimensional $\cN = 8$ theory has the unexpected exceptional group $E_{7(7)}$ as a symmetry. In fact, unexpected symmetries of that type also appear in the reduction to dimensions other than four. They are usually referred to as ``hidden symmetries", since they are not visible in the original eleven-dimensional theory. These classical symmetries of supergravity are broken to a discrete subgroup by quantum effects in string theory, which is conjectured to be an exact symmetry of the full quantum string theory \cite{Hull:1994ys,Witten:1995ex}. \\

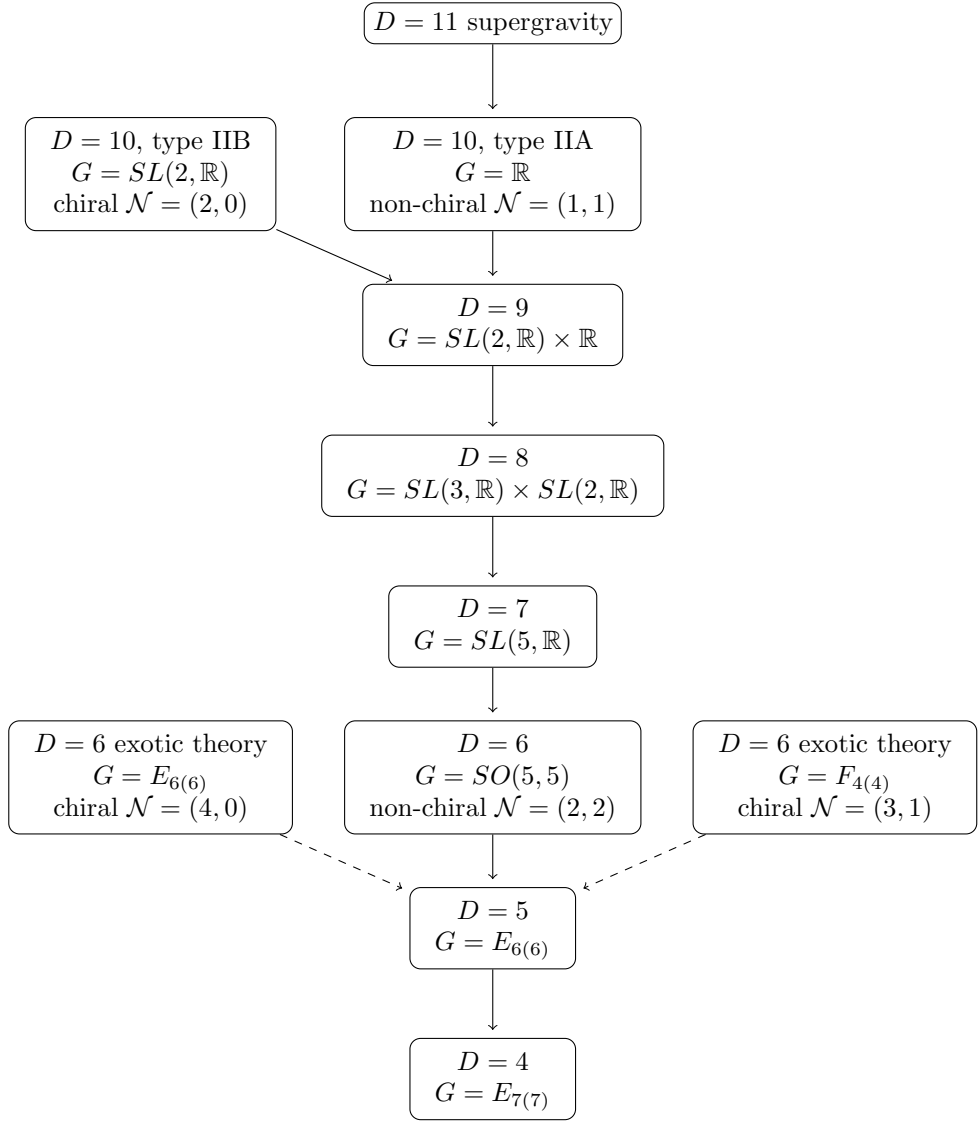
\begin{figure}[p]
\centering
\begin{tikzpicture}
\tikzset{r/.style={rectangle, rounded corners, text centered, draw}}
\node[r] (11) at (0,16) {$D = 11$ supergravity};
\node[r] (A) at (0,14) {\onetab{$D=10$, type IIA\\$G=\R$\\non-chiral $\cN = (1,1)$}};
\node[r] (9) at (0,12) {\onetab{$D=9$\\$G = SL(2,\R) \times \R$}};
\node[r] (8) at (0,10) {\onetab{$D=8$\\$G = SL(3,\R) \times SL(2,\R)$}};
\node[r] (7) at (0,8) {\onetab{$D=7$\\$G = SL(5,\R)$}};
\node[r] (6) at (0,6) {\onetab{$D=6$\\$G = SO(5,5)$\\non-chiral $\cN = (2,2)$}};
\node[r] (5) at (0,4) {\onetab{$D=5$\\$G=E_{6(6)}$}};
\node[r] (4) at (0,2) {\onetab{$D=4$\\$G=E_{7(7)}$}};
\node[r] (B) at (-4.5,14) {\onetab{$D=10$, type IIB\\ $G=SL(2,\R)$\\ chiral $\cN = (2,0)$}};
\node[r] (40) at (-4.5,6) {\onetab{$D=6$ exotic theory\\$G=E_{6(6)}$\\chiral $\cN = (4,0)$}};
\node[r] (31) at (4.5,6) {\onetab{$D=6$ exotic theory\\$G=F_{4(4)}$\\ chiral $\cN = (3,1)$}};
\foreach \from/\to in {11/A, A/9, B/9, 9/8, 8/7, 7/6, 6/5, 5/4}
	\draw [->, shorten >=3pt] (\from) -- (\to);
\foreach \from/\to in {40/5, 31/5}
	\draw [->, dashed, shorten >=3pt] (\from) -- (\to);
\end{tikzpicture}
\vspace{0.5cm}
\caption[Maximal supergravity, from eleven to four dimensions]{\label{fig:oxidation}Higher-dimensional parents of $\cN = 8$ supergravity in four dimensions, with their global symmetry group $G$. We have indicated the chirality of the supersymmetry algebra in dimensions ten and six; the main vertical line (from $D=11$ to $D=4$) is non-chiral. The conjectured $\cN = (4,0)$ and $\cN = (3,1)$ theories in six dimensions of \cite{Hull:2000zn,Hull:2000rr} are also shown.}
\end{figure}

It can happen that there are several higher-dimensional origins for the same theory; this is the case for example in ten and six dimensions, where several maximal supergravities exist since one can assign different chiralities to the supercharges \cite{Nahm:1977tg,Strathdee:1986jr}. This tree of possible higher-dimensional parents of $\cN = 8$ supergravity in four dimensions is depicted in figure \ref{fig:oxidation}, with their various global symmetry groups. On the other hand, starting from the same higher-dimensional theory, various four-dimensional models can be engineered by changing the internal manifold $\cK$; for example, the $SO(8)$ gauged supergravity of \cite{deWit:1981sst} can be understood as the reduction of eleven-dimensional supergravity on the seven-sphere $S^7$ \cite{deWit:1986oxb}.

\subsection*{Electric-magnetic dualities}

The $E_{7(7)}$ symmetry group of maximal supergravity in four dimensions includes \emph{electric-magnetic duality symmetries}, which exchange electric and magnetic fields. They are generalisations of the well-known symmetry of the vacuum Maxwell equations
\begin{align*}
\vec{\nabla} \cdot \vec{E} &= 0 &\quad \vec{\nabla} \cdot \vec{B} &= 0 \\
\vec{\nabla} \times \vec{E} &= - \frac{\pd \vec{B}}{\pd t} &\quad \vec{\nabla} \times \vec{B} &= \frac{\pd \vec{E}}{\pd t}
\end{align*}
under the replacement $\vec{E} \rightarrow \vec{B}$, $\vec{B} \rightarrow -\vec{E}$ of electric and magnetic fields or, more generally, under the rotations
\[ \col{\vec{E}}{\vec{B}} \rightarrow \begin{pmatrix}
\cos \theta & \sin \theta \\ - \sin \theta & \cos \theta
\end{pmatrix} \col{\vec{E}}{\vec{B}}\, . \]
These symmetries are of fundamental importance in the modern study of gauge theories, supergravity and string theory. In extended supergravities, they form a subgroup of the symplectic group $Sp(2 n_v, \R)$ (where $n_v$ is the number of vector fields), as was first proved in \cite{Gaillard:1981rj}. There is also evidence that its interacting and supersymmetric extension (non-abelian super-Yang-Mills theory) enjoys this symmetry at the quantum level \cite{Montonen:1977sn,Seiberg:1994rs,Vafa:1994tf,Seiberg:1994aj}; this can be understood geometrically from the existence of the elusive $\cN = (2,0)$ superconformal theory in six dimensions \cite{Verlinde:1995mz}, which contains chiral forms in its spectrum.\\

This duality symmetry has a direct counterpart in linearized gravity, where it exchanges the Riemann tensor and its dual or, equivalently, the gravitational analogues of the electric and magnetic fields \cite{Hull:2000zn,Hull:2001iu,Hull:2000rr,West:2001as,Nieto:1999pn,Bekaert:2002jn,Boulanger:2003vs,Henneaux:2004jw}. This can be understood from the existence of some exotic chiral fields in six dimensions, and the extension of this fact to the maximally supersymmetric case is conjectured to involve the superconformal $\cN = (4,0)$ theory in six dimensions \cite{Hull:2000zn,Hull:2000rr} in complete analogy with the case of super-Yang-Mills. This is by now well understood at the free level, but the status of this duality symmetry in a fully interacting gravitational theory remains unclear. However, given the success of its electromagnetic counterpart, it is certainly very intriguing and deserves closer study.

\section*{Structure and results of the thesis}

We present now the organisation of the thesis, highlighting the original results of \cite{Henneaux:2015opa,Compere:2015roa,Henneaux:2016opm,Henneaux:2017kbx,Henneaux:2017xsb,Barnich:2017nty,Lekeu:2018kul,Henneaux:2018rub} from the parts that are reviews of existing literature.

\subsection*{Part \ref{PART:SUGRA}: Symmetries of maximal supergravity}

The first part of the thesis is concerned with the global symmetries of maximal supergravity in various dimensions.\\

We start in chapter \ref{chap:maxsugra} by reviewing the $E$-series exceptional group symmetries that appear upon toroidal reduction of eleven-dimensional supergravity. The most celebrated example of those is the $E_{7(7)}$ symmetry of $\cN = 8$ supergravity in four dimensions discovered forty years ago by Cremmer and Julia \cite{Cremmer:1978ds,Cremmer:1979up}. We then recall the possibility of chiral supersymmetry in six and ten dimensions \cite{Nahm:1977tg,Strathdee:1986jr}. In ten dimensions, there are two possibilities: the well known type IIA and type IIB supergravities, which are the low-energy limits of the corresponding superstring theories. In six dimensions, in addition to the maximal supergravity which comes directly from eleven dimensions, the exotic $\cN = (4,0)$ and $\cN = (3,1)$ supergravity theories which are the subject of part \ref{PART:6D} appear. Those theories, first considered by Hull in \cite{Hull:2000zn,Hull:2000rr}, contain self-dual tensors with mixed Young symmetry instead of a metric.
We also comment briefly on the infinite-dimensional Kac-Moody symmetries $E_{10}$ and $E_{11}$ which are conjectured to be fundamental symmetries of supergravity and $M$-theory \cite{West:2001as,Damour:2002cu}.\\

Chapter \ref{chap:cscouplings} presents the results of \cite{Henneaux:2015opa}. We prove that the exceptional symmetries of maximal supergravity appear if and only if the Chern-Simons coupling takes a specific value in the eleven-dimensional theory. This value is exactly the one imposed by supersymmetry of the full supergravity Lagrangian \cite{Cremmer:1978km}. Indeed, the bosonic theory in eleven dimensions makes sense as a classical theory for any value of this coupling, but only a smaller group of symmetries (which is expected from toroidal reduction) remains in lower dimensions if this coupling does not have the value fixed by supersymmetry. We carry out this proof for the reduction to three dimensions, where the only fields are scalars and the symmetry structure is more transparent. A toy model for this phenomenon is minimal supergravity in five dimensions, which has the smallest exceptional group $G_{2(2)}$ in three dimensions \cite{Mizoguchi:1998wv}. We then turn to the $E_{8(8)}$ symmetry of $D = 3$ maximal supergravity.\\

Chapter \ref{chap:emduality} is again only a review, devoted to the duality properties of vector-scalar models in four dimensions which are typical of extended supergravities. As mentioned above, those models contain duality symmetries that generally mix the electric and magnetic fields or, equivalently, field equations and Bianchi identities \cite{Gaillard:1981rj}. Those symmetries, when written in terms of the fundamental variables (vector potentials), act non-locally at the level of the action \cite{Deser:1976iy,Deser:1981fr}. The subgroup of symmetries that only transforms the electric fields among themselves acts locally at the level of the Lagrangian and constitutes what is called the electric group of that Lagrangian. The electric group depends of course on the choice of duality frame, i.e., on the choice of electric or magnetic potentials as fundamental variables. This is of fundamental importance in the study of gauged supergravities, as we review in part \ref{PART:GAUGINGS}. The presentation follows that of \cite{Henneaux:2017kbx,Bunster:2011aw} and uses the unconstrained first-order (Hamiltonian) formalism, where all the duality symmetries act in a local way and these facts have a clear symplectic interpretation. We finish the chapter by contrasting the duality groups in $4m$ and $4m + 2$ dimensions.\\

Chapter \ref{chap:blackholes} presents the results of \cite{Compere:2015roa}. In it, we derive a manifestly $E_{7(7)}$-invariant entropy formula for the most general non-extremal, asymptotically flat black hole of $\cN = 8$ supergravity in four dimensions \cite{Chow:2013tia,Chow:2014cca}. This derivation is based on the fact that these black hole solutions can be obtained by acting with $E_{7(7)}$ symmetries on solutions of the $STU$ model \cite{Sen:1994eb,Cvetic:1996zq}. Therefore, once invariants of the $STU$ model are constructed (following \cite{Sarosi:2015nja}), it is sufficient to extend them to $E_{7(7)}$ invariants.

\subsection*{Part \ref{PART:GAUGINGS}: Four-dimensional gaugings}

This part contains a systematic study of the gauging deformations of coupled vector-scalar models, with a view towards extended supergravities in four dimensions.\\

Chapter \ref{chap:YMgaugings} starts with a review of Yang-Mills gaugings of four-dimensional models, in which the original abelian gauge group is deformed to a non-abelian one. In that procedure, some global symmetries of the Lagrangian are promoted to local symmetries (with space-time dependent parameters) in the usual way: for the vector fields, abelian fields strengths are replaced by the non-abelian ones, and ordinary derivatives by covariant derivatives (extra topological terms may also appear \cite{deWit:1984rvr,deWit:1987ph}). The symmetries for which this is well understood are only the electric ones (in the sense of chapter \ref{chap:emduality}); therefore, the possible gaugings depend on the choice of electric frame. There are two action principles which allow for an arbitrary choice of duality frame: those are the embedding tensor formalism \cite{deWit:2002vt,deWit:2005ub,deWit:2007kvg} and the action of chapter \ref{chap:emduality}. We show that neither of those allow for qualitatively new gaugings, which could not be obtained by starting from the conventional action in a definite duality frame. This statement is the original result of \cite{Henneaux:2017kbx}.\\

In chapter \ref{chap:introbrst}, we review the BRST-BV field-antifield formalism and its application to the problem of finding all local and consistent deformations of gauge theories \cite{Batalin:1981jr,Batalin:1984jr,Barnich:1993vg}, of which the gaugings of the Yang-Mills type are a specific class. It allows for a cohomological reformulation of that problem, thereby providing a strong mathematical foundation enabling complete proofs of existence and/or unicity theorems without \emph{a priori} assumptions on the form of the deformations. The main results reviewed in that chapter are 1) first-order deformations are classified by elements of the local BRST cohomology group $H^0(s|d)$ of the undeformed theory; and 2) global symmetries of the action are classified by elements of $H^{-1}(s|d)$.\\

The results of \cite{Barnich:2017nty} are then presented in chapter \ref{chap:vsgaugings}. We classify completely the local BRST cohomology groups of a general class of vector-scalar models in four dimensions. This contains the deformations of ungauged supergravity models: we find that, besides the obvious deformations that consist in adding gauge-invariant terms to the action, the only deformations that deform the gauge transformations are indeed gaugings of the Yang-Mills type. In particular, the link between $H^{-1}(s|d)$ (symmetries) and $H^0(s|d)$ (deformations) is explicitly done. Combining these results with those of chapter \ref{chap:YMgaugings}, this proves that the embedding tensor formalism captures the most general local deformations of ungauged supergravity models that deform the gauge transformations.

\subsection*{Part \ref{PART:6D}: Six-dimensional exotic fields}

The subject of this part is the study of chiral tensors of mixed Young symmetry in six space-time dimensions and their relation to gravitational electric-magnetic duality.\\

In chapter \ref{chap:gravduality}, we review the first motivation for their introduction: the duality symmetry of linearized gravity in four dimensions \cite{Hull:2000zn,Hull:2001iu,Hull:2000rr,West:2001as,Nieto:1999pn,Bekaert:2002jn,Boulanger:2003vs,Henneaux:2004jw}, which is the exact analogue of the one studied before for vector fields. Chiral tensors in six dimensions allow for a geometric realization of this symmetry as rotations in the internal space \cite{Hull:2000rr}; this is the direct generalization of the well known situation for vector fields and chiral $2$-forms in six dimensions \cite{Verlinde:1995mz}.\\

Chapter \ref{chap:twisted} contains the original results of \cite{Lekeu:2018kul}. We derive an action principle where the graviton and its dual appear on the same footing, generalizing the results of \cite{Henneaux:2004jw,Bunster:2013oaa} to arbitrary dimension. This is done by going to the Hamiltonian formalism and solving the constraints, thereby expressing the fields as derivatives of some more primitive variables, called the prepotentials. As was first noticed in \cite{Henneaux:2004jw}, the action is invariant under local Weyl transformations of the prepotentials. Control of this invariance is done through the systematic construction of conformal invariants, which are called the Cotton tensors of the prepotentials by analogy with three-dimensional gravity. We then write an analogue action for the gravitino and explain how the usual supersymmetry transformations are translated in this formalism, following \cite{Bunster:2012jp}.\\

In chapter \ref{chap:selfdual}, we also use prepotential techniques to generalize the procedure of \cite{Henneaux:1988gg} and write actions for the chiral tensors in six dimensions considered previously. These results were presented in the papers \cite{Henneaux:2016opm,Henneaux:2017xsb,Henneaux:2018rub} for the various types of fields. In contrast to the cases considered in chapter \ref{chap:twisted}, however, there is no quadratic, manifestly Lorentz-invariant action for those fields \cite{Marcus:1982yu}. The actions are thus ``intrinsically Hamiltonian". They are nevertheless Lorentz-invariant, even though not manifestly, just like the Hamiltonian actions of more familiar relativistic field theories. Upon dimensional reduction, they yield the actions of chapter \ref{chap:twisted} where duality is manifest; therefore, the conclusions of chapter \ref{chap:gravduality} are also valid at the level of the action and not only at the level of equations of motion.\\

Finally, chapter \ref{chap:6dsusy} contains the second motivation for the chiral tensors of mixed symmetry, which was already alluded to in chapter \ref{chap:maxsugra}: they appear in the maximally supersymmetric $\cN = (4,0)$ and $\cN = (3,1)$ chiral multiplets in six dimensions \cite{Hull:2000zn,Hull:2000rr}. The results of the two previous chapters can be used straightforwardly to write the free action for these theories. We also check explicitly the supersymmetry invariance, and the fact that they both reduce to (linearized) maximal supergravity in five dimensions. This completes at the level of the action the picture of all the higher-dimensional parents of $\cN = 8$ supergravity in four dimensions of figure \ref{fig:oxidation}, at least for the linearized theory. These results appeared in \cite{Henneaux:2017xsb} (for the $\cN = (4,0)$ theory) and in \cite{Henneaux:2018rub} (for the $\cN = (3,1)$ theory).\\
\vspace{1em}

\noindent
We then conclude with open questions and perspectives for future work. Appendices containing conventions, useful formulas and technical derivations for each of the three parts close the thesis.

\part{Symmetries of maximal supergravity}\label{PART:SUGRA}
\chapter{Maximal supergravity in various dimensions}
\label{chap:maxsugra}

The maximally supersymmetric theory in four dimensions is the $\cN = 8$ supergravity\footnote{We will not consider particles with spin greater than $2$ in this thesis.} first constructed by Cremmer and Julia in \cite{Cremmer:1978ds,Cremmer:1979up}, with the remarkable $E_{7(7)}$ exceptional group as global ``hidden" symmetry. It was constructed by dimensional reduction of the unique supergravity theory in eleven dimensions \cite{Cremmer:1978km}, whose bosonic Lagrangian is
\begin{align} \label{eq:11lag}
\cL\up{11} &= R \star \1 - \frac{1}{2} \star F\dwn{4} \wedge F\dwn{4} + \frac{1}{6} F\dwn{4} \wedge F\dwn{4} \wedge A\dwn{3}
\end{align}
in differential form notation\footnote{In components, this is
\begin{equation*}
\cL\up{11} = \sqrt{-g} \,d^{11}\!x \left( R - \frac{1}{48} F_{ABCD} F^{ABCD} - \frac{1}{12^4} \varepsilon^{M_1 \dots M_{11}} F_{M_1 \dots M_4} F_{M_5 \dots M_8} A_{M_9 M_{10} M_{11}} \right)\, .
\end{equation*}
Our conventions for differential forms are collected in appendix \ref{app:diffgeo}. We use the normalizations of \cite{Cremmer:1997ct}.}. Exceptional groups of $E$-type also appear in the toroidal reduction of this Lagrangian to dimensions other than four. Those symmetries were conjectured by simple counting in \cite{Cremmer:1979up,Cremmer:1980zb,Cremmer:1980gs,Julia:1980gr} and proved by direct construction of the maximal theory in various dimensions (see \cite{Salam:1989fm} for references). They were only later derived in an unified way from eleven dimensions in \cite{Cremmer:1997ct}.

The goal of this chapter is to review these exceptional symmetries of maximal supergravity. We start by analysing the toroidal reduction of \eqref{eq:11lag} down to three dimensions, following \cite{Cremmer:1997ct}. We then comment on the cases of ten and six dimensions, where supersymmetry is chiral \cite{Nahm:1977tg,Strathdee:1986jr} and several maximal supersymmetry algebras are therefore available. In ten dimensions, those are the well known type IIA \cite{Campbell:1984zc,Giani:1984wc} and IIB \cite{Schwarz:1983qr,Schwarz:1983wa,Howe:1983sra} supergravity theories, while in six dimensions, one is led to the exotic theories first considered by Hull \cite{Hull:2000zn,Hull:2000rr}. Finally, we review briefly the conjectured infinite-dimensional version of these symmetries \cite{West:2001as,Damour:2002cu}. Technical prerequisites can be found in appendix \ref{chap:app1}.

\section{Field content and exceptional symmetries}

\subsection{Field content and dualizations}

The bosonic field content of eleven-dimensional supergravity reduced on a $n$-dimensional torus $T^n$ can be understood by simply counting the number of internal indices on the fields. From the eleven-dimensional metric $\hat{g}_{MN}$, one gets
\begin{itemize}
\item one metric $g_{\mu\nu} \sim \hat{g}_{\mu\nu}$
\item $n$ vector fields $\cA_{i\, \mu} \sim \hat{g}_{\mu\,i}$
\item $n(n+1)/2$ scalar fields $\phi_{ij} \sim \hat{g}_{ij}$
\end{itemize}
and from the eleven-dimensional three-form $\hat{A}_{MNP}$,
\begin{itemize}
\item one three-form $A_{\mu\nu\rho} \sim \hat{A}_{\mu\nu\rho}$
\item $n$ two-forms $A_{i\, \mu\nu} \sim \hat{A}_{\mu\nu\,i}$
\item $n(n-1)/2$ vector fields $A_{ij\,\mu} \sim \hat{A}_{\mu\, ij}$
\item $n(n-1)(n-2)/6$ scalar fields $\chi_{ijk} \sim \hat{A}_{ijk}$ .
\end{itemize}
Therefore, dimensional reduction of eleven-dimensional supergravity to $D$ dimensions on a torus $T^n$ ($n = 11-D$) gives a theory with the field content presented in table \ref{table:fieldcontent}.
\begin{table}
\centering
\begin{tabular}{c|ccccc}
$D$ & metric & $3$-forms & $2$-forms & vectors & scalars \\ \midrule
11 & 1 & 1 & 0 & 0 & 0 \\
10 & 1 & 1 & 1 & 1 & 1 \\
9 & 1 & 1 & 2 & 3 & 3 \\
8 & 1 & 1 & 3 & 6 & 7 \\
7 & 1 & 1 & 4 & 10 & 14 \\
6 & 1 & 1 & 5 & 15 & 25 \\
5 & 1 & 1 & 6 & 21 & 41 \\
4 & 1 & 1 & 7 & 28 & 63 \\
3 & 1 & 1 & 8 & 36 & 92
\end{tabular}
\caption[Field content of maximal supergravity, without dualization]{\label{table:fieldcontent}The field content of eleven-dimensional supergravity reduced to $D$ dimensions, without dualization of fields.}
\end{table}
The explicit form of the Lagrangian can be derived from \eqref{eq:11lag} by iterating $n$ times the rules for dimensional reduction on a circle given in appendix \ref{app:dimred}; this is done in reference \cite{Cremmer:1997ct}.

In $D$ dimensions, $p$-form fields and $(D-p-2)$-form fields are equivalent, as reviewed in appendix \ref{app:pforms}. For $D \leq 7$, this can be used to replace some of the $p$-forms by lower-degree forms:
\begin{itemize}
\item in $D = 7$, a three-form can be dualized to a two-form;
\item in $D = 6$, a three-form can be dualized to a vector;
\item in $D = 5$, a three-form can be dualized to a scalar and a two-form to a vector;
\item in $D = 4$, a three-form has no degree of freedom and a two-form can be dualized to a scalar;
\item in $D = 3$, the three-form and the two-forms have no degree of freedom, and vectors can be dualized to scalars. 
\end{itemize}
For those dimensions, the field content simplifies and is summarized in table \ref{table:fieldcontentdual}.
\begin{table}
\centering
\begin{tabular}{c|cccc}
$D$ & $2$-forms & vectors & scalars \\ \midrule
7 & 5 & 10 & 14 \\
6 & 5 & 16 & 25 \\
5 & 0 & 27 & 42 \\
4 & 0 & 28 & 70 \\
3 & 0 & 0 & 128
\end{tabular}
\caption[Field content of maximal supergravity, with dualization]{\label{table:fieldcontentdual}The field content of eleven-dimensional supergravity reduced to $D$ dimensions, where fields are dualized to get the lower form degree possible.}
\end{table}

\subsection{Global $E_{7(7)}$ symmetry in four dimensions}

As was noticed in \cite{Cremmer:1978ds,Cremmer:1979up}, after the dualizations of the two-forms are performed, the number of scalars exactly matches the dimension of the homogeneous space $E_{7(7)}/SU(8)$ (indeed, $70 = 133 - 63$). This is no coincidence; in fact, the whole theory has the exceptional group $E_{7(7)}$ as a global symmetry. The unexpected appearance of $E_{7(7)}$ here is similar to that of the Ehlers group $\SL$ in the dimensional reduction of pure gravity to three dimensions \cite{Ehlers} (see also \cite{Julia:1980gr} and appendix \ref{app:dimred}).

The symmetry assignments of the various fields are:
\begin{itemize}
\item The metric is inert.
\item The vector fields and their duals transform in the $56$-dimensional representation of $E_{7(7)}$.
\item The $70$ scalar fields transform in the standard non-linear realisation of $E_{7(7)}$ (see appendix \ref{app:cosetlagrangians}).
\item Fermionic fields transform in representations of the maximal compact subgroup $SU(8)$: the $8$ gravitinos are in the fundamental and the $56$ spin $1/2$'s are in the rank $3$ antisymmetric tensor representation.
\end{itemize}
Explicit formulas can be found in \cite{Cremmer:1979up,Cremmer:1997ct} and in appendix \ref{app:N8E7}.

\subsection{$E$-series global symmetries}

It turns out that in other dimensions the scalar fields also parametrize homogeneous spaces of the form $G/K$, where the groups $G$ and $K$ are given by table \ref{table:scalarcosets} (after the necessary dualizations are performed for $D \leq 7$). In all cases, $G$ is a semi-simple group in its maximally non-compact real form and $K$ is the maximal compact subgroup of $G$.
\begin{table}
\centering
\begin{tabular}{c|cc|ccc}
$D$ & $G$ & $H$ & $\dim(G)$ & $\dim(H)$ & $\dim(G/H)$ \\ \midrule
10 & $\R$ & - & 1 & 0 & 1 \\
9 & $SL(2,\R)\times \R$ & $SO(2)$ & 4 & 1 & 3 \\
8 & $SL(3,\R) \times SL(2,\R)$ & $SO(3) \times SO(2)$  & 11 & 4 & 7 \\
7 & $SL(5,\R)$ & $SO(5)$ & 24 & 10 & 14 \\
6 & $SO(5,5)$ & $SO(5)\times SO(5)$ & 45 & 20 & 25 \\
5 & $E_{6(6)}$ & $USp(8)$ & 78 & 36 & 42 \\
4 & $E_{7(7)}$ & $SU(8)$ & 133 & 63 & 70 \\
3 & $E_{8(8)}$ & $SO(16)$ & 248 & 120 & 128 
\end{tabular}
\caption[Scalar cosets of maximal supergravity]{\label{table:scalarcosets}The scalar cosets of maximal supergravity in various dimensions. The dimension $\dim(G/H) = \dim(G) - \dim(H)$ of the coset space matches the number of scalars in table \ref{table:fieldcontent} (for $D \geq 6$) or \ref{table:fieldcontentdual} (for $D \leq 7$). The group $G$ extends to a symmetry of the full bosonic sector.}
\end{table}

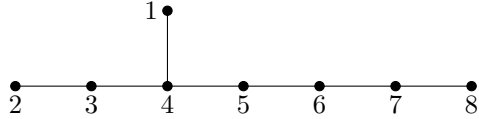
\begin{figure}
\centering
\begin{tikzpicture}
\tikzset{v/.style={circle,fill,inner sep=0pt,minimum size=3.5pt,draw}}
\draw (0,0) node[v]{} node[below]{$2$} -- (1,0) node[v]{} node[below]{$3$} -- (2,0) node[v]{} node[below]{$4$} -- (3,0) node[v]{} node[below]{$5$} -- (4,0) node[v]{} node[below]{$6$} -- (5,0) node[v]{} node[below]{$7$} -- (6,0) node[v]{} node[below]{$8$};
\draw (2,0) -- (2,1) node[v]{} node[left]{$1$};
\end{tikzpicture}
\caption[Dynkin diagram of $E_8$]{\label{fig:E8dynkin}The Dynkin diagram of $E_8$. The Dynkin diagrams of the $E_n$ algebras for $n \geq 3$ appearing in the dimensional reduction on a $n$-torus are obtained by keeping only the vertices numbered $1$ to $n$.}
\end{figure}

In particular, the three exceptional groups $G = E_{6(6)}$, $E_{7(7)}$ and $E_{8(8)}$ appear naturally in dimensions five, four and three respectively. For $n \leq 5$, it is conventional to define $E_{0(0)} =$ trivial, $E_{1(1)} = \R$, $E_{2(2)} = SL(2,\R)\times \R$, $E_{3(3)} = SL(3,\R) \times SL(2,\R)$, $E_{4(4)} = SL(5,\R)$ and $E_{5(5)} = SO(5,5)$. With this convention, the coset spaces appearing after the reduction on a torus can be written as $E_{n(n)}/K(E_{n(n)})$, where $n = 11 - D$ and $K(E_{n(n)})$ is the maximal compact subgroup of $E_{n(n)}$. For $n \geq 3$, this chain of groups can be visualized from the Dynkin diagram of $E_{8(8)}$ by removing the appropriate roots at each step of the reduction; see figure \ref{fig:E8dynkin}.

As before, these symmetries of the scalar sector do not act on the metric, while the fermionic fields transform under the compact sugroup $K(E_{n(n)})$. The $E_{n(n)}$ symmetry acts linearly on the various $p$-forms of the theory as follows:
\begin{itemize}
\item For $D=10$, the shift of the scalar is accompanied by the appropriate rescalings of the $p$-forms.
\item For $D=9$, the $SL(2,\R)$ factor acts on the internal indices. Therefore, the two vector fields coming from the metric form a doublet, and the same goes for the two two-forms coming from the $D=11$ three-form. The three-form and the last vector are singlets. The commuting $\R$ factor acts as a shift of one of the scalars, again accompanied by the appropriate rescalings of the other fields.
\item For $D=8$, the three-form and its dual form an $SL(2,\R)$ doublet, and are inert under the $SL(3,\R)$ factor. The two-forms are in the fundamental representation of $SL(3,\R)$ and are inert under $SL(2,\R)$. The six vectors are in the tensor product of those representations.
\item For $D=7$, the five two-forms transform in the fundamental representation of $SL(5,\R)$. The ten vectors are in the antisymmetric representation.
\item For $D=6$, the five two-forms and their duals transform in the fundamental representation of $SO(5,5)$. The sixteen vectors are in the irreducible $16$-dimensional representation of $SO(5,5)$ (Majorana-Weyl spinor representation).
\item For $D=5$, the vector fields are in the fundamental (28-dimensional) representation of $E_{6(6)}$. 
\item For $D=4$, the $28$ vector fields and their duals are in the fundamental (56-dimensional) representation of $E_{7(7)}$, as mentioned above and reviewed in appendix \ref{app:N8E7}.
\item For $D=3$, there are only scalar fields.
\end{itemize}

In even dimensions $D = 4$, $6$ and $8$, there are in the spectrum some $p$-forms whose field strength is of degree $D/2$. In those cases, the global symmetry mixes those field strengths and their dual (or, equivalently, the $p$-forms themselves and their duals). Those symmetries are therefore symmetries of the equations of motion and Bianchi identities. Since the actions obtained by direct reduction of \eqref{eq:11lag} involve either a $p$-form or its dual, but not both at the same time, they cannot be realized as symmetries of the action acting in a local manner. Other action principles where these duality symmetries can be realized locally will be presented in chapters \ref{chap:emduality} and \ref{chap:selfdual}.

\section{Chiral supersymmetry}

\subsection{Ten-dimensional type II supergravities}

In dimension ten, supersymmetry is chiral, i.e., can have both left and right generators. The supersymmmetry algebra $\cN = (\cN_+, \cN_-)$ is maximal when $\cN_+ +\cN_- = 2$: the two possibilities are then $\cN = (1,1)$ or $\cN = (2,0)$. The non-chiral $\cN = (1,1)$ theory is the type IIA supergravity, which can be obtained from the reduction of (non-chiral) eleven-dimensional supergravity on a circle \cite{Campbell:1984zc,Giani:1984wc,Huq:1983im}. The $\cN = (2,0)$ theory is the type IIB supergravity, which has no higher-dimensional origin \cite{Schwarz:1983qr,Schwarz:1983wa,Howe:1983sra}. We review here briefly the field content and symmetries of those theories, and their common reduction to nine dimensions \cite{Bergshoeff:1995as}, following the conventions of \cite{Cremmer:1997ct}.

\subsubsection{Type IIA supergravity}

The type IIA supergravity in ten dimensions is obtained by direct reduction of eleven-dimensional supergravity on a circle. In doing so, the eleven-dimensional metric gives a ten-dimensional metric $g$, a scalar field $\phi$ (the dilaton), and a vector field $A\dwn{1}$, and the eleven-dimensional three-form gives a ten-dimensional three-form $A\dwn{3}$ and a two-form $A\dwn{2}$. (This agrees with table \ref{table:fieldcontent}.) Its Lagrangian is
\begin{align}\label{eq:IIAlag}
\cL\up{10}_\text{IIA} = R \star \1 &- \frac{1}{2} \star d\phi \wedge d\phi - \frac{1}{2} e^{\frac{3}{2}\phi} \star F\dwn{2} \wedge F\dwn{2} - \frac{1}{2} e^{-\phi} \star F\dwn{3} \wedge F\dwn{3} \nn \\
&- \frac{1}{2} e^{\frac{1}{2}\phi} \star F\dwn{4} \wedge F\dwn{4} + \frac{1}{2} dA\dwn{3} \wedge dA\dwn{3} \wedge A\dwn{2} ,
\end{align}
where the field strengths of the various $p$-forms are given by
\begin{align}
F\dwn{2} = dA\dwn{1}, \quad F\dwn{3} = dA\dwn{2}, \quad F\dwn{4} = dA\dwn{3} - dA\dwn{2} \wedge A\dwn{1}.
\end{align}
As announced in table \ref{table:scalarcosets},  this Lagrangian has a $G = \R$ global symmetry which acts as a shift of the dilaton and a rescaling of the $p$-forms,
\begin{equation}
\phi' = \phi + c, \quad A\dwn{1}' = e^{-\frac{3}{4} c} A\dwn{1}, \quad A\dwn{2}' = e^{\frac{1}{2} c} A\dwn{2}, \quad A\dwn{3}' = e^{-\frac{1}{4} c} A\dwn{3}
\end{equation}
(notice that the exponents conspire to give a homogeneous scaling of $F\dwn{4}$).

\subsubsection{Type IIB supergravity}

The type IIB supergravity is the other theory of maximal supergravity in ten dimensions, which does not arise from dimensional reduction of eleven-dimensional supergravity. Its bosonic field content is the following:
\begin{itemize}
\item two scalar fields $\phi$, $\chi$ (dilaton and axion);
\item two $2$-forms $A_{(2)\,i}$ ($i = 1$, $2$); 
\item one self-dual $4$-form $C\dwn{4}$.
\end{itemize}
It has a global $SL(2,\R)$ symmetry, with the two scalar fields parametrizing the $SL(2,\R)/SO(2)$ coset space. The two-forms are in a doublet, and the self-dual $4$-form is inert.

Because of the self-dual $4$-form, there is no simple manifestly Lorentz-invariant action for type IIB supergravity \cite{Marcus:1982yu}. There is, however, a convenient action that gives the correct equations of motion except the self-duality equation, which has to be imposed by hand after varying the action to obtain the equations of motion \cite{Bergshoeff:1995sq}. This ``pseudo-action" is given by the integral of
\begin{align}\label{eq:IIBlagcov}
\cL\up{10}_\text{IIB} = R \star \1 &- \frac{1}{2} \star d\phi \wedge d\phi - \frac{1}{2} e^{2\phi} \star d\chi \wedge d\chi \nn \\
&- \frac{1}{2} (\cM^{-1})^{ij} \star F_{(3)\,i} \wedge F_{(3)\,j} - \frac{1}{2} \star F\dwn{5} \wedge F\dwn{5} \\
&- \frac{1}{2\sqrt{2}} \varepsilon^{ij} C\dwn{4} \wedge F_{(3)\,i} \wedge F_{(3)\,j}\, , \nn
\end{align}
where the field strengths are
\begin{equation}
F_{(3)\,i} = d A_{(2)\,i}\, , \quad F\dwn{5} = d C\dwn{4} + \frac{1}{2\sqrt{2}} \varepsilon^{ij} A_{(2)\,i} \wedge dA_{(2)\,j}\, .
\end{equation}
Here, $\cM$ is the usual $\SL/SO(2)$ scalar matrix \eqref{eq:cosetSL2SO2M}, which transforms as
\begin{equation}
\cM \rightarrow S^T \cM S
\end{equation}
under $S\in \SL$ (see appendix \ref{app:cosetlagrangians}). This Lagrangian has therefore a manifest $SL(2,\R)$ symmetry, provided the two-forms transform as a doublet
\begin{equation}
A\dwn{2} \rightarrow S^T A\dwn{2}
\end{equation}
and the metric and four-form do not transform.

Let us now comment on the self-duality condition. The equation of motion coming from the variation of $C\dwn{4}$ in \eqref{eq:IIBlagcov} and the Bianchi identity for $F\dwn{5}$ are
\begin{equation}
d \star F\dwn{5} = \frac{1}{2\sqrt{2}} \varepsilon^{ij} F_{(3)i} \wedge F_{(3)j}\, , \quad d F\dwn{5} = \frac{1}{2\sqrt{2}} \varepsilon^{ij} F_{(3)i} \wedge F_{(3)j}\, ,
\end{equation}
respectively. They have the same right-hand side and are compatible with the self-duality equation
\begin{equation}
\star F\dwn{5} = F\dwn{5}
\end{equation}
but do not imply it. The Lagrangian \eqref{eq:IIBlagcov}, even though it is convenient for the discussion of symmetries and for packaging all the other equations of motion, is therefore incomplete in the sense that it does not give all the equations of the theory. A complete Lagrangian for type IIB was first written in \cite{DallAgata:1997gnw,DallAgata:1998ahf} following the methods of \cite{Pasti:1995ii,Pasti:1995tn,Pasti:1996vs}, but it involves extra fields and non-polynomial interactions. An alternative, which is quadratic and does not contain extra fields, is given in \cite{Bekaert:1999sq} (following \cite{Henneaux:1988gg,Deser:1997se,Bekaert:1998yp}). The drawback of this formulation is that it is not manifestly Lorentz invariant (see also \cite{Henneaux:2015gya} for a similar situation in six dimensions). A similar feature (non-standard transformations under spacetime diffeomorphisms) is also encountered in the variational principle proposed in \cite{Sen:2015nph}. These issues will be reviewed in part \ref{PART:6D} (see also section \ref{sec:4m4m2}).

\subsubsection{Reduction to $D=9$}

The reduction of type IIA supergravity is equivalent to the reduction of eleven-dimensional supergravity on $T^2$ and gives nine-dimensional supergravity, as presented above. On the other hand, the reduction of type IIB supergravity to nine dimensions gives the following field content:
\begin{itemize}
\item The metric gives a metric, a scalar, and a vector;
\item The two scalars give two scalars;
\item The two $2$-forms give two $2$-forms and two vectors;
\item The $4$-form gives a priori a $4$-form and a $3$-form. However, because of the self-duality condition in ten dimensions, one is dual to the other and we can choose to have only one $3$-form in nine dimensions.
\end{itemize}
All in all, this gives one metric, one $3$-form, two $2$-forms, three vectors and three scalars: exactly the field content of maximal supergravity in nine dimensions, as displayed in table \ref{table:fieldcontent}. In fact, the global symmetries and the details of the interactions also match \cite{Bergshoeff:1995as}.

Therefore, nine-dimensional maximal supergravity can be seen in two different ways: coming either from type IIA or type IIB supergravity in ten dimensions reduced on a circle. This confirms what is expected from the fact that there is only one maximal supergravity theory in nine dimensions.

\subsection{Six-dimensional conjectures}

Just as in ten dimensions, supersymmetry in six dimensions is chiral, $\cN = (\cN_+, \cN_-)$. It is maximal when $\cN_+ +\cN_- = 4$ and there are now three possibilities: the non-chiral $\cN = (2,2)$ algebra, and the chiral $\cN = (3,1)$ and $\cN = (4,0)$ algebras.

The non-chiral $\cN = (2,2)$ theory is well known; it is simply the reduction of eleven-dimensional supergravity on $T^5$.
The chiral $\cN = (3,1)$ and $\cN = (4,0)$ theories, however, are quite mysterious: their spectrum contains ``exotic" self-dual fields which are the object of part \ref{PART:6D} \cite{Strathdee:1986jr,Hull:2000zn,Hull:2000rr}. Since there is only one maximal supersymmetry algebra in five dimensions, one would expect that these exotic theories also give maximal supergravity upon dimensional reduction. This is easily seen at the level of the free theory \cite{Hull:2000zn,Henneaux:2017xsb,Henneaux:2018rub}, but is still a conjecture at the interacting level since no interactions are known for these exotic theories.

This completes the picture of the various higher-dimensional origins of maximal supergravity, as depicted in figure \ref{fig:oxidation} of the introduction.

\section{Infinite-dimensional symmetries}

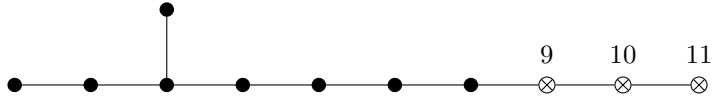
\begin{figure}
\centering
\begin{tikzpicture}
\tikzset{v/.style={circle,fill,inner sep=0pt,minimum size=5pt,draw}}
\tikzset{cr/.style={circle,inner sep=2pt, minimum size=6pt,draw,path picture={\draw (path picture bounding box.south east) -- (path picture bounding box.north west) (path picture bounding box.south west) -- (path picture bounding box.north east);}}}

\draw (0,0) node[v]{} -- (1,0) node[v]{} -- (2,0) node[v]{} -- (3,0) node[v]{} -- (4,0) node[v]{} -- (5,0) node[v]{} -- (6,0) node[v]{};
\draw (2,0) -- (2,1) node[v]{};

\node[cr] (9) at (7,0) {};
\node[cr] (10) at (8,0) {};
\node[cr] (11) at (9,0) {};
\draw (6,0) -- (9) -- (10) -- (11);

\node at (7,0.4) {$9$};
\node at (8,0.4) {$10$};
\node at (9,0.4) {$11$};
\end{tikzpicture}
\caption[Dynkin diagram of $E_{11}$]{\label{fig:E11dynkin}The Dynkin diagram of $E_{11}$. The black nodes constitute the diagram of $E_{8}$; adding the crossed nodes one by one, one gets the Dynkin diagrams of the infinite-dimensional Kac-Moody algebras $E_{9}$, $E_{10}$ and $E_{11}$ successively.}
\end{figure}

Even though the Cartan classification of finite-dimensional simple Lie algebras stops at $E_8$, the algebras $E_9$, $E_{10}$ and $E_{11}$ can nevertheless be defined. They are simple but infinite-dimensional Lie algebras, defined from $E_8$ by a systematic extension procedure (see \cite{Henneaux:2007ej} for a review). Their Dynkin diagram is pictured in figure \ref{fig:E11dynkin}.

By looking at table \ref{table:scalarcosets} or figure \ref{fig:oxidation}, it is natural to postulate that the $E$-series of exceptional groups will continue to $E_9$, $E_{10}$ and $E_{11}$ in dimensions $2$, $1$ and $0$ respectively \cite{Julia:1980gr,Julia:1982gx,Julia:1997cy}. We now briefly review how these ideas have been realized so far.

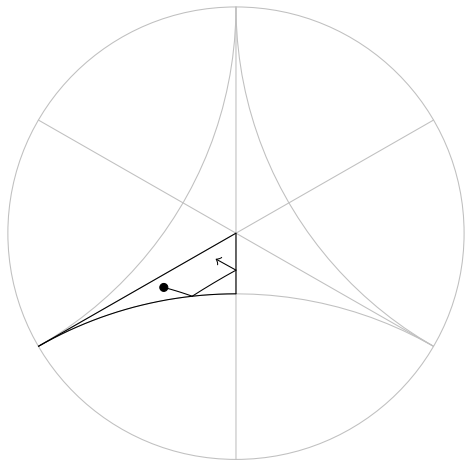
\begin{figure}
\centering
\definecolor{cqcqcq}{rgb}{0.7529411764705882,0.7529411764705882,0.7529411764705882}
\begin{tikzpicture}[line cap=round,line join=round,x=1.0cm,y=1.0cm]
\clip(-5.282285868157842,-3) rectangle (5.319090193058347,3);
\draw [color=cqcqcq] (0.0,-0.0) circle (3.0cm);
\draw [color=cqcqcq] (0.0,3.0)-- (0.0,-3.0);
\draw [color=cqcqcq] (-2.598076211353316,-1.4999999999999993)-- (2.598076211353316,1.4999999999999993);
\draw [color=cqcqcq] (2.598076211353315,-1.5000000000000007)-- (-2.598076211353315,1.5000000000000007);
\draw [shift={(0.0,-6.0)},color=cqcqcq]  plot[domain=1.0471975511965979:2.0943951023931957,variable=\t]({1.0*5.19615242270663*cos(\t r)+-0.0*5.19615242270663*sin(\t r)},{0.0*5.19615242270663*cos(\t r)+1.0*5.19615242270663*sin(\t r)});
\draw [shift={(5.196152422706632,2.9999999999999987)},color=cqcqcq]  plot[domain=1.0471975511965979:2.0943951023931957,variable=\t]({-0.4999999999999998*5.19615242270663*cos(\t r)+-0.8660254037844387*5.19615242270663*sin(\t r)},{0.8660254037844387*5.19615242270663*cos(\t r)+-0.4999999999999998*5.19615242270663*sin(\t r)});
\draw [shift={(-5.19615242270663,3.0000000000000013)},color=cqcqcq]  plot[domain=1.0471975511965979:2.0943951023931957,variable=\t]({-0.5000000000000003*5.19615242270663*cos(\t r)+0.8660254037844384*5.19615242270663*sin(\t r)},{-0.8660254037844384*5.19615242270663*cos(\t r)+-0.5000000000000003*5.19615242270663*sin(\t r)});
\draw [shift={(4.981665387570458,-9.43815404942545)}] plot[domain=2.0787837523173627:2.1444152151180305,variable=\t]({1.0*10.241493927149392*cos(\t r)+-0.0*10.241493927149392*sin(\t r)},{0.0*10.241493927149392*cos(\t r)+1.0*10.241493927149392*sin(\t r)});
\draw [shift={(0.0,-6.0)}] plot[domain=1.5707963267948966:2.0943951023931957,variable=\t]({1.0*5.19615242270663*cos(\t r)+-0.0*5.19615242270663*sin(\t r)},{0.0*5.19615242270663*cos(\t r)+1.0*5.19615242270663*sin(\t r)});
\draw [->] (0.0,-0.4899023863283305) -- (-0.266263313638047,-0.3416412527710819);
\draw [shift={(-1.9713187898984863,-4.641257211032341)}] plot[domain=1.2193801507424396:1.3158542329315148,variable=\t]({1.0*4.053068747303275*cos(\t r)+-0.0*4.053068747303275*sin(\t r)},{0.0*4.053068747303275*cos(\t r)+1.0*4.053068747303275*sin(\t r)});
\draw (0.0,-0.0)-- (0.0,-0.80384757729337);
\draw (0.0,-0.0)-- (-2.598076211353316,-1.4999999999999993);
\begin{scriptsize}
\draw [fill=black] (-0.9491779174329227,-0.7191921545599382) circle (1.5pt);
\end{scriptsize}
\end{tikzpicture}
\caption[Hyperbolic billiard of pure Einstein gravity]{Chaotic motion of a billiard ball in a region of hyperbolic space. The ``billiard table" is the fundamental Weyl chamber of the Kac-Moody algebra $\SL^{++}$. The motion of the ball is equivalent to the (BKL-type) chaotic dynamics of Einstein gravity near a spacelike singularity.}\label{fig:BKL}
\end{figure}

\begin{itemize}
\item[$E_9$:] The appearance of the group $E_9$ has been established for maximal supergravity in two dimensions in reference \cite{Nicolai:1987kz}. It has the same origin as the Geroch group \cite{Geroch:1972yt,Breitenlohner:1986um} in pure Einstein gravity (see also \cite{Katsimpouri:2012ky} and references therein). The idea is that, in the reduction from four to two dimensions, the theory has a $\SL$ symmetry acting on the two internal directions (this is called the Matzner–Misner group). It does not commute with the hidden symmetry $G_E$ which is already present in three dimensions (as we saw above, $G_E$ is the Ehlers group $\SL$ in the case of Einstein gravity, or $E_{8(8)}$ for maximal supergravity); instead, it combines with $G_E$ to generate the infinite-dimensional affine extension $G_E^+$. This is the Geroch group $\SL^+$ for pure gravity, or $E_9$ in maximal supergravity.

\item[$E_{10}$:] The Kac-Moody algebra $E_{10}$ appears in the dynamics of maximal supergravity in the vicinity of a spacelike singularity (see \cite{Damour:2002et,Henneaux:2007ej} for reviews)\footnote{An important difference with the previous symmetries is that this is the case already in the eleven-dimensional theory. However, in the limit where one is asymptotically close to the singularity, spatial gradients become negligible and the dynamics become effectively one-dimensional.}. There, the theory exhibits chaotic behaviour, much like the BKL-type singularities of pure gravity in four dimensions \cite{Belinsky:1970ew,Belinsky:1982pk}. It turns out that this chaotic behaviour is equivalent to the motion of a billiard ball in a closed region of hyperbolic space; moreover, the billiard table is precisely the Weyl chamber of the $E_{10}$ algebra (or $\SL^{++}$ for pure gravity, see figure \ref{fig:BKL}).

Inspired by this phenomenon, it was conjectured by Damour, Henneaux and Nicolai \cite{Damour:2002cu} that the \emph{complete} eleven-dimensional Lagrangian \eqref{eq:11lag} might be equivalent to the motion of a particle in the $E_{10}/K(E_{10})$ coset space. This idea is remarkably successful up to some level in the expansion of the infinite-dimensional $E_{10}$ algebra. However, its interpretation becomes difficult when the ``dual graviton" (see part \ref{PART:6D}) appears.

\item[$E_{11}$:] Before the $E_{10}$ conjectures of \cite{Damour:2002cu}, a similar idea involving the $E_{11}$ algebra was put forward by West in \cite{West:2001as}. The idea is also to establish an equivalence with a coset model based on the Kac-Moody algebra $E_{11}$, but it is even more radical in the sense that it is ``zero-dimensional", the space-time coordinates being already contained in the $E_{11}$ structure itself. This brings a lot of extra structure which has not yet been unraveled; it experiences similar difficulties as the $E_{10}$ model when the dual graviton is involved.
\end{itemize}
Even though their ultimate realization is yet unknown, it is clear that the infinite-dimensional $E_{10}$ and $E_{11}$ algebras contain a lot of information about maximal supergravity. They also motivate further work on the intriguing electric-magnetic and gravitational duality properties of supergravity and their interplay with supersymmetry, which are recurrent themes in this thesis.

\chapter{Enhancement of symmetries and Chern-Simons couplings}
\label{chap:cscouplings}

Quite generically, the global symmetry group that is expected to arise in the lower-dimensional theory from the reduction on a $n$-torus is
\begin{equation}\label{eq:parabolic}
GL(n,\R) \ltimes \R^k \, .
\end{equation}
The $GL(n,\R)$ factor comes from diffeomorphisms in the internal space and the $\R^k$ factor comes from shifts of the scalar fields (which do not necessarily commute) associated with the reduction and/or dualisation of $p$-form gauge potentials. (This is nicely reviewed in \cite{Cremmer:1997ct,PopeKKLectures}; see also \cite{Henneaux:2015opa} and appendix \ref{app:dimred}.) As we saw in the previous chapter, the group that appears in maximal supergravity is not \eqref{eq:parabolic} but the much larger $E_{n(n)}$, of which \eqref{eq:parabolic} is a parabolic subgroup. For this reason, $E_{n(n)}$ is often referred as the ``hidden symmetries" of the theory, since they are neither manifest in eleven dimensions nor expected from dimensional reduction.

It is clear that the surprising appearance of hidden symmetries must somehow be linked to the presence of maximal supersymmetry. However, the discussion of the previous chapter only dealt with the bosonic sector of maximal supergravity. The signature of supersymmetry in the bosonic Lagrangian \eqref{eq:11lag} is the value of its Chern-Simons coupling, which is indeed fixed by supersymmetry in the full supergravity theory \cite{Cremmer:1978km} (see also \cite{freedman_vanproeyen_2012} for a pedagogical derivation). In other words, the theory is \emph{not} supersymmetric if this coefficient is anything else than $1/6$. (Diffeomorphism and gauge invariance do not place any constraint on this coupling.) 

In this chapter, we prove in detail that the exceptional symmetries encountered in the previous chapter only appear if the Chern-Simons coupling take the ``correct" value $1/6$. Away from this critical value, only the parabolic subgroup \eqref{eq:parabolic} of symmetries remains. To avoid complications related to the presence of $p$-form potentials, we show this for the three-dimensional Lagrangian obtained by a reduction of \eqref{eq:11lag} on $T^8$, which only contains scalar fields in this case. Before going to maximal supergravity and $E_{8(8)}$, we also start with the analogue but simpler case of minimal supergravity in five dimensions in which $E_{8(8)}$ is replaced by the much smaller $G_{2(2)}$ \cite{Mizoguchi:1998wv}. We then indicate how our results are connected with rigidity theorems of Borel-like algebras \cite{LegerLuks}.

This chapter is based on the paper \cite{Henneaux:2015opa}, written in collaboration with M. Henneaux and A. Kleinschmidt.

\section{Minimal \texorpdfstring{$D=5$}{D=5} supergravity and \texorpdfstring{$G_{2(2)}$}{G2(2)}}
\label{G2sec}

We start by considering variations of minimal supergravity in $D=5$, in the conventions of~\cite{Compere:2009zh}. The bosonic Lagrangian density is given by
\begin{align}
\label{SUGRA5}
\mathcal{L}_{(5)} = R \star \1 - \frac12 \star F \wedge F + \frac{1}{3\sqrt{3}} F\wedge F \wedge A.
\end{align}
It is similar to the Lagrangian \eqref{eq:11lag} of eleven-dimensional supergravity, but now the gauge potential $A$ is a one-form and hence $F$ is a two-form.

The dimensional reduction of~\eqref{SUGRA5} to three dimensions gives rise to a total of $8$ scalar fields:
\begin{itemize}
\item Three scalars coming directly from the metric, associated with the coset $GL(2,\R)/SO(2)$. Two out of the three scalars are of dilatonic type because they come from diagonal components of the metric. We will call the dilatons $\phi_1$ and $\phi_2$ and the third metric scalar $\chi_1$.
\item Two scalars from the direct reduction of the one-form gauge potential, $\chi_2$ and $\chi_3$.
\item One scalar from dualising the vector coming from $A$ in five dimensions, $\chi_4$.
\item Two scalars from dualising the two Kaluza-Klein vectors, $\chi_5$ and $\chi_6$.
\end{itemize}
Remarkably, the eight scalars parametrize the coset space $G_{2(2)}/SO(4)$ (the dimensions are correct: $14 - 6 = 8$). The exceptional group $G_2$ plays the same role here as $E_8$ for maximal supergravity \cite{Mizoguchi:1998wv}.

\subsection{The $G_2$ algebra}

\begin{figure}
\centering
\begin{tikzpicture}
\begin{scope}[shift={(-3,0)},scale=1.5]
\tikzset{v/.style={circle,fill,inner sep=0pt,minimum size=7pt,draw}}
\node at (0,-0.3) {$1$};
\node at (1,-0.3) {$2$};
\draw (0,0) node[v]{} -- (1,0) node[v]{};
\draw (0, 0.07) -- (1, 0.07);
\draw (0, -0.07) -- (1, -0.07);
\draw (0.4, 0.15) -- (0.6, 0) -- (0.4, -0.15);
\end{scope}
\begin{scope}[shift={(3,0)}]
\node[left] at (-{sqrt(3)},1) {$\vec{\alpha}_1$};
\node[right] at ({2/sqrt(3)},0) {$\vec{\alpha}_2$};
\pgfmathsetmacro{\t}{deg(pi/3)}
\foreach \x in {0,...,6}{
	\draw[->, rotate=\t*\x] (0,0) -- (-{sqrt(3)},1);
	\draw[->, rotate=\t*\x] (0,0) -- ({2/sqrt(3)},0); }
\end{scope}
\end{tikzpicture}
\caption[Dynkin diagram and root system of $G_2$]{The Dynkin diagram and root system of $G_2$. The roots $\alpha_1$ and $\alpha_2$ are the simple roots.}\label{fig:G2}
\end{figure}
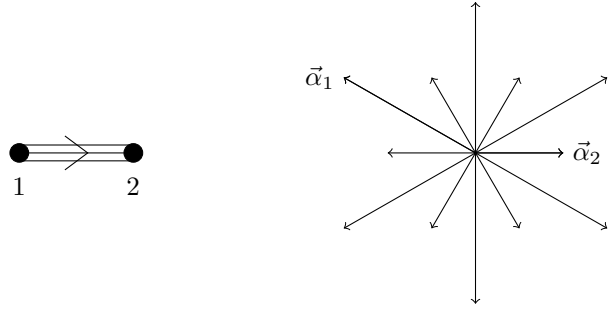

The $G_2$ algebra is of rank two and is specified by the Cartan matrix
\begin{equation}
A(G_2) = \begin{pmatrix}
2 & -1 \\ -3 & 2
\end{pmatrix} .
\end{equation}
Its Dynkin diagram and root system are displayed in figure \ref{fig:G2}. 
We follow the notations of \cite{Compere:2009zh} for a basis of $G_2$, which consists of two Cartan generators $h_1$, $h_2$, six positive generators $e_i$ ($i = 1, \dots, 6$) and six negative generators $f_i$.\footnote{Note that these are \emph{not} the canonical Chevalley-Serre generators as defined in appendix \ref{app:liealg}. The necessary change of basis is given in \cite{Compere:2009zh}.} The non-zero commutators between the $e_i$ are
\begin{align}\label{eq:g2plus}
[e_1,e_2] &= e_3 ,&
[e_2,e_3] &= -\frac{2}{\sqrt 3} e_4,&
[e_2,e_4] &= -e_5,& 
[e_1,e_5] &= e_6 ,&
[e_3,e_4] &= -e_6.
\end{align}
The commutators between the Cartan generators and the $e_i$ are
\begin{align}
\lb \, k_1 h_1 + k_2 h_2 \, , e_i \rb = (\vec{\alpha}_i \cdot \vec{k}) \, e_i ,
\end{align}
where
\begin{align}
\vec{\alpha}_1 = \left(-\sqrt{3},1\right), \quad\vec{\alpha}_2 = \left( \frac{2}{\sqrt 3}, 0 \right)
\end{align}
and
\begin{align}
\vec{\alpha}_3=\vec{\alpha}_1+\vec{\alpha}_2,\quad \vec{\alpha}_4=\vec{\alpha}_1 + 2\vec{\alpha}_2,\quad \vec{\alpha}_5=\vec{\alpha}_1+3\vec{\alpha}_2,\quad \vec{\alpha}_6=2\vec{\alpha}_1+3\vec{\alpha}_2 .
\end{align}
The vectors $\vec{\alpha}_i$ are the positive roots of $G_{2(2)}$, $\vec{\alpha}_1$ and $\vec{\alpha}_2$ being the simple ones. The remaining commutators involving only the $f_i$ and/or the $h_i$ can be obtained from those by using the Chevalley involution. The commutators involving both $e_i$ and $f_i$ can be worked out using the previous commutators, the relations
\begin{equation}
[e_1, f_1] = \frac{1}{2} \left( h_2 - \sqrt{3} h_1 \right), \quad [e_2, f_2] = \frac{1}{\sqrt{3}} h_1, \quad [e_1, f_2] = [e_2, f_1 ] = 0
\end{equation}
and the Jacobi identity. A faithful matrix representation of $G_2$ by $7 \times 7$ matrices is also provided in appendix A of \cite{Compere:2009zh}, which is useful for explicit computations.

\subsection{Scalar manifold and global symmetries}
\label{sec:varg2}

The $G_{2(2)}/SO(4)$ coset can be parametrized by
\begin{align}\label{eq:G2coset}
\cV = e^{\frac{1}{2}\phi_1 h_1+\frac{1}{2}\phi_2 h_2}e^{\chi_1 e_1}e^{-\chi_2 e_2 + \chi_3 e_3}e^{\chi_6 e_6}e^{\chi_4 e_4-\chi_5 e_5}.
\end{align}
Direct computation then shows that
\begin{equation}\label{eq:g2MC}
d\cV \cV^{-1} = \frac{1}{2} d\phi_1 \, h_1 + \frac{1}{2} d\phi_2 \, h_2 + \sum_{i=1}^6 \omega_i \, e_i,
\end{equation}
where the one-forms $\omega_i$ are given by
\begin{align}
\omega_1 &= e^{\vec{\alpha}_1 \cdot \vec{\phi}/2}d\chi_1, \label{oneforms5} \\
\omega_2 &= -e^{\vec{\alpha}_2 \cdot \vec{\phi}/2}d\chi_2, \\
\omega_3 &= e^{\vec{\alpha}_3 \cdot \vec{\phi}/2}\left( d\chi_3 - \chi_1 d\chi_2 \right), \\
\omega_4 &= e^{\vec{\alpha}_4 \cdot \vec{\phi}/2}\left( d\chi_4 + \frac{1}{\sqrt 3}(\chi_2 d\chi_3 - \chi_3 d\chi_2) \right), \\
\omega_5 &= -e^{\vec{\alpha}_5 \cdot \vec{\phi}/2}\left( d\chi_5 - \chi_2 d\chi_4 + \frac{1}{3 \sqrt 3}\chi_2(\chi_3 d\chi_2 - \chi_2 d\chi_3) \right), \\
\omega_6 &= e^{\vec{\alpha}_6 \cdot \vec{\phi}/2}\big( d\chi_6 - \chi_1 d\chi_5 + (\chi_1\chi_2 - \chi_3) d\chi_4 \nn \\
&\qquad\qquad + \frac{1}{3 \sqrt 3}(-\chi_1\chi_2+\chi_3)(\chi_3 d\chi_2 - \chi_2 d\chi_3) \big). \label{oneforms5fin}
\end{align}
The metric on the scalar manifold is then
\begin{align}
\label{scalars5}
ds^2=d\phi_1^2+d\phi_2^2+\sum_{i=1}^6 \omega_i^2 .
\end{align}
This matches the Lagrangian of the dimensionally reduced theory, and the scalar manifold is isometric to the standard Borel subgroup of $G_{2(2)}$ (i.e., the subgroup generated by the Cartan and raising operators $h_i$ and $e_i$ only).

The Borel subgroup of $G_{2(2)}$ leaves the one-forms $d\phi_1$, $d\phi_2$, $\omega_i$ invariant, and therefore also the metric \eqref{scalars5}. We now investigate the action of the rest of $G_{2(2)}$ on this metric. It is more convenient to consider the compact generators $k_i = e_i-f_i$ instead of the lowering generators $f_i$ themselves. Since the scalar manifold is an homogeneous space (the Borel subgroup acts transitively on it), we can moreover look only at the variations at the special point $\phi=0$, $\chi=0$. 

It is also sufficient to determine the action of $k_1$ and $k_2$ at zero, since the other $k_i$ can be obtained from these by commutation. Performing the standard non-linear realisation one finds for $k_1$:
\begin{align}
\delta d\phi_1 &= \sqrt{3}\omega_1,&  \delta d\phi_2 &= -\omega_1 ,& \label{k13} \\
\delta \omega_1 &= -\sqrt{3} d\phi_1+d\phi_2,&  \delta \omega_2 &= \omega_3, & \delta \omega_3 &= -\omega_2,& \\
\delta \omega_4 &= 0,& \delta \omega_5 &= \omega_6,& \delta \omega_6 &= -\omega_5& \label{k13fin}
\end{align}
and for $k_2$:
\begin{align}
\delta d\phi_1 &= -\frac{2}{\sqrt{3}}\omega_2, &\delta d\phi_2 &= 0 ,& \label{k23}\\
\delta \omega_1 &= -\omega_3, & \delta \omega_2 &= \frac{2}{\sqrt{3}} d\phi_1,& \delta \omega_3 &= \omega_1 -\frac{2}{\sqrt{3}}\omega_4 ,&\\
\delta \omega_4 &= \frac{2}{\sqrt{3}}\omega_3-\omega_5, & \delta \omega_5 &= \omega_4,& \delta \omega_6& = 0.& \label{k23fin}
\end{align}
We note that $k_2$ mixes the groups $(\omega_1,\omega_2,\omega_3)$ and $(\omega_4,\omega_5,\omega_6)$, while $k_1$ does not. Plugging these explicit transformations into the coset metric~\eqref{scalars5} one checks that the transformations indeed leave the metric invariant. This recovers the well-known fact that minimal supergravity has a hidden global $G_{2(2)}$ symmetry.

Another and more efficient way to perform this computation is to work directly with the Maurer-Cartan form \eqref{eq:g2MC}: the variations of the one-forms $d\phi_1$, $d\phi_2$, $\omega_i$ are simply the coefficients of the generators $h_1$, $h_2$ and $e_i$ in the variation of $d\cV \cV^{-1}$. Moreover, the action of a compact generator $k_i$ at the origin is simply given by
\begin{equation}
\delta \left( d\cV \cV^{-1} \right) = [d\cV \cV^{-1}, k_i] + c_i,
\end{equation}
where the compensator $c_i$ is an element of the compact subalgebra, uniquely determined from the requirement that $d\cV \cV^{-1}$ contains no lowering $f_i$ generator. This approach is most useful in the case of $E_{8(8)}$, whose matrix representations are less convenient to manipulate than those of $G_{2(2)}$.

\subsection{The role of the Chern--Simons coupling}
\label{CSG2}

Our main interest is to see the role of the Chern--Simons coupling. To this end we modify the Lagrangian~\eqref{SUGRA5} to
\begin{align}
\label{Lk5}
\mathcal{L}_\kappa = R \star \1 - \frac12 \star F \wedge F + \frac{\kappa}{3\sqrt{3}} F\wedge F \wedge A\, .
\end{align}
The value $\kappa=1$ corresponds to the theory considered above. The presence of $\kappa$ influences the scalar Lagrangian one obtains after reduction: the reduced theory in three dimensions is a non-linear sigma model with scalar metric given by
\begin{align}
ds^2=d\phi_1^2+d\phi_2^2+\sum_{i=1}^6 \tilde{\omega}_i^2 ,
\end{align}
where the invariant forms are now
\begin{align}
\tilde{\omega}_1 &= e^{\vec{\alpha}_1 \cdot \vec{\phi}/2}d\chi_1, \label{invoneG2} \\
\tilde{\omega}_2 &= -e^{\vec{\alpha}_2 \cdot \vec{\phi}/2}d\chi_2, \\
\tilde{\omega}_3 &= e^{\vec{\alpha}_3 \cdot \vec{\phi}/2}\left( d\chi_3 - \chi_1 d\chi_2 \right), \\
\tilde{\omega}_4 &= e^{\vec{\alpha}_4 \cdot \vec{\phi}/2}\left( d\chi_4 + \frac{\kappa}{\sqrt 3}(\chi_2 d\chi_3 - \chi_3 d\chi_2) \right), \\
\tilde{\omega}_5 &= -e^{\vec{\alpha}_5 \cdot \vec{\phi}/2}\left( d\chi_5 - \chi_2 d\chi_4 + \frac{\kappa}{3 \sqrt 3}\chi_2(\chi_3 d\chi_2 - \chi_2 d\chi_3) \right), \\
\tilde{\omega}_6 &= e^{\vec{\alpha}_6 \cdot \vec{\phi}/2}\big( d\chi_6 - \chi_1 d\chi_5 + (\chi_1\chi_2 - \chi_3) d\chi_4 \nn\\
&\qquad\qquad + \frac{\kappa}{3 \sqrt 3}(-\chi_1\chi_2+\chi_3)(\chi_3 d\chi_2 - \chi_2 d\chi_3) \big), \label{invoneG2fin}
\end{align}
This has to be compared with~\eqref{oneforms5} -- \eqref{oneforms5fin}: we see that only $\tilde{\omega}_4$, $\tilde{\omega}_5$ and $\tilde{\omega}_6$ differ.

Rather than investigating the symmetries of this metric, we perform a field redefinition in order to use the previous results. As long as $\kappa\ne 0$, we can redefine
\begin{align}
\label{redG2}
\chi_4 \rightarrow \kappa \chi_4, \quad \chi_5 \rightarrow \kappa \chi_5, \quad \chi_6 \rightarrow \kappa \chi_6
\end{align}
to obtain
\begin{align}
\tilde{\omega}_i = \kappa \omega_i \quad\textrm{for $i=4,5,6$},
\end{align}
while the first three $\omega_i$ were identical already before. This means that the scalar metric of the $\kappa$-deformed model can be written as
\begin{align}
ds^2=d\phi_1^2+d\phi_2^2+\omega_1^2+\omega_2^2+\omega_3^2+\kappa^2(\omega_4^2+\omega_5^2+\omega_6^2)\, .
\end{align}
This metric is therefore also invariant under the Borel subgroup of $G_{2(2)}$, since this subgroup leaves all the one-forms $d\phi_1$, $d\phi_2$, $\omega_i$ invariant by themselves.

Applying now the transformations $k_1$ and $k_2$ from equations~\eqref{k13}--\eqref{k13fin} and~\eqref{k23}--\eqref{k23fin}, one finds that for $\kappa^2 \neq 1$ only $k_1$ leaves this deformed metric invariant whereas $k_2$ does not. The $k_1$ symmetry has to be there because of the $GL(2,\R)$ part of the parabolic symmetry that is always present from the toroidal reduction. The enhancement due to the $k_2$ symmetry, however, is not present for values of the Chern--Simons coupling different from the value of minimal supergravity. This is the claimed result that the requirement of an enhanced symmetry implies the same constraints on the Chern--Simons coupling as supersymmetry would.

The fact that the Borel subalgebra is always a symmetry can be rephrased more clearly in terms of commutators as follows. The non-linear shift invariances of the one-forms \eqref{invoneG2} -- \eqref{invoneG2fin} are
\begin{align}
\delta \chi_1 &= c_1 \label{eq:g2shift} \\
\delta \chi_2 &= -c_2 \\
\delta \chi_3 &= c_3 + c_1 \chi_2 \\
\delta \chi_4 &= c_4 + \frac{c_3 \kappa}{\sqrt 3} \chi_2 + \frac{c_2 \kappa}{\sqrt 3} \chi_3 \\
\delta \chi_5 &= -c_5 + \frac{c_3 \kappa}{3 \sqrt 3} \chi_2^2 - c_2 \left( \chi_4 - \frac{\kappa}{3\sqrt 3} \chi_2 \chi_3 \right) \\
\delta \chi_6 &= c_6 + c_3 \left( \chi_4 + \frac{\kappa}{3\sqrt 3} \chi_2\chi_3 \right) + \frac{c_2 \kappa}{3\sqrt 3} \chi_3^2 + c_1 \chi_5, \label{eq:g2shiftfin}
\end{align}
where the $c_i$ are constant parameters. The finite transformations, which contain terms quadratic and cubic in the $c_i$, can be found in \cite{Henneaux:2015opa}; they are not necessary for this discussion.

These shift symmetries commute according to
\begin{align}\label{eq:g2deformed}
[\bar{e}_1, \bar{e}_2] &= \bar{e}_3 ,&
[\bar{e}_2, \bar{e}_3] &= -\frac{2\kappa}{\sqrt 3} \bar{e}_4,&
[\bar{e}_2, \bar{e}_4] &= -\bar{e}_5,& 
[\bar{e}_1, \bar{e}_5] &= \bar{e}_6 ,&
[\bar{e}_3, \bar{e}_4] &= -\bar{e}_6,
\end{align}
where the $\bar{e}_i$ generator is associated to the $c_i$ parameter in \eqref{eq:g2shift} -- \eqref{eq:g2shiftfin} (the other commutators are zero). When $\kappa=1$, this is exactly the upper triangular subalgebra of $G_{2(2)}$, but it is a continuous deformation thereof when $\kappa \neq 1$. However, when $\kappa \neq 0$, one can do the redefinition
\begin{equation}
e_i = \kappa \bar{e}_i \quad (i=4,5,6)
\end{equation}
and $e_i = \bar{e}_i$ for $i = 1, 2, 3$. This redefinition brings the commutation relations \eqref{eq:g2deformed} to those of $G_{2(2)}$ (equations \eqref{eq:g2plus}). In other words, the deformation \eqref{eq:g2deformed} is trivial in the sense that it can be undone by a redefinition of the generators of the algebra. Moreover, the action of the Cartan generators of $G_{2(2)}$ is independent on the value of $\kappa$: this shows that the Borel subgroup of $G_{2(2)}$ is always a symmetry of the model.

For the value $\kappa=0$, the Chern--Simons term is absent and the redefinition~\eqref{redG2} above is not allowed. The structure of the shift algebra simplifies to
\begin{align}
[\bar{e}_1, \bar{e}_2] &= \bar{e}_3 ,&
[\bar{e}_2, \bar{e}_3] &= 0&
[\bar{e}_2, \bar{e}_4] &= -\bar{e}_5,& 
[\bar{e}_1, \bar{e}_5] &= \bar{e}_6 ,&
[\bar{e}_3, \bar{e}_4] &= -\bar{e}_6
\end{align}
(the remaining commutators are all zero). This is a contraction of the previous shift algebra. 

A final comment concerns the enhanced symmetry that exists in pure $D=5$ gravity, i.e., without the gauge potential $A$. In this case it is known that there is an enhancement of the global symmetry to $SL(3,\R)$. One might wonder whether at least this enhancement survives for arbitrary $\kappa$.
Since it is generated by $\{ h_1, h_2, e_1, e_5, e_6, f_1, f_5, f_6 \}$, it is also broken by the ``wrong" value of $\kappa$.

\section{\texorpdfstring{$D=11$}{D=11} supergravity and \texorpdfstring{$E_{8(8)}$}{E8(8)}}
\label{E8sec}

We start with the following bosonic Lagrangian density in $D=11$
\begin{align}
\label{Lk11}
\mathcal{L}_\kappa = R \star \1 - \frac12 \star F \wedge F + \frac{\kappa}{6} F\wedge F \wedge A
\end{align}
and follow the same procedure as in section \ref{G2sec}.
Compared to the bosonic part of $D=11$ supergravity in~\eqref{eq:11lag}, we have introduced a free parameter $\kappa$ that controls the strength of the Chern--Simons self-interaction of the three-form $A$.

\subsection{The Borel subalgebra of $E_{8(8)}$}

We use the notations of \cite{Cremmer:1997ct}. The Borel subalgebra of $E_{8(8)}$ contains 8 Cartan generators, packed in a vector $\vec{H}$, and 120 ``raising" operators, which are
\begin{itemize}
\item the raising operators $E\indices{_i^j}$ of the $SL(8,\R)$ subalgebra (defined for $i < j$);
\item a three-form $E^{ijk}$;
\item a two-form $D_{ij}$; and
\item a one-form $D_i$.
\end{itemize}
All indices take values from one to eight. The $SL(8,\R)$ subalgebra coresponds to the nodes numbered $2$ to $8$ in figure \ref{fig:E8dynkin}. From the perspective of dimensional reduction, these generators correspond non non-linear shift invariances of the scalar fields in three dimensions as follows:
\begin{itemize}
\item $E\indices{_i^j}$ corresponds to shifts of the  axionic scalars coming from the direct reduction of the metric;
\item $E^{ijk}$ to the scalars from the three-form;
\item $D_{ij}$ to the scalars coming from the dualization of the vectors that arise from the reduction of the three-form; and 
\item $D_i$ to the scalars coming from the dualization of the Kaluza-Klein (metric) vectors.
\end{itemize}
Commutation relations can be found in section 4 of \cite{Cremmer:1997ct}. The important one for our purposes is
\begin{equation}\label{eq:commE8}
[E^{ijk}, E^{lmn}] = - \frac{1}{2} \varepsilon^{ijklmnpq} D_{pq} .
\end{equation}

\subsection{Global symmetries}

Taking the coset representative $\cV$ of \cite{Cremmer:1997ct}, the invariant forms on the scalar manifold are defined by
\begin{equation}
d\mathcal{V} \mathcal{V}^{-1} = \frac{1}{2} d\vec{\phi} \cdot \vec{H} + \sum_{i < j} \omega\indices{^i_j} E\indices{_i^j} + \sum_{i<j<k} \omega_{ijk} E^{ijk} + \sum_{i<j} \omega^{ij} D_{ij} + \sum_i \omega_i D^i
\end{equation}
(see section 4 and appendix A of that reference for explicit formulas).
Each one-form is invariant under the Borel subalgebra of $E_{8(8)}$. As long as the Chern-Simons coefficient $\kappa$ is not zero, the metric on the scalar manifold obtained by the reduction of \eqref{Lk11} can be expressed in terms of those one-forms as
\begin{align}
ds^2 = \sum_{i=1}^8 d\phi^2_i + \sum_{i < j} (\omega\indices{^i_j})^2 + \sum_{i<j<k} (\omega_{ijk})^2 + \kappa^2 \left( \sum_{i<j} (\omega^{ij})^2 + \sum_i (\omega_i)^2 \right)
\end{align}
by a suitable rescaling of the axion fields.

This form of the metric is preserved by all the generators of the $GL(8,\R)$ subalgebra and the symmetry is
\begin{align}\label{eq:parE8}
GL(8,\R) \ltimes\left( \R^{56} + \R^{28} + \R^{8}\right) .
\end{align}
The compact generator associated to the exceptional node (labelled ``$1$" in figure \ref{fig:E8dynkin}), however, rotates the $\omega_{ijk}$ and $\omega^{ij}$ among themselves: it is a symmetry of this metric if and only if $\kappa^2 = 1$, in which case the symmetry \eqref{eq:parE8} is enhanced to the full $E_{8(8)}$ hidden symmetry of Cremmer and Julia. This result follows from the commutation relations of $E_{8(8)}$, using the strategy outlined at the end of section \ref{sec:varg2} for computing the variations of the Borel-invariant one-forms.

It can also be understood as follows: without any redefinition of the fields, the scalar shift symmetries commute according to
\begin{align}
[E^{ijk}, E^{lmn}] = - \frac{\kappa}{2} \varepsilon^{ijklmnpq} D_{pq}\, .
\end{align}
As long as $\kappa\neq 0$, we can introduce a rescaled shift generator $\tilde{D}_{pq} = \kappa D_{pq}$ that brings the above commutation relation into the standard form \eqref{eq:commE8}. We also see that for $\kappa=0$, when there is no self-interaction of the three-form, shift symmetries abelianise.

\section{Scalar manifold and rigidity theorems}
\label{Rigidity}

We have seen in the previous examples that except for $\kappa=0$, one could always absorb $\kappa$ through redefinitions in the structure constants of the smaller symmetry algebra.  This smaller symmetry algebra present for all values of $\kappa$ has therefore always the same structure provided that $\kappa$ does not vanish. This is not an accident as we now explain.

The situation is the following. Consider a theory in $D$ spacetime dimensions that has hidden symmetry algebra $E$ (which can be any simple Lie algebra \cite{Breitenlohner:1987dg,Cremmer:1999du}) upon dimensional reduction to three dimensions.  This theory involves $p$-form fields and Chern-Simons couplings.  For the critical values of the Chern-Simons coefficient for which $E$ appears, the complete scalar manifold in three dimensions can be identified with the group manifold of the Borel subgroup $B(E)$, which is part of the symmetry. When deforming away from the critical point, the symmetry algebra is reduced and contains, as we have seen,  $B(GL(d,\R)) \ltimes U$ where $U$ is the exponential of some nilpotent algebra, which has same dimension as $B(E)$. The structure constants of the subalgebra $B(GL(d,\R)) \ltimes U$ depend continuously on the Chern-Simons coefficients, and so $B(GL(d,\R)) \ltimes U$ is a deformation of $B(E)$.  But by the rigidity theorems of \cite{LegerLuks}, the algebra $B(E)$ admits no non-trivial deformation.  Hence, $B(GL(d,\R)) \ltimes U$ is isomorphic with $B(E)$.

Of course the argument is valid in the vicinity of the critical values and does not eliminate the possibility of having contractions of $B(E)$ under deformations going out of that vicinity, just as the rigidity of simple Lie algebras does not eliminate the possibility to contract them to abelian algebras.
\chapter{Off-shell electric-magnetic duality}
\label{chap:emduality}

As we saw in chapter \ref{chap:maxsugra}, the $E_{7(7)}$ symmetry of $\cN = 8$ supergravity in four dimensions contains electric-magnetic duality transformations that mix the vector fields with their duals or, equivalently, equations of motion with Bianchi identities. These are generalizations of the usual electric-magnetic duality of Maxwell's equations. More generally, this situation can arise when the theory contains $p$-form fields whose field strengths are of degree exactly equal to half the space-time dimension: some symmetries of the theory can mix the $p$-forms with their duals (see chapter \ref{chap:maxsugra} in dimensions $4$, $6$ and $8$ for examples in maximal supergravity).

These symmetries define what is called the ``duality group" $G$ of the theory. Not all symmetries of the duality group act in a local way at the level of the usual (second-order) action\footnote{However, contrary to ``popular folklore", all symmetries of $G$ are symmetries of the action. It is just that some of them have a non-local expression (in space) when written in terms of the fundamental variables of the variational principle \cite{Deser:1976iy,Deser:1981fr,Bunster:2011aw}.}; those that do constitute the electric group $G_e \subset G$. This electric group depends on the choice of duality frame, i.e., the choice of whether the usual potentials or their duals are taken as fundamental variables in the action. This is an important point in the study of supergravity gaugings, in which only local symmetries of the action can be gauged: the possible gauge groups $G_\text{gauge} \subset G_e$ then depend on the choice of duality frame. (This is the subject of part \ref{PART:GAUGINGS}.)

The goal of this chapter is to review those facts. The easiest way to do so is through the construction of a first-order action in which all duality symmetries act locally. The presentation is based on the paper \cite{Henneaux:2017kbx}, written in collaboration with M. Henneaux, B. Julia and A. Ranjbar. The originals results of that paper will be presented in chapter \ref{chap:YMgaugings}. We neglect gravity in this chapter, since it does not play a crucial role in this discussion.

\section{Bianchi identities and equations of motion}

In this section, we review the duality symmetries at the level of the Bianchi identities and equations of motion. We first consider a simple example with duality group $G = \SL$ before going to the general case of coupled vector and scalar fields.

\subsection{Maxwell-dilaton-axion theory}

We begin with a Lagrangian containing a single vector field $A$ with field strength $F = dA$ and two scalar fields $\phi$, $\chi$ parametrizing the $SL(2,\R)/SO(2)$ coset space,
\begin{equation}\label{eq:N4simplified}
\cL = - \frac{1}{2} \star d\phi \wedge d\phi - \frac{1}{2}e^{2 \phi} \star d\chi \wedge d \chi - \frac{1}{2} e^{-\phi} \star F \wedge F - \frac{1}{2} \chi \, F \wedge F \, .
\end{equation}
It has the following symmetry features, which illustrate well the general case when one has field strengths of degree equal to half the dimension.
\begin{itemize}
\item The model has $SL(2,\R)$ symmetry, acting in the usual non-linear way on the scalar fields (see appendix \ref{app:cosetlagrangians}) and mixing the vector field with its dual.
\item All the $SL(2,\R)$ symmetries do not act in a local way at the level of the Lagrangian \eqref{eq:N4simplified}, since it contains the vector field $A$ but not its dual. This Lagrangian has only a $B^+$ symmetry acting locally, where $B^+$ is the Borel subgroup of $\SL$ generated by upper triangular matrices.
\item By dualizing the vector field, one can get a Lagrangian with a different group of local symmetries, namely the subgroup $B^-$ of $SL(2,\R)$ generated by lower triangular matrices. These Lagrangians give the same equations of motion and Bianchi identities with the full $SL(2,\R)$ symmetry. (They both have the full $SL(2,\R)$ symmetry at the level of the action, but some of the transformations act in a non-local way.)
\end{itemize}
The equations of motion and Bianchi identity for the vector field are\footnote{Those are exactly the Lagrangian and equations of motion that one gets by trucating $D=8$ maximal supergravity to its $SL(3,\R)$ singlet sector, as discussed in \cite{Cremmer:1997ct}, except that the role of the $3$-form in $D=8$ is played here by a $1$-form in $D=4$.}
\begin{equation}
d(e^{-\phi} \star F + \chi F) = 0, \quad d F = 0 .
\end{equation}
They can be written more compactly as
\begin{equation}
dG = 0, \quad dF = 0,
\end{equation}
where
\begin{equation}\label{eq:N4GF}
G = e^{-\phi} \star F + \chi F .
\end{equation}
If $F$ and $G$ were independent, the set of equations of motion and Bianchi identities would be invariant under any linear $GL(2,\R)$ transformation of $F$ and $G$. However, the fact that such a linear transformation must preserve the relation \eqref{eq:N4GF} puts restrictions on the possible linear transformations.
The easiest way to see this is to observe that, using the fact that $\star^2 = -1$ in this case, \eqref{eq:N4GF} is equivalent to
\begin{equation}\label{eq:N4twisted}
\star \cG = M(\phi,\chi)\, \epsilon \, \cG
\end{equation}
where
\begin{equation}
\cG = \begin{pmatrix}
F \\ G
\end{pmatrix}, \qquad \epsilon = \begin{pmatrix}
0 & 1 \\ -1 & 0
\end{pmatrix}, \qquad M(\phi,\chi) = \begin{pmatrix} e^\phi & \chi e^\phi \\ \chi e^\phi & e^{-\phi} + \chi^2 e^\phi \end{pmatrix} .
\end{equation}
($M$ is the standard $SL(2,\R)/SO(2)$ coset matrix, see eq. \eqref{eq:cosetSL2SO2M}). The equation \eqref{eq:N4twisted} is known as the ``twisted self-duality" equation. In this notation, the equations of motion and Bianchi identities are simply $d\cG = 0$, and they are invariant under linear transformations
\begin{equation}
\cG \rightarrow g^T \cG, \quad g \in GL(2,\R).
\end{equation}
One gets further restrictions on $g$ from the fact that it must be compatible with the twisted self-duality equation \eqref{eq:N4twisted}. This is possible when $g \in SL(2,\R)$: then, there exists a symmetry of the scalar sector such that
\begin{equation}
M \rightarrow g^T M g\, .
\end{equation}
The right-hand side of \eqref{eq:N4twisted} then transforms as
\begin{equation}
M \epsilon \cG \rightarrow g^T M g \epsilon g^T \cG = g^T M \epsilon \cG,
\end{equation}
where we used the property $g \epsilon g^T = \epsilon$ which holds since $g\in SL(2,\R)$. This shows that the vector equations of motion and Bianchi identities have a $SL(2,\R)$ symmetry. If one introduces the dual vector field $\tilde{A}$ through $G = d\tilde{A}$, this can be seen equivalently as a $SL(2,\R)$ symmetry acting on the original vector and its dual as a doublet,
\begin{equation}
\begin{pmatrix}
A \\ \tilde{A}
\end{pmatrix} \rightarrow g^T \begin{pmatrix}
A \\ \tilde{A}
\end{pmatrix}, \qquad g \in SL(2,\R).
\end{equation}
In infinitesimal form, the generators $h$, $e$, $f$ of $\SL$ act on the various fields as
\begin{equation}\label{eq:N4sym}
\begin{array}{lllll}
h: \quad & \delta \phi = 2 \varepsilon, & \delta \chi = - 2 \varepsilon \chi, &\delta A = \varepsilon A, &\delta \tilde{A} = - \varepsilon\tilde{A},\\
e: \quad & \delta \phi = 0, & \delta \chi = \varepsilon, &\delta A = 0, &\delta \tilde{A} = \varepsilon A,\\
f: \quad & \delta \phi = 2 \varepsilon \chi, & \delta \chi = \varepsilon ( e^{-2\phi} - \chi^2), &\delta A = \varepsilon \tilde{A}, &\delta \tilde{A} = 0 \\
\end{array}
\end{equation}
with $\varepsilon$ an infinitesimal parameter.
Only the transformations in which the variation of $A$ does not involve $\tilde{A}$ can act in a local way at the level of the Lagrangian \eqref{eq:N4simplified} (indeed, the expression of $\tilde{A}$ in terms of $A$ is non-local, since the Poincaré lemma involves a space-time integration). This leaves $h$ and $e$, which can indeed easily be checked to give symmetries of \eqref{eq:N4simplified}\footnote{We define symmetries as invariance of the action, which is invariance of the Lagrangian up to a total derivative. This happens for the $e$ symmetry, under which \eqref{eq:N4simplified} is not invariant but varies to the total derivative $\delta \cL = \epsilon F\wedge F/2 = d (\epsilon A \wedge F /2)$. This kind of symmetries will play an important role in part \ref{PART:GAUGINGS}.}. They generate the $B^+$ subgroup of $\SL$.

Let us now replace $A$ by its dual $\tilde{A}$ in \eqref{eq:N4simplified}, following the procedure outlined in appendix \ref{app:pformduality}. This is done by treating $F$ as the elementary field and introducing $\tilde{A}$ as a Lagrange multiplier for the Bianchi identity $dF = 0$,
\begin{equation}\label{eq:N4simpaux}
\cL_V^\text{aux} = - \frac{1}{2} e^{-\phi} \star F \wedge F - \frac{1}{2} \chi \, F \wedge F + \tilde{A} \wedge dF .
\end{equation}
Solving for $\tilde{A}$ leads back to \eqref{eq:N4simplified}. The equation of motion for $F$ is of course $d\tilde{A} = e^{-\phi} \star F + \chi F$, which can be solved algebraically for $F$ as
\begin{equation}
F = \frac{1}{(e^{-\phi} + \chi^2 e^{\phi})} (- \star G + \chi e^\phi G)
\end{equation}
where $G = d\tilde{A}$. Plugging this back in the auxiliary Lagrangian \eqref{eq:N4simpaux} gives a Lagrangian with $\tilde{A}$ only,
\begin{equation}
\tilde{\cL}_V = - \frac{1}{2(e^{-\phi} + \chi^2 e^{\phi})} \star G \wedge G - \frac{\chi e^\phi}{2(e^{-\phi} + \chi^2 e^{\phi})} \, G \wedge G .
\end{equation}
Its local symmetries are now those of \eqref{eq:N4sym} in which the variation of $\tilde{A}$ do not involve $A$, i.e., $h$ and $f$ which generate the subgroup $B^-$ of $\SL$. This shows that the group $G_e$ of local symmetries of the second-order Lagrangian depends on the choice of duality frame.

\subsection{The general case}

The above discussion generalizes to any scalar-vector Lagrangian of the form
\begin{equation}
\cL^{(2)} = \cL^{(2)}_S + \cL_V^{(2)}, 
\end{equation}
where the vector Lagrangian is
\begin{align} 
\cL^{(2)}_V &= - \frac{1}{2} \mathcal{I}_{IJ}(\phi) \star F^I \wedge F^{J} - \frac{1}{2} \mathcal{R}_{IJ}(\phi)\, F^I \wedge F^J \label{eq:lag4dvectors}
\end{align}
and the explicit form of the scalar Lagrangian is not crucial in what follows. The indices $I,J$ run from $1$ to the number $n_v$ of vector fields. The symmetric matrices $\cI$ and $\cR$ give the non-minimal couplings between the scalars and the abelian vectors. In the free Maxwell case, one has $\cI_{IJ} = \delta_{IJ}$ and $\cR_{IJ} = 0$, while in the example \eqref{eq:N4simplified} of the previous section, $n_v = 1$, $\cI = e^{2\phi}$ and $\cR = \chi$.

We follow the review \cite{Trigiante:2016mnt}, except for the sign of $\cI$ to be consistent with \cite{Henneaux:2017kbx,Barnich:2017nty}.\footnote{Note also that the sign of the $\varepsilon_{\mu\nu\rho\sigma}$ tensor is different in \cite{Barnich:2017nty}.} Neglecting gravity, this is the generic bosonic sector of ungauged supergravities in four dimensions.

The equations of motion and Bianchi identities can be written as
\begin{equation}
d G_I = 0, \quad dF^I = 0,
\end{equation}
where $G_I$ is defined by
\begin{equation}
G_I = \cI_{IJ} \star F^J + \cR_{IJ} F^J .
\end{equation}
As before, those equations are equivalent to
\begin{equation}\label{eq:dG}
d \cG = 0, \qquad \cG = \begin{pmatrix}
F^I \\ G_I
\end{pmatrix} ,
\end{equation}
and the relation between $F^I$ and $G_I$ can be written as the twisted self-duality condition
\begin{equation}\label{eq:twisted}
\star \cG = \Omega \cM(\phi) \, \cG\, ,
\end{equation}
where the matrices $\Omega$ and $\mathcal{M}(\phi)$ are the $2n_v \times 2n_v$ matrices
\begin{equation} \label{eq:OMdef}
\Omega = \begin{pmatrix}
0 & I \\ -I & 0
\end{pmatrix}, \qquad
\mathcal{M} = \begin{pmatrix}
\mathcal{I} + \mathcal{R}\mathcal{I}^{-1}\mathcal{R} & - \mathcal{R} \mathcal{I}^{-1} \\
- \mathcal{I}^{-1} \mathcal{R} & \mathcal{I}^{-1}
\end{pmatrix},
\end{equation}
each block being $n_v\times n_v$ and $0$, $I$ being the $n_v \times n_v$ zero and identity matrices respectively.

Now, a linear transformation
\begin{equation}\label{eq:transfG}
\cG \rightarrow g\, \cG
\end{equation}
of $F^I$ and $G_I$, where $g$ is an invertible $2n_v \times 2n_v$ matrix, is a symmetry of the equations \eqref{eq:dG} and \eqref{eq:twisted} if and only if the following two conditions are met, as was discovered in \cite{Gaillard:1981rj}:
\begin{enumerate}
\item The matrix $g$ is symplectic,
\begin{equation} \label{eq:duality1}
g^T \Omega g = \Omega,
\end{equation}
\item There is a symmetry $\phi \rightarrow \phi'$ of the scalar Lagrangian such that the matrix $\cM$ transforms as
\begin{equation}\label{eq:duality2}
\cM(\phi') = g^{-T} \cM(\phi) g^{-1} .
\end{equation}
\end{enumerate}
The group of transformations satisfying these conditions is called the ``(electric-magnetic) duality group" $G$.

As we have just indicated, the duality group is a subgroup of $Sp(2n_v, \mathbb{R})$.  This symplectic condition holds irrespective of the scalar sector and its couplings to the vectors.  Which transformations among those of $Sp(2n_v, \mathbb{R})$ are actual symmetries depends, however, on the number of scalar fields, their internal manifold and their couplings to the vectors \cite{Ferrara:1976iq,Cremmer:1977zt,Cremmer:1977tc,Cremmer:1977tt,Gaillard:1981rj,deWit:2001pz}.

For example, if there is no scalar field, a duality transformation should also belong to $SO(2n_v)$ to leave $\cM_{MN} = \delta_{MN}$ invariant, and the duality group is then the unitary group $U(n_v) = Sp(2n_v, \mathbb{R}) \cap SO(2n_v)$. If there are scalars forming the coset manifold $Sp(2n_v, \mathbb{R})/U(n_v)$, with appropriate couplings to the vectors, the duality group is the full symplectic group $Sp(2n_v, \mathbb{R})$ \cite{Bunster:2011aw}. In the case of maximal supergravity toroidally reduced to four dimensions, the scalars are such that the duality group is $E_{7(7)} \subset Sp(56,\R)$ as discovered in \cite{Cremmer:1978ds,Cremmer:1979up}. For the example of the previous section, one has $G = SL(2,\R) = Sp(2,\R)$.

The easiest way to see this fact, as well as the generalization of the other phenomenons encountered in the example \eqref{eq:N4simplified}, is to go to the Hamiltonian formalism.

\section{Off-shell invariance: first order action}

The canonical momenta conjugate to the $A^I$ are given by
\begin{equation}
\pi^i_I = \frac{\pd \mathcal{L}}{\pd \dot{A}^I_i} = \mathcal{I}_{IJ}\, F\indices{^J_0^i} -\frac{1}{2} \mathcal{R}_{IJ}\,\varepsilon^{ijk}F^J_{jk}, 
\end{equation}
along with the constraint $\pi^0_I=0$. This relation can be inverted to get
\begin{equation}
\dot{A}^{Ii} = (\mathcal{I}^{-1})^{IJ} \,\pi^i_J + \partial^i A^I_0 + \frac{1}{2} (\mathcal{I}^{-1} \mathcal{R})\indices{^I_J} \,\varepsilon^{ijk} F^J_{jk} ,
\end{equation}
from which we can compute the first-order Hamiltonian action
\begin{equation}
S_H =  \int\!d^4x\, \left( \pi^i_I \dot{A}^I_i - \mathcal{H} -  A^I_0 \, \mathcal{G}_I \right),
\end{equation}
where
\begin{align}
\mathcal{H} &= \frac{1}{2} (\mathcal{I}^{-1})^{IJ} \pi^i_I \pi_{J i} + \frac{1}{4} (\mathcal{I} + \mathcal{R}\mathcal{I}^{-1}\mathcal{R})_{IJ} F^I_{ij} F^{Jij} + \frac{1}{2} (\mathcal{I}^{-1} \mathcal{R})\indices{^I_J} \,\varepsilon^{ijk} \pi_{Ii} F^J_{jk} \\
\mathcal{G}_I &= - \partial_i \pi^i_I .
\end{align}
The time components $A^I_0$ appears in the action as Lagrange multipliers for the constraints
\begin{equation}
\partial_i \pi^i_I = 0 .
\end{equation}
These constraints can be solved by introducing new (dual) potentials $Z_{I i}$ through the equation
\begin{equation}\label{eq:momentumZ}
\pi^i_I = -  \varepsilon^{ijk} \partial_j Z_{I k},
\end{equation}
which determines $Z_{I}$ up to a gauge transformation $Z_{I i} \rightarrow Z_{I i} + \partial_i \tilde{\epsilon}_I$. Note that the introduction of these potentials is non-local but permitted in (flat) contractible space. 
Putting this back in the action gives
\begin{equation} \label{eq:symlag}
S = \frac{1}{2} \int \!d^4x \left(  \Omega_{MN} \mathcal{B}^{Mi} \dot{\mathcal{A}}^N_i -   \mathcal{M}_{MN}(\phi) \mathcal{B}^M_i \mathcal{B}^{Ni} \right),
\end{equation}
where the doubled potentials are packed into a vector
\begin{equation}
(\mathcal{A}^M) = \begin{pmatrix} A^I \\ Z_I \end{pmatrix}, \quad M= 1, \dots, 2n_v
\end{equation}
and their curls  $\mathcal{B}^{Mi}$ are
\begin{equation}
\mathcal{B}^{Mi} =  \varepsilon^{ijk} \partial_j \mathcal{A}^M_k .
\end{equation}
The matrices $\Omega$ and $\mathcal{M}(\phi)$ are those defined in \eqref{eq:OMdef}. The dual vector potentials are canonically conjugate to the magnetic fields. The action (\ref{eq:symlag}), which puts electric and magnetic potentials on the same footing, is called the ``first-order action" in this chapter.

Now, in order to be a symmetry of this action, a linear transformation
\begin{equation}
\cA \rightarrow g\, \cA
\end{equation}
of the vector potentials, possibly accompanied by a scalar transformation $\phi \rightarrow \phi'$, must preserve the kinetic term and the (scalar-dependent) energy density term separately. This gives the two conditions \eqref{eq:duality1} and \eqref{eq:duality2} respectively. This transformation induces on-shell the transformation \eqref{eq:transfG} of the field strengths. In this way, the full duality group $G$ acts as local, off-shell symmetries of the first-order action \eqref{eq:symlag}. We stress that the duality transformations are defined here in terms of the fundamental variables of the theory that are varied in the action principle, namely the potentials, following \cite{Deser:1976iy}. In the case of $\cN = 8$ supergravity and $E_{7(7)}$, this action was first written in \cite{Hillmann:2009zf}.

\section{Choice of symplectic frames and the electric group}

While the first-order (action \eqref{eq:symlag}) and second-order (action \eqref{eq:lag4dvectors}) formalisms are equivalent in terms of symmetries (any symmetry in one formalism is also a symmetry of the other), they are not equivalent in what concerns the concept of locality (specifically, locality in space).  A local function in one formulation may not be local in the other.  The origin of this difference is made more precise in this section.

\subsection{Choice of symplectic frame and locality restriction}\label{sec:symplecticchoice}

In order to go from the first-order formalism to the standard second-order formalism, one has to tell what are the ``positions" $q$'s (to be kept) and the ``conjugate momenta" $p$'s (to be eliminated in favour of the velocities through the inverse Legendre transformation). In symplectic geometry, this is called a choice of ``Darboux" or ``symplectic" frame. In our case, since the electric and magnetic fields are conjugate, a choice of $q$'s and $p$'s is equivalent to  choosing what are the ``electric potentials" and what are the conjugate ``magnetic potentials". For that reason, a choice of Darboux frame is also called a choice of duality frame. Since we consider here only linear canonical transformations, a choice of canonical coordinates amounts to a choice of an element of the symplectic group $Sp(2n_v, \mathbb{R})$ relating that symplectic frame to  a reference symplectic frame.

Once a choice of symplectic frame has been made, one goes from \eqref{eq:symlag} to the second-order formalism by following the  steps reverse to those that led to the first-order formalism.
\begin{itemize}
\item One keeps the first half of the vector potentials $\mathcal{A}^I = A^I$, the ``electric ones" in the new frame ($I = 1, \cdots, n_v$).
\item One replaces the second half of the vector potentials $\mathcal{A}^{I +n_v}= Z_I$ (the ``magnetic ones") by the momenta
\begin{equation}
\pi^i_I = - \varepsilon^{ijk} \partial_j Z_{I k}
\end{equation}
subject to the constraints 
\begin{equation}
\partial_i \pi^i_I = 0 
\end{equation}
which one enforces by introducing the Lagrange multipliers $A_0^I$.  This is the non-local step (in space).
\item One finally eliminates the momenta $\pi^i_I$, which can be viewed as auxiliary fields, through their equations of motion which are nothing else but the inverse Legendre transformation expressing $\dot{A}^I_i$ in terms of $\pi^i_I$.
\end{itemize}

We can encode the choice of symplectic frame through the symplectic transformation that relates the chosen frame to a  reference symplectic frame,
\begin{equation} \label{eq:Aprime}
\mathcal{A} = E \mathcal{A}',
\end{equation}
where $E$ is a symplectic matrix and $\mathcal{A}$, $\mathcal{A}'$ are respectively the potentials in the reference and new duality frames. The symplectic property $E^T \Omega E = \Omega$ ensures that the kinetic term in \eqref{eq:symlag} remains invariant. The first-order action therefore takes the same form, but with matrices $\mathcal{M}$ and $\mathcal{M}'$  related by 
\begin{equation}\label{eq:Mprime}
\mathcal{M}' = E^T \mathcal{M} E .
\end{equation}
Using the fact that $E$ and $E^T$ are symplectic, a straightforward but not very illuminating computation shows that $\mathcal{M}'$ determines matrices $\mathcal{I}'$, $\mathcal{R}'$ uniquely such that equation \eqref{eq:OMdef} with primes holds,  i.e, there exist unique $\mathcal{I}'$ and $\mathcal{R}'$ such that
\begin{equation} \label{eq:OMdefBis}
\mathcal{M}' = \begin{pmatrix}
\mathcal{I}' + \mathcal{R}'\mathcal{I}'^{-1}\mathcal{R}' & - \mathcal{R}' \mathcal{I}'^{-1} \\
- \mathcal{I}'^{-1} \mathcal{R'} & \mathcal{I}'^{-1}
\end{pmatrix}.
\end{equation}
The matrices $\mathcal{I}'$ and $\mathcal{R}'$ depend on the scalar fields, but also on the symplectic matrix $E$.
Performing in reverse order the steps described above,  one then gets the second-order Lagrangian  
\begin{equation} \label{eq:lprime}
\cL^{(2)\prime}_V = - \frac{1}{4} \mathcal{I}'_{IJ}(\phi) F'^I_{\mu\nu} F'^{J\mu\nu} + \frac{1}{8} \mathcal{R}_{IJ}'(\phi)\, \varepsilon^{\mu\nu\rho\sigma} F'^I_{\mu\nu} F'^J_{\rho\sigma} .
\end{equation}
The new Lagrangian depends on the parameters of the symplectic transformation $E$ used in equation \eqref{eq:Mprime}; therefore, we have a family of Lagrangians labelled by an $Sp(2n_v, \mathbb{R})$ element $E$.  By construction, these Lagrangians differ from one another by a change of variables that is in general non-local in space, but give the same set of equations of motion and Bianchi identities.

There are of course redundancies in this description.
\begin{itemize}
\item First, it is clear that different symplectic transformations can lead to the same final $n_v$-dimensional ``Lagrangian subspace" of the $q$'s upon elimination of the momenta.  [The choice of the $q$'s is equivalent to a choice of Lagrangian linear subspace, or linear polarization, because the $q$'s form  a complete set of commuting variables (in the Poisson bracket). The same is true for a choice of $p$'s.] The stability subgroup $T$ of a Lagrangian subspace consists of the block lower triangular symplectic transformations, yielding second-order Lagrangians that differ by a redefinition of the $q$'s (when $E$ is a canonical ``point" transformation) and the addition of a total derivative (when $E$ is a canonical ``phase" transformation).

More explicitly, when $E$ is lower-triangular, i.e., of the form (since it must be symplectic)
\begin{equation}
E = \begin{pmatrix}
P & 0 \\ SP & P^{-T}
\end{pmatrix}, \qquad P \in GL(n_v, \R), \quad S^T = S,
\end{equation}
a short computation then shows that $\mathcal{I}'$ and $\mathcal{R}'$ are given by
\begin{align}
\mathcal{I}' &= P^T \mathcal{I} P \\
\mathcal{R}' &= P^T \mathcal{R} P + P^T S P .
\end{align}
By using these formulas in \eqref{eq:lprime}, we recover the original Lagrangian \eqref{eq:lag4dvectors} by the change of variables
\begin{equation}
A'^I_\mu = P\indices{^I_J} A^J_\mu
\end{equation}
(the second term of $\cR'$ corresponds to the addition of a total derivative).

\item Second, duality symmetries correspond also to redundancies in the transition to the second-order formalism since they do not modify the first-order Lagrangian (when the scalars are transformed appropriately) and hence clearly lead to the same final second-order Lagrangian.
\end{itemize}
One can characterize the redundancies as follows. The symplectic transformations \eqref{eq:Aprime} with matrices $E \in Sp(2 n_v, \mathbb{R})$ and $g E t$, where $t$ belongs to the stability subgroup $T$ of the final Lagrangian subspace and $g$ belongs to the duality group $G$, are equivalent. Indeed, the first-order Lagrangian is left invariant under the transformation  $\mathcal{A} \rightarrow g^{-1} \mathcal{A}$ (provided the scalars are transformed appropriately), and the two sets of new symplectic coordinates $\mathcal{A}'$ and $ t \mathcal{A}'$ yield equivalent second-order Lagrangians after elimination of the momenta\footnote{One could also consider what happens when the coordinates $\mathcal{A}'$ defined by \eqref{eq:Aprime} are taken as the reference frame. Then, the changes of frame $\mathcal{A}' = S \mathcal{A}''$ and $\mathcal{A}' = g' S t' \mathcal{A}''$ are equivalent, where $g' = E^{-1} g E$ is the matrix associated with the duality symmetry $g$ in the $\mathcal{A}'$--frame and $t'$ is an element of the stability subgroup of the Lagrangian subspace in the $\mathcal{A}''$--frame. This gives an equivalent description of the redundancies.}. The relevant space is thus the quotient $G \backslash Sp(2n_v, \mathbb{R})/ T$. It should be noted that the authors of \cite{deWit:2002vt} considered a smaller stability subgroup $T = GL(n_v,\mathbb{R})$ by requiring  invariance of the Lagrangian itself (and not invariance up to a total derivative that does not matter classically).  

\subsection{Electric group}

The duality transformations
$\mathcal{A} \rightarrow \mathcal{A}' = g \mathcal{A}$, $\phi \rightarrow \phi'$, where the symplectic matrix $g$ and the isometry $\phi \rightarrow \phi'$ of the scalar manifold are such that the condition \eqref{eq:duality2} holds, 
generically mix the electric and magnetic potentials.  Accordingly, when expressed in terms of the variables of the second-order formalism, they will generically take a non-local form, since the magnetic potentials become non-local functions of the electric potentials and their time derivatives in the second-order formalism \cite{Deser:1976iy,Deser:1981fr,Bunster:2011aw}\footnote{Given a duality symmetry of the action $S[\mathcal{A}^I_k, \phi^i]$, one gets the corresponding duality symmetry of the first-order action $S[A^I_k, \pi_I^k, A_0^I]$ by (i) expressing in the variation $\delta A^I_k$ the magnetic potentials $Z^I_k$ (if they occur) in terms of $\pi_I^k$, which is a non-local expression, determined up to a gauge transformation that can be absorbed in a gauge transformation of the electric variables;  (ii) computing the variation $\delta \pi^k_I$ from  $\pi^i_I = - \varepsilon^{ijk} \partial_j Z_{I k}$; and (iii) determining the variation of the  Lagrange multipliers $A_0^I$ so that the terms proportional to the constraint terms $\partial_i \pi^i_I$ cancel in $\delta S[A^I_k, \pi_I^k, A_0^I]$.  One gets the symmetry of the second order action by expressing the auxiliary fields $\pi^i_I$ that are eliminated in terms of the retained variables.}. The transformations that do not mix the electric and magnetic potentials are called ``electric symmetries".
 
Thus, an electric symmetry transformation is characterized by the property that the matrix $g$ is lower-triangular,
\begin{equation} \label{eq:Ge0}
g = \begin{pmatrix} M & 0 \\ BM & M^{-T} \end{pmatrix} , \quad M \in GL(n_v,\R), \quad B^T = B.
\end{equation}
An electric symmetry is therefore a linear transformation $\bar{A}^I_\mu = M\indices{^I_J} A^J_\mu$ of the electric potentials for which there is a symmetric matrix $B$ and an isometry $\phi \rightarrow \bar{\phi}$ of the scalar manifold such that
\begin{equation} \label{eq:Ge}
\mathcal{M}(\bar{\phi}) = g^{-T} \mathcal{M}(\phi) g^{-1}
\end{equation}
with $g$ given by \eqref{eq:Ge0}. The transformations with $B \not=0$ involve transformations of Peccei-Quinn type (axion shift symmetry) \cite{Peccei:1977hh,Peccei:1977ur}.

The electric group $G_e$ in a given frame depends of course on the chosen duality frame. Indeed, going to another duality frame with the symplectic matrix $E$ as in \eqref{eq:Aprime} will replace the matrices $g$ by their conjugates $g' = E^{-1} g E$, and these might not be lower-triangular.
The condition that $g'$ has the lower-triangular form \eqref{eq:Ge0} therefore depends on the choice of $E$.

\section{\texorpdfstring{$4m$}{4m} versus \texorpdfstring{$4m + 2$}{4m+2} space-time dimensions}
\label{sec:4m4m2}

For a $p$-form in $D= 2p + 2$ spacetime dimensions, the field strength and its dual are forms of same rank $p+1=D/2$ and one may consider electric-magnetic duality transformations that mix them while leaving the theory invariant. It turns out that the cases $p$ odd ($p= 2m - 1$, $D= 4m$) and $p$ even ($p=2m$, $D=4m +2$) lead to different duality groups. The electric-magnetic duality group is a subgroup of $Sp(2n, \mathbb{R})$ in $4m$ spacetime dimensions and $O(n, n)$ in $4m+2$ spacetime dimensions, where $n$ is the number of such $p$-forms. These results were established in \cite{Deser:1997mz,Deser:1997se,Julia:1997cy} and already anticipated in \cite{Julia:1980gr} (see also \cite{Cremmer:1997ct,Bremer:1997qb,Deser:1998vc,Julia:2005wg} for independent developments).

As before, the easiest way to see this is to solve the constraint that appears in the Hamiltonian formalism, following \cite{Bunster:2011qp}. The constraint is in this case
\begin{equation}
\pd_{i_1} \pi_I^{i_1 i_2 \dots i_p} = 0,
\end{equation}
where $\pi_I^{i_1 i_2 \dots i_p}$ is the conjugate momentum to $A^I_{i_1 i_2 \dots i_p}$ and the index $I$ runs from $1$ to $n$. It is solved in $2p+1$ spatial dimensions in terms of another $p$-form potential $Z_I$ as
\begin{equation}
\pi_I^{i_1 i_2 \dots i_p} = \cB^{i_1 i_2 \dots i_p}[Z_I] = \frac{1}{p!} \varepsilon^{i_1 \dots i_p j k_1 \dots k_p} \pd_j Z_{I\, k_1 \dots k_p},
\end{equation}
where we introduced the definition of the magnetic field $\cB$.
Then, the kinetic term in the Hamiltonian action is (up to integration by parts)
\begin{equation}\label{eq:pformkinetic}
\pi_I^{i_1 i_2 \dots i_p} \dot{A}^I_{i_1 i_2 \dots i_p} = \frac{1}{2} \rho_{MN} \cB^{M \, i_1 i_2 \dots i_p} \dot{\cA}^N_{i_1 i_2 \dots i_p},
\end{equation}
where the vector $\cA^M$ is $(\cA^M) = (A^I, Z_I)$ and the $2n \times 2n$ matrix $\rho_{MN}$ is
\begin{equation}
\rho = \begin{pmatrix}
0 & I \\ (-1)^p I & 0
\end{pmatrix}.
\end{equation}
This matrix is symmetric when $p$ is even (with signature $(n,n)$), and antisymmetric when $p$ is odd. Those cases correspond to dimensions $4m+2$ or $4m$ respectively. In doing so for the whole Hamiltonian action, one gets the actions first written in \cite{Schwarz:1993vs}.

Now, the requirement that a linear transformation $\cA \rightarrow g \cA$ of the potentials is an invariance of the kinetic term \eqref{eq:pformkinetic} constrains the matrix $g \in GL(2n, \R)$ to satisfy
\begin{equation}
g^T \rho \, g = \rho .
\end{equation}
In other terms, the matrix $g$ must belong to $O(n,n)$ in dimensions $4m + 2$, but to $Sp(2n,\R)$ in dimensions $4m$, as announced. (The invariance of the Hamiltonian itself may impose further restrictions on $g$, analog to \eqref{eq:duality2}, that depend on the precise theory at hand.)

Let us finally mention that in dimension $4m + 2$, the diagonalization of $\rho$ amounts to the split of the $p$-forms into chiral (self-dual) and anti-chiral (anti-self-dual) parts \cite{Deser:1997se,Bekaert:1998yp}, recovering the actions for (anti-)chiral forms of \cite{Henneaux:1988gg}. This will be reviewed in part \ref{PART:6D}.
\chapter{\texorpdfstring{$E_{7(7)}$}{E7(7)}-invariant black hole entropy}
\label{chap:blackholes}

The most general non-extremal black hole solution of four-dimensional $\cN=8$ supergravity has been found recently by Chow and Compère in \cite{Chow:2013tia,Chow:2014cca}. In this chapter, we write a manifestly $E_{7(7)}$-invariant formula for the entropy of those black holes (i.e., the area of the event horizon divided by $4G$, according to Bekenstein and Hawking's celebrated work).

This chapter is based on the paper \cite{Compere:2015roa}, written in collaboration with G. Compère. I would also like to thank G. Sárosi for interesting discussions, both before and after the publication of \cite{Compere:2015roa}.

\section{The \texorpdfstring{$F$}{F}-invariant}

Finding a $E_{7(7)}$-invariant formula for the entropy is equivalent to finding such a formula for the so-called ``$F$-invariant", as we review now.

The construction of the black hole solutions of \cite{Chow:2013tia,Chow:2014cca} was done in two steps, which rely both on the existence of hidden symmetries.
\begin{enumerate}
\item First, they considered the $STU$ truncation of maximal supergravity, with duality group $\SL^3 \subset E_{7(7)}$. Once the most general black hole of the $STU$ model is constructed, one then gets the most general black hole of $\cN = 8$ supergravity by acting on that solution with the whole of $E_{7(7)}$ \cite{Sen:1994eb,Cvetic:1996zq}.
\item To construct the most general black hole of the $STU$ model, they reduced it to three dimensions, where there are only scalar fields parametrizing the coset $SO(4,4)/SL(2,\R)^4$. Acting with all $SO(4,4)$ symmetries on a well-chosen ``seed" solution and uplifting back to four dimensions yields the result.
\end{enumerate}
The entropy  of the $STU$ model black holes is then found to take the remarkably simple form
\begin{equation}\label{eq:entropy}
S= 2\pi \left( \sqrt{\Delta + F} + \sqrt{- J^2 + F} \right) .
\end{equation}
In this formula, $\Delta$ is Cayley's hyperdeterminant \cite{Duff:2006uz}, which is a $\SL^3$-invariant quartic function of the electromagnetic charges $Q_I$ and $P^I$ only, and $J$ is the angular momentum of the black hole. The quantity
\begin{equation}
F = F(M,Q_I,P^I,z_i^\infty)
\end{equation}
depends on the mass $M$ of the black hole, the electromagnetic charges, and the values $z_i^\infty$ of the scalar fields at spatial infinity. Since also $J$ and $S$ are invariant under the action of the duality group $\SL^3$ (since the metric does not transform), $F$ must also be $\SL^3$ invariant. This $F$-invariant was written in terms of auxiliary parameters (the parameters of the $SO(4,4)$ generating transformation) in \cite{Chow:2013tia,Chow:2014cca}, but a manifestly duality-invariant expression was missing in the case of non-extremal black holes.

In general, the $F$-invariant is not a polynomial function of $M$, $Q_I$, $P^I$ and $z_i^\infty$, as illustrated by explicit subcases \cite{Chow:2014cca}. In the paper \cite{Sarosi:2015nja}, Sárosi was nonetheless able to conveniently write the $F$-invariant as a polynomial expression in terms of $\SL^3$ invariants of the $STU$ model. His strategy was to use the scalar charges which appear as the first subleading term in the asymptotic expansion of
the scalar fields at radial infinity,
\begin{equation}
z_i = z_i^{\infty} + \frac{q_i}{r} + \mathcal{O}(r^{-2}) .
\end{equation}
These auxiliary variables absorb most of the algebraic complexity. It is in principle possible to write the scalar charges in terms of the physical quantities $M$, $Q_I$, $P^I$ and $z_i^\infty$, but this process is algebraically complicated. For the example of the Kaluza-Klein black hole, an explicit formula in terms of the physical charges could be obtained but involved trigonometric functions \cite{Chow:2014cca}. For the four-charge Cveti{c}-Youm black hole \cite{Cvetic:1996kv}, it has been stated  in \cite{Sarosi:2015nja} that one encounters a fifth order polynomial in the inversion algorithm, which therefore cannot be solved by radicals.

In section \ref{sec:STUinv}, we review the results of \cite{Sarosi:2015nja} and also present them in a direct $\SL^3$ fashion, whereas \cite{Sarosi:2015nja} used a $SL(6,\R)$ embedding of the duality group. In section \ref{sec:E7inv}, we show how those $STU$ model invariants extend to $E_{7(7)}$-invariants of the whole $\cN = 8$ supergravity theory\footnote{A formula in the symplectic formalism of $\cN = 2$ supergravities was also written in \cite{Compere:2015roa}, but we will not present it here.}.

\section{The \texorpdfstring{$STU$}{STU} model}

This section provides a short review of the $STU$ model and its embedding in maximal supergravity. The $STU$ model in four dimensions is a model of $\cN = 2$ supergravity coupled to three vector multiplets \cite{Duff:1995sm}. The bosonic field content is therefore given by one metric, four vector fields (three from the vector multiplets and one from the gravity multiplet) and six scalar fields (two from each of the vector multiplets). We follow the conventions of \cite{Chow:2014cca}.

\subsection{Lagrangian and $\SL^3$ symmetries}

We write the fields coming from the vector multiplets with an index $i$, ($i = 1, 2, 3$). They are the six scalar fields $\varphi_i$, $\chi_i$, parametrizing three copies of the $\SL/SO(2)$ coset space, and three vector fields $\tilde{A}_i$. The vector field of the gravity multiplet is written $A^4$.

The Lagrangian is
\begin{align}
\cL_\text{STU} = R \star \1 &- \frac{1}{2}\sum_{i=1}^3 \left( \star d\varphi_i \wedge d\varphi_i + e^{2 \varphi_i} \star d\chi_i \wedge d \chi_i \right) \label{eq:STUlag}\\
&- \frac{1}{2} e^{-\varphi_1 -\varphi_2-\varphi_3} \star F^4 \wedge F^4 - \frac{1}{2} \sum_{i=1}^3 e^{2\varphi_i-\varphi_1-\varphi_2-\varphi_3} \star \tilde{\cF}_i \wedge \tilde{\cF}_i \nn \\
& - \left(\chi_1\chi_2 \,\tilde{F}_3 +\chi_2\chi_3 \,\tilde{F}_1 +\chi_3\chi_1\, \tilde{F}_2 \right) \wedge F^4 \nn \\
&+ \chi_1\, \tilde{F}_2 \wedge \tilde{F}_3 + \chi_2\, \tilde{F}_3 \wedge \tilde{F}_1 + \chi_3 \,\tilde{F}_1 \wedge \tilde{F}_2 + \chi_1 \chi_2 \chi_3 F^4 \wedge F^4, \nn
\end{align}
where the field strengths are
\begin{align}
\tilde{F}_i = d\tilde{A}_i, \qquad \tilde{\cF}_i = d\tilde{A}_i - \chi_i dA^4, \qquad F^4 = dA^4 .
\end{align}
This Lagrangian is invariant under permutations of the $1$, $2$, $3$ indices.
The $SL(2,\R)^3$ symmetry of the scalar sector extends to the full theory: the four vector fields and their duals transform in the tensor product of the defining representations of the three $\SL$ factors, as we will see below.

The equations of motion and Bianchi identities for the vector fields $\tilde{A}_i$ can be written as
\begin{equation}
dF^i = 0, \quad d \tilde{F}_i = 0,
\end{equation}
where
\begin{align}
F^1 &= e^{\varphi_1-\varphi_2-\varphi_3} \star \tilde{\cF}_1 + \chi_2\chi_3 F^4 - \chi_2 \tilde{F}_3 - \chi_3 \tilde{F}_2, \nn\\
F^2 &= e^{\varphi_2-\varphi_3-\varphi_1} \star \tilde{\cF}_2 + \chi_3\chi_1 F^4 - \chi_3 \tilde{F}_1 - \chi_1 \tilde{F}_3, \label{eq:STUFi}\\
F^3 &= e^{\varphi_3-\varphi_1-\varphi_2} \star \tilde{\cF}_3 + \chi_1\chi_2 F^4 - \chi_1 \tilde{F}_2 - \chi_2 \tilde{F}_1 \nn
\end{align}
($F^2$ and $F^3$ are obtained from $F^1$ by cyclic permutation of the indices).
For $A^4$, they are
\begin{equation}\label{eq:STUeom4}
dF^4 = 0,\quad d\tilde{F}_4 = 0,
\end{equation}
where
\begin{align}
\tilde{F}_4 = &- e^{-\varphi_1-\varphi_2-\varphi_3} \star F^4 + \sum_{i=1}^3 e^{2\varphi_i-\varphi_1-\varphi_2-\varphi_3} \chi_i \star \tilde{\cF}_i \label{eq:STUtF4}\\
&- \chi_1\chi_2 \,\tilde{F}_3 -\chi_2\chi_3 \,\tilde{F}_1 - \chi_3\chi_1\, \tilde{F}_2 + 2 \chi_1\chi_2\chi_3 F^4 . \nn
\end{align}
As usual, the equations of motion imply the existence of dual vector fields $A^i$, $\tilde{A}_4$, defined by
\begin{equation}
F^i = dA^i, \qquad \tilde{F}_4 = d\tilde{A}_4 .
\end{equation}
A convenient way to package the eight vector fields is to define the $1$-form-valued tensor $\cA_{aa'a''}$, where the indices can take the two values $0$ or $1$, by
\begin{align}
(\cA_{000}, \cA_{111}) & = (\tilde{A}_4,-A^4) , & (\cA_{100}, \cA_{011}) & = (A^1,-\tilde{A}_1) , \nn \\
(\cA_{010}, \cA_{101}) & = (A^2,-\tilde{A}_2) , & (\cA_{001}, \cA_{110}) & = (A^3,-\tilde{A}_3) . \label{eq:STUvectors}
\end{align}
The equations of motion and Bianchi identities are then simply $d\cA_{aa'a''} = 0$. The twisted self-duality equations \eqref{eq:STUFi}, \eqref{eq:STUtF4} are equivalent to
\begin{equation}\label{eq:STUtsd}
\star \,d\cA_{aa'a''} = - (\epsilon \wt{\cM}_1)\indices{_a^b} \,(\epsilon \wt{\cM}_2)\indices{_{a'}^{b'}} \,(\epsilon \wt{\cM}_3)\indices{_{a''}^{b''}} \,d\cA_{bb'b''},
\end{equation}
where the matrices $\wt{\cM}_i$ are defined from the usual coset matrices $\cM_i$ as
\begin{equation}\label{eq:tildeM}
\wt{\cM}_i = h \cM_i h =
\begin{pmatrix} e^{\varphi_i} & - \chi_i e^{\varphi_i} \\ - \chi_i e^{\varphi_i} & e^{-\varphi_i} + \chi_i^2 e^{\varphi_i} \end{pmatrix}.
\end{equation}
They differ from $\cM_i$ by a sign flip of the axion $\chi_i$. Note that we write the indices corresponding to the first, second and third $\SL$ factor with zero, one and two primes respectively. Under a $\SL^3$ transformation, these objects transform as
\begin{equation}\label{eq:cosetSTUtransf}
\wt{\cM}_i \rightarrow (S_i)^{-T} \wt{\cM}_i (S_i)^{-1}\, ,
\end{equation}
and
\begin{equation}
\cA_{aa'a''} \mapsto (S_1)\indices{_a^b}(S_2)\indices{_{a'}^{b'}}(S_3)\indices{_{a''}^{b''}}\cA_{bb'b''}\, ,
\end{equation}
where the $S_i$ are elements of the $i$-th $\SL$. These transformations preserve the twisted self-duality equation \eqref{eq:STUtsd}. 

The transformation law of $\cA_{aa'a''}$ can be rephrased as follows. Under the first $SL(2,\R)$, the following $2$-component vectors transform as doublets $v \rightarrow S_1 v$,
\begin{equation}
\SL_1:\quad (\tilde{A}_4, A^1), \quad  (A^3, -\tilde{A}_2), \quad (A^2, -\tilde{A}_3),\quad (-\tilde{A}_1, -A^4) .
\end{equation}
They are equal to $\cA_{a00}$, $\cA_{a01}$, $\cA_{a10}$ and $\cA_{a11}$ respectively.
Under the second and third $\SL$ factors, the same reasoning give the doublets
\begin{align}
&\SL_2:\quad &(\tilde{A}_4, A^2),\quad &(A^3, -\tilde{A}_1),\quad  &(A^1, -\tilde{A}_3),\quad &(-\tilde{A}_2, -A^4), \\
&\SL_3:\quad &(\tilde{A}_4, A^3),\quad &(A^2, -\tilde{A}_1),\quad  &(A^1, -\tilde{A}_2),\quad &(-\tilde{A}_3, -A^4)
\end{align}
which can also be obtained from the previous ones by cyclic permutations of $1$, $2$, $3$. From this, one sees that none of the $\SL$'s are local symmetries of the Lagrangian \eqref{eq:STUlag}. However, by dualizing one of the $\tilde{A}_i$ vector fields to $A^i$, one can get from \eqref{eq:STUlag} three Lagrangians in which two of the $\SL$ factors act as local symmetries, while the third one is non-local. For example, writing the Lagrangian with $(A^1, \tilde{A}_2, \tilde{A}_3, A^4)$, the second and third $\SL$ factors are local symmetries and the first is non-local. The advantage of \eqref{eq:STUlag} is that it treats the first three vector fields on an equal footing.

As explained in chapter \ref{chap:emduality}, one can also write another, first-order Lagrangian in which all the $SL(2,\R)^3$ symmetries act in a local way, but we will not use it in this chapter.

\subsection{Higher-dimensional origin}

The higher-dimensional origin of the $STU$ model is reviewed in \cite{Chow:2014cca}. We reproduce here the relevant formulas.

\subsubsection*{Five dimensions}

The Lagrangian \eqref{eq:STUlag} can be obtained from the reduction of the following five-dimensional theory:
\begin{equation}\label{eq:STUlag5}
\cL\up{5}_\text{STU} = R\star \1 - \frac{1}{2} \sum_{i=1}^3 h_i^{-2}\left( \star dh_i \wedge d h_i + \star \hat{F}_i \wedge \hat{F}_i \right) + \hat{F}_1 \wedge \hat{F}_2 \wedge \hat{A}_3 .
\end{equation}
It contains three vector fields $\hat{A}_i$ with field strengths $\hat{F}_i = d\hat{A}_i$, and scalar fields $h_1$, $h_2$, $h_3$ subject to the constraint $h_1 h_2 h_3 = 1$. Since one can permute the indices in the Chern-Simons term up to integration by parts, this Lagrangain is invariant under any permutation of the $1$, $2$, $3$ indices.

The reduction ansatz for the metric and the vector fields is
\begin{equation}\label{eq:STUansatzfive}
ds^2_5 = f^{-1} ds_4^2 + f^2(dz - A^4)^2, \qquad \hat{A}_i = \tilde{A}_i + \chi_i (dz - A^4) ,
\end{equation}
and the four-dimensional scalar fields $\varphi_i$ are given by $e^{-\varphi_i} = f h_i$. A convenient parametrization of the $h_i$ in terms of two real scalars $\varphi'_1$, $\varphi'_2$ is
\begin{equation}
h_1 = e^{-\varphi'_1/\sqrt{6} - \varphi'_2/\sqrt{2}}, \quad h_2 = e^{-\varphi'_1/\sqrt{6} + \varphi'_2/\sqrt{2}}, \quad h_3 = e^{2\varphi'_1/\sqrt{6}}.
\end{equation}
Writing $f = e^{-\phi/\sqrt{3}}$ as in \eqref{eq:KKansatz}, this gives for the four-dimensional dilatons
\begin{equation}
\varphi_1 = \frac{\phi}{\sqrt{3}} + \frac{\varphi_1'}{\sqrt{6}} + \frac{\varphi_2'}{\sqrt{2}}, \quad \varphi_2 = \frac{\phi}{\sqrt{3}} + \frac{\varphi_1'}{\sqrt{6}} - \frac{\varphi_2'}{\sqrt{2}}, \quad \varphi_3 = \frac{\phi}{\sqrt{3}} - \frac{2\varphi_1'}{\sqrt{6}} .
\end{equation}
These relations can be inverted to 
get the scalar factors $f$, $h_i$ explicitly in terms of the four dimensional dilatons as
\begin{align}
f &= e^{-\frac{1}{3}\left( \varphi_1 + \varphi_2 + \varphi_3 \right)} ,\\
h_1 &= e^{\frac{1}{3}\left( - 2 \varphi_1 + \varphi_2 + \varphi_3 \right)}, \quad h_2 = e^{\frac{1}{3}\left( \varphi_1 - 2 \varphi_2 + \varphi_3 \right)}, \quad h_3 = e^{\frac{1}{3}\left( \varphi_1 + \varphi_2 - 2 \varphi_3 \right)} .
\end{align}

\subsubsection*{Eleven dimensions}

The Lagrangian \eqref{eq:STUlag5} can be obtained from the Lagrangian \eqref{eq:11lag} of eleven-dimensional supergravity by reducing on a six-torus $T^6$ and taking the fields as
\begin{align}
ds^2_{11} &= ds^2_5 + h_1 (dX_1^2 + dX_2^2) + h_2 (dX_3^2 + dX_4^2) + h_3 (dX_5^2 + dX_6^2), \label{eq:STU11metric}\\
A\up{3} &= \hat{A}_1 \wedge dX_1 \wedge dX_2 + \hat{A}_2 \wedge dX_3 \wedge dX_4 + \hat{A}_3 \wedge dX_5 \wedge dX_6. \label{eq:STU113form}
\end{align}
The constraint $h_1 h_2 h_3 = 1$ is such that the internal $T^6$ has constant volume. The permutation symmetry of \eqref{eq:STUlag5} in $1$, $2$, $3$ comes from permutations of the $T^2$ factors in $T^6 = T^2\times T^2 \times T^2$.

Putting this together with the ansatz \eqref{eq:STUansatzfive} for going from five to four dimensions gives the expression of the eleven-dimensional fields in the $STU$ truncation.

\subsection{Embedding in $D=4$, $\cN = 8$ supergravity}

The eleven-dimensional origin of the Lagrangian \eqref{eq:STUlag} naturally gives an embedding of the $STU$ model in $D=4$ maximal supergravity. The fields of $D=4$, $\cN = 8$ supergravity are defined by the eleven-dimensional fields through equations \eqref{eq:ansatzmetric} and \eqref{eq:ansatz3form} of appendix \ref{app:N8E7}; the $STU$ fields are then identified by comparing those with \eqref{eq:STU11metric} and \eqref{eq:STU113form} respectively. With respect to \ref{app:N8E7}, the coordinates on the internal space are $dz = dz^7$ and $dX^i = dz^i$ for $i=1, \dots, 6$.

\subsubsection*{Metric ansatz}

For the vector field $A^4$, we identify directly
\begin{equation}\label{eq:STUA4}
\cA\dwn{1}^7 = - A^4
\end{equation}
since only $\omega^7$ is non-trivial. For the dilatons, we get the various relations
\begin{align}
f &= e^{-\frac{1}{3} \vec{g} \cdot \vec{\phi}} = e^{\vec{\gamma}_7 \cdot \vec{\phi}}, \\
h_1 &= e^{2\vec{\gamma}_1 \cdot \vec{\phi}} = e^{2\vec{\gamma}_2 \cdot \vec{\phi}}, \quad
h_2 = e^{2\vec{\gamma}_3 \cdot \vec{\phi}} = e^{2\vec{\gamma}_4 \cdot \vec{\phi}}, \quad
h_3 = e^{2\vec{\gamma}_5 \cdot \vec{\phi}} = e^{2\vec{\gamma}_6 \cdot \vec{\phi}}
\end{align}
which can be solved to express the seven $\cN=8$ dilatons in terms of the three $STU$ dilatons. One then obtains the relation
\begin{equation}
\vec{\phi} = \sum_{i=1}^3 \varphi_i \vec{v}_i\, ,
\end{equation}
where the $\vec{v}_i$ are the three orthonormal vectors
\begin{align}
\vec{v}_1 &= \left(\frac{1}{2},\frac{3}{2 \sqrt{7}},-\frac{1}{2\sqrt{21}},-\frac{1}{2 \sqrt{15}},-\frac{1}{2 \sqrt{10}},-\frac{1}{2 \sqrt{6}},\frac{1}{\sqrt{3}}\right), \nn\\
\vec{v}_2 &= \left(-\frac{1}{4},-\frac{3}{4 \sqrt{7}},\frac{2}{\sqrt{21}},\frac{2}{\sqrt{15}},-\frac{1}{2 \sqrt{10}},-\frac{1}{2 \sqrt{6}},\frac{1}{\sqrt{3}}\right), \nn \\
\vec{v}_3 &= \left(-\frac{1}{4},-\frac{3}{4 \sqrt{7}},-\frac{\sqrt{3}}{2 \sqrt{7}},-\frac{\sqrt{3}}{2 \sqrt{5}},\frac{1}{\sqrt{10}},\frac{1}{\sqrt{6}},\frac{1}{\sqrt{3}}\right).
\end{align}

\subsubsection*{Three-form ansatz}

In the $STU$ truncation, the eleven-dimensional three-form is
\begin{align}
A\up{3} = &\quad (\tilde{A}_1 - \chi_1 A^4) \wedge dX_1 \wedge dX_2 \\
&+ (\tilde{A}_2 - \chi_2 A^4) \wedge dX_3 \wedge dX_4 \nn\\
&+ (\tilde{A}_3 - \chi_3 A^4) \wedge dX_5 \wedge dX_6 \nn\\
&+ \chi_1 \,dX_1 \wedge dX_2 \wedge dz + \chi_2 \,dX_3 \wedge dX_4 \wedge dz + \chi_3 \,dX_5 \wedge dX_6 \wedge dz \nn.
\end{align}
This gives for the vector fields of $\cN = 8$ supergravity
\begin{equation}\label{eq:STUAi}
A_{(1)12} = - \tilde{A}_1 + \chi_1 A^4,\quad A_{(1)34} = - \tilde{A}_2 + \chi_2 A^4,\quad
A_{(1)56} = - \tilde{A}_3 + \chi_3 A^4
\end{equation}
and for the axions
\begin{equation}
A_{(0)127} = - \chi_1,\quad A_{(0)347} = - \chi_2,\quad A_{(0)567} = - \chi_3.
\end{equation}

\subsubsection*{Group embedding $\SL^3 \subset E_{7(7)}$}

We can identify the $\SL^3$ generators of the $STU$ model by comparing the scalar coset representatives. In the $STU$ truncation, the coset matrix $\cV$ of $\cN = 8$ supergravity (see \eqref{eq:N8coset}) reduces to
\begin{equation}\label{eq:STUcoset}
\cV=\exp\left[\frac{1}{2}\sum_{i=1}^3 \varphi_i \,\vec{v}_i \cdot \vec{H}\right]\exp\left[-\chi_1 E^{127}-\chi_2 E^{347} -\chi_3 E^{567}\right] .
\end{equation}
We deduce that the relevant $SL(2,\R)^3$ subalgebra of $E_{7(7)}$ is generated by
\begin{align}
h_1 &= \vec{v}_1 \cdot \vec{H},\quad e_1 = -E^{127},\quad f_1 = - (E^{127})^\#, \\
h_2 &= \vec{v}_2 \cdot \vec{H},\quad e_2 = -E^{347},\quad f_2 = - (E^{347})^\#, \\
h_3 &= \vec{v}_3 \cdot \vec{H},\quad e_3 = -E^{567},\quad f_3 = - (E^{567})^\#.
\end{align}
(Each $\SL_i$ factor is generated by $\{h_i, e_i, f_i\}$.) Indeed, using the $E_{7(7)}$ commutators provided in \cite{Cremmer:1997ct}, one can check that the $\SL^3$ commutation relations
\begin{equation}
[h_i, e_i] = 2 e_i, \quad [h_i, f_i] = - 2 f_i, \quad [e_i, f_i] = h_i
\end{equation}
hold. All unwritten commutators vanish; in particular, the different $\SL_i$ factors commute, as they should.

\section{\texorpdfstring{$STU$}{STU} black holes}
\label{sec:STUinv}

In order to define $\SL^3$ invariants that are relevant for describing the $F$-invariant of the $STU$ model, the first step is to write down covariant tensors in terms of the electromagnetic charges, scalar moduli and scalar charges which transform naturally under the duality group. We will do so in section \ref{covtSTU}. We then describe the construction of invariants in section \ref{consSTU}, compare them with those of \cite{Sarosi:2015nja} in section \ref{SL6STU} and conclude with the result for the $F$-invariant in section \ref{FinvSTU}.

\subsection{Covariant tensors}
\label{covtSTU}

According to \eqref{eq:STUvectors}, the electromagnetic charges $Q^I$, $P_I$ ($I = 1, 2, 3, 4$) are organized in the \emph{charge tensor} $\gamma_{aa'a''}$, with components\footnote{The charge tensor defined in (6.14) in \cite{Chow:2014cca} was incorrect since it does not transform covariantly under $SL(2,\mathbb R)^3$ transformations. With the present correction, no other formula of \cite{Chow:2014cca} is affected.}
\begin{align}
(\gamma_{000}, \gamma_{111}) & = (P^4,-Q_4) , & (\gamma_{100},\gamma_{011}) & = (Q_1,-P^1) , \nn \\
(\gamma_{010}, \gamma_{101}) & = (Q_2,-P^2) , & (\gamma_{001}, \gamma_{110}) & = (Q_3,-P^3) .
\end{align}
It transforms as
\begin{equation} \label{SL2charges}
\gamma_{aa'a''} \mapsto (S_1)\indices{_a^b}(S_2)\indices{_{a'}^{b'}}(S_3)\indices{_{a''}^{b''}}\gamma_{bb'b''}
\end{equation}
under $SL(2,\R)^3$.

From the scalar fields, we also have the three coset matrices $\wt{\cM}_i$ defined in \eqref{eq:STUcoset}. At spatial infinity, they admit the asymptotic expansion
\be \wt{\cM}_i = \wt{\cM}_i^{(0)} + \frac{\wt{\cM}_i^{(1)}}{r} + \mathcal{O}\left(\frac{1}{r^2}\right). \ee
Here $\wt{\cM}_i^{(0)}$ encodes the scalar moduli at infinity $\varphi_i^{\infty}$, $\chi_i^{\infty}$, while $\wt{\cM}_i^{(1)}$ encodes the scalar charges $\Sigma_i$, $\Xi_i$ defined as
\begin{align}
\varphi_i &= \varphi_i^{\infty} + \frac{\Sigma_i}{r} + \mathcal{O}(r^{-2}), & \chi_i &= \chi_i^{\infty} + \frac{\Xi_i}{r} + \mathcal{O}(r^{-2}).
\end{align}
We then define the \emph{dressed scalar charge tensor} as
\be R_i = (\wt{\cM}_i^{(0)})^{-1} \wt{\cM}_i^{(1)} .\label{ds}\ee
This tensor is invariant under $SL(2,\mathbb R)_j$ with $j \neq i$ and transforms under $SL(2,\mathbb R)_i$ as
\be R_i \mapsto S_i R_i S_i^{-1} .\ee

Finally, we also have the invariant tensor for each copy of $SL(2,\mathbb R)_i$
\be \varepsilon = \begin{pmatrix} 0 & 1 \\ -1 & 0 \end{pmatrix}, \ee
that satisfies
\be S^T \varepsilon S = S \varepsilon S^T = \varepsilon \ee
for any $S$ in $SL(2,\R)$.

Let us now summarize our ingredients and their transformation laws under $S_1 \otimes S_2 \otimes S_3 \in SL(2,\R)^3$, in index notation. To avoid notational clutter, we will write $\wt{\cM}_i^{(0)}=M_i$ from now on. We have the following objects:
\begin{itemize}
\item the charge tensor $\gamma_{aa'a''} \mapsto (S_1)\indices{_a^b}(S_2)\indices{_{a'}^{b'}}(S_3)\indices{_{a''}^{b''}}\gamma_{bb'b''}$;
\item the asymptotic coset tensors $(M_i)^{ab} \mapsto (M_i)^{cd} (S_i^{-1})\indices{_c^a}(S_i^{-1})\indices{_d^b}$;
\item the dressed scalar charge tensors $(R_i)\indices{_a^b} \mapsto (S_i)\indices{_a^c} (R_i)\indices{_c^d} (S_i^{-1})\indices{_d^b}$.
\end{itemize}
We can also use the invariant epsilon tensor $\varepsilon^{ab}$.

\subsection{Construction of invariants}
\label{consSTU}

We proceed in two steps.
\begin{enumerate}
\item First, we make $SL(2,\R)^3$ invariants by contracting all indices, with the constraint that only indices corresponding to the same $SL(2,\R)$ (i.e., with the same number of primes) can be contracted together.
\item Second, we implement invariance under permutations of the three $SL(2,\R)$ factors by summing the expression with all others obtained by permuting its different $SL(2,\mathbb R)$ internal indices. In general there are $6$ terms but symmetries might reduce them to $3$ or only $1$ term. For example, from the $SL(2,\R)^3$-invariant expression
\be \varepsilon^{ab}\varepsilon^{a'b'}M_3^{a''b''} \gamma_{aa'a''}\gamma_{bb'b''} \ee
we make
\be \left( M_1^{ab}\varepsilon^{a'b'}\varepsilon^{a''b''} + \varepsilon^{ab}M_2^{a'b'}\varepsilon^{a''b''} +\varepsilon^{ab}\varepsilon^{a'b'}M_3^{a''b''}  \right) \gamma_{aa'a''}\gamma_{bb'b''} . \ee
There are only three terms in this example because the permutations of two $\varepsilon^{ab}$ tensors give identical terms. 
\end{enumerate}

Let us also define a degree as follows: the mass, NUT charge (that we include for completeness), electromagnetic charges and scalar charges have degree 1 while the scalar moduli have degree 0. Therefore, $M_i^{ab}$ and $\varepsilon^{ab}$ have degree 0 while $\gamma_{aa'a''}$ and $(R_i)\indices{_a^b}$ have degree 1. Inspection reveals that the $F$-invariant is a homogeneous function of degree 4. Restricting to degree $\leq 4$, we find the following independent invariants:
\begin{itemize}
\item Degree 1:
\begin{align}
M, \, N.
\end{align}
\item Degree 2:
\begin{align}
L_1 &= M_1^{ab}M_2^{a'b'}M_3^{a''b''} \gamma_{aa'a''}\gamma_{bb'b''}, \\
L_2 &= \frac{1}{3}  \left( \Tr{R_1^2}+\Tr{R_2^2}+\Tr{R_3^2} \right) .
\end{align}
\item Degree 3:
\begin{align}
C_1 &= \frac{1}{3} \sum \varepsilon^{ac} R\indices{_{1c}^b} \varepsilon^{a'b'}\varepsilon^{a''b''} \gamma_{aa'a''}\gamma_{bb'b''} ,\\
C_2 &= \frac{1}{3} \sum M_1^{ac} R\indices{_{1c}^b} M_2^{a'b'}M_3^{a''b''} \gamma_{aa'a''}\gamma_{bb'b''} .
\end{align}
\item Degree 4:
\begin{align}
\Delta &= \frac{1}{32} \varepsilon^{ac} \varepsilon^{a'b'} \varepsilon^{a''b''} \varepsilon^{bd} \varepsilon^{c'd'} \varepsilon^{c''d''} \gamma_{aa'a''}\gamma_{bb'b''} \gamma_{cc'c''}\gamma_{dd'd''} ,\\
\Delta_2 &= \frac{1}{96} \sum M_1^{ac} \varepsilon^{a'b'} \varepsilon^{a''b''} M_1^{bd} \varepsilon^{c'd'} \varepsilon^{c''d''} \gamma_{aa'a''}\gamma_{bb'b''} \gamma_{cc'c''}\gamma_{dd'd''} ,\\
\Delta_3 &=\frac{1}{96}  \left( \Tr{R_1^4}+\Tr{R_2^4}+\Tr{R_3^4} \right).
\end{align}
\end{itemize}
Here, each sum is over the three cyclic permutations of the $SL(2,\mathbb R)$ indices. The familiar quartic invariant (Cayley's hyperdeterminant) is $\Delta$. Many more invariants can be formulated, but we will not classify them here: this list will be sufficient to express the $F$-invariant below. 

\subsection{Reformulation in a $SL(6,\mathbb R)$ embedding}
\label{SL6STU}

In the paper \cite{Sarosi:2015nja}, S\'arosi constructed duality invariants using an embedding of the duality group into $SL(6,\mathbb R)$. The $SL(2,\mathbb R)^3$ transformations are expressed as $S \in SL(6,\mathbb R)$ with 
\be
S = \begin{pmatrix} S_1&&\\ & S_2& \\&& S_3 \end{pmatrix},
\ee
while the 6 permutations of the $\SL$ factors are represented by block permutation matrices. 

One starts from the pair of tensors
\be
(\psi_1)_{aa'a''} = -\gamma_{aa'a''},\qquad (\psi_2)_{aa'a''} = \widetilde \gamma_{aa'a''},
\ee
where $\widetilde \gamma_{aa'a''}$ is obtained from $\gamma_{aa'a''}$ by electromagnetic duality,
\be\label{tQP}
\widetilde Q^I  =  P^I ,\qquad \widetilde P_I  = - Q_I.
\ee
We denote them as $(\psi_\alpha)_{aa'a''} $ with $\alpha=1,2$.
One can then construct an antisymmetric $SL(6,\mathbb R)$-covariant tensor $(P_{\psi_\alpha})_{ABC}$ ($A,B,C=1,\dots, 6$) according to
\be (P_{\psi_\alpha})_{a+1, a'+3, a''+5} = (\psi_\alpha)_{a a' a''} \qquad (a,a',a''=0,1).  \ee
The other components of $P_{\psi_\alpha}$ are either obtained from those by antisymmetry or are zero.
It transforms as
\be
(P_{\psi_\alpha})_{ABC} \mapsto S_A^{\;\; A'}S_B^{\;\; B'}S_C^{\;\; C'} (P_{\psi_\alpha})_{A'B'C'}.
\ee
Finally, one builds the four $SL(6,\mathbb R)$-covariant tensors 
\be
(K_{\alpha\beta})^A_{\; B} = \frac{1}{2!3!} \epsilon^{A C_1C_2C_3C_4C_5} (P_{\psi_\alpha})_{B C_1C_2}(P_{\psi_\beta})_{C_3C_4C_5}
\ee
which transform as
\be
K_{\alpha\beta} \mapsto (S^{-1})^T K_{\alpha\beta} S^T.
\ee
One can also construct a block-diagonal matrix $R$ whose blocks are $R^T_i$, where $R_i$ ($i=1,2,3$) is defined in \eqref{ds}. It transforms as 
\be
R \mapsto (S^{-1})^T R S^T. 
\ee

Restricting to degree $\leq 4$, the independent invariants constructed in \cite{Sarosi:2015nja} from these objects are related to those defined in the previous section as follows:
\begin{itemize}
\item Degree 1:
\begin{align}
M,\, N.
\end{align}
\item Degree 2:
\begin{align}
\Tr(K_{12}) &= -3 L_1 ,\\
\Tr(R^2) &= 3 L_2.
\end{align}
\item Degree 3:
\begin{align}
\Tr(K_{11}R) &= 3 C_1 ,\\
\Tr(K_{12}R) &= - 3 C_2.
\end{align}
\item Degree 4:
\begin{align}
\Tr(K_{11}^2) &= - 96 \Delta, \\
\Tr(K_{11}K_{22}) &= - 96 \Delta_2, \\
\Tr(R^4) &= 96 \Delta_3.\label{TrR4}
\end{align}
\end{itemize}

\subsection{The $F$-invariant}
\label{FinvSTU}

The area over $4G$ of the outer and inner horizons of the general non-extremal black hole of the $STU$ model, with NUT charge included, takes the form 
\be
S_{\pm} =2\pi \left(  \sqrt{F+\Delta} \pm \sqrt{F - J^2}  \right)
\ee
where the $F$-invariant is given in terms of the auxiliary parameters $m,n,\nu_1,\nu_2$ defined in \cite{Chow:2014cca} as
\be
F = \frac{(m^2+n^2)}{G^2}(-n \nu_1 + m \nu_2)^2.\label{iF}
\ee
In  \cite{Sarosi:2015nja}, this formula was matched after extensive and convincing numerical tests to the following formula written in terms of triality invariants built from the  $SL(6,\mathbb R)$ embedding of the triality group,
\begin{align}
\label{eq:Finvsarosi}
F &= M^4+M^2 N^2+\frac{ M^2}{12}\Tr  K_{12}-\frac{M }{24}\Tr ( K_{12}R)+\frac{N^2}{24}\Tr(R^2)\nn \\ 
&-\frac{N}{24}\Tr(K_{11} R)+\frac{1}{192}\left(\Tr(K_{11}^2)-\Tr(K_{11}K_{22})-\frac{1}{2}(\Tr ²R^2)^2+\Tr(R^4)\right).
\end{align}
We checked independently that formulae \eqref{iF} and \eqref{eq:Finvsarosi} numerically agree for around a hundred random values of the parameters.

Using this and the dictionary above, we obtain the formula for the $F$-invariant in terms of triality invariants which are functions of the mass, the NUT charge, the electromagnetic charges, the moduli and the scalar charges:
\be\label{Finv1}
F = M^4 +M^2 N^2-\frac{M^2}{4} L_1 +\frac{N^2}{8}L_2 + \frac{M}{8}C_2 -\frac{N}{8}C_1+ \frac{- \Delta +\Delta_2 +\Delta_3}{2}  -\frac{3}{128}(L_2)^2.
\ee
One can obviously set the NUT charge to zero, $N=0$, to find physical configurations. 

An important comment is that formula \eqref{iF} is valid for trivial scalar asymptotics, $\varphi_i^{\infty} = 0 =\chi_i^{\infty}$, which corresponds to the special forms
\begin{align}
M_i &= \begin{pmatrix} 1 & 0 \\ 0 & 1 \end{pmatrix}, & R_i &= \begin{pmatrix} \Sigma_i & -\Xi_i \\ -\Xi_i & -\Sigma_i \end{pmatrix}\label{trivscal}
\end{align}
of the scalar matrices which we used to compare \eqref{Finv1} with \eqref{iF}. Non-trivial scalar moduli at infinity are generated by acting with the $SL(2,\R)^3$ transformations\footnote{In the $\mathfrak{so}(4,4)$ formalism of \cite{Chow:2014cca}, this corresponds to acting on the three-dimensional coset matrix of the solution with trivial moduli with
\[ g_\text{mod} = \exp\left(\frac{1}{2}\varphi_1^\infty H_1 + \frac{1}{2}\varphi_2^\infty H_2 + \frac{1}{2}\varphi_3^\infty H_3 \right) \exp\left( -\chi_1^\infty E_1 - \chi_2^\infty E_2 - \chi_3^\infty E_3 \right), \]
where $H_i$, $E_i$ are the $\mathfrak{so}(4,4)$ generators defined in \cite{Chow:2014cca}.}
\be S_i = \begin{pmatrix} e^{-\varphi_i^\infty/2} & \chi_i^\infty e^{\varphi_i^\infty/2} \\ 0 & e^{\varphi_i^\infty/2} \end{pmatrix}, \ee
which act on the scalars as
\be \varphi_i \mapsto \varphi_i + \varphi_i^\infty, \qquad \chi_i \mapsto \chi_i e^{-\varphi_i^\infty} + \chi_i^\infty .\ee
This transformation leaves formula \eqref{Finv1} invariant; it is therefore also valid for any value of the scalars at infinity.

\section{\texorpdfstring{$\cN = 8$}{N = 8} black holes}
\label{sec:E7inv}

As was already mentioned, the generic non-extremal black hole of the $STU$ model can be used as a seed for the generic non-extremal black hole of $\mathcal N = 8$ supergravity \cite{Sen:1994eb,Cvetic:1996zq,Chow:2013tia,Chow:2014cca}. We derived the black hole entropy for the $STU$ model in terms of duality invariants in the previous section. We will now reformulate the black hole entropy in terms of $E_{7(7)}$ invariants by matching individual invariants in $\mathcal N = 8$ supergravity with their corresponding invariants in the $STU$ truncation. This will give the correct formula for the most general non-extremal black hole of $\mathcal N = 8$ supergravity. In the static extremal case, it reduces to the well known formula of \cite{Kallosh:1996uy} in terms of the $E_{7(7)}$ quartic invariant.

\subsection{Symplectic and quartic invariants}

The $56$ electromagnetic charges of $\cN=8$ supergravity transform in the fundamental representation of $E_{7(7)}$ reviewed in appendix \ref{app:E77properties}. Accordingly, we can pack them into a $56$-component vector $X$ that transforms (infinitesimally) as
\be \delta X = g X \ee
under $g \in \mathfrak{e}_{7(7)}$. The vector $X$ is equivalently written as a pair of antisymmetric $SL(8,\R)$ tensors, $X = (X_{ij},X^{ij})$.
According to appendix \ref{app:E77properties}, we can directly identify two invariants:
\begin{itemize}
\item the symplectic invariant, which is a quadratic form over two distinct fundamental representations
\be \label{eq:symplecticE7} X^T \Omega Y = X^{ij}Y_{ij} - X_{ij}Y^{ij}, \ee
\item the quartic invariant $\cI_4(X)$ defined in \eqref{eq:I4E7}.
\end{itemize}
We could also switch from a $SL(8,\R)$-covariant basis to a $SU(8)$-covariant basis using formula \eqref{eq:E7SL8SU8}.  Since the $E_{7(7)}$ transformation laws in those two bases are formally identical (cf. equations \eqref{eq:E7transf} and \eqref{eq:E7transfSU8}), we can also construct invariants in the $SU(8)$ basis by replacing $i,j,\dots$ indices by $A,B,\dots$ indices in the previous invariants. Therefore, the two quantities
\begin{align}
\diamondsuit(X) &= \bar{X}^{AB}X_{BC}\bar{X}^{CD}X_{DA} -\frac{1}{4}(\bar{X}^{AB}X_{AB})^2 \nn \\
&\quad +\frac{1}{96} \varepsilon^{ABCDEFGH} X_{AB}X_{CD}X_{EF}X_{GH} +\frac{1}{96} \varepsilon_{ABCDEFGH} \bar{X}^{AB}\bar{X}^{CD}\bar{X}^{EF}\bar{X}^{GH}, \\
(X,Y)_\Omega &= \bar{X}^{AB}Y_{AB} - X_{AB} \bar{Y}^{AB} 
\end{align}
are $E_{7(7)}$-invariant. However, the invariants constructed in the two bases are proportional to each other: we have the relations $\diamondsuit(X)=-\mathcal{I}_4(X)$ \cite{Gunaydin:2000xr} and $X^T \Omega Y  = -i (X,Y)_\Omega$, as we checked using the explicit change of basis of appendix \ref{app:E77properties}.

\subsection{Construction of additional invariants}

The $70$ scalar fields of $\cN = 8$ supergravity parametrize the coset $E_{7(7)}/SU(8)$ and are therefore contained in a matrix $\cV$ that transforms as
\be \delta \cV = k \cV - \cV g \ee under $(k,g) \in \mathfrak{su}(8)_{\text{local}} \times \mathfrak{e}_{7(7)\,\text{global}}$. The local transformation $k$ is the compensator required to keep $\cV$ in the Borel gauge. Under the action of the group, these transformations are
\begin{align}
X &\mapsto G X, \\
\cV &\mapsto K\cV G^{-1}.
\end{align}
where $K \in SU(8)$, $G \in E_{7(7)}$. In particular, the object $\cV X$ only transforms under $SU(8)$.
From $\cV$, we define the usual matrix\footnote{We use the faithful $56 \times 56$ matrix representation of $E_{7(7)}$ presented in appendix \ref{app:E77properties}, in which the generalised transpose and usual matrix transpose coincide.}
\be \label{eq:E7M} \cM = \cV^T \cV, \ee which transforms as
\be \cM \mapsto (G^{-1})^T \cM G^{-1}. \ee
Again, from the asymptotic expansion
\be \cM = \cM^{(0)} + \frac{\cM^{(1)}}{r} + \mathcal{O}\left(\frac{1}{r^2}\right), \ee
we define the dressed charge matrix
\be \cR = (\cM^{(0)})^{-1} \cM^{(1)} \ee
which transforms in the adjoint representation of $E_{7(7)}$,
\be \cR \mapsto G \cR G^{-1} .\ee
In building invariants, we will also make use of $\Omega$, which has the property
\be G^T \Omega G = \Omega \ee
for any $G$ in $E_{7(7)}\subset Sp(56,\R)$.

We can now construct several additional invariants.
\begin{itemize}
\item Since both $X$ and $\cR X$ transform in the fundamental, we can make the following invariants of order two and three:
\be X^T \cM^{(0)} X, \quad X^T \cM^{(0)} \cR X, \quad X^T \Omega \cR X .\ee
\item As noted above, $\cV X$ transforms only under $SU(8)$. Switching to the $SU(8)$ notation (with indices $A,B,\dots$) using equation \eqref{eq:E7SL8SU8}, we can make invariants simply by contracting indices, e.g.,
\begin{align}
T_2 &= (\cV X)_{AB} \overline{(\cV X)}^{BA} ,\\
T_4 &= (\cV X)_{AB} \overline{(\cV X)}^{BC} (\cV X)_{CD} \overline{(\cV X)}^{DA}
\end{align}
where $\overline{(\cV X)}^{AB} = ((\cV X)_{AB})^*$. Higher order invariants can also be constructed in the same fashion, but we will not need them here. We have the relation
\be T_2 = - X^T \cM^{(0)} X, \ee
so we discard $T_2$ from our list of invariants.
\item Since $\cR$ transforms in the adjoint, all traces
\be \Tr(\cR^k) \ee
of powers of $\cR$ are invariant. These invariants are not all independent; in fact, those with odd $k$ vanish identically. We checked that $\Tr \cR^2$, $\Tr \cR^6$ and $\Tr \cR^8$ are independent, while for $\Tr \cR^4$ we have the relation
\be 
\Tr \cR^4 = \frac{1}{24} (\Tr \cR^2 )^2 . \label{rel}
\ee
In fact, it is known by mathematicians \cite{Lee:1974:IPW,Berdjis:1981:CCC,Berdjis:1981:CO} (see also the textbooks \cite{azcarraga_izquierdo_1995,fuchs2003symmetries}) that the only independent ones are those with
\be
k = 2, 6, 8, 10, 12, 14 \text{ and } 18 .
\ee
This will introduce a subtlety in identifying the $E_{7(7)}$ generalization of the invariant $\text{Tr}(R^4)=96 \Delta_3$ \eqref{TrR4}: it will be a non-polynomial expression in $\Tr \cR^2$, $\Tr \cR^6$ and $\Tr \cR^8$ as we will describe below\footnote{This fact was not properly recognized in the first version of \cite{Sarosi:2015nja}, which makes the generalization of \eqref{eq:Finvsarosi} proposed there incorrect. Our final formula \eqref{FinvE7} solves this problem.}.
\end{itemize}

\subsection{$STU$ truncation}

In this notation, the embedding is the following.
\begin{itemize}
\item The electromagnetic charges are given by
\begin{align}
(X^{12}, X^{34}, X^{56}, X^{78}) &= (Q_1,Q_2,Q_3,-Q_4) ,\nn \\
(X_{12}, X_{34}, X_{56}, X_{78}) &= (P^1,P^2,P^3,-P^4) ,
\end{align}
the other $X^{ij}$, $X_{ij}$ being zero. This follows from the identification \eqref{eq:STUA4}, \eqref{eq:STUAi} of the vector fields in the $STU$ truncation (and the fact that the scalar fields vanish at infinity, which is not a restriction as discussed at the end of section \ref{FinvSTU}).
\item The coset matrix $\cV$ was given in equation \eqref{eq:STUcoset}. We used the identification \eqref{eq:E7dict} to get the explicit form of $\cV$ as a $56\times 56$ matrix.
\end{itemize}
We can now compare our $E_{7(7)}$ invariants computed in the $STU$ truncation with those of the previous sections. We find
\begin{itemize}
\item Order two:
\begin{align}
L_1&=  X^T \cM^{(0)} X, \\
L_2 &= \frac{1}{36} \Tr(\cR^2).
\end{align}
\item Order three:
\begin{align}
C_1 &=\frac{1}{3} X^T \Omega \cR X, \\
C_2&= \frac{1}{3} X^T \cM^{(0)} \cR X.
\end{align}
\item Order four:
\begin{align}
\Delta &= - \frac{1}{16} \cI_4(X), \label{I4} \\
\Delta_2 &= \frac{1}{96} \left(  8 T_4 + 6 \cI_4 - (X^T \cM^{(0)} X)^2\right)
\end{align}
and
\begin{align}
0 &= 2^{17} 3^7 5 (\Delta_3 )^2 - 2^8 3^3 5 \Delta_3 (\Tr(\cR^2))^2 - 5 (\Tr(\cR^2))^4 \nn \\
&\quad + 2^5 3^2 11 \Tr(\cR^2) \Tr(\cR^6) - 2^6 3^5 \Tr(\cR^8). \label{eq:TrR4}
\end{align}
\end{itemize}
As announced, we find a non-polynomial expression for $\Delta_3$,
\be 
\Delta_3 = \frac{1}{2^{10} 3^4} \left[ \Tr(\cR^2)^2 + 5 \sqrt{\Tr(\cR^2)^4 - (2^8 3^3 5^{-3} 11) \Tr(\cR^2) \Tr(\cR^6) + (2^9 3^6 5^{-3}) \Tr(\cR^8)}\, \right].
\ee
(only this root of \eqref{eq:TrR4} correctly reproduces $\Tr(R^4)$).

\subsection{The $F$-invariant}

Using formula \eqref{Finv1} and the dictionary above, we find the following formula for the $F$-invariant in terms of $E_{7(7)}$ invariants:
\begin{align}
F &= M^4+M^2 N^2 - \frac{M^2}{4}X^T \cM^{(0)} X + \frac{N^2}{288} \Tr(\cR^2)+\frac{M}{24} X^T \cM^{(0)} \cR X- \frac{N}{24}  X^T \Omega \cR X \nn \\
&\quad +\frac{1}{16}\mathcal I_4(X)+ \frac{1}{24}T_4 - \frac{1}{192} (X^T \mathcal M^{(0)} X)^2 - \frac{1}{2^{10}3^4}\Tr(\cR^2)^2 \nn \\
&\quad +  \frac{5}{2^{11}3^4}  \sqrt{\Tr(\cR^2)^4 - (2^8 3^3 5^{-3} 11) \Tr(\cR^2) \Tr(\cR^6) + (2^9 3^6 5^{-3}) \Tr(\cR^8)}.
\label{FinvE7}
\end{align}
As was emphasized before, the entropy of the $STU$ black hole does not change under $E_{7(7)}$ dualities (since the four-dimensional metric is invariant), and the $STU$ black hole is a seed for the generic $\mathcal N = 8$ black hole. Therefore, although \eqref{FinvE7} was derived in the $STU$ truncation, it is the correct formula for the most general non-extremal black hole of $\mathcal N = 8$ supergravity. The entropy of such a black hole takes the form \eqref{eq:entropy}, where the quartic invariant $\Delta$ is given by \eqref{I4} and \eqref{eq:I4E7} and the $F$-invariant is given by \eqref{FinvE7}\footnote{Our final formula differs from \cite{Sarosi:2015nja}, as discussed above.}.

\part{Four-dimensional gaugings}\label{PART:GAUGINGS}
\chapter{Supergravity gaugings}
\label{chap:YMgaugings}

\begin{figure}
\centering
\begin{tikzpicture}
\tikzset{r/.style={rectangle, rounded corners, text centered, draw}}

\node[r] (11) at (0,10) {$D = 11$ supergravity};
\node[r] (U) at (0,0) {\onetab{Ungauged supergravity\\$G_g=U(1)^{28}$}};
\node[r] (G) at (10,0) {\onetab{Gauged supergravity\\$G_g = SO(8)$}};

\node (l) at (5,0.4) {gauging $SO(8) \subset SL(8,\R)$};
\draw[->, shorten >=10pt] (U) -- (G);

\pgfmathsetmacro{\x}{0.8}
\pgfmathsetmacro{\off}{0.1}
\draw (11) -- (0,6);
\draw (0,5) circle [x radius = 1.5, y radius = 0.75];
\draw (- \x , 5+\off) arc[x radius=\x, y radius=0.2, start angle=180, end angle=360];
\begin{scope}
\clip (0 , 5+\off) circle [x radius=\x, y radius=0.2];
\draw (0 , 5-\off) circle [x radius=\x, y radius=0.2];
\end{scope}
\draw[->, shorten >=10pt] (0,4) -- (U);
\node (t) at (-1,6) {$T^7$};

\draw (11) -- (4,6);
\draw (5,5) circle [radius=1];
\draw[dashed] (6,5)  arc[x radius=1, y radius=0.25, start angle=0, end angle=180];
\draw (4,5)  arc[x radius=1, y radius=0.25, start angle=180, end angle=360];
\draw[->, shorten >=10pt] (6,4) -- (G);
\node (s) at (6,6) {$S^7$};
\end{tikzpicture}
\vspace{0.5cm}
\caption[Two ways to $SO(8)$ supergravity]{\label{fig:SO8}Two ways to the gauged $SO(8)$ maximal supergravity in four dimensions: 1) start from the ungauged theory, which is eleven-dimensional supergravity reduced on $T^7$, and ``gauge" the group $SO(8) \subset SL(8,\R)$; or 2) directly reduce the eleven-dimensional theory on the sphere $S^7$. Here, we will be interested in the possible ``horizontal arrows", i.e., the possible deformations of supergravity. We will not do the link with the ``vertical arrows", i.e., how to get the gauged theories from compactifications of higher-dimensional models. We refer to the review \cite{Trigiante:2016mnt} (and references therein) for this equally interesting problem.}
\end{figure}
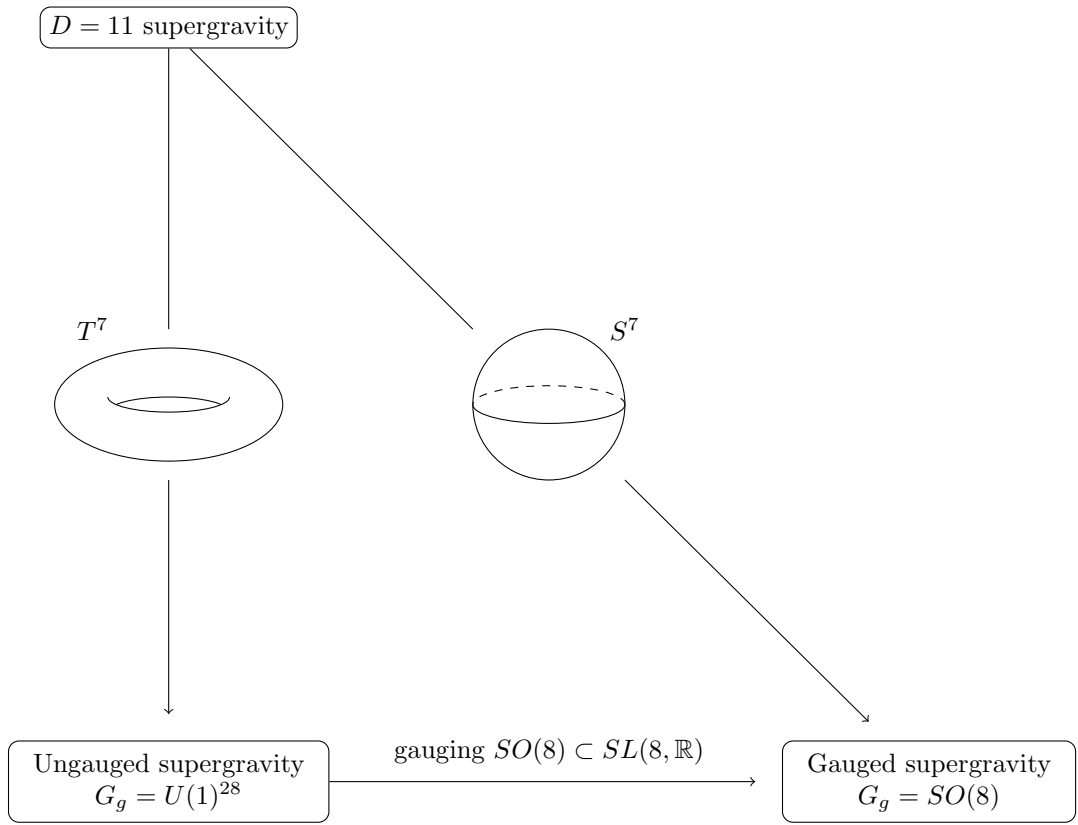

Yang-Mills type gaugings in extended supergravities, in which a subgroup of the rigid symmetries of the Lagrangian is promoted to a local symmetry (i.e., with space-time dependent parameters), have a long history that goes back
to \cite{Freedman:1976aw,Fradkin:1976xz,Freedman:1978ra,Zachos:1978iw}. For maximal supergravity, the first gauging has been performed in \cite{deWit:1981sst} with a $SO(8)$ gauge group in the Lagrangian formulation of \cite{Cremmer:1979up}, which involves a specific choice of duality frame with $SL(8,\R)$ rigid symmetry (see part \ref{PART:SUGRA} and appendix \ref{app:N8E7}). It was later understood as arising from the compactification of eleven-dimensional supergravity on a seven-sphere \cite{deWit:1986oxb} (see figure \ref{fig:SO8}). Other gaugings in the $SL(8,\R)$ frame have been constructed in \cite{Hull:1984yy,Hull:1984vg,Hull:1984qz} and later completely classified in \cite{Cordaro:1998tx}.

In this chapter, we first review these gaugings in a definite electric frame. They are smooth deformation of the theory, and the original (ungauged) theory is recovered as the couplings constants go to zero. As was emphasized in chapter \ref{chap:emduality}, the group of rigid symmetries of the usual second-order vector models depends on the choice of duality frame. Therefore, the same is true for the available gaugings. In the case of maximal supergravity, new gaugings were obtained in this way (outside the $SL(8,\R)$ frame) in \cite{Andrianopoli:2002mf,Hull:2002cv} (see also \cite{DallAgata:2012mfj} for a more recent example, and the analysis in \cite{DallAgata:2014tph,Inverso:2015viq}).

There are two formalisms in the literature in which an arbitrary duality frame is handled: the first is the embedding tensor formalism, which contains extra two-forms and vector potentials \cite{deWit:2002vt,deWit:2005ub,deWit:2007kvg} (see \cite{Samtleben:2008pe,Trigiante:2016mnt} for reviews), and the second is the two-potential first order action of chapter \ref{chap:emduality} (see \cite{Hillmann:2009zf} for the case of $\cN = 8$ supergravity). A natural question, motivated by the quest for new gaugings, is then whether these extended actions allow for new deformations, which cannot be realized in the usual second-order action in a definite duality frame. We will show that the answer is, unfortunately, negative. Another natural question is then whether the only deformations of the second-order action in a fixed duality frame are gaugings of the Yang-Mills type. It is answered in the next chapter.

The original results of this chapter were presented in the paper \cite{Henneaux:2017kbx}, written in collaboration with M. Henneaux, B. Julia and A. Ranjbar. I would also like to thank G. Barnich and N. Boulanger for useful discussions and collaborations during the early stages of that project.

\section{Gaugings of the Yang-Mills type}

A gauging of the Yang-Mills type refers to the procedure in which a subgroup of the rigid symmetries of the Lagrangian is promoted to a local symmetry, using some of the vector fields of the theory as non-abelian gauge fields. In doing so, ordinary derivatives of the matter fields are converted to covariant derivatives, and the abelian field strengths of the vector fields are replaced by the non-abelian ones. Some extra topological terms (or obstructions to the gauging) may also arise when there are symmetries of the original Lagrangian only up to a total derivative (``Peccei-Quinn" symmetries)\footnote{Consistency with supersymmetry brings other features (scalar potential, gravitino masses) which are important for applications but that we will not consider here. In any case, supersymmetry imposes no new constraint on the gauging procedure \cite{Trigiante:2016mnt}.}.

We begin by reviewing this type of deformations, following the notation of \cite{Barnich:2017nty}. We also comment briefly on the embedding tensor formalism that was developed to handle them in an arbitrary duality frame. We will neglect gravity and fermions in this part of the thesis.

\subsection{Elementary examples}

We begin with a few simple examples, in which the ``gauged" theory is simply obtained by using non-abelian field strengths and covariant derivatives. We also see an example where a symmetry of the Lagrangian only up to a total derivative brings an obstruction in the gauging procedure.

\paragraph{Pure Yang-Mills theory.}

The ungauged theory is a collection of $n_v$ free (abelian) vector fields, with action and gauge invariances
\begin{align}
S[A^I] &= - \frac{1}{4} \int \dx[n]\; \delta_{IJ} F^I_{\mu\nu} F^{J \mu\nu}, \quad F^I_{\mu\nu} = \pd_\mu A^I_\nu - \pd_\nu A^I_\mu, \\
\delta A^I_\mu &= \pd_\mu \epsilon^I \, .
\end{align}
The ``gauged" theory is pure Yang-Mills, with action and gauge invariances
\begin{align}
S[A^I] &= - \frac{1}{4} \int \dx[n]\; \delta_{IJ} \cF^I_{\mu\nu} \cF^{J \mu\nu}, \quad \cF^I_{\mu\nu} = \pd_\mu A^I_\nu - \pd_\nu A^I_\mu + g \, f\indices{^I_{JK}} A^J_\mu A^K_\nu, \label{eq:YMaction}\\
\delta A^I_\mu &= \pd_\mu \epsilon^I + g\, f\indices{^I_{JK}} A^J_\mu \epsilon^K \, . \label{eq:YMgauge}
\end{align}
Here, the $f\indices{^I_{JK}}$ are the structure constants of a compact Lie algebra, i.e. satisfying the two conditions 1) $f_{IJK} \equiv \delta_{IL}f\indices{^L_{JK}}$ is totally antisymmetric and 2) The Jacobi identity. These two conditions are necessary for consistency (invariance of \eqref{eq:YMaction} under \eqref{eq:YMgauge}). It is an interacting theory, with cubic (order $g$) and quartic (order $g^2$) interactions between the vector fields. Note that the free action and gauge transformations are smoothly recovered as the coupling constant $g$ goes to zero. Because the matrices $t\indices{^I_J} \equiv f\indices{^I_{JK}} \epsilon^K$ are antisymmetric in $IJ$, the transformations \eqref{eq:YMgauge} can be understood as a gauging of the rigid $SO(n_v)$ symmetry of the free theory.

\paragraph{Scalar shift symmetry.}

We start with the free action for a scalar and a vector field,
\begin{equation}\label{eq:freeactionsv}
S[\phi,A] = \int \dx[n] \left( - \frac{1}{2} \pd_\mu \phi \,\pd^\mu \phi - \frac{1}{4} F_{\mu\nu} F^{\mu\nu} \right),
\end{equation}
with gauge invariance
\begin{equation}
\delta \phi = 0, \quad \delta A_\mu = \pd_\mu \epsilon .
\end{equation}
This action has the global shift symmetry $\phi \rightarrow \phi + c$. We want to make it local by using the gauge parameter of the vector field, so the gauge transformations of the interacting theory will be
\begin{equation}\label{eq:gaugeshift}
\delta \phi = g \,\epsilon, \quad \delta A_\mu = \pd_\mu \epsilon .
\end{equation}
The action is easily written down by noticing that the covariant derivative
\begin{equation}
D_\mu \phi = \pd_\mu \phi - g A_\mu
\end{equation}
is invariant under the gauge transformations \eqref{eq:gaugeshift}. The interacting action is therefore simply taken as
\begin{align}
S[\phi,A] &= \int \dx[n] \left( - \frac{1}{2} D_\mu \phi \,D^\mu \phi - \frac{1}{4} F_{\mu\nu} F^{\mu\nu} \right) \\
&= \int \dx[n] \left( - \frac{1}{2} \pd_\mu \phi \,\pd^\mu \phi + g A^\mu \pd_\mu \phi - \frac{1}{2} g^2 A^\mu A_\mu - \frac{1}{4} F_{\mu\nu} F^{\mu\nu} \right).
\end{align}
It involves an interaction vertex between the scalar and vector fields, which can be written as $gA^\mu j_\mu$, with $j_\mu = \pd_\mu \phi$ the conserved current associated to the rigid symmetry of the original theory. A mass term is also produced for the vector field.  Again, the action and gauge invariances of the free theory are smoothly recovered as $g$ goes to zero.

\paragraph{Axion-vector model.}

In four dimensions, we can add a term to the action \eqref{eq:freeactionsv} and consider instead
\begin{equation}\label{eq:axionaction}
S[\phi,A] = \int \dx[4] \left( - \frac{1}{2} \pd_\mu \phi \,\pd^\mu \phi - \frac{1}{4} F_{\mu\nu} F^{\mu\nu} + \frac{1}{8} \phi \,\varepsilon^{\mu\nu\rho\sigma} F_{\mu\nu} F_{\rho\sigma} \right).
\end{equation}
This action still has the rigid shift symmetry $\phi \rightarrow \phi + c$, since the variation of the last term is a total derivative. However, this symmetry cannot be gauged. Indeed, when the parameter depends on space-time, the variation
\begin{equation}
\delta \cL = \frac{1}{8}\, c(x) \,\varepsilon^{\mu\nu\rho\sigma} F_{\mu\nu} F_{\rho\sigma}
\end{equation}
of the Lagrangian density is no longer a total derivative, and cannot be canceled by a new term.

\subsection{Vector-scalar models}

We now consider the non-minimally coupled vector-scalar models $\cL = \cL_S + \cL_V$ of chapter \ref{chap:emduality}, with vector Lagrangian
\begin{equation}
  \cL_V = - \frac{1}{4}\, \mathcal{I}_{IJ}(\phi) F^I_{\mu\nu}
          F^{J\mu\nu}
          + \frac{1}{8} \,\mathcal{R}_{IJ}(\phi)\,
          \varepsilon^{\mu\nu\rho\sigma}
          F^I_{\mu\nu} F^J_{\rho\sigma} . \label{eq:lag}
\end{equation}
The matrices $\cI$ and $\cR$ depend on the chosen duality frame, as discussed there.

\paragraph{Electric group symmetries.} This Lagrangian has the ``electric group symmetries"
\begin{equation}
  \label{eq:64}
\delta A^I_{\mu} = f\indices{^I_J} A^J_{\mu}, \quad
\delta\phi^i = \Phi^i(\phi),
\end{equation}
where the $f\indices{^I_J}$ are constants and $\Phi^i(\phi)$ are functions of the (undifferentiated) scalar fields only\footnote{Strictly speaking, the electric group is the subgroup of those transformations with $f \neq 0$. We will also include purely scalar symmetries in this part of the thesis.}.
These transformations are symmetries of the
action if and only if
\begin{enumerate}
\item the scalar
variations leave the scalar action invariant separately and
\item the quantities $f\indices{^I_J}$, $\Phi^i(\phi)$ satisfy
\begin{align}
  \frac{\pd \cI}{\pd \phi^i} \Phi^i &= - f^T \mathcal{I} - \mathcal{I} f,\label{var-I}\\
  \frac{\pd \cR}{\pd \phi^i} \Phi^i &= - f^T \mathcal{R} - \mathcal{R} f - 2 h, \label{var-R-theta}
\end{align}
where the $h$ are constant symmetric matrices which correspond to a total derivative in the variation of \eqref{eq:lag}.
\end{enumerate}
In particular, when the scalar Lagrangian is given by
$\cL_S = - \frac{1}{2} g_{ij}(\phi) \pd_\mu \phi^i \pd^\mu \phi^j$, the first condition means that $\Phi^i$ must be a Killing vector of the
metric $g_{ij}$.

We label a basis of $G_e$ by a capital greek index,
\begin{align}
\delta_\Gamma A^I_{\mu} &= (f_{\Gamma})\indices{^I_J}
                          A^J_{\mu}, \label{global-sym-A}\\ 
\delta_\Gamma \phi^i &= \Phi_{\Gamma}^i(\phi). \label{global-sym-phi}
\end{align}
Closure of the algebra then implies
\begin{align}
[f_\Delta, f_\Gamma] &= -C\indices{^\Sigma_{\Delta\Gamma}} f_\Sigma , \label{X-alg}\\
f_\Gamma^T h_\Delta - f_\Delta^T h_\Gamma + h_\Delta f_\Gamma -
  h_\Gamma f_\Delta &
= - C\indices{^\Sigma_{\Delta\Gamma}} \,h_\Sigma , \label{5idGlobal} \\
  \frac{\partial \Phi^i_\Delta}{\partial \phi^j} \, \Phi^j_\Gamma
  - \frac{\partial \Phi^i_\Gamma}{\partial \phi^j} \, \Phi^j_\Delta &
= -C\indices{^\Sigma_{\Delta\Gamma}} \Phi^i_\Sigma , \label{F-alg}
\end{align}
where the $C\indices{^\Sigma_{\Delta\Gamma}}$ are the structure constants of the electric group. Equation \eqref{5idGlobal} follows by commutation of the variations \eqref{var-R-theta}.

\paragraph{Gauging the rigid symmetries.} The gauge invariances of the undeformed theory are
\begin{equation}
\delta \phi^i = 0, \quad \delta A^I_\mu = \pd_\mu \epsilon^I .
\end{equation}
In the gauged theory, they are deformed to
\begin{align}
  \delta A^I_{\mu} &= \partial_\mu \epsilon^I + f\indices{^I_{JK}}
 A^J_{\mu} \epsilon^K , \label{gauge-sym-A}\\
  \delta \phi^i &= \epsilon^I \Phi_I^i(\phi). \label{gauge-sym-phi}
\end{align}
The link with the original rigid symmetries goes as follows: the gauged theory is characterized by a $\dim(G_e)\times n_v$ matrix $k\indices{^\Gamma_I}$, in terms of which we have
\begin{align}
f\indices{^I_{JK}} &= (f_\Gamma)\indices{^I_J} k\indices{^\Gamma_K} , \label{eq:f-k} \\
\Phi^i_I(\phi) &= \Phi^i_\Gamma(\phi) k\indices{^\Gamma_I} . \label{eq:phi-k}
\end{align}
This deformation of the gauge symmetries corresponds therefore to gauging the underlying rigid symmetries by using local parameters $\eta^\Gamma (x) = k\indices{^\Gamma_I} \epsilon^I (x)$ (we absorb the coupling constants in $k$). 
For the gauging to be possible, the matrix $k\indices{^\Gamma_I}$ must satisfy the two linear constraints
\begin{align}
(f_\Gamma)\indices{^I_J} k\indices{^\Gamma_K} + (f_\Gamma)\indices{^I_K}
  k\indices{^\Gamma_J} &= 0 ,
\label{eq:constr-antisymmetry}\\
h_{\Gamma \, (IJ} k\indices{^\Gamma_{K)}} &= 0, \label{eq:constr-chernsimons}
\end{align}
and the quadratic constraint
\begin{equation} \label{eq:quadraticconstraint}
  k\indices{^{\Gamma}_I} k\indices{^{\Delta}_J}
  C\indices{^{\Sigma}_{\Gamma\Delta}} -
  (f_{\Gamma})\indices{^K_I} k\indices{^\Gamma_J}
  k\indices{^{\Sigma}_K} = 0.
\end{equation}
The first linear constraint is the antisymmetry of the structure constants $f\indices{^I_{JK}}$ of the gauge algebra in their last two indices. The second linear constraint appears when $h_\Gamma \neq 0$: it corresponds to the obstruction encountered when there is a symmetry of the Lagrangian only up to a total derivative (cf. example \eqref{eq:axionaction} of the previous section). Finally, the quadratic constraint corresponds to the Jacobi identity for the $f\indices{^I_{JK}}$ and closure of the gauge algebra, as we will see below.

Then, the complete Lagrangian after gauging reads
\begin{align}\label{Full-Lagrangian}
  \mathcal{L} = \mathcal{L}_S(\phi^i, D_\mu \phi^i) &
  - \frac{1}{4}\,\mathcal{I}_{IJ}(\phi)
   \mathcal{F}^I_{\mu\nu} \mathcal{F}^{J\mu\nu} + \frac{1}{8}\,
   \mathcal{R}_{IJ}(\phi)\,
    \varepsilon^{\mu\nu\rho\sigma}
 \mathcal{F}^I_{\mu\nu} \mathcal{F}^J_{\rho\sigma} \nonumber \\
    &+ \frac{2}{3}\, X_{IJ,K}\, \varepsilon^{\mu\nu\rho\sigma} A^J_\mu
      A^K_\nu \left( \partial_\rho A^I_\sigma + \frac{3}{8}\,
      f\indices{^I_{LM}} A^L_\rho A^M_\sigma \right).
\end{align}
In this equation, the covariant derivatives and non-abelian field strengths are
\begin{align}
  \mathcal{F}^I_{\mu\nu} &= \partial_\mu A^I_\nu - \partial_\nu A^I_\mu
 + f\indices{^I_{JK}} A^J_\mu A^K_\nu  , \\
D_\mu \phi^i &= \partial_\mu \phi^i - \Phi^i_I(\phi) A^I_\mu ,
\end{align}
and the tensor $X_{IJ,K}$ is defined by
\begin{equation}
X_{IJ,K} = (h_\Gamma)_{IJ} k^\Gamma_K .
\end{equation}
This Lagrangian is obtained from the ungauged one by replacing derivatives by covariant derivatives, abelian field strengths by non-abelian ones and, when $h_\Gamma \neq 0$, adding the topological term of the second line. This term was first discussed in \cite{deWit:1984rvr,deWit:1987ph}.

Checking directly the invariance of this action under
\eqref{gauge-sym-A} and \eqref{gauge-sym-phi} without first
parametrizing $f\indices{^I_{JK}}$, $\Phi_I^i(\phi)$ and $X_{IJ,K}$
through symmetries requires the use of the linear identities
\begin{equation}
f\indices{^I_{JK}} = f\indices{^I_{[JK]}}, \quad X_{(IJ,K)} = 0
\end{equation}
and of the quadratic ones
\begin{align}
f\indices{^I_{J[K_1}}f\indices{^J_{K_2K_3]}} &= 0, \\
  f\indices{^K_{I[L}} X_{M]J,K} + f\indices{^K_{J[L}} X_{M]I,K}
  - \frac{1}{2}\, X_{IJ,K} f\indices{^K_{LM}} &= 0, \\
  \frac{\pd \Phi^i_I}{\pd \phi^j} \Phi^j_J -
  \frac{\pd \Phi^i_J}{\pd \phi^j} \Phi^j_I + f\indices{^K_{IJ}} \Phi^i_K &= 0.
\end{align}
In terms of $k\indices{^\Gamma_I}$, the two linear identities are equivalent to \eqref{eq:constr-antisymmetry} and \eqref{eq:constr-chernsimons} respectively. The three quadratic identities all come
from the single quadratic constraint \eqref{eq:quadraticconstraint}
once the algebra of global symmetries \eqref{X-alg} -- \eqref{F-alg}
is taken into account.

\section{Embedding tensor formalism} \label{sec:embeddingtensor}

As we mentioned before, the electric symmetry group depends on the choice of duality frame. Therefore, the same is true for the available gaugings. A formalism has been developed in the literature that allows to work independently of the specific duality frame: it is the embedding tensor formalism of
\cite{deWit:2002vt,deWit:2005ub,deWit:2007kvg} (see also \cite{Samtleben:2008pe,Trigiante:2016mnt} for reviews). This method has proven very powerful for the investigation of gaugings of the Yang-Mills type; in particular, it enabled the discovery of a one-parameter family of gaugings that involve a continuous change of the duality frame \cite{DallAgata:2012mfj}.

In this formalism, the gauging is characterized by the embedding tensor
\begin{equation}
\Theta\indices{_M^\alpha} .
\end{equation}
The index $M$ goes from $1$ to $2 n_v$ (twice the number of vector fields), and $\alpha$ from $1$ to $\dim(G)$, where $G$ is the full duality group (in which one can also include pure scalar symmetries). It is the analogue of the matrix $k\indices{^\Lambda_I}$, where both indices have an increased range. This embedding tensor satisfies a number of linear and quadratic constraints.
\begin{enumerate}
\item One of the quadratic constraints is the ``locality constraint"
\begin{equation}\label{eq:localityconstraint}
\Omega^{MN} \Theta\indices{_M^\alpha} \Theta\indices{_N^\beta} = 0,
\end{equation}
where $\Omega^{MN}$ is the usual $Sp(2n_v, \R)$-invariant antisymmetric matrix. This constraint ensures that one can always go to a definite duality frame, as shown in \cite{deWit:2005ub,Trigiante:2016mnt,Henneaux:2017kbx}. In that frame, $\Theta$ reduces to $k$ and the gauge group is a subgroup of an electric subgroup of $G$.
\item The other linear and quadratic constraints on $\Theta$ are just ``covariantizations" of the constraints \eqref{eq:constr-antisymmetry}, \eqref{eq:constr-chernsimons} and \eqref{eq:quadraticconstraint} on $k$, to which they reduce once the locality constraint has been used \cite{Trigiante:2016mnt}.
\end{enumerate}
Therefore, the embeding tensor contains information both about 1) the choice of duality frame in which the gauging is effected, and 2) the choice of gauge group $G_g \subset G_g$ in that duality frame. This corresponds to 1) the symplectic matrix $E$ of chapter \ref{chap:emduality} encoding the choice of duality frame, and 2) the matrix $k$ encoding the gauging of the previous section, respectively.

\subsection{Lagrangian - ungauged limit}

In addition to the usual $n_v$ ``electric" vector fields $A^I_\mu$ and the embedding tensor $\Theta\indices{_M^\alpha}$, the Lagrangian of the embedding tensor formalism also contains a collection of extra fields: $n_v$ ``magnetic" vector fields $\tilde{A}_{I\mu}$ and $\dim(G)$ two-forms $B_{\alpha \mu \nu}$. It reads  \cite{deWit:2005ub}
\begin{align}
\mathcal{L}^{\Theta, \text{int.}}(A,\tilde{A},B) = &- \frac{1}{4} \mathcal{I}_{IJ}(\phi) \mathcal{H}^I_{\mu\nu} \mathcal{H}^{J\mu\nu} + \frac{1}{8} \mathcal{R}_{IJ}(\phi)\, \varepsilon^{\mu\nu\rho\sigma} \mathcal{H}^I_{\mu\nu} \mathcal{H}^J_{\rho\sigma} \nonumber \\
&- \frac{g}{8} \varepsilon^{\mu\nu\rho\sigma} \Theta^{I\alpha}B_{\alpha \mu\nu} \left( \tilde{\mathcal{F}}_{I \rho\sigma} - \frac{g}{4} \Theta\indices{_I^\beta} B_{\beta \rho\sigma}\right) + g \mathcal{L}^\text{extra} (A,\tilde{A}, gB ) \, ,
\label{eq:19}
\end{align}
where
\begin{equation}
\mathcal{H}^I_{\mu\nu} = \mathcal{F}^I_{\mu\nu} + \frac{g}{2} \Theta^{I \alpha} B_{\alpha \mu\nu} 
\end{equation}
and $\mathcal{F}^I_{\mu\nu}$, $\tilde{\mathcal{F}}_{I\mu\nu}$ are the Yang-Mills curvatures, differing from the abelian ones by $O(g)$-terms. We also split the index $M$ of the embedding tensor,
\begin{equation}\label{eq:embeddingsplit}
\Theta\indices{_M^\alpha} = (\Theta\indices{_I^\alpha},\Theta\indices{^{I\alpha}})
\end{equation}
The ``extra" terms in \eqref{eq:19} are terms necessary to secure gauge invariance but vanish in the limit taken below. They are not written for that reason and their explicit form may be found in \cite{deWit:2005ub}. As we emphasized in \eqref{eq:19}, they depend on the two-forms $B_\alpha$ only through the combination $B{''}_\alpha = g B_\alpha$ and carry at least an extra factor of $g$. The authors of \cite{deWit:2005ub} also consider matter couplings that we do not consider here.

Note that the electric an magnetic fields are not treated equally: the electric fields have a proper kinetic term, while the magnetic ones appear only through a topological term with the $2$-forms. This ensures that the number of degrees of freedom is not changed by the addition of these fields (see below)\footnote{Similar models inspired by this construction have been studied in \cite{Boulanger:2008nd}. They are 1) a toy model in $D = 3$ with scalars and their dual vectors, and 2) gravity with the dual graviton in arbitrary dimensions. In those cases also, only one field of the pair has a kinetic terms, while the other appears through a topological coupling with some auxiliary field to avoid the doubling of degrees of freedom. We thank N. Boulanger for pointing out this reference.}.
 
To view the gauged Lagrangian \eqref{eq:19} as a deformation of an ungauged Lagrangian with the same field content, so as to phrase the gauging problem as a deformation problem, we observe that the straightforward free limit $g\rightarrow 0$ of \eqref{eq:19} is just the original Lagrangian \eqref{eq:lag}, without any additional fields and without the embedding tensor. But if we first redefine the two-forms appearing in \eqref{eq:19} as $B{''}_\alpha = g B_\alpha$ and then take the limit $g \rightarrow 0$, one gets
\begin{equation} \label{lagtwoforms}
\mathcal{L}^\Theta = - \frac{1}{4} \mathcal{I}_{IJ}(\phi) H^I_{\mu\nu} H^{J\mu\nu} + \frac{1}{8} \mathcal{R}_{IJ}(\phi)\, \varepsilon^{\mu\nu\rho\sigma} H^I_{\mu\nu} H^J_{\rho\sigma} - \frac{1}{8} \varepsilon^{\mu\nu\rho\sigma} \Theta^{I\alpha}B{''}_{\alpha \mu\nu} \left( \tilde{F}_{I \rho\sigma} - \frac{1}{4} \Theta\indices{_I^\beta} B{''}_{\beta \rho\sigma}\right),
\end{equation}
with
\begin{equation}
H^I_{\mu\nu} = F^I_{\mu\nu} + \frac{1}{2} \Theta^{I \alpha} B{''}_{\alpha \mu\nu} 
\end{equation}
and $F^I_{\mu\nu}$, $\tilde{F}_{I \mu\nu}$ the abelian curvatures.  This Lagrangian is invariant under the $2n_v$ gauge transformations
\begin{align}
\delta A^I_\mu &= \partial_\mu \lambda^I - \frac{1}{2} \Theta^{I \alpha} \Xi{''}_{\alpha \mu}\, , \\
\delta \tilde{A}_{I\mu} &= \partial_\mu \tilde{\lambda}_I + \frac{1}{2} \Theta\indices{_I^\alpha} \Xi{''}_{\alpha \mu}
\end{align}
and the $\dim G$ gauge transformations of the two-forms
\begin{align}
\delta B{''}_{\alpha\mu\nu} &= \partial_{\mu}\Xi{''}_{\alpha \nu} - \partial_{\nu}\Xi{''}_{\alpha \mu},
\end{align}
where we redefined the gauge parameter as $\Xi{''}_\alpha = g \Xi_\alpha$ with respect to \cite{deWit:2005ub}. The parameters $\lambda^I$, $\tilde{\lambda}_I$ and $\Xi{''}_{\alpha \nu}$ are arbitrary functions.  The $\lambda$'s  and $\tilde{\lambda}$'s correspond to standard $U(1)$ gauge symmetries of the associated one-form potentials, while the $\Xi''$'s define ordinary abelian two-form gauge symmetries and also appear as shift transformations of the one-form potentials. This set of gauge symmetries is reducible since adding a gradient to $\Xi''$ and shifting simultaneously the $\lambda^I$ and $\tilde{\lambda}_I$, i.e.,
\begin{equation}
\Xi{''}_{\alpha \mu} \rightarrow \Xi{''}_{\alpha \mu} + \partial_\mu \xi_{\alpha}, \qquad \lambda^I \rightarrow \lambda^I + \frac{1}{2} \Theta^{I\alpha} \xi_{\alpha}, \qquad \tilde{\lambda}_I \rightarrow \tilde{\lambda}_I - \frac{1}{2} \Theta\indices{_I^\alpha} \xi_\alpha,
\end{equation}
leads to no modification of the gauge transformations.

The Lagrangian \eqref{lagtwoforms} contains the extra fields and still has the ability to cover generic symplectic frames through  the embedding tensor components $\Theta^{I\alpha}$ and $\Theta\indices{_I^\alpha}$, as we will see in the next subsection. These satisfy the locality constraint \eqref{eq:localityconstraint} written above, that we will use in the form
\begin{equation}
\Theta^{I [\alpha} \Theta\indices{_I^{\beta]}} = 0 \label{eq:ThetaCons}
\end{equation}
which follows from \eqref{eq:localityconstraint} using the split of indices \eqref{eq:embeddingsplit}\footnote{As mentioned above, the embedding tensor satisfies other constraints in addition to \eqref{eq:ThetaCons}. Those have to do with the gauging itself and not with the choice of duality frame, so we will not use them here.}. By construction, the non-abelian Lagrangian \eqref{eq:19} can be viewed as a consistent local deformation of \eqref{lagtwoforms}. The abelian Lagrangian \eqref{lagtwoforms} is thus a sensible starting point for the deformation procedure, since the space of consistent local deformations of \eqref{lagtwoforms} will necessarily include \eqref{eq:19}.

\subsection{Local deformations}

A natural question to be asked is whether the space of local deformations  of \eqref{lagtwoforms} is isomorphic to the space of local deformations of the conventional Lagrangian \eqref{eq:lag} in an appropriate symplectic frame. We now show that this is indeed the case.

To that end, we will perform a sequence of local field redefinitions and show that the Lagrangian \eqref{lagtwoforms} with a given definite (arbitrary) choice of embedding tensor differs from the Lagrangian \eqref{eq:lag} in a related symplectic frame by the presence of two kinds of fields:
\begin{enumerate}
\item pure gauge fields (i.e., with gauge transformations that are pure shifts), which therefore do not appear at all in the Lagrangian;
\item algebraic auxiliary fields, which can be eliminated from the Lagrangian using their (algebraic) equations of motion.
\end{enumerate}
We follow closely the steps given in \cite{deWit:2005ub} to prove that the extra fields appearing in the embedding tensor formalism do not add new degrees of freedom with respect to conventional gaugings (see also \cite{Trigiante:2016mnt} for another proof). An important difference with those references is that we do not fix the gauge at any stage. We also closely monitor locality. Indeed, gauge-fixing and nonlocalities change the space of available local deformations. However, using the tools of BRST cohomology reviewed in the next chapter, it is proved in appendix \ref{app:auxfields} (following \cite{Barnich:1994db}) that the presence of the extra fields described above does not.

The magnetic components $\Theta^{I\alpha}$ of the embedding tensor form a rectangular $n_v \times (\dim G)$ matrix. Let $r \leq n_v$ be its rank.
Then, there exists an $n_v \times n_v$ invertible matrix $Y$ and a $(\dim G) \times (\dim G)$ invertible matrix $Z$ such that
\begin{equation} \label{eltheta}
\Theta^{\prime I \alpha} = Y\indices{^I_J} \Theta^{J \beta} Z\indices{_\beta^\alpha}
\end{equation}
is of the form
\begin{equation} \label{formthetaprime}
(\Theta^{\prime I \alpha}) = \left(\begin{array}{c|c}
\theta & 0  \\ \hline
0 & 0
\end{array}\right),
\end{equation}
where $\theta$ is an $r\times r$ invertible matrix. (Indeed, the matrices $Y$ and $Z$ can be constructed as a product of the matrices that implement the familiar ``elementary operations" on the rows and columns of $\Theta^{I \alpha}$.)
We also define the new electric components by
\begin{equation} \label{magtheta}
\Theta\indices{^\prime_I^\alpha} = (Y^{-1})\indices{^J_I} \Theta\indices{_J^\beta} Z\indices{_\beta^\alpha},
\end{equation}
so that the new components still satisfy the constraint \eqref{eq:ThetaCons},
\begin{equation} \label{primedconstraint}
\Theta^{\prime I [\alpha} \Theta\indices{^\prime_I^{\beta]}} = 0 .
\end{equation}
As in section 5.1 of \cite{deWit:2005ub}, let us split the indices as $I = (\hat{I}, \hat{U})$ and $\alpha = (i, m)$, where $\hat{I}, i = 1, \dots, r$, $\hat{U} = r+1, \dots, n_v$ and $m = r+1, \dots, \dim G$. We also define
\begin{equation}
\tilde{\theta}\indices{_{\hat{I}}^i} = \Theta\indices{^\prime_{\hat{I}}^i} .
\end{equation}
With this split, equations \eqref{formthetaprime} and \eqref{primedconstraint} become
\begin{align} \label{thetasplit}
\Theta^{\prime\hat{I}i} &= \theta^{\hat{I}i} \quad \text{(invertible)}\, , \nn \\
\Theta^{\prime\hat{I}m} &= \Theta^{\prime\hat{U}i} = \Theta^{\prime\hat{U}m} = \Theta\indices{^\prime_{\hat{I}}^m} = 0\, , \nn \\
\theta^{\hat{I}i} \tilde{\theta}\indices{_{\hat{I}}^j} &= \theta^{\hat{I}j} \tilde{\theta}\indices{_{\hat{I}}^i} \, .
\end{align}

Now, we make the field redefinitions
\begin{equation} \label{eq:change1}
B{''}_\alpha = Z\indices{_\alpha^\beta} B'_\beta, \qquad A^I = (Y^{-1})\indices{^I_J} A^{\prime J}, \qquad \tilde{A}_I = Y\indices{^J_I} \tilde{A}'_J .
\end{equation}
In those variables, the Lagrangian takes the same form \eqref{lagtwoforms} but with  primed quantities everywhere. The new matrices $\mathcal{I}'$ and $\mathcal{R}'$ are given by
\begin{equation}
\mathcal{I}' = Y^{-T} \mathcal{I} Y^{-1}, \qquad
\mathcal{R}' = Y^{-T} \mathcal{R} Y^{-1}.
\end{equation}
These field  redefinitions are local, so any local function of the old set of variables is also a local function of the new set of variables. Using equations \eqref{thetasplit}, it can be seen that the magnetic vector fields $\tilde{A}'_{\hat{U}}$ and the two-forms $B'_m$ do not appear at all in the Lagrangian; this means that their gauge symmetries are in fact pure shift symmetries\footnote{In particular, a complete description of the gauge symmetries of the  two-forms $B'_m$ is actually given by $\delta B'_m = \epsilon'_m$ rather than $\delta B'_m = d \Xi'_m$, implying that the above set of gauge transformations is not complete.  This is correctly taken into account in the BRST discussion of the appendix \ref{app:auxfields}.  Similar features (shift symmetries for some of the $2$-forms) hold when $g$ is turned on.}.

Let us also redefine the gauge parameters as $\Xi'_\alpha = (Z^{-1})\indices{_\alpha^\beta} \Xi{''}_\beta$, $\lambda'^I = Y\indices{^I_J} \lambda^J$ and $\tilde{\lambda}'_I=(Y^{-1})\indices{^J_I} \tilde{\lambda}_J$. Ignoring temporarily $A'^{\hat{U}}$,  the gauge variations of $A'^{\hat{I}}$, $\tilde{A}'_{\hat{I}}$ and $B'_i$ are then
\begin{equation}
\delta A'^{\hat{I}}_\mu = \partial_\mu \lambda'^{\hat{I}} - \frac{1}{2} \theta^{\hat{I}i} \Xi'_{i\mu}, \quad \delta \tilde{A}^\prime_{\hat{I}\mu} = \partial_\mu \tilde{\lambda}'_{\hat{I}} + \frac{1}{2} \tilde{\theta}\indices{_{\hat{I}}^i} \Xi'_{i\mu}, \quad
\delta B'_{i\mu\nu} = 2 \partial_{[\mu}\Xi'_{i \nu]} .
\end{equation}
 They suggest the further changes of variables
\begin{align} \label{eq:change2}
\bar{A}^i_{\mu} &= \theta^{\hat{I}i} \tilde{A}'_{\hat{I}\mu} + \tilde{\theta}\indices{_{\hat{I}}^i} A'^{\hat{I}}_{\mu} \, ,\\
\Delta_{i\mu\nu} &= B'_{i\mu\nu} + 2 (\theta^{-1})_{i\hat{I}} F'^{\hat{I}}_{\mu\nu} \, ,
\end{align}
which we complete in the $A$-sector by taking other independent linear combinations of the $A'^{\hat{I}}$, $\tilde{A}'_{\hat{I}}$, which can be taken to be for instance the $A'^{\hat{I}}_\mu$.   The change of variables is again such that any local function of the old set of variables is also a local function of the new set of variables.  Using the constraint $\theta^{\hat{I}i}\tilde{\theta}\indices{_{\hat{I}}^j} = \theta^{\hat{I}j}\tilde{\theta}\indices{_{\hat{I}}^i}$, one finds that the variation of the new variables simplifies to
\begin{equation}
\delta \bar{A}^i_\mu = \partial_\mu \eta^i, \quad \delta \Delta_{i\mu\nu} = 0, \quad \delta A'^{\hat{I}}_\mu = \epsilon^{\hat{I}}_\mu
\end{equation}
with $\eta^i = \theta^{\hat{I}i} \tilde{\lambda}'_{\hat{I}} + \tilde{\theta}\indices{_{\hat{I}}^i} \lambda'^{\hat{I}}$ and $\epsilon^{\hat{I}}_\mu = \partial_\mu \lambda'^{\hat{I}} - \frac{1}{2} \theta^{\hat{I}i} \Xi'_{i\mu}$. We have used a different symbol $\Delta_{i\mu\nu}$ to emphasize that it does not transform as a $2$-form gauge potential anymore. Because $\theta^{\hat{I}i}$ is invertible, the gauge parameters  $\eta^i$ and $ \epsilon^{\hat{I}}_\mu$ provide an equivalent description of the gauge symmetries but, contrary to that given by $\lambda'^{\hat{I}}$, $\tilde{\lambda}'_{\hat{I}}$ and $\Xi'_{i\mu}$, it is an irreducible one.

Written in those variables, the Lagrangian only depends on $n_v$ vector fields $A'^{\hat{U}}$ and $\bar{A}^i$ and on $r$ two-forms $\Delta_i$ (the variables $A'^{\hat{I}}_\mu$ drop out, in agreement with the shift symmetry $\delta A'^{\hat{I}}_\mu = \epsilon^{\hat{I}}_\mu$).
The Lagrangian is explicitly
\begin{equation} \label{eq:lbarB}
\mathcal{L} = - \frac{1}{4} \mathcal{I}'_{IJ}(\phi) \bar{H}^I_{\mu\nu} \bar{H}^{J\mu\nu} + \frac{1}{8} \mathcal{R}'_{IJ}(\phi)\, \varepsilon^{\mu\nu\rho\sigma} \bar{H}^I_{\mu\nu} \bar{H}^J_{\rho\sigma} - \frac{1}{8} \varepsilon^{\mu\nu\rho\sigma} \Delta_{i \mu\nu} \left( \bar{F}^i_{\rho\sigma} - \frac{1}{4} \tilde{\theta}\indices{_{\hat{I}}^i} \theta^{\hat{I}j} \Delta_{j \rho\sigma}\right),
\end{equation}
where the $\bar{H}^I$ are
\begin{equation}
\bar{H}^{\hat{I}} = \frac{1}{2} \theta^{\hat{I}i} \Delta_i, \quad \bar{H}^{\hat{U}} = F'^{\hat{U}} .
\end{equation}
The gauge variations of the fields are
\begin{equation}
\delta A'^{\hat{U}}_\mu = \partial_\mu \lambda'^{\hat{U}}, \quad \delta \bar{A}^i_\mu = \partial_\mu \eta^i, \quad \delta \Delta_{i\mu\nu} = 0.
\end{equation}
The two-forms $\Delta_i$ are auxiliary fields and can be eliminated from the Lagrangian \eqref{eq:lbarB}, yielding a Lagrangian of the form \eqref{eq:lag} in a definite symplectic frame. This is because the relevant quadratic form is invertible (see \cite{deWit:2005ub}, section 5.1, where the final Lagrangian may also be found).

One can get this Lagrangian more directly as follows.  After the auxiliary fields are eliminated, the variables that remain are the scalar fields and the $n_v$ vector fields $A'^{\hat{U}}$ and $\bar{A}^i$.  We thus see that the embedding tensor determines a symplectic frame. The matrix $E \in Sp(2n_v, \mathbb{R})$ defining this symplectic frame can be viewed in this approach as the function $E(\Theta)$ of the embedding tensor obtained through the above successive steps that lead to the final Lagrangian where only half of the potentials remain\footnote{Of course, there are ambiguities in the derivation of $E$ from the embedding tensor, since choices were involved at various stages in the construction.  One gets $E$ up to a transformation of the stability subgroup of the Lagrangian subspace of the $n_v$ electric potentials  $A'^{\hat{U}}$ and $\bar{A}^i$.}.

More explicitly, one gets the matrix $E(\Theta)$ from the above construction as follows: the change of variables \eqref{eq:change1} can be written as
\begin{equation} \label{E1}
\begin{pmatrix}
A^I \\ \tilde{A}_I
\end{pmatrix} = \begin{pmatrix}
(Y^{-1})\indices{^I_J} & 0 \\ 0 & Y\indices{^J_I}
\end{pmatrix} \begin{pmatrix}
A'^J \\ \tilde{A}'_J
\end{pmatrix}
\end{equation}
and the full change \eqref{eq:change2} (including the trivial redefinitions of the other vector fields) as
\begin{equation} \label{E2}
\begin{pmatrix}
A'^{\hat{I}} \\ A'^{\hat{U}} \\\tilde{A}'_{\hat{I}} \vspace{3pt} \\ \tilde{A}'_{\hat{U}}
\end{pmatrix} =
\begin{pmatrix}
0 & 0 & \delta^{\hat{I}}_{\hat{J}} & 0 \\
0 & \delta^{\hat{U}}_{\hat{V}} & 0 & 0 \\
\delta^{\hat{J}}_{\hat{I}} & 0 & - (\theta^{-1})_{i \hat{I}} \tilde{\theta}\indices{_{\hat{J}}^i} & 0 \\
0 & 0 & 0 & \delta^{\hat{V}}_{\hat{U}}
\end{pmatrix}
\begin{pmatrix}
\bar{A}_{\hat{J}} \vspace{3pt} \\ \bar{A}^{\hat{V}} \\ \bar{A}^{\hat{J}} \\ \bar{A}_{\hat{V}}
\end{pmatrix},
\end{equation}
where we defined
\begin{equation}
\bar{A}_{\hat{I}} = (\theta^{-1})_{i \hat{I}} \bar{A}^i
\end{equation}
with respect to \eqref{eq:change2} in order to have the same kind of indices. The matrix $E(\Theta)$ is therefore simply given by
\begin{equation}
E(\Theta) = E_1 E_2,
\end{equation}
where $E_1$ and $E_2$ are the matrices appearing in equations \eqref{E1} and \eqref{E2} respectively. Using the property
\begin{equation}
(\theta^{-1})_{i[\hat{I}} \tilde{\theta}\indices{_{\hat{J}]}^i} = 0,
\end{equation}
which follows from the last equation of \eqref{thetasplit} upon contractions with $\theta^{-1}$, one can show that $E(\Theta)$ defined this way is indeed a symplectic matrix. Once $E(\Theta)$ is known, the final Lagrangian in this symplectic frame (i.e., the Lagrangian that follows from the elimination of $\Delta_i$ in \eqref{eq:lbarB}) is then simply the one of section \ref{sec:symplecticchoice}, that is, the Lagrangian \eqref{eq:lag} with $\mathcal{I}'$ and $\mathcal{R}'$ determined from \eqref{eq:OMdefBis} where $\mathcal{M}'$ is given by 
\begin{equation}\label{eq:MprimeTheta}
\mathcal{M}' = E(\Theta)^T \mathcal{M} E(\Theta) .
\end{equation}
(see \eqref{eq:Mprime}). In general, the matrix $E$ will have an upper triangular part, and hence, will not belong to the stability subgroup of the original electric frame.

\section{Deformations of the first-order action}

Another formalism covering arbitrary choices of duality frames is of course the first-order action of chapter \ref{chap:emduality}. Again, one can wonder if this allows for new gauging possibilities. Unfortunately, this is not the case, as we show in this section.

It was shown in \cite{Bunster:2010wv} that the first order action \eqref{eq:symlag} without scalar fields,
\be
S = \frac12\int \dtdx[3] \left(\Omega_{MN} \mathcal{B}^{Mi} \dot{\mathcal{A}}^N_i -  \delta_{MN} \mathcal{B}^M_i \mathcal{B}^{Ni}\right) \label{eq:LV1bis}
\ee
does not admit non-abelian deformations of the Yang-Mills type. A much stronger result was actually derived earlier in \cite{Bekaert:2001wa} through the BRST formalism, namely, that the action \eqref{eq:LV1bis} admits no local deformation that deforms the gauge algebra at all.

In fact, the obstruction described in \cite{Bunster:2010wv}  does not depend on the scalar sector and obstructs Yang-Mills deformations even when scalar fields are present.  The clash comes from the incompatibility of an adjoint action (as required by the Yang-Mills construction) and the symplectic condition (as required by the invariance of the scalar-independent kinetic term). The persistence of the obstruction even in the presence of scalar fields was announced in the conclusion of \cite{Bunster:2011aw}, but the proof was only sketched there.  For completeness, we give the details here.

To be more specific, consider the Yang-Mills deformation 
\be
\delta \mathcal{A}^M_i = \partial_i \mu^M + g \, C\indices{^M_{NP}} \mathcal{A}^N_i \mu^P 
\ee
of the original abelian gauge symmetry $\delta \mathcal{A}^M_i = \partial_i \mu^M$ of \eqref{eq:LV1bis} with gauge parameters $\mu^M$.   Here, the $C\indices{^M_{NP}} = - C\indices{^M_{PN}}$ are the structure constants of the gauge group $G_g$ of dimension $2n_v$ into which the original abelian gauge group  is deformed\footnote{As context should make clear, these are not the structure constants of the electric group introduced above. This is an unfortunate clash of notations between \cite{Henneaux:2017kbx} and \cite{Barnich:2017nty}, whose notations we keep.} and $g$ is the deformation parameter.
To make the abelian gauge invariance of the starting point more manifest, one can introduce the temporal component $\mathcal{A}^M_0$ of the potentials and rewrite the action in terms of the abelian curvatures $f_{\mu \nu}^M = \partial_\mu  \mathcal{A}^M_\nu - \partial_\nu  \mathcal{A}^M_\mu$.  One can replace $\dot{\mathcal{A}}^N_i$ by $f_{0i}^M$ in \eqref{eq:LV1bis} because the magnetic fields are identically transverse.

Under the Yang-Mills deformation, the abelian curvatures are replaced by the non-abelian ones,
\be
\cF_{\mu \nu}^M = f_{\mu \nu}^M + g C\indices{^M_{NP}} \mathcal{A}^N_\mu \mathcal{A}^P_\nu,
\ee
which transform in the adjoint representation as
\be
\delta \cF_{\mu \nu}^M = g C\indices{^M_{NP}} \cF^N_{\mu \nu} \mu^P \label{eq:adjoint}
\ee
and the ordinary derivatives $\partial_\mu \phi^i$ of the scalar fields are replaced by the covariant derivatives $D_\mu \phi^i$.  These contain linearly the undifferentiated vector potentials $\mathcal{A}^N_\mu$.  

The deformed action reads
\be
S = - \frac{1}{2} \int \dx[4] \, g_{ij}(\phi) D_\mu \phi^i D^\mu \phi^j + \frac{1}{4} \int \dx[4] \left(\Omega_{MN}\, \varepsilon^{ijk}\cF^M_{ij} \cF^N_{0k} - \mathcal{M}_{MN}(\phi) \cF^M_{ij} \cF^{Nij} \right)
\ee
The key observation now is that the kinetic term for the vector potentials is the same as in the absence of scalar fields.  Invariance of the kinetic term under the Yang-Mills gauge transformations  yields therefore the same conditions as when there is no $\phi^i$ (the kinetic term must be invariant by itself since the scalar Lagrangian is invariant and there can be no compensation with the energy density $\mathcal{M}_{MN}(\phi) \cF^M_{ij} \cF^{Nij}$ that contains no time derivative).  These conditions are that the adjoint representation \eqref{eq:adjoint} of the gauge group $G_g$ should preserve the symplectic form $\Omega_{MN}$ in internal electric-magnetic space \cite{Bunster:2010wv} and read explicitly
\be
C_{NPM} = C_{MPN} \label{eq:InvSymp}
\ee
with $C_{MNP} \equiv \Omega_{MQ}C\indices{^Q_{NP}}$. The symmetry of $C_{MNP}$ under the exchange of its first and last indices, and its antisymmetry in its last two indices, force it to vanish.  Hence, the structure constants $C\indices{^M_{NP}}$ must also vanish and there can be no non-abelian deformation of the gauge symmetries.  The Yang-Mills construction in which the potentials become non-abelian connections is unavailable in the manifestly duality invariant first-order formulation \cite{Bunster:2010wv}\footnote{Other types of deformations, for example adding functions of the abelian curvatures, are not excluded by the argument since they preserve the abelian nature of the gauge group.}.  In order to have access to these deformations, one must avoid the condition \eqref{eq:InvSymp} expressing the invariance of the symplectic form.  This is what is effectively achieved when one goes to the second-order formalism by choosing a Lagrangian submanifold, on which the pull-back of $\Omega_{MN}$ is by definition zero.

We also note that one may add extra scalar fields to the first-order action \eqref{eq:symlag} in order to have covariant gauge transformations $\delta \cA^M_\mu = \pd_\mu \mu^M$ \cite{Barnich:2017nty}, along the lines of \cite{Barnich:2007uu}. This does not help, however: it is proved in \cite{Barnich:2017nty} using the BRST formalism that non-abelian deformations are still obstructed.

\chapter{Introduction to BRST-BV deformation theory}
\label{chap:introbrst}

The gaugings considered above are a special case of smooth deformations, in which the original theory is recovered by sending the coupling constants to zero. One systematic way to explore deformations of theories with a gauge
freedom is provided by the BV-BRST field-antifield formalism \cite{Batalin:1981jr,Batalin:1984jr,Barnich:1993vg,Henneaux:1997bm}. This point of view has proved very powerful and has enabled the proof of a variety of uniqueness and no-go theorems (see \cite{Barnich:1993vg,Barnich:1993pa,Barnich:1994mt,Bautier:1997yp,Henneaux:1997ha,Bekaert:2000qx,Boulanger:2000rq,Bekaert:2001wa,Boulanger:2001wq,Bekaert:2002uh,Bekaert:2004dz,Boulanger:2008tg,Boulanger:2018fei} for a non-exhaustive list).

We briefly review this formalism in this chapter, following \cite{Barnich:1993vg,Henneaux:1997bm,Barnich:2017nty,Gomis:1994he}.
We focus on the following results on the local cohomology groups $H^k(s|d)$ of the BRST operator at various ghost numbers:
\begin{itemize}
\item consistent first-order deformations of the action are elements of $H^0(s|d)$;
\item obstructions to completing these to second-order are controlled by $H^1(s|d)$; and
\item global symmetries of the action are elements of $H^{-1}(s|d)$.
\end{itemize}
They will be used in the next chapter.

\section{Consistent deformations}

We begin with an action $I\up{0}$, depending on a collection of fields $\{ \varphi^a \}$.\footnote{We do not specify the type of fields at this stage: the $\varphi^a$ may include gauge fields or matter fields, gravity, etc. Although this is not necessary in the general case, we will nevertheless assume that the fields are bosonic and that there are no reducibility identities among the gauge transformations.} The action is invariant under some gauge transformations, written in a condensed notation as
\begin{equation}
\delta_\epsilon^{(0)} \varphi^a = \epsilon^\alpha R\indices{^{(0)a}_\alpha},
\end{equation}
where the parameters $\epsilon^\alpha$ are arbitrary functions of space-time. Invariance of the action means that it satisfies
\begin{equation}\label{eq:inv0}
\frac{\delta I^{(0)}}{\delta \varphi^a} R\indices{^{(0)a}_\alpha} = 0 \, .
\end{equation}
We now want to deform smoothly the action and gauge transformations,
\begin{align}
I\up{0} &\rightarrow I = I\up{0} + g\, I\up{1} + g^2\, I\up{2} + \dots, \label{eq:actiondef}\\
R\indices{^{(0)a}_\alpha} &\rightarrow R\indices{^a_\alpha} = R\indices{^{(0)a}_\alpha} + g\, R\indices{^{(1)a}_\alpha} + g^2\, R\indices{^{(2)a}_\alpha} + \dots, \label{eq:gaugedef}
\end{align}
in such a way that the deformed action is invariant under the deformed gauge symmetries, i.e.,
\begin{equation}\label{eq:fullinvariance}
\frac{\delta I}{\delta \varphi^a} R\indices{^a_\alpha} = 0 .
\end{equation}
This condition is then analysed order by order in the deformation parameter (coupling constant) $g$. Moreover, there are two kinds of trivial deformations, which do not really lead to a new theory:
\begin{enumerate}
\item The first comes from $g$-dependent gauge transformations, $\varphi^a \rightarrow \varphi'^a = \varphi^a + g F^a + O(g^2)$. This induces a deformation of the action,
\begin{equation}\label{eq:actionredef}
I\up{0}[\varphi'^a] = I\up{0}[\varphi^a] + g \frac{\delta I\up{0}}{\delta \varphi^a} F^a + O(g^2),
\end{equation}
and similar deformations for the gauge transformations if they were field-dependent (such as in the case of Yang-Mills, see \eqref{eq:YMgauge}).
\item The second one comes from redefinitions of the gauge transformations,
\begin{equation}\label{eq:trivgauge}
R\indices{^a_\alpha} = M\indices{_\alpha^\beta} R\indices{^{(0)a}_\beta} + t^{ab}_\alpha \frac{\delta I^{(0)}}{\delta \varphi^b},
\end{equation}
where $M$ is invertible and $t$ is antisymmetric in its $a,b$ indices. This produces a deformation of the type \eqref{eq:gaugedef} when $M$ and $t$ are $g$-dependent.
\end{enumerate}
The goal of deformation theory is to find all the solutions of \eqref{eq:fullinvariance}, order by order in $g$, while at the same time excluding the trivial deformations.

At order zero, \eqref{eq:fullinvariance} is just \eqref{eq:inv0}, i.e., the invariance of the original action under the original gauge transformations. It is satisfied by assumption. The first condition comes at order one in $g$: it is
\begin{equation}
\frac{\delta I^{(1)}}{\delta \varphi^a} R\indices{^{(0)a}_\alpha} + \frac{\delta I^{(0)}}{\delta \varphi^a} R\indices{^{(1)a}_\alpha} = 0.
\end{equation}
This condition states that the first-order deformation $I\up{1}$ of the action is invariant \emph{on-shell} under the undeformed gauge transformations $R\indices{^{(0)a}_\alpha}$,
\begin{equation}
\delta_\epsilon\up{0} I^{(1)} \approx 0,
\end{equation}
where $\approx$ means on-shell equality, i.e., equality up to a combination of the equations of motion\footnote{We assume the necessary regularity conditions on the action that ensure that any on-shell vanishing quantity is indeed a combination of the equations of motion, see \cite{Henneaux:1992ig,Gomis:1994he,Barnich:2000zw}.}. Moreover, $I\up{1}$ is trivial if and only if it vanishes on-shell (see \eqref{eq:actionredef}),
\begin{equation}
I\up{1} \approx 0 \qquad \text{(trivial solution)}.
\end{equation}
The solution at first order is therefore given by the classification of the non-trivial ``observables" of the undeformed theory, i.e., on-shell gauge-invariant quantities. This is is provided by the ghost number zero cohomology of the BRST operator $H^0(s)$, as we will see below.

\section{BRST differential}

The BRST differential $s$ acts in the extended space of the original fields $\varphi^a$ and several extra fields: ghosts $C^\alpha$ and antifields $\varphi^*_a$, $C^*_\alpha$. As the indices suggest, there are as many ghosts (and antighosts) as gauge transformations (in the case of reducible theories, there are also ghosts for ghosts, etc.). The ghosts have the opposite Grasmmann parity as their corresponding gauge parameters. Similarly, all antifields have the opposite parity of their conjugates.

We will now describe this differential for our case of interest, with $n_s$ scalar fields and $n_v$ abelian vectors in four dimensions. This will be sufficient for the purposes of chapter \ref{chap:vsgaugings}. Lots of other examples can be found in the review \cite{Gomis:1994he}.

\subsection{Structure of the models}
\label{sec:vsmodels}

We take the action and gauge invariances
\begin{equation}\label{eq:simplebrst}
I[\phi^i, A^I] = \int \dx[4] \, \cL, \quad \delta_\epsilon \phi^i = 0, \quad \delta_\epsilon A^I_\mu = \pd_\mu \epsilon^I ,
\end{equation}
where $\cL$ is a local function of the fields and a finite number of their derivatives that we will write as $\mathcal{L} = \mathcal{L}_S[\phi^i]+\mathcal{L}_V[A^I_\mu,\phi^i]$. In four spacetime dimensions, there is no Chern-Simons term in the Lagrangian, which can be assumed to be strictly gauge invariant and not just invariant up to a total derivative. Therefore, $\mathcal{L}_V[A^I_\mu,\phi^i]$ depends on the vector potentials $A^I_\mu$ only through $F^I_{\mu\nu}=\d_\mu A_\nu^I-\d_\nu A^I_\mu$ and their derivatives.
We define 
\begin{equation}
  \label{eq:47}
  \vddl{\mathcal L_V}{F^I_{\mu\nu}} = \half (\star G_I)^{\mu\nu}\, ,
\end{equation}
where the $(\star G_I)^{\mu\nu}$ are also manifestly gauge invariant functions.
The equations of motion for the vector fields can then be written as 
\begin{equation}
  \label{eq:4bis}
  \vddl{\mathcal L_0}{A_\mu^I}=\d_\nu(\star G_I)^{\mu\nu} .
\end{equation}
Note that the Lagrangian \eqref{eq:lag} falls into this general class of models, with the gauge invariant two-form $G_I=\mathcal I_{IJ}\star F^J+\mathcal R_{IJ} F^J$ and $d^4x\, \mathcal{L}_V = \frac{1}{2} G_I F^I$.

The spectrum of fields and antifields is thus
\begin{equation}
\{ \; \phi^i, \; A^I_\mu, \; C^I, \; \phi^*_i, \; A^{*\mu}_I, \; C^*_I \; \}\, ,
\end{equation}
where the indices run as $i = 1, \dots, n_s$, $I = 1, \dots, n_v$. Among these, the Grassmann odd variables are the $C^I$, $\phi^*_i$ and $A^{*\mu}_I$; the other are ordinary (commuting) variables.  In this simple case, the differential $s$ is given by the sum of two pieces,
\begin{equation}
s = \gamma + \delta .
\end{equation}
They satisfy
\begin{equation}
\gamma^2 = 0, \quad \delta^2 = 0
\end{equation}
(they are also differentials), and they anticommute,
\begin{equation}
\gamma \delta + \delta \gamma = 0 .
\end{equation}
This last property implies that $s$ is indeed a differential,
\begin{equation}
s^2 = 0 .
\end{equation}
The differential $\gamma$ is called the ``derivative along the gauge orbits" \cite{Henneaux:1992ig} and contains the information about the gauge transformations. It is defined on the fields and antifields by
\begin{align}
&\gamma \phi^i = 0,& &\gamma A^I_\mu = \pd_\mu C^I,& &\gamma C^I = 0 , \nn \\
&\gamma \phi^*_i = 0,& &\gamma A^{*\mu}_I = 0,& &\gamma C^*_I = 0
\end{align}
and takes the form ``gauge transformation with parameters replaced by ghosts".
The differential $\delta$ is called the ``Koszul-Tate differential" \cite{Fisch:1989rp,Henneaux:1992ig} and contains the information about the equations of motion. It is defined on the basic variables as
\begin{align}
&\delta \phi^i = 0,& &\delta A^I_\mu = 0,& &\delta C^I = 0, \nn \\ &\delta \phi^*_i = \frac{\delta \cL}{\delta \phi^i},& &\delta A^{*\mu}_I = \frac{\delta \cL}{\delta A^I_\mu},& &\delta C^*_I = - \d_\mu A^{*\mu} .
\end{align}
These definitions are extended to arbitrary functions of all the fields and their derivatives by requiring that $\gamma$ and $\delta$
\begin{enumerate}
\item act as (graded) derivations, and
\item commute with the ordinary derivative, $[\gamma, \pd_\mu] = 0 = [\delta, \pd_\mu]$.
\end{enumerate}
Since the one-forms $dx^\mu$ are Grassmann-odd, the second property implies that $\gamma$ and $\delta$ anticommute with the exterior derivative,
\begin{equation}
d \gamma + \gamma d = 0 = \delta d + d \delta .
\end{equation}
The same properties are satisfied by $s$.
\begin{table}
\centering
\begin{tabular}{c|cccc}
& pgh & afd & gh & gr \\ \midrule
$\phi^i$, $A^I_\mu$ & 0 & 0 & 0 & even \\
$C^I$ & 1 & 0 & 1 & odd \\ \midrule
$\phi^*_i$, $A^{*\mu}_I$ & 0 & 1 & -1 & odd \\
$C^*_I$ & 0 & 2 & -2 & even \\ \midrule
$\gamma$ & 1 & 0 & 1 & odd \\
$\delta$ & 0 & -1 & 1 & odd \\
$s$ & - & - & 1 & odd \\
$d$ & 0 & 0 & 0 & odd
\end{tabular}
\caption[Gradings of BRST-BV fields, antifields and differentials]{Pure ghost number, antifield number, ghost number and Grassmann parity of the fields, antifields and differentials appearing in the BRST-BV formalism.} \label{tab:gradings}
\end{table}
Besides Grassmann parity, it is useful to introduce the following gradings:
\begin{itemize}
\item the pure ghost number, which is one for the ghosts $C^I$ and zero otherwise,
\item the antifield number (also called antighost number), which is one for $\phi^*_i$, $A^{*\mu}_I$, two for $C^*_I$ and zero otherwise, and
\item the ghost number, which is the difference between the two.
\end{itemize}
The various gradings of the fields, ghosts, their antifields and the differentials are collected in table \ref{tab:gradings}. An important property of the BRST differential $s$ is that increases the ghost number by one.

For all these differentials, the (co)homology groups are defined in the usual way. We will also use the following notation: if $d_1$ and $d_2$ are two anticommuting differentials, i.e., $d_1^2 = d_2^2 = d_1 d_2 + d_2 d_1 = 0$, the cohomology of $d_1$ ``modulo $d_2$" is written $H(d_1|d_2)$ and is defined by
\begin{equation}
d_1 \omega + d_2 \rho = 0, \quad \omega \sim \omega + d_1 \alpha + d_2 \beta .
\end{equation}
The main example will be the cohomology groups $H(s|d)$.

Finally, for functions $f[\varphi]$ that depend only on the usual fields and their derivatives (but not on the ghosts and antifields), the definitions above imply the following useful reformulations:
\begin{enumerate}
\item $f$ is gauge-invariant if and only if it is $\gamma$-closed,
\begin{equation}
\delta_\epsilon f = 0 \; \Leftrightarrow \; \gamma f = 0 ;
\end{equation}
\item the regularity assumptions on the action ensure that any on-shell vanishing quantity is a combination of the equations of motion and is therefore $\delta$-exact,
\begin{equation}\label{eq:onshelldelta}
f \approx 0 \; \Leftrightarrow \; f = t^a \frac{\delta \cL}{\delta \varphi^a} \; \Leftrightarrow \; f = \delta \!\left( t^a \varphi^*_a \right) .
\end{equation}
\end{enumerate}

\subsection{Local deformations and $H^0(s|d)$}

Let us come back to the deformation problem. As we have seen, the first-order deformation $I\up{1}$ to the action should be gauge-invariant on-shell. With the above definitions, this is equivalent to
\begin{equation}\label{eq:gammas1}
\gamma I\up{1} + \delta A_1 = 0
\end{equation}
for some $A_1$ linear in the antifields $\phi^*_i$ and $A^{*\mu}_I$. Moreover, this solution is trivial if and only if $I\up{1}$ vanishes on-shell, i.e., can be written as
\begin{equation}\label{eq:deltab1}
I\up{1} = \delta A'_1 \qquad \text{(trivial solution)}
\end{equation}
for some $A'_1$ of antifield number one. Now, it is a standard result of the BRST formalism \cite{Henneaux:1992ig,Barnich:1994db} that any solution to \eqref{eq:gammas1} determines uniquely quantities $A_2$, $A_3$, $\dots$ of increasing antifield number such that
\begin{equation}
S\up{1} = I\up{1} + A_1 + A_2 + \dots
\end{equation}
is a BRST-cocycle, i.e., satisfies
\begin{equation}
s S\up{1} = 0 .
\end{equation}
On the other hand, trivial solutions of the form \eqref{eq:deltab1} correspond to BRST-exact terms of the form
\begin{equation}
S\up{1} = s B .
\end{equation}
Expanding those equations in antifield number, we recover equations \eqref{eq:gammas1} and \eqref{eq:deltab1} at antifield number zero. As announced, consistent first-order deformations correspond to elements of the cohomology of $s$ at ghost number zero,
\begin{equation}
[S\up{1}] \in H^0(s)\, ,
\end{equation}
the superscript denoting ghost number\footnote{Note also that ghost number zero cohomology is also relevant to the study of renormalisability of gauge theories, $\hbar$ playing the role of the deformation parameter.}. If one imposes locality of the deformation, i.e., that $I\up{1}$ is the integral of some local functional $\cL\up{1}$, the relevant space becomes $H^0(s|d)$ \cite{Barnich:1993vg} (cohomology of $s$ modulo $d$ at ghost number zero).
This space is defined by
\begin{equation}
s a + d b = 0, \quad a \sim a + s a' + db',
\end{equation}
where we wrote $S\up{1} = \int a$ (we ignore surface terms). This is what we will consider here.

\subsection{Symmetries and $H^{-1}(s|d)$}
\label{sec:symmetriesH-1}

In the previous chapter, we have seen that Yang-Mills type deformations are closely linked to the global symmetries of the theory. It will therefore be important to understand how the global symmetries fall into the BRST formalism. It this section, we explain that they are elements of the cohomology $H^{-1}(s|d)$ at ghost number $-1$ and form degree $n$, and present the cohomological version of Noether's theorem. We follow \cite{Barnich:1994db,Barnich:2000zw}, to which we refer for proofs and further details.

Inequivalent global symmetries $\delta_Q \varphi^a = Q^a(\varphi)$ of the action are defined by functions $Q^a$ such that
\begin{equation}\label{eq:globalsym}
Q^a \frac{\delta \cL}{\delta \varphi^a} + \pd_\mu j^\mu = 0 .
\end{equation}
They are defined only up to a (possibly trivial) gauge transformation (see \eqref{eq:trivgauge}),
\begin{equation}\label{eq:globalsymambiguity}
Q^a \sim Q^a + M\indices{^a_b^\alpha} R\indices{^b_\alpha} + t^{[ab]} \frac{\delta \cL}{\delta \varphi^b} .
\end{equation}
We now introduce the $n$-form of antifield number $1$ (independent of the ghosts)
\begin{equation}\label{eq:omegaQ}
\omega_1^n = d^n\!x\, Q^a \varphi^*_a .
\end{equation}
The condition \eqref{eq:globalsym} on the global symmetries is then
\begin{equation}\label{eq:deltaomega}
\delta \omega_1^n + d \omega^{n-1}_0 = 0 ,
\end{equation}
where $\omega^{n-1}_0$ is determined by the current $j^\mu$ as $\omega_0^{n-1} = (dx^{n-1})_\mu j^\mu$ (see appendix \ref{app:conv2} for $p$-form conventions). At the level of $\omega_1^n$, the ambiguity \eqref{eq:globalsymambiguity} is
\begin{equation}
\omega_1^n \sim \omega_1^n + \delta \rho_2^n + d\omega_1^{n-1}.
\end{equation}
Thus, global symmetries determine equivalence classes of $H^n_1(\delta | d)$, where the superscript is the form degree and the subscript the antifield number. The converse is also true: any cocycle $\omega_1^n$ of this group can be brought in the form \eqref{eq:omegaQ} and therefore determines a global symmetry of the action \cite{Barnich:1994db,Barnich:2000zw}.

Then, the isomorphism
\begin{equation}
H^n_1(\delta | d) \simeq H^{-1}(s|d)
\end{equation}
shows that global symmetries correspond to elements of $H^{-1}(s|d)$ at form degree $n$ \cite{Barnich:1994db}. This isomorphism is proved using the same kind of reasoning (``homological perturbation theory") as in the previous section: $\omega_1^n$ determines uniquely $n$-forms $\omega_2$, $\omega_3$, $\dots$ of increasing antifield number such that
\begin{equation}
a = \omega^n_1 + \omega_2 + \omega_3 + \dots 
\end{equation}
satisfies $sa + db = 0$, i.e., is an element of $H^{-1}(s|d)$ (the ghost number of $a$ is indeed minus one, see \eqref{eq:omegaQ} and table \ref{tab:gradings}). Equation \eqref{eq:deltaomega} is then recovered from $sa + db = 0$ at lowest antifield number.

\subsubsection*{Cohomological version of Noether's theorem}

The inequivalent conserved (Noether) currents of the theory are defined by vectors $j_\mu$, which are functions of the fields and their derivatives, such that
\begin{equation}\label{eq:conservedcurrent}
\pd_\mu j^\mu \approx 0
\end{equation}
up to the trivial redefinitions
\begin{equation}\label{eq:conservedcurrentambiguity}
j^\mu \sim j^\mu + \pd_\nu k^{[\mu\nu]} + t^\mu \quad (t^\mu \approx 0)
\end{equation}
(the symbol $\approx$ denotes on-shell equality, as before). The condition $\pd_\mu j^\mu \approx 0$ is of course the same equation as the symmetry equation \eqref{eq:globalsym}, since the first term of \eqref{eq:globalsym} is a combination of the equations of motion. In fact, (equivalence classes of) global symmetries are in one-to-one correspondence with (equivalence classes of) conserved currents: this is the well-known theorem of Noether. Let us now rephrase this discussion in cohomological terms.

We first introduce the characteristic cohomology spaces $H^p_{\rm char}(d)$. They are defined by $p$-forms $\omega^p$, depending on the original fields $\varphi^a$ and their derivatives, such that
\begin{equation}
  \label{eq:10}
  d\omega^{p}\approx 0,\quad 
  \omega^{p} \sim \omega^{p} + d \eta^{p-1} + t^{p},
\end{equation}
with $t^{p}\approx 0$ and where the superscript denotes form degree. In dual notations, this is exactly the conditions \eqref{eq:conservedcurrent}, \eqref{eq:conservedcurrentambiguity} for the $(n-1)$-form $\omega_0^{n-1} = (dx^{n-1})_\mu j^\mu$. Therefore, conserved currents are determined by elements of $H^{n-1}_{\rm char}(d)$. In the BRST language, equations \eqref{eq:conservedcurrent} and \eqref{eq:conservedcurrentambiguity} are, using property \eqref{eq:onshelldelta} above,
\begin{equation}
d \omega_0^{n-1} + \delta \rho_1^{n-2} = 0, \quad \omega \sim \omega + d\eta + \delta \rho',
\end{equation}
for some $\rho$, $\rho'$ of antifield number $1$. Conserved currents are then classified by $H^{n-1}_{\rm char}(d) \simeq H^{n-1}_0(d|\delta)$. In the latter group, superscript refers to form degree and subscript to antifield number. In this language, Noether's theorem is reformulated as the isomorphism \cite{Barnich:1994db,Barnich:2000zw}
\begin{equation}
H^{n-1}_0(d|\delta) \simeq H^n_1(\delta | d) .
\end{equation}

\section{Antibracket and obstructions}
\label{sec:antibracket}

The structure of a gauge system (Lagrangian, gauge transformations, gauge algebra, ...) is completely captured by the
Batalin-Vilkovisky (BV) master action $S$ \cite{Batalin:1981jr,Batalin:1984jr}. It is a ghost number zero functional
\begin{equation}
  \label{eq:masteraction}
  S[\varphi^a, C^\alpha, \varphi^*_a, C^*_\alpha] = \int \!d^n\!x\, \left[ \cL + \varphi^*_{a}
    R\indices{^{a}_\alpha} \,  
    C^\alpha + \frac{1}{2}C^*_\alpha
    f\indices{^\alpha_{\beta\gamma}} C^\beta C^\gamma + \dots  \right].
\end{equation}
It starts with the usual action $I$, which is of course independent of ghosts and antifields, but also contains higher-order terms that contain explicit information about the gauge structure. In the case of \eqref{eq:simplebrst}, the master action is simply
\begin{equation}
S = \int \!d^n\!x\, \left( \cL + A^{*\mu}_I \pd_\mu C^I \right) .
\end{equation}
In general, the original action and gauge transformations are recovered from the master action by setting the antifields to zero,
\begin{equation}
I = S[\varphi^a, \varphi^*_{a} = 0 ], \quad \delta_\epsilon \varphi^a = \frac{\delta S}{\delta \varphi^*_a}[\varphi^a, \varphi^*_{a} = 0,\, C^\alpha \rightarrow \epsilon^\alpha ] .
\end{equation}
The master action satisfies the ``master equation"
\begin{equation}
\label{eq:masterequation}
(S,S)=0\, ,
\end{equation}
which completely encodes the consistency of the gauge structure. In this equation, $(\cdot, \cdot)$ (the ``antibracket") is the odd graded Lie bracket defined by
\begin{equation}
  \label{eq:antibracket}
  (X,Y)= \int\!d^n\!x\, \left[ \frac{\delta^R X}{\delta \Phi^A(x)}
\frac{\delta^L Y}{\delta \Phi^*_A(x)}-\frac{\delta^R X}{\delta 
\Phi^*_A(x)}\frac{\delta^L Y}{\delta \Phi^A(x)} \right],
\end{equation}
where we wrote $\Phi^A=\{ \varphi^{a}, \, C^\alpha \}$ for the
original fields and ghosts and $\Phi^*_A$ for their antifields. With these definitions, the BRST differential can be written as
\begin{equation}
s = (S, \cdot )
\end{equation}
and the nilpotency of $s$ (i.e., $s^2 = 0$) follows from the graded Jacobi identity satisfied by the antibracket.

Moreover, the antibracket gives rise to a well defined map in cohomology,
\begin{equation}
  \label{eq:31a}
  (\cdot,\cdot): H^{g_1}(s)\times H^{g_2}(s) \longrightarrow H^{g_1+g_2+1}(s).
\end{equation}
For cocycles $C_i$ with $[C_i]\in H^{g_i}(s)$, it is explicitly given
by
\begin{equation}
  \label{eq:44}
  ([C_1],[C_2])=[(C_1,C_2)] \in H^{g_1+g_2+1}(s). 
\end{equation}
When $g_1=-1=g_2$, we have $(\cdot,\cdot) : H^{-1} \times H^{-1}\to H^{-1}$ and the antibracket map encodes the Lie algebra structure of the
inequivalent global symmetries \cite{Barnich:1996mr}. More generally,
it follows from $(\cdot,\cdot) : H^{-1} \times H^{g}\to H^{g}$ that,
for any ghost number $g$, the BRST cohomology classes form a
representation of the Lie algebra of inequivalent global symmetries. This will be crucial in chapter \ref{chap:vsgaugings}.

The deformation problem can be compactly rephrased as follows: we start with the master action $S\up{0}$ of the undeformed theory and deform it to
\begin{equation}
S\up{0} \rightarrow S = S\up{0} + g\, S\up{1} + g^2\, S\up{2} + \dots
\end{equation}
in such a way that the master equation $(S,S) = 0$ is satisfied. The (usual) action $I$ and gauge transformations $R\indices{^a_\alpha}$ of the deformed theory (as in \eqref{eq:actiondef} and \eqref{eq:gaugedef}) can then be read off the final master action $S$. Expanding the master equation in powers of $g$ gives the equations
\begin{align}
(S\up{0}, S\up{0}) &= 0, \\
(S\up{0}, S\up{1}) &= 0, \\
(S\up{0}, S\up{2}) + \frac{1}{2} (S\up{1}, S\up{1}) &= 0, \label{eq:43} \\
&\dots \, .\nn
\end{align}
The first equation is satified by assumption (it is just the consistency equation for the starting point). The second equation tells us the $S\up{1}$ is an element of $H^0(s\up{0})$, where $s\up{0}$ is the BRST operator of the undeformed theory, as we saw above. The third constrains the infinitesimal
deformation $S^{(1)}$ to satisfy
\begin{equation}
  \label{eq:45}
  ([S^{(1)}],[S^{(1)}]) = [0] \in H^1(s\up{0}).
\end{equation}
If this is the case, $S^{(2)}$ in \eqref{eq:43} is defined up to a
cocycle in ghost number $0$. If this is not the case, there is an obstruction in extending the first-order deformation to a full one. Thus, it is the group $H^1(s\up{0})$ that controls the obstructions. Actually, it was proved in \cite{Barnich:1993vg} that the condition \eqref{eq:45} is always satisfied: there never is an obstruction to completing the first-order deformation $S\up{1}$. Nevertheless, this completion is in general highly non-local: obstructions can appear when one imposes locality (as we will do here), i.e., when the groups $H^k(s\up{0}|d)$ are considered instead of $H^k(s\up{0})$. In the following, we will omit the superscript on the BRST operator $s\up{0}$, since we work only with the operator of the undeformed theory.

\section{Length and depth}
\label{sec:lengthdepth}

There are two main approaches to solving the equation
\begin{equation}\label{eq:cocycleintro}
s a + d b = 0
\end{equation}
characterizing an element $a$ of the local BRST cohomology $H^k(s|d)$ at ghost number $k$. They define two integers associated with $a$, the length and the depth.

The first one arises by decomposing $a$ according to an expansion in antifield number,
\begin{equation}
a = a_0 + a_1 + \dots + a_L ,
\end{equation}
where each $a_i$ has antifield number $i$. (In the cases considered above, it can be proved that this expansion indeed terminates and that, moreover, we always have $L \leq 2$. See \cite{Barnich:1994mt}.) We will call the integer $L$ the \emph{length} of the cocycle $a$. Doing a similar expansion for $b$ and writing $s = \delta + \gamma$, the cocycle condition $s a + db = 0$ becomes
\begin{align}
\gamma a_L + d b_L &= 0, \label{eq:lengtheqs} \\
\delta a_L + \gamma a_{L-1} + d b_{L-1} &= 0, \\
&\dots \nn\\
\delta a_1 + \gamma a_0 +  d b_0 &= 0.
\end{align}
To solve these equations, one starts from the top to determine the general solution for $a_L$, then works their way down iteratively. In the case of Yang-Mills theories, this is the method followed in \cite{Barnich:1994mt}. It is also the one we will follow in the next chapter.

Another method comes by noticing that, acting with $s$ on \eqref{eq:cocycleintro}, we have $s (d b) = - d ( sb ) = 0$. Therefore, the algebraic Poicaré lemma \cite{Henneaux:1992ig} implies that there exists a $c$ such that $s b + d c = 0$. The form degree of $c$ is one less than that of $b$, which is itself one less than that of $a$. Therefore, indicating the form degree by a superscript, equation \eqref{eq:cocycleintro} is equivalent to the chain of ``descent equations"
\begin{align}
s a^p + d a^{p-1} &= 0, \\
s a^{p-1} + d a^{p-2} &= 0, \\
&\dots \nn \\
s a^{p-d} &= 0 .
\end{align}
The integer $d$ characterizing the shortest non-trivial descent defines the \emph{depth} of the cocycle $a$ (it is bounded by the space-time dimension). To solve these equations, one now starts form the bottom and goes upwards; this is the method followed in \cite{Barnich:2000zw} for Yang-Mills theories. A useful theorem involving the relation between the antibracket map and the depths of its argument is proved in appendix \ref{sec:antibr-maps-desc}. It states that the depth of the image of the antibracket map is always inferior or equal to the depth of its most shallow argument,
\begin{equation}
\text{depth} (C_1, C_2) \leq \min \{ \text{depth}(C_1), \, \text{depth}(C_2) \}.
\end{equation}
This will be useful in chapter \ref{chap:vsgaugings}.

\chapter{Deformations in a definite symplectic frame}
\label{chap:vsgaugings}

In this chapter, we answer the second question raised in the introduction of chapter \ref{chap:YMgaugings}, i.e., whether the gaugings of the Yang-Mills type are the most general deformations of the supergravity-inspired Lagrangian \eqref{eq:lag}. The uniqueness of the Yang-Mills coupling has of course been proven long ago (see for example \cite{Deser:1963zzc,Ramond:1986vt,Barnich:1993pa}) but couplings to nonlinear scalar fields were not considered in these early works. We show that the answer to this question is, essentially, affirmative: besides the obvious deformations that
consist in adding gauge invariant terms to the Lagrangian without
changing the gauge symmetries, all deformations are gaugings of the Yang-Mills type related to the
global symmetries of the original action.

In so doing, we completely characterize the BRST cohomology classes (with non trivial antifield dependence) for a general class of vector-scalar Lagrangians in four dimensions (see section \ref{sec:vsmodels}), of which \eqref{eq:lag} is a particular case. The above result is obtained at ghost number zero. The link with conserved currents and rigid symmetries is explicitely done; a classification of the global symmetries in $U$, $V$ and $W$ type (defined below) arises naturally in the study of the cohomology.

The results of this chapter were presented in the paper \cite{Barnich:2017nty}, written in collaboration with G. Barnich, N. Boulanger, B. Julia, M. Henneaux and A. Ranjbar.

\section{Consistent deformations: ghost number zero}

We illustrate explicitly the procedure for $H^0(s \vert d)$ in maximum
form degree, which defines the consistent local deformations, as reviewed in the previous chapter. We consider the case of general ghost number in the next section.

The main equation to be solved for $a$ is 
\begin{equation}
\label{eq:cocycle}
sa + db = 0,
\end{equation}
 where $a$ has form degree $4$ and ghost number $0$.  To solve it, we expand the cocycle $a$
according to the antifield number, \be a = a_0 + a_1 + a_2. \ee 
As shown in \cite{Barnich:1994mt}, the expansion stops at most at
antifield number $2$. Because $a$ has total ghost number zero, each term $a_n$ has antifield number $n$ and pure ghost number $n$ as well. The interpretation of these three terms is the following: $a_0$ is the (first order) deformation
of the Lagrangian, a non-vanishing $a_1$ corresponds to a deformation of the gauge variations, and a non-vanishing $a_2$ corresponds to a deformation of the gauge algebra. All three terms are related by the cocycle condition \eqref{eq:cocycle}.

\subsection{Solutions of $U$-type ($a_2$ non trivial)}

The first case to consider is when $a_2$ is non-trivial.  This defines
``class I'' solutions in the terminology of \cite{Barnich:1994mt},
which we call here ``$U$-type'' solutions to comply with the general
terminology introduced below.  One has from general theorems on the invariant characteristic
cohomology \cite{Barnich:1994db,Barnich:1994mt} that \be a_2 = d^4\!x\,C^*_I\,  \Theta^I \ee with \be
\Theta^I=\frac{1}{2!}{f^I}_{J_1J_{2}} C^{J_1} C^{J_{2}}\, .  \ee 
Here
${f^I}_{J_1J_{2}}$ are some constants, antisymmetric in $J_1$, $J_2$.
The reason why the coefficient $d^4 x\,C^*_I $ of the ghosts in $a_2$
is determined by the characteristic cohomology follows from the
equation $\delta a_2 + \gamma a_1 + db_1 =0$ that $a_2$ must fulfill
in order for $a$ to be a cocycle of $H(s \vert d)$.  Given that $a_2$
has antifield number equal to $2$, it is the 
characteristic cohomology
in form degree $n-2=2$ that is relevant. (We refer to
\cite{Barnich:1994db,Barnich:1994mt} for the further details.)  The emergence
of the characteristic cohomology in the computation of $H(s \vert d)$
will be observed again for $a_1$ below, where it will be the conserved
currents that appear.

We now consider the equation $\delta a_2 + \gamma a_1 + d b_1 = 0$ to find $a_1$. By the argument of \cite{Barnich:1994mt}, the term $a_1$ is then
\be a_1 = \star A^*_I A^{K}\d_{K}\Theta^I + m_1 \, , \ee
where $\gamma m_1 =0$ and $\d_K=\ddl{}{C^K}$.  
The term $m_1$ (to be determined by the next
equation) is linear in $C^I$ and can be taken to be linear in the
undifferentiated antifields $A^*_I$ and $\phi^*_i$ since derivatives
of these antifields, which can occur only linearly, can be redefined
away through trivial terms.  We thus write \be m_1 = \hat K=\star
A^*_I \hat g^I -\star \phi^*_i \hat \Phi^i\, , \ee with \be \hat
g^I=dx^\mu g^{I}_{\mu K}C^{K}\,, \quad~ \hat \Phi^i=\Phi^{i}_{K}C^{K}\, .
\ee Here $g^{I}_{\mu K}$ and $\Phi^{i}_{K}$ are gauge invariant
functions, arbitrary at this stage, but which will be constrained by
the requirement that $a_0$ exists.

We must now consider the equation $\delta a_1 + \gamma a_0 + d b_0 = 0$
that determines $a_0$ up to a solution of $\gamma a'_0 + db'_0 =0$.
This equation is equivalent to
 \begin{equation}
 \label{eq:00}
  \left(\vddl{\mathcal L_0}{A_\mu^I}\delta_{K} A_\mu^I
  +\vddl{\mathcal
    L_0}{\phi^i}\delta_{K}\phi^i \right) C^K+\gamma \alpha_0 + \d_\mu \beta_0^\mu=0, 
\end{equation}
where we have passed to dual notations ($a_0 = d^4 x \,\alpha_0$,
$db_0 = d^4 x \,\d_\mu \beta_0^\mu$) and where we have set
\begin{equation}
 \delta_{K} A_\mu^I=A^J_\mu {f^I}_{JK}+g_{\mu
    K}^I  ,\quad \delta_{K}
  \phi^i=\Phi^i_{K}.  \label{eq:TypeISymmb}
\end{equation}
Writing $\beta_0^\mu = j^\mu_K C^K + $ ``terms containing derivatives
of the ghosts'', we read from (\ref{eq:00}), by comparing the
coefficients of the undifferentiated ghosts, that \be \vddl{\mathcal
  L_0}{A_\mu^I}\delta_{K} A_\mu^I +\vddl{\mathcal
  L_0}{\phi^i}\delta_{K}\phi^i + \d_\mu j_K^\mu=0 .  \label{eq:3.19} \ee A necessary
condition for $a_0$ (and thus $a$) to exist is therefore that
$ \delta_{K} A_\mu^I$ and $ \delta_K \phi^i$ define symmetries.
 
 To proceed further and determine $a_0$, we observe that the non-gauge
 invariant term $ \vddl{\mathcal L_0}{A_\mu^I} A^J_\mu {f^I}_{JK}$ in
 $\vddl{\mathcal L_0}{A_\mu^I}\delta_{K} A_\mu^I$ can be written as
 $ \partial_\mu \left( \star G^{\nu\mu}_I A_\nu^J {f^I}_{JK}\right) $
 plus a gauge invariant term, so that
 $j^\mu_K - \star G^{\mu\nu}_I A_\nu^J {f^I}_{JK}$ has a
 gauge invariant divergence. Results on the invariant cohomology of
 $d$ \cite{Brandt:1989gy,DuboisViolette:1992ye} imply then that the
 non-gauge invariant part of such an object can only be a Chern-Simons
 form, i.e.,
 $j^\mu_K - \star G^{\mu\nu}_I A_\nu^J {f^I}_{JK} = J^\mu_{K} + \half
 \epsilon^{\mu\nu\rho\sigma}A_\nu^I F_{\rho\sigma}^J h_{I|JK}$ or
\begin{equation}
  j^\mu_{K}
  =J^\mu_{K}+\star G^{\mu\nu}_I
  A_\nu^J {f^I}_{JK} + \half
  \epsilon^{\mu\nu\rho\sigma}A_\nu^I F_{\rho\sigma}^J 
  h_{I|JK}\, ,
\end{equation}  
where $J^\mu_{K}$ is gauge invariant and where the symmetries of the
constants $h_{I|JK}$ will be discussed in a moment.  It is useful to
point out that one can switch the indices $I$ and $J$ modulo a trivial
term.

The equation (\ref{eq:00}) becomes
$-(\partial_\mu j^\mu_K) \, C^K + \gamma \alpha_0 + \partial_\mu
\beta_0^\mu = 0$, i.e.,
$j^\mu_K \, (\gamma A_\mu^K) + \gamma \alpha_0 + \partial_\mu
{\beta'}_0^\mu = 0$.  The first two terms in the current yield
manifestly $\gamma$-exact terms,
\begin{equation}
J^\mu_K \,  (\gamma A_\mu^K) = \gamma(J^\mu_K \,   A_\mu^K), \; \;
\; \star G^{\mu\nu}_I 
A_\nu^J {f^I}_{JK} \, (\gamma A_\mu^K)= \frac12 \gamma (\star
G^{\mu\nu}_I A_\nu^J {f^I}_{JK} \, A_\mu^K)
\end{equation} 
and so $h_{I|JK}$ must be such that the term $A^I F^J dC^K h_{I|JK}$
is by itself $\gamma$-exact modulo $d$. This is a problem that has
been much studied in the literature through descent equations (see
e.g. \cite{DuboisViolette:1985cj}).  It has been shown that
$h_{I|JK}$ must be antisymmetric in $J$, $K$ and should have vanishing
totally antisymmetric part in order to be ``liftable" to $a_0$ and
non-trivial,
\begin{equation}
  h_{I|JK}=h_{I|[JK]},\quad  h_{[I|JK]}=0. \label{eq:Symmh}
\end{equation}
Putting things together, one finds for $a_0$
\be
a_0 = A^I\d_I\hat J + \half G_I
  A^{K}A^{L}\d_{L}\d_{K}\Theta^I + \half F^I
  A^KA^L\d_L\d_K\Theta'_I \,,
\ee
where
\be
\hat J=\star dx^\mu J_{\mu K}
    C^{K}\;,\quad \Theta'_I=\frac{1}{2}h_{I|J_1 J_2}C^{J_1} C^{ J_{2}} .
\ee

A non-trivial $U$-solution
modifies the gauge
algebra. It is characterized by constants ${f^I}_{J_1J_{2}}$ which
are antisymmetric in $J_1$, $J_2$. These constants must be such that
there exist gauge invariant functions $g^{I}_{\mu K}$ and
$\Phi^{i}_{K}$ such that $ \delta_{K} A_\mu^I$ and $ \delta_K \phi^i$
define symmetries of the undeformed Lagrangian.  Here
$ \delta_{K} A_\mu^I$ and $ \delta_K \phi^i$ are given by
\eqref{eq:TypeISymmb}.  Furthermore, the $h$-term in the corresponding
conserved current (if any) must fulfill \eqref{eq:Symmh}.

Given the ``head'' $a_2$ of a $U$-type solution, characterized by a set of ${f^I}_{J_1J_{2}}$'s,  the lower terms $a_1$ and $a_0$, and in particular the $h$-piece, are not uniquely determined.  One can always add solutions of $W$, $V$ or $I$-types described below, which have the property that they have no $a_2$-piece.  
Hence one may require that the completion of the ``head'' $a_2$ of a $U$-type solution should be chosen to vanish when $a_2$ itself vanishes. But this leaves some freedom in the completion of $a_2$, since for instance any $W$-type solution multiplied by a component of ${f^I}_{J_1J_{2}}$ will vanish when the ${f^I}_{J_1J_{2}}$'s are set to zero. The situation has a triangular nature since two $U$-type solutions with the same $a_2$ differ by solutions of ``lower'' types, for which there might not be a canonical choice.
 
Further constraints on ${f^I}_{J_1J_{2}}$ (notably the
 Jacobi identity) will arise at second order in the deformation parameter.

\subsection{Solutions of $W$ and $V$-type (vanishing $a_2$ but
   $a_1$ non trivial)}

These solutions are called ``class II'' solutions in \cite{Barnich:1994mt}.

We now have \be a = a_0 + a_1 \ee and $a_1$ can be taken to be gauge
invariant, i.e., annihilated by $\gamma$ \cite{Barnich:1994mt}.  We
thus have \be a_1 = \hat K=\star A^*_I \hat g^I -\star \phi^*_i \hat
\Phi^i \,, \ee with \be \hat g^I=dx^\mu g^{I}_{\mu K}C^{K}\;, \quad~ \hat
\Phi^i=\Phi^{i}_{K}C^{K} \, . \ee Here $g^{I}_{\mu K}$ and $\Phi^{i}_{K}$
are again gauge invariant functions, which we still denote by the same
letters as above, although they are independent from the similar
functions related to the constants ${f^I}_{J_1J_{2}}$.  We also set
\begin{equation}
 \delta_{K} A_\mu^I=g_{\mu
    K}^I  ,\quad \delta_{K}
  \phi^i=\Phi^i_{K}.  \label{eq:DefKbb} 
\end{equation}

The equation $\delta a_1 + \gamma a_0 + d b_0 = 0$ implies then, as
above,
\begin{equation} \vddl{\mathcal L_0}{A_\mu^I}\delta_{K} A_\mu^I
  +\vddl{\mathcal L_0}{\phi^i}\delta_{K}\phi^i + \d_\mu j_K^\mu=0\,. \label{eq:3.28}
\end{equation}
A necessary condition for $a_0$ (and thus $a$) to exist is
therefore that $ \delta_{K} A_\mu^I$ and $ \delta_K \phi^i$ given by
\eqref{eq:DefKbb} define symmetries of the original action. Equation \eqref{eq:3.28} take the same form as \eqref{eq:3.19}, but there is one important difference: the divergence of the current  $j^\mu_{K}$ is now gauge invariant, while it is {\em not} in \eqref{eq:3.19} due to the contribution coming from $a_2$.

The current takes this time the form
 \begin{equation}
  j^\mu_{K}
  =J^\mu_{K}+ \half
  \epsilon^{\mu\nu\rho\sigma}A_\nu^I F_{\rho\sigma}^J 
  h_{I|JK}
\end{equation}
(with $h_{I|JK}$ fulfilling the symmetry properties \eqref{eq:Symmh} as above), yielding
\be a_0 = A^I\d_I\hat J + \half F^I A^KA^L\d_L\d_K\Theta'_I \,, \ee where still
\be \hat J=\star dx^\mu J_{\mu K} C^{K}\;,\quad
\Theta'_I=\frac{1}{2}h_{I|J_1 J_2}C^{J_1} C^{ J_{2}} .  \ee We define
$W$-type solutions to have $h_{I|JK} \neq 0$, while $V$-type have
$h_{I|JK}=0$.  Both these types deform the gauge transformations but
not their algebra (to first order in the deformation).  They are
determined by rigid symmetries of the undeformed Lagrangian 
with gauge invariant variations (\ref{eq:DefKbb}). The $V$-type have gauge invariant currents, while the
currents of the $W$-type contain a non-gauge invariant piece.

Note that again, the solutions of $W$ and $V$-types are determined up to a solution of lower type with no $a_1$-``head'', and that there might not be a canonical choice. In fact one may require similarly that $W$-type transformations become trivial when 
$h_{I|JK}$ tends to zero.

\subsection{Solutions of $I$-type (vanishing $a_2$ and $a_1$)}

In that case, 
\be
a = a_0
\ee
with $\gamma a_0 + db_0 =0$.

Since there is no Chern-Simons term in four dimensions, one can assume
that $b_0 =0$.  The deformation $a_0$ is therefore a gauge invariant function, i.e., a function only of the abelian curvatures $F_{\mu \nu}^I$, the scalar fields, and their derivatives. The $I$-type deformations neither deform the gauge transformations nor (a fortiori) the gauge algebra.  Born-Infeld deformations belong to this type. They are called ``class III'' solutions in \cite{Barnich:1994mt}.

\section{Local BRST cohomology at other ghost numbers}
\label{sec:locBRST}

By applying the above method, one finds that the local BRST cohomology of the models of section \ref{sec:vsmodels} can be described along exactly the same lines. The tools necessary to
handle the ``$h$-term'' in the non gauge invariant currents have
been generalized to higher ghost numbers in
\cite{DuboisViolette:1985hc,DuboisViolette:1985jb,DuboisViolette:1985cj} (see also \cite{Barnich:2017nty,Barnich:2018nqa} for further details). Note also that the cohomology at
negative ghost numbers reflect general properties of the
characteristic cohomology that go beyond the models considered
here \cite{Barnich:1994db}.

The local BRST cohomology in all ghost numbers is described as follows.
\begin{enumerate}[1)]
\item $H^g(s|d)$ is empty for $g\leq -3$.  

\item $H^{-2}(s|d)$ is represented by the $4$-forms 
  \begin{equation}
    U^{-2}=\mu^I d^4x\, C^*_I\label{eq:11a}. 
\end{equation}
If $A^*_I=dx^\mu A^*_{I\mu}$, the associated descent equations
are
  \begin{equation}
s\, d^4x\, C^*_I+d \star A^*_I=0,\quad s \star A^*_I+d
G_I=0,\quad sG_I=0.\label{eq:14}
\end{equation}
Characteristic cohomology $H^{n-2}_{\rm char}(d)$ is then
represented by the 2-forms $\mu^IG_I$.

\item Several types of cohomology classes in ghost numbers $g \geq
  -1$, which we call $U$, $W$ and $V$-type, can be described by
  constants ${f^I}_{JK_1\dots K_{g+1}}$ which are antisymmetric in the
  last $g+2$ indices,
\begin{equation}
    \label{sk}
   {f^I}_{JK_1\dots K_{g+1}}={f^I}_{[JK_1\dots K_{g+1}]}, 
\end{equation}
and constants $h_{I|JK_1\dots K_{g+1}}$ that are antisymmetric in the last $g+2$
indices but without any totally antisymmetric part\footnote{We will write
  $h_{IJ} \equiv h_{I|J}$ for $g=-1$.}, 
\begin{equation}
    \label{Y1}
  h_{I|JK_1\dots K_{g+1}}=h_{I|[JK_1\dots K_{g+1}]},\quad  h_{[I|JK_1\dots K_{g+1}]}=0, 
\end{equation}
together with gauge invariant
functions $g^I_{\mu K_1\dots K_{g+1}},\Phi^i_{K_1\dots K_{g+1}}$ 
that are antisymmetric in the last $g+1$ indices. They are constrained by
the requirement that the transformations
\begin{equation}
  \label{eq:16}
 \delta_{K_1\dots K_{g+1}} A_\mu^I=A^J_\mu {f^I}_{JK_1\dots K_{g+1}}+g_{\mu
    K_1\dots K_{g+1}}^I  ,\quad \delta_{K_1\dots K_{g+1}}
  \phi^i=\Phi^i_{K_1\dots K_{g+1}}, 
\end{equation}
define symmetries of the action in the sense that 
 \begin{equation}
  \label{eq:47A}
  \vddl{\mathcal L_0}{A_\mu^I}\delta_{K_1\dots K_{g+1}} A_\mu^I
  +\vddl{\mathcal
    L_0}{\phi^i}\delta_{K_1\dots K_{g+1}}\phi^i+\d_\mu j^\mu_{K_1\dots K_{g+1}}=0,
\end{equation}
with currents $j^\mu_{K_1\dots K_{g+1}}$ that are antisymmetric in the
last $g+1$ indices. This can be made more precise by making the gauge
(non-)invariance properties of these currents manifest: one finds
\begin{equation}\label{eq:current}
  j^\mu_{K_1\dots K_{g+1}}
  =J^\mu_{K_1\dots K_{g+1}}+\star G^{\mu\nu}_I
  A_\nu^J {f^I}_{JK_1\dots K_{g+1}} + \half
  \epsilon^{\mu\nu\rho\sigma}A_\nu^I F_{\rho\sigma}^J 
  h_{I|JK_1\dots K_{g+1}},
\end{equation}
where $J^\mu_{K_1\dots K_{g+1}}$ is gauge invariant and
antisymmetric in the lower $g+1$ indices. When taking into account that
\begin{equation}
  \label{eq:15}
  G_I F^J= d(G_I A^J+\star A^*_I C^J)+s(\star A^*_IA^J+d^4x C^*_IC^J) ,
  \quad F^IF^J=d(A^I F^J),  
\end{equation}
and defining 
\begin{align}
C^{{K_1}\dots K_g} &= C^{K_1}\dots C^{K_g} & & \nn \\
\Theta^I &= \frac{1}{(g+2)!}{f^I}_{J_1\dots J_{g+2}}C^{J_1\dots
      J_{g+2}}, &\Theta'_I&=\frac{1}{(g+2)!}h_{I|J_1\dots
        J_{g+2}}C^{J_1\dots J_{g+2}}\;,\nn\\
\hat J &= \star dx^\mu J_{\mu K_1\dots K_{g+1}}
    \frac{1}{(g+1)!}C^{K_1\dots K_{g+1}}\;, &\hat K&=(\star A^*_I
    \hat g^I -\star \phi^*_i \hat \Phi^i)\;,\label{eq:17b}\\ 
\hat g^I &= \frac{1}{(g+1)!}dx^\mu g^{I}_{\mu K_1\dots
      K_{g+1}}C^{K_1\dots K_{g+1}}\;, &\hat
    \Phi^i& =\frac{1}{(g+1)!}\Phi^{i}_{K_1\dots
      K_{g+1}}C^{K_1\dots K_{g+1}}\;,\nn
\end{align} 
the ``global symmetry" condition \eqref{eq:47A} is equivalent to a 
$(s,d)$-obstruction equation, 
\begin{equation}
  G_I
  F^J\d_J\Theta^I+F^IF^J\d_{J}\Theta'_{I}+s(\hat K +A^I\d_I \hat J) + d\hat J
 = 0,\label{eq:16b}
\end{equation}
with $\d_I=\ddl{}{C^I}$. Note that the last two terms combine into
\be d[\star dx^\mu J_{\mu K_1\dots K_{g+1}}]
  \frac{1}{(g+1)!}C^{K_1}\dots C^{K_{g+1}},\ee so that this equation
involves gauge invariant quantities only. It is this form that arises
in a systematic analysis of the descent equations along the lines of \cite{Barnich:2000zw} (see \cite{Barnich:2018nqa}). One can now
distinguish the three types of solutions.
\begin{enumerate}[a)]
\item $U$-type corresponds to solutions with non vanishing
  \be {f^I}_{JK_1\dots K_{g+1}} \ee and particular
  \be {}^U h_{I|JK_1\dots K_{g+1}}, \; {}^U g^I_{\mu K_1\dots K_{g+1}}, \; {}^U\Phi^i_{K_1\dots K_{g+1}}, \; {}^U J_{\mu K_1\dots K_{g+1}} \ee that
  vanish when the $f$'s vanish (and that may be vanishing even when
  the $f$'s do not\footnote{As we explained above, different choices of the particular completion ${}^U h$, ${}^U g$, ${}^U
  \Phi$, ${}^U J$ of $a_2$ exist and there might not be a canonical one, but a completion exists if the $U$-type solution is indeed a solution. Similar ambiguity holds for the solutions of $W$ and $V$-types described below.}).  A $U$-type solution is trivial if and only
  if it vanishes. Denoting by $\hat K_U,\hat J_U,(\Theta'_U)_I$, the
  expressions as in \eqref{eq:17b} but involving the particular
  solutions, the associated BRST cohomology classes are represented by
\begin{align}
  \label{eq:18b}
  U &= (d^4\!x\, C^*_I +\star A^*_IA^{K}\d_{K}+ \half G_I
  A^{K}A^{L}\d_{L}\d_{K})\Theta^I \\
  &\qquad +\hat K_U+\half F^I
  A^KA^L\d_L\d_K(\Theta'_U)_I+A^I\d_I\hat J_U,\nn
\end{align}
with $sU+d(\star
A^*_I\Theta^I+G_IA^J\d_J\Theta^I+F^IA^J\d_J(\Theta'_U)_I+\hat J_U)=0\,$.

\item $W$-type corresponds to solutions with vanishing $f$'s but non
vanishing $h_{I|J K_1\dots K_{g+1}}$ and particular ${}^W g^I_{\mu
  K_1\dots K_{g+1}},{}^W 
\Phi^i_{K_1\dots K_{g+1}}, {}^W J_{\mu K_1\dots K_{g+1}}$ that may be chosen to vanish
when the $h$'s vanish. Such solutions are trivial when the $h$'s
vanish. With the obvious notation, the associated BRST
cohomology classes are represented by 
\begin{equation}
  \label{eq:19b}
  W=\hat K_W+\half F^IA^KA^L\d_L\d_K\Theta'_I+A^I\d_I\hat J_W,
\end{equation}
with $sW+d(F^IA^J\d_J\Theta'_I+\hat J_W)=0\,$.

\item $V$-type corresponds to solutions with vanishing $f$'s and
  $h$'s. They are represented by
\begin{equation}
  \label{eq:20}
  V=\hat K_V+A^I\d_I \hat J_V,
\end{equation}
with $s V+d\hat J_V=0\,$ and $s \hat J_V=0\,$.

\end{enumerate}

\item Lastly, $I$-type cohomology classes exist in ghost numbers
  $g\geq 0$ and are described by
\begin{equation}
\hat I=d^4x\, \frac{1}{g!} I_{K_1\dots K_g}C^{K_1}\dots C^{K_g}\, ,\label{eq:12}
\end{equation}
with $s\hat I=0$, i.e., gauge invariant $I_{K_1\dots K_g}$ that are
completely antisymmetric in the $K$ indices. Such classes are to be
considered trivial if the $I_{K_1\dots K_s}$ vanish on-shell up to a
total derivative. This can again be made more precise by making the
gauge (non-)invariance properties manifest: an element of type $I$ is
trivial if and only if
\begin{equation}
  \label{eq:13}
  d^4x\, I_{K_1\dots K_g}\approx dJ_{K_1\dots
    K_g}+{m^I}_{JK_1\dots K_g}G_IF^J+\half F^IF^J m'_{IJK_1\dots K_g},
\end{equation}
where $J_{K_1\dots K_g}$ are gauge invariant $3$-forms that are
completely antisymmetric in the $K$ indices, while
${m^I}_{JK_1\dots K_g},m'_{IJK_1\dots K_g}$ are constants that are
completely antisymmetric in the last $g+1$ indices. Note also that the
on-shell vanishing terms in \eqref{eq:13} need to be gauge invariant.
When there are suitable restrictions on the space of gauge invariant
functions (such as for instance $x^\mu$ independent, Lorentz invariant
polynomials with power counting restrictions) one may sometimes
construct an explicit basis of non-trivial gauge invariant $4$-forms,
in the sense that if $d^4x I\approx \rho^{\cA}I_{\cA}+d\omega^{3}$ and
$\rho^{\cA}I_{\cA}\approx d\omega^{3}$, then $\rho^{\cA}=0$ \cite{Barnich:2000zw}. The
associated BRST cohomology classes are then parametrized by constants
${\rho^{\cA}}_{K_1\dots K_g}$.

\end{enumerate}
At a given ghost number $g \geq -1$, the cohomology is the direct sum
of elements of type $U$, $W$, $V$ (and also $I$ when $g\geq 0$).

This completes our general discussion of the local BRST cohomology. One can extend the above results to cover simple factors in
addition to the abelian factors, as well as any spacetime dimension $\geq 3$: this was done by G. Barnich and N. Boulanger in \cite{Barnich:2018nqa}, by following the different route (``depth" instead of ``length") adopted in \cite{Barnich:2000zw}.

\section{Antibracket map and structure of symmetries}
\label{sec:AntiMap}

In this section, we link the above cohomology classes to the global symmetries of the action, i.e., to elements of $H^{-1}(s|d)$. In particular, for $g = 0$, this provides the link between deformations and symmetries which appears crucially in the Yang-Mills gaugings.

\subsection{Antibracket map in cohomology}

We now investigate the antibracket map
$H^g \times H^{g'} \to H^{g+g'+1}$ for the different types of
cohomology classes described above. It follows from the detailed
discussion of the cohomology of the previous section that the
shortest non trivial length of descents, the ``depth'', of elements of
type $U$, $W$, $V$ and $I$ is $2$, $2$, $1$ and $0$ respectively. Indeed, the $U$-type and $W$-type solutions have depth $2$ because they involve $A_\mu j^\mu$ with a non-gauge invariant current.  The $V$-type solutions have depth $1$ because the Noether term
$A_\mu j^\mu$ involves for them a gauge invariant current.  Finally, $I$-type solutions clearly have depth $0$. The antibracket map is sensitive to the depth of its arguments: the depth of the image is less than or equal to the depth of its most shallow element, see appendix \ref{sec:antibr-maps-desc}.

The antibracket map involving $U^{-2}=\mu^I d^4x\, C^*_I$ in $H^{-2}$ is given by
\begin{equation}
(\,\cdot\,,U^{-2}) : H^g \to H^{g-1}, \quad \omega^{g,n}\mapsto
\vddl{{}^R\omega^{g,n}}{C^I}\mu^I.
\label{eq:53A}
\end{equation}
More explicitly, it is trivial for $g=-2$. It is also trivial for
$g=-1$, except for $U$-type where it is described by
$f\indices{^I_J}\mapsto f\indices{^I_J}\mu^J$. For $g> 0$, it is
described by
$\rho^{\cA}_{K_1\dots K_g}\mapsto \rho^{\cA}_{K_1\dots K_g}\mu^{K_g}$
for $I$-type,
$k\indices{^{v_1}_{K_1\dots K_{g+1}}}\mapsto
k\indices{^{v_1}_{K_1\dots K_{g+1}}}\mu^{K_{g+1}}$ for $V$-type, 
$h_{IJK_1\dots K_{g+1}}\mapsto h_{IJK_1\dots K_{g+1}}\mu^{K_{g+1}}$
and
$f\indices{^I_{JK_1\dots K_{g+1}}} \mapsto f\indices{^I_{JK_1\dots
    K_{g+1}}}\mu^{K_{g+1}}$ for $U$- and $W$-type.

The antibracket map for $g, g' \geq -1$ has the triangular structure presented in table \ref{tab:triangular}.
\begin{table}
\centering
$\begin{array}{c|c c c c}
  (\cdot,\cdot)  & U & W & V & I \\
  \midrule
  U & U\oplus W\oplus V\oplus I & W\oplus V\oplus I  & V\oplus I & I
  \\
  W & W\oplus V \oplus I & W\oplus V\oplus I & V\oplus I & I 
  \\
  V & V\oplus I & V\oplus I & V\oplus I & I
\\
  I & I & I & I & 0
\end{array}$
\caption[Triangular structure of the antibracket map]{Triangular structure of the antibracket map.}
\label{tab:triangular}
\end{table}
Indeed, $(\hat I,\hat I')=0$ because $I$-type cocycles can be chosen to be
antifield independent. For all other brackets involving $I$-type
cocycles, it follows from appendix \ref{sec:antibr-maps-desc} that the
result must have depth $0$ and the only such
classes are of $I$ type. Alternatively, since all cocycles can be
chosen to be at most linear in antifields, the result will be a
cocycle that is antifield independent and only classes of $I$-type
have trivial antifield dependence. It thus follows that $I$-type
cohomology forms an abelian ideal. According to appendix \ref{sec:antibr-maps-desc}, the depth of the
antibracket map of $V$-type cohomology classes with $V$, $W$, or $U$-type classes is less or equal to $1$, so it must be of $V$ or $I$-type. Finally, the remaining structure follows from the fact that only
brackets of $U$-type cocycles with themselves may give rise to terms
that involve $C^*_I$'s.  

\subsection{Structure of the global symmetry
  algebra} \label{sec:globalsymmetries}

Let us now concentrate on brackets 
between two elements that
have both ghost number $-1$, i.e., on the detailed structure of the
Lie algebra of inequivalent global symmetries when taking into account
their different types.

In this case, one may use the table above supplemented by the fact
that $I = 0$ at ghost number $g = -1$. Let then
\begin{equation}
U_{u}, \quad W_{w},\quad V_{v},
\end{equation} 
be bases of symmetries of $U$, $W$ and $V$-type respectively\footnote{These are bases in the
  cohomological sense, i.e., $\sum_u \lambda^u [U_u] = [0]$
  $\Rightarrow$ $\lambda^u = 0$ (and similarly for $W_w$ and
  $V_v$). In terms of the representatives, this becomes
  $\sum_{u} \lambda^u U_u = s a + db$ $\Rightarrow$ $\lambda^u =
  0$.}. At ghost number $g=-1$, equations \eqref{eq:18b} -- \eqref{eq:20} give
\begin{equation}
  \label{eq:5bis}
  V_{v}=K_{v},\quad W_{w}=K_{w},\quad
  U_{u}= (f_{u})\indices{^I_J}[d^4x\, C^*_I C^J+\star
  A^*_I A^J]+K_{u}. 
\end{equation}
It follows from table \ref{tab:triangular} that $V$-type symmetries and the direct sum of $V$ and $W$-type symmetries form ideals in the Lie algebra of inequivalent global symmetries.

We define the symmetry algebra $\mathfrak{g}_U$ as the quotient of all inequivalent global symmetries by the ideal of
$V\oplus W$-type symmetries. (In particular, if $U$-type symmetries form a subalgebra, it is isomorphic to $\mathfrak g_U$.)

First, $V$-type symmetries are parametrized by constants $k^v$,
$V^{-1}=k^v V_{v}$. The gauge invariant symmetry transformations on the
original fields then are
\begin{equation}
  \delta_{v}
  A^I_\mu=-(V_{v},A^I_\mu)=g\indices{_{{v}\mu}^I},\quad 
  \delta_{v} \phi^i=
  -(V_{v},\phi^i)=\Phi^i_{v}\label{eq:50}. 
\end{equation}
Furthermore, there
exist constants ${C^{v_3}}_{{v_1}{v_2}}$ such that
\begin{equation}
  \label{eq:25}
  ([V_{v_1}],[V_{v_2}]) = -{C^{v_3}}_{{v_1}{v_2}} [V_{v_3}]
\end{equation}
holds for the cohomology classes.
We choose the minus sign because 
\begin{equation}
  \label{eq:28a}
  (V_{v_1},V_{v_2}) = - d^4x (A^{*\mu}_I
  [\delta_{v_1},\delta_{v_2}] A_\mu^I+\phi^*_i
  [\delta_{v_1},\delta_{v_2}]\phi^i),
\end{equation}
so that the ${C^{v_3}}_{{v_1}{v_2}}$ are the structure constants of
the commutator algebra of the $V$-type symmetries,
$[\delta_{v_1},\delta_{v_2}] = {C^{v_3}}_{{v_1}{v_2}}
\delta_{v_3}$. For the functions $g\indices{_{v\mu}^I}$ and
$\Phi^i_v$, this gives
\begin{equation} \label{eq:commsymV}
  \begin{split}
    \delta_{v_1} g\indices{_{v_2\mu}^I} - \delta_{v_2}
    g\indices{_{v_1\mu}^I} &
    = {C^{v_3}}_{{v_1}{v_2}} g\indices{_{v_3\mu}^I} + (\text{trivial}) \\
    \delta_{v_1} \Phi^i_{v_2} - \delta_{v_2} \Phi^i_{v_1} &=
    {C^{v_3}}_{{v_1}{v_2}} \Phi^i_{v_3} + (\text{trivial}).
  \end{split}
\end{equation}
The ``trivial" terms on the right hand side take the form ``(gauge
transformation) $+$ (antisymmetric combination of the equations of
motion)" which is the usual ambiguity in the form of global
symmetries (see section \ref{sec:symmetriesH-1}). They come
from the fact that equation \eqref{eq:25} holds for classes: for the
representatives $V_v$ themselves, \eqref{eq:25} is
$(V_{v_1},V_{v_2}) = -{C^{v_3}}_{{v_1}{v_2}} V_{v_3} + sa + db$. The
trivial terms in \eqref{eq:commsymV} are then the symmetries generated
by the extra term $sa + db$, which is zero in cohomology.  The graded
Jacobi identity for the antibracket map implies the ordinary Jacobi
identity for these structure constants,
\begin{equation}
  \label{eq:26}
  {C^{v_1}}_{v_2[v_3}{C^{v_2}}_{v_4v_5]}=0. 
\end{equation}

Next, $W$-type symmetries are parametrized by constants $k^w$,
$W^{-1}=k^w W_{w}$ and encode the gauge invariant symmetry
transformations
\begin{equation}
  \delta_{w}
  A^I_\mu=-(W_{w},A^I_\mu)=g\indices{_{{w}\mu}^I},\quad 
  \delta_{w} \phi^i=
  -(W_{w},\phi^i)=\Phi^i_{w}\label{eq:49a} 
\end{equation}
with associated Noether $3$-forms
$j_W=k^w (h_w)_{IJ}F^{(I}A^{J)} + k^w J_{Ww}$. There then exist
${C^{v_2}}_{w v_1}$, $C\indices{^{w_3}_{w_1 w_2}}$, $C\indices{^{v}_{w_1 w_2}}$
such that
\begin{equation}
  \label{eq:27a}
  \begin{split}
([W_w],[V_{v}]) &= -{C^{v_2}}_{w v} [V_{v_2}],\\
([W_{w_1}], [W_{w_2}]) &= - C\indices{^{w_3}_{w_1 w_2}} [W_{w_3}] -
C\indices{^{v}_{w_1 w_2}} [V_v],
\end{split}
\end{equation}
with associated Jacobi identities that we do not spell out. For the functions $g\indices{_{w\mu}^I}$ and $\Phi^i_w$, this implies
\begin{align}
    \delta_{w} g\indices{_{v\mu}^I} - \delta_{v} g\indices{_{w\mu}^I} &= {C^{v_2}}_{{w}{v}} g\indices{_{v_2\mu}^I}, \\
    \delta_{w} \Phi^i_{v} - \delta_{v} \Phi^i_{w} &= {C^{v_2}}_{{w}{v}} \Phi^i_{v_2}, \\
    \delta_{w_1} g\indices{_{w_2\mu}^I} - \delta_{w_2}
    g\indices{_{w_1\mu}^I} &= {C^{w_3}}_{{w_1}{w_2}}
    g\indices{_{w_3\mu}^I} + {C^{v}}_{{w_1}{w_2}}
    g\indices{_{v\mu}^I}, \label{eq:435}
    \\
    \delta_{w_1} \Phi^i_{w_2} - \delta_{w_2} \Phi^i_{w_1} &=
    {C^{w_3}}_{{w_1}{w_2}} \Phi^i_{w_3} + {C^{v}}_{{w_1}{w_2}}
    \Phi^i_{v}\,,
    \label{eq:436}
  \end{align}
up to trivial terms, see the discussion below \eqref{eq:commsymV}.

Finally, $U$-type symmetries are parametrized by $k^{u}$,
$U^{-1}=k^{u} U_{u}$ and encode the symmetry
transformations
\begin{align}
  \label{eq:51a}
  \delta_{u} A^I_\mu &= -(U_{u},A^I_\mu)=(f_{u})\indices{^I_J}A^J_\mu
  + g\indices{_{{u}\mu}^I}, &\delta_{u} \phi^i &=
  -(U_{u},\phi^i)=\Phi^i_{u}, \\
  \delta_{u} A^{*\mu}_I &= -(f_{u})\indices{^K_I} A^{*\mu}_K
  - \frac{\delta}{\delta A^I_\mu} ( A^{*\nu}_K g\indices{_{u\nu}^K} +
  \phi^*_i \Phi^i_u ), &\delta_{u} \phi^*_i &= - \frac{\delta}{\delta \phi^i}
  ( A^{*\nu}_K g\indices{_{u\nu}^K} + \phi^*_j \Phi^j_u ),  \nn \\
  \delta_{u} C^I &= (f_{u})\indices{^I_J} C^J, &\delta_{u} C^*_I &=
  - (f_{u})\indices{^K_I} C^*_K. \nn
\end{align}
Again, there exist constants $C$ with various types of indices such that
\begin{align}
    ([U_u],[V_{v}]) &= -C\indices{^{v_2}_{u v}} [V_{v_2}],
    \label{eq:UVbracket}
    \\
([U_u],[W_{w}]) &= -C\indices{^{w_2}_{u w}} [W_{w_2}] -C\indices{^{v}_{u w}} [V_{v}],
\label{eq:UWbracket}
\\
([U_{u_1}], [U_{u_2}]) &= -C\indices{^{v}_{u_1 u_2}} [V_{v}]
-C\indices{^{w}_{u_1 u_2}} [W_{w}] -C\indices{^{u_3}_{u_1 u_2}} [U_{u_3}],
                     \label{eq:29b}
\end{align}
with associated Jacobi identities. Working
out the term proportional to $C^*_I$ in
$(U_{u_1},U_{u_2})$ gives the commutation relations for
the $(f_u)\indices{^I_J}$ matrices,
\begin{equation}\label{eq:commU}
[f_{u_1},f_{u_2}] = -C\indices{^{u_3}_{u_1u_2}}
f_{u_3} .
\end{equation}
In turn, this implies Jacobi identities for this type of structure
constants alone,
\begin{equation}
  \label{eq:26a}
  {C^{u_1}}_{u_2[u_3}{C^{u_2}}_{u_4u_5]}=0. 
\end{equation}
The $C\indices{^{u_3}_{u_1u_2}}$ are the structure constants of $\mathfrak{g}_U$. 

From equation \eqref{eq:UVbracket}, we get the identities
\begin{align}\label{eq:422}
  \delta_{u} g\indices{_{v\mu}^I} - \delta_{v} g\indices{_{u\mu}^I}
  - (f_u)\indices{^I_J} g\indices{_{v\mu}^J} &= {C^{v_2}}_{{u}{v}}
  g\indices{_{v_2\mu}^I}, \quad \delta_{u} \Phi^i_{v}
 - \delta_{v} \Phi^i_{u} = {C^{v_2}}_{{u}{v}} \Phi^i_{v_2} .
\end{align}
Equation \eqref{eq:UWbracket} gives the same identities with the
right-hand side replaced by the appropriate sum, as in
\eqref{eq:435}--\eqref{eq:436}. The last relation \eqref{eq:29b} gives
\begin{equation}\label{eq:423}
  \begin{split}
  \delta_{u_1} g\indices{_{u_2\mu}^I} - (f_{u_1})\indices{^I_J}
  g\indices{_{u_2\mu}^J}
  - (u_1 \leftrightarrow u_2) &= {C^{u_3}}_{{u_1}{u_2}}
  g\indices{_{u_3\mu}^I}
  + {C^{w}}_{{u_1}{u_2}} g\indices{_{w\mu}^I} + {C^{v}}_{{u_1}{u_2}}
  g\indices{_{v\mu}^I}, \\
  \delta_{u_1} \Phi^i_{u_2} - \delta_{u_2} \Phi^i_{u_1} &=
  {C^{u_3}}_{{u_1}{u_2}} \Phi^i_{u_3}
  + {C^{w}}_{{u_1}{u_2}} \Phi^i_{w} + {C^{v}}_{{u_1}{u_2}} \Phi^i_{v} .
\end{split}
\end{equation}
Equations \eqref{eq:422} and \eqref{eq:423} are again valid only
  up to trivial symmetries.

Let us now concentrate on identities containing the $h_{IJ}$, which
appear in the currents of $U$ and $W$-type. We first consider
$(U_u,W_w)$ projected to $W$-type. As in appendix
\ref{sec:antibr-maps-desc}, we have
\begin{align}
s(U_u,W_w)_{{\rm alt}} &=-d(U_u,(h_{w})_{IJ}F^{I}A^{J}+J_{w})_{{\rm
    alt}} \nn \\
    &= d\{(h_{w})_{IJ} [(f_u)\indices{^I_K} F^K A^J +F^I
(f_u)\indices{^J_K} A^K ]+{\rm invariant}\}\, .
\end{align}
When comparing this to
$s$ applied to the right hand side of \eqref{eq:UWbracket} and using
the fact that $W$-type cohomology is characterized by the Chern-Simons
term in its Noether current, we get
\begin{equation} \label{eq:hwfu}
(h_{w})_{IN}(f_u)\indices{^I_M}+(h_{w})_{MI}(f_u)\indices{^I_N}=
C\indices{^{w_2}_{u w}}(h_{w_2})_{MN}. 
\end{equation}
This computation amounts to identifying the Chern-Simons term in
  the $U$-variation $\delta_u j_w$ of a current of $W$-type.  The same
  computation applied to $(W_{w_1}, W_{w_2})$ shows that
  $C\indices{^{w_3}_{w_1 w_2}}(h_{w_3})_{MN} = 0$, which implies
\begin{equation} \label{eq:hwfw}
  C\indices{^{w_3}_{w_1 w_2}} = 0
\end{equation}
since the matrices $h_w$ are linearly independent (otherwise, the
$W_w$ would not form a basis). In other words, the $W$-variation
$\delta_{w_1} j_{w_2}$ of a current of $W$-type is gauge invariant up
to trivial terms, i.e., is of $V$-type. 

In order to work out
$(U_{u_1},U_{u_2})$ projected to $W$-type, a slightly involved
reasoning gives 
\begin{align}
\label{EQ:UTRASF}
\delta_u G_I + (f_u)\indices{^J_I} G_J \approx  - 2 (h_u)_{IJ} F^J + \lambda^w_u (h_w)_{IJ} F^J + d(\text{invariant})
\end{align}
for some constants $\lambda^w_u$. This is proved in Appendix \ref{app:derivation}  in the case where $G_I$ does not depend on derivatives of $F^I$ (but can have otherwise arbitrary dependence of $F^I$). We were not able to find the analog of \eqref{EQ:UTRASF} in the higher derivative case.

Applying then $(U_{u_1},\cdot)_{\rm alt}$
to the chain of descent equations for $U_{u_2}$ and adding the chain
of descent equations for $C^{u_3}_{u_1u_2} U_{u_3}$ yields
\begin{align}
  (h_{u_2})_{IN}(f_{u_1})\indices{^I_M} &+(h_{u_2})_{MI}(f_{u_1})\indices{^I_N}
  -
    (h_{u_1})_{IN}(f_{u_2})\indices{^I_M}-(h_{u_1})_{MI}(f_{u_2})\indices{^I_N}
    \nonumber \\ &+ \frac{1}{2} \left[ (h_w)_{IN} (f_{u_2})\indices{^I_M} +  (h_w)_{IM} (f_{u_2})\indices{^I_N}  \right]\lambda^w_{u_1} \nonumber \\
&= C\indices{^{u_3}_{u_1 u_2}}(h_{u_3})_{MN} + C\indices{^{w}_{u_1
                   u_2}}(h_{w})_{MN}.
\end{align}
Again, this amounts to identifying the Chern-Simons terms in the
$U$-variation $\delta_{u_1} j_{u_2}$ of a $U$-type current. Equation
\eqref{EQ:UTRASF} is crucial for this computation since $U$-type
currents contain $G_I$.
Using \eqref{eq:hwfu}, this becomes
\begin{align}\label{eq:hufu}
  (h_{u_2})_{IN}(f_{u_1})\indices{^I_M} &+(h_{u_2})_{MI}(f_{u_1})\indices{^I_N}
  -
    (h_{u_1})_{IN}(f_{u_2})\indices{^I_M}-(h_{u_1})_{MI}(f_{u_2})\indices{^I_N}
    \nonumber \\ &= C\indices{^{u_3}_{u_1 u_2}}(h_{u_3})_{MN} + \left[ C\indices{^{w}_{u_1 u_2}} - \frac{1}{2} C\indices{^w_{u_2 w_2}} \lambda^{w_2}_{u_1} \right] (h_{w})_{MN}.
\end{align}
We see that the effect of the $\lambda^w_u$ is to shift the structure constants of type $C\indices{^{w}_{u_1 u_2}}$. The constants $\lambda^w_u$ vanish for the explicit models considered below; it would be interesting to find an explicit example where this is not the case. As a last comment, we note that antisymmetry of equation \eqref{eq:hufu} in $u_1$ and $u_2$ imposes the constraint
\begin{equation}
C\indices{^w_{u_2 w_2}} \lambda^{w_2}_{u_1} + C\indices{^w_{u_1 w_2}} \lambda^{w_2}_{u_2} = 0
\end{equation}
on the constants $\lambda^w_u$.

\subsection{Parametrization through symmetries} \label{sec:parametrization}

It follows from the discussion of the explicit form of the antibracket map involving $H^{-2}$ (after \eqref{eq:53A}) that cohomologies of $U,W,V$-type in
ghost numbers $g\geq 0$ can be parametrized by symmetries of the
corresponding type with suitably constrained coefficients
\begin{equation}
  \label{gd}k\indices{^{u}_{K_1\dots K_{g+1}}},
  \quad k\indices{^{v}_{K_1\dots K_{g+1}}},\quad k\indices{^{w}_{K_1\dots K_{g+1}}}.
\end{equation}
In this way, for $g=0$, the problem of finding all infinitesimal
deformations can be reformulated as the question of which of these
symmetries can be gauged.

In order to do this, it is useful to first rewrite the constants
$h_{I|JK_1\dots K_{g+1}}$ appearing in the cohomology classes of $U$
and $W$-types in the equivalent symmetric convention
\begin{equation}
    X_{IJ,K_1\dots K_{g+1}}:=h_{(I|J)K_1\dots K_{g+1}}\iff 
    h_{I|JK_1\dots K_{g+1}}=\frac{2(g+2)}{g+3}X_{I[J,K_1\dots K_{g+1}]}\, , 
\label{Xintermsofh}
\end{equation}
where \eqref{Y1} is now replaced by
\begin{equation}\label{YX}
  X_{IJ,K_1 \dots K_{g+1}} = X_{(IJ),[K_1 \dots K_{g+1}]}, \quad
  X_{(IJ,K_1) K_2 \dots K_{g+1}} = 0\;. \end{equation}
Note that for $g=-1$, $h_{IJ} = X_{IJ}$.

For cohomology classes of $U,W$-type, we can write
\begin{align}
  \label{eq:352}
  {f^I}_{JK_1\dots K_{g+1}} &= (f_{u})\indices{^I_{J}}
                              \, k\indices{^{u}_{K_1\dots K_{g+1}}}, \\
  {}^U X_{IJ,K_1\dots K_{g+1}} &=(h_{u})_{IJ} \, k\indices{^{u}_{K_1\dots K_{g+1}}}, \\
  X_{IJ,K_1\dots K_{g+1}} &= (h_{w})_{IJ} \, k\indices{^{w}_{K_1\dots K_{g+1}}},
\end{align}
where $(f_{u})\indices{^I_{J}}$, $(h_{u})_{IJ}$ and $(h_{w})_{IJ}$
appear in the basis elements $U_{u}$ and $W_{w}$. (One has similar
parametrizations for the quantities $g^I_{\mu K_1\dots K_{g+1}}$,
$\Phi^i_{K_1\dots K_{g+1}}$, $J_{\mu K_1\dots K_{g+1}}$ in the
cohomology classes of the various types.)  This guarantees that
condition \eqref{eq:47A} (or \eqref{eq:16b}) is automatically
satisfied.

However, the symmetry properties \eqref{sk} and \eqref{YX} imply the
following linear constraints on the parameters:
\begin{align}
(f_{u})\indices{^I_{(J}} \, k\indices{^{u}_{K_1)K_2\dots K_{g+1}}} &= 0, \label{linf}\\
(h_{u})_{(IJ} \, k\indices{^{u}_{K_1)K_2 \dots K_{g+1}}} &= 0, \label{358}\\
(h_{w})_{(IJ} \, k\indices{^{w}_{K_1) K_2 \dots K_{g+1}}} &= 0. \label{465}
\end{align}
From the discussion of the cohomology, it also follows that $V$-type
cohomology classes are entirely determined by $V$-type symmetries in
terms of $k\indices{^{v}_{K_1\dots K_{g+1}}}$ without any additional
constraints.

\subsection{Second-order constraints on deformations and gauge algebra}
\label{sec:parametrization-2nd-order}

The most general infinitesimal gauging is given by
$S^{(1)} = \int ( U^0+W^0+V^0+I^0 )$. To extend it to second order, we have to compute
\begin{equation}
\half (S^{(1)}, S^{(1)}) = \int \left( U^1+W^1+V^1+I^1 \right) \label{eq:22a}. 
\end{equation}
The infinitesimal deformation $S^{(1)}$ can be extended to second
order whenever the right hand side vanishes in cohomology, resulting in quadratic
constraints on the constants $k\indices{^{u_1}_K}$,
$k\indices{^{w_1}_K}$, $k\indices{^{v_1}_K}$ and $\rho^{\cA}$.
Working all of them out explicitly requires computing all brackets between $U^0$, $W^0$, $V^0$ and $I^0$.

It follows from the previous section that the only
contribution to $U^1$ comes from $\half (U^0,U^0)$. The vanishing of the terms containing the antighosts $C^*_I$ requires
\begin{equation}
  \label{eq:23}
  f\indices{^I_{J[K_1}}f\indices{^J_{K_2K_3]}}=0,
\end{equation}
i.e., the Jacobi identity for the $f\indices{^I_{JK}}$. The
  associated $n_v$-dimensional Lie algebra is the gauge
  algebra and is denoted by $\mathfrak g_g$.
Using
$f\indices{^I_{JK}} = (f_{u_1})\indices{^I_J} k\indices{^{u_1}_K}$ and equation \eqref{eq:commU}, the Jacobi identity reduces to the
following quadratic constraint on $k\indices{^{u_1}_K}$:
\begin{equation} \label{eq:quadu}
 k\indices{^{u_1}_I} k\indices{^{u_2}_J} C\indices{^{u_3}_{u_1u_2}} -
  (f_{u_4})\indices{^K_I} k\indices{^{u_4}_J} k\indices{^{u_3}_K} = 0 .
\end{equation}
Note that the antisymmetry in $IJ$ of the second term is guaranteed by
the linear constraint \eqref{linf}. 

The terms at antifield number one give the constraints 
\begin{align}
 \delta_I g^K_J + f\indices{^K_{MJ}} g^M_I - (I\leftrightarrow J) &= f\indices{^L_{IJ}} g^K_L \\
 \delta_I \Phi^i_J  - (I\leftrightarrow J) &= f\indices{^L_{IJ}} \Phi^i_L.
\end{align}
Expressed with $k$'s, this gives
\begin{equation}
    k\indices{^{\Gamma}_I} k\indices{^{\Delta}_J} C\indices{^{\Sigma}_{\Gamma\Delta}} -
  (f_{u})\indices{^K_I} k\indices{^{u}_J} k\indices{^{\Sigma}_K} = 0,
\end{equation}
where the capital Greek indices take all values $u,w,v$.  This gives
three constraints, according to the type of the free index
$\Sigma$. When $\Sigma = u$, we get the constraint $\eqref{eq:quadu}$,
because the only non-vanishing structure constants with an upper $u$
index are the $C\indices{^{u_3}_{u_1u_2}}$. When $\Sigma = w$, the
possible structure constants are $C\indices{^{w_3}_{w_1w_2}}$,
$C\indices{^{w_3}_{u_1w_2}} = -C\indices{^{w_3}_{w_2u_1}}$ and
$C\indices{^{w_3}_{u_1u_2}}$, giving the constraint
\begin{equation}
  k\indices{^{w_1}_I} k\indices{^{w_2}_J} C\indices{^{w_3}_{w_1w_2}}
  + 2 k\indices{^{u_1}_{[I}} k\indices{^{w_2}_{J]}}
  C\indices{^{w_3}_{u_1w_2}}
  + k\indices{^{u_1}_I} k\indices{^{u_2}_J} C\indices{^{w_3}_{u_1u_2}}
  - (f_{u_4})\indices{^K_I} k\indices{^{u_4}_J} k\indices{^{w_3}_K} = 0 .
\end{equation}
Finally, when the free index $\Sigma$ is of type $v$, one gets a similar
identity with all possible types of values in the lower indices of the
structure constants,
\begin{align}
  k\indices{^{v_1}_I} k\indices{^{v_2}_J} C\indices{^{v_3}_{v_1v_2}} &+
  2 k\indices{^{w_1}_{[I}} k\indices{^{v_2}_{J]}}
  C\indices{^{v_3}_{w_1v_2}} + k\indices{^{w_1}_I} k\indices{^{w_2}_J}
  C\indices{^{v_3}_{w_1w_2}} + 2 k\indices{^{u_1}_{[I}}
  k\indices{^{v_2}_{J]}} C\indices{^{v_3}_{u_1v_2}} \nn \\
  &+ 2
  k\indices{^{u_1}_{[I}} k\indices{^{w_2}_{J]}}
  C\indices{^{v_3}_{u_1w_2}} + k\indices{^{u_1}_I} k\indices{^{u_2}_J}
  C\indices{^{v_3}_{u_1u_2}} - (f_{u_4})\indices{^K_I}
  k\indices{^{u_4}_J} k\indices{^{v_3}_K} = 0 . 
\end{align}

\section{Quadratic vector models}
\label{sec:second order}

To go further, one needs to specialize the form of the Lagrangian,
which has been assumed to be quite general so far.  In this section, we focus on the usual second order Lagrangian \eqref{eq:lag} arising in the context of extended
supergravities.

\subsection{Restriction to electric symmetries}

An important result of our general analysis is that the symmetries of the action that can lead to consistent deformations may have a term that is not gauge invariant. This term is present only in the variation of the vector potential and is restricted to be linear in the undifferentiated vector potential, i.e.,
$\delta A^I_{\mu} = f\indices{^I_J} A^J_{\mu} + g^I_\mu$,
$\delta \phi^i = \Phi^i$.  Here $f\indices{^I_J}$ are constants and $g^I_\mu$, $\Phi^i$ are gauge invariant functions; the symbol
$\delta$ represents the variation of the fields and not
the Koszul-Tate differential (no confusion should arise as the
context is clear).

It is of interest to investigate a subalgebra of these general symmetries, obtained by restricting oneself from the outset to
transformations of the gauge potentials that are linear and
homogeneous in the undifferentiated potentials and to transformations of the scalars that depend on undifferentiated scalars alone. This means that one takes $g_\mu^I = 0$ and that the functions $\Phi^i$ only depend on the undifferentiated scalar fields, leading to the electric symmetries \eqref{eq:64} considered in chapter \ref{chap:YMgaugings}. It generically does not exhaust all symmetries and does not contain for example the conformal symmetries of free electromagnetism. However, in many examples (with particular forms of the matrices $\cI$ and $\cR$), $U$ and $W$-type symmetries can be explicitely computed, using the assumption that there is no explicit space-time dependence\footnote{This assumption is necessary if one wants to couple these theories to gravity \cite{Barnich:1995ap}.}. They are then indeed found to be of ``electric form" (see section 5 of \cite{Barnich:2017nty} for details).

Finally, decomposing the indices into $U$, $W$ and $V$ type, the electric symmetry relations \eqref{X-alg} -- \eqref{F-alg} are consistent with the relations of section \ref{sec:globalsymmetries} with $\lambda_u^w=0$. In order to simplify formulas, we will no longer make the distinction between $U$, $W$ and $V$-type which can easily be recovered.

\subsection{Deformations}

We now limit ourselves to deformations of the master
action with the condition that all infinitesimal deformations come from symmetries that belong to the electric symmetry algebra above.

According to section \ref{sec:parametrization}, the deformations are parametrized through electric symmetries by a matrix $k\indices{^\Gamma_I}$ as in \eqref{eq:f-k} and \eqref{eq:phi-k}. 
The linear constraints \eqref{linf} -- \eqref{465} on the matrix
$k\indices{^\Gamma_I}$ become \eqref{eq:constr-antisymmetry} and \eqref{eq:constr-chernsimons}. They guarantee that the first order deformation of the master action
is given by
\begin{equation}
S^{(1)} = \int \!d^4\!x\,\left( a_2 + a_1 + a_0 \right),
\end{equation}
where 
\begin{equation}
  a_2 = \frac{1}{2} C^*_I f\indices{^I_{JK}} C^J C^K
\end{equation}
encodes the first order deformation of the gauge algebra and
\begin{equation}
  a_1 = A^{*\mu}_I f\indices{^I_{JK}} A^J_\mu C^K + \phi^*_i \Phi^i_K C^K
\end{equation}
encodes the first order deformation of the gauge symmetries. The deformation $a_0$ of the Lagrangian is given by the sum of three terms:
\begin{align}
  a^\text{(YM)}_0 &= \frac{1}{2} (\star G_I)^{ \mu\nu}
 f\indices{^I_{JK}} A^J_\mu A^K_\nu , \\
a^\text{(CD)}_0 &= J^\mu_K A^K_\mu ,  \\
  a^\text{(CS)}_0 &= \frac{1}{3} X_{IJ,K} \epsilon^{\mu\nu\rho\sigma}
F^I_{\mu\nu} A^J_\rho A^K_\sigma.
\end{align}
The terms $a^\text{(YM)}$ and $a^\text{(CD)}$ are exactly those
necessary to complete the abelian field strengths and ordinary
derivatives of the scalars into covariant quantities. The ``Chern-Simons" term $a_0^\text{(CS)}$ appears when $h_\Gamma \neq 0$, as emphasized in chapter \ref{chap:YMgaugings}.

The second order deformation $S^{(2)}$ to the master action is then determined by the first order deformation through equation
\eqref{eq:43}. As discussed in section
\ref{sec:parametrization-2nd-order}, the existence of $S^{(2)}$
imposes the additional quadratic constraint \eqref{eq:quadraticconstraint} on the matrix $k\indices{^\Gamma_I}$. It yields the various constraints of section \ref{sec:parametrization-2nd-order} upon splitting the indices in $V$, $W$, $U$ symmetries.

Explicit computation shows that $S^{(2)}$ can be chosen such that
there is no further deformation of the gauge symmetries or of their
algebra. The second order terms in the Lagrangian are exactly those
necessary to complete abelian field strengths and
ordinary derivatives of the scalars to non-abelian field strengths and covariant derivatives. One also finds a non-abelian completion of the Chern-Simons term $a_0^\text{(CS)}$. 
Putting everything together, the full Lagrangian after adding the second order deformation is precisely \eqref{Full-Lagrangian}. Since that Lagrangian is fully gauge invariant, the deformation stops at second
order, i.e., $S = S^{(0)} + S^{(1)} + S^{(2)}$ gives a solution to
the master equation $(S,S)=0$ (we absorb $g$ in $k\indices{^\Gamma_I}$).

\subsection{Comparison with the embedding tensor constraints}

As explained in section \ref{sec:embeddingtensor}, one can always
go to a duality frame in which the magnetic components of the
embedding tensor vanish, $\Th^{I\a} = 0$. Moreover, only the
components $\Th\indices{_I^\Gamma}$ survive, where $\Gamma$ runs over
the generators of the electric subgroup $G_e \subset G$ that act as
local symmetries of the Lagrangian in that frame. Then, the gauged
Lagrangian in the electric frame is exactly the Lagrangian
\eqref{Full-Lagrangian}, where the matrix $k$ is identified with the
remaining electric components of the embedding tensor,
$k\indices{^\Gamma_I} = \Th\indices{_I^\Gamma}$ (or
$\Th\indices{_{\hat{I}}^\Gamma}$ in the notation of
\cite{Trigiante:2016mnt}). The linear and quadratic constraints on the
embedding tensor then agree with the constraints on $k$. More
precisely, the constraints (3.11), (3.12) and (3.39) of
\cite{Trigiante:2016mnt} in the electric frame correspond to our
\eqref{eq:quadraticconstraint}, \eqref{eq:constr-antisymmetry} and
\eqref{eq:constr-chernsimons} respectively. As explained in sections
\ref{sec:parametrization} and \ref{sec:parametrization-2nd-order}, the
constraints can be refined using the split corresponding to the
various ($U$, $W$, $V$) types of symmetry.

It was shown in section \ref{sec:embeddingtensor} (and appendix \ref{app:auxfields}) that the embedding tensor
formalism does not allow for more general deformations than those of the second-order Lagrangian \eqref{eq:lag} studied in this chapter. Indeed, their
BRST cohomologies are isomorphic even though the field content and
gauge transformations are different. 

Conversely, as long as one
restricts the attention to the symmetries of \eqref{eq:lag} that are of the electric type \eqref{eq:64}, we showed that the embedding tensor formalism captures all consistent deformations that deform the gauge transformations of the fields.

\part{Six-dimensional exotic fields}\label{PART:6D}
\chapter{Gravitational duality and dimensional reduction}
\label{chap:gravduality}

As was reviewed in part \ref{PART:SUGRA}, Maxwell's theory of a free abelian gauge field can equivalently be described by a $(D-3)$-form field. Moreover, when $D = 4$, the field and its dual have the same index structure and one actually finds a continuous $SO(2)$ symmetry of the theory mixing the field and its dual\footnote{There is also a commuting $\R_0$ factor, which simply acts a constant rescaling of the field and its dual. We will not consider it here.}.

Remarkably, a similar duality arises for linearized gravity \cite{Hull:2000zn,Hull:2001iu,Hull:2000rr,West:2001as,Nieto:1999pn,Bekaert:2002jn,Boulanger:2003vs,Henneaux:2004jw}. In space-time dimension $D$, the ``dual graviton" is a tensor of mixed Young symmetry type $[D-3,1]$,
\begin{equation}
T_{\mu_1 \dots \mu_{D-3} \nu} \sim \begin{ytableau}
{} & {} \\
{} \\
\none[\vdots] \\
\none \\ 
{} \\
\end{ytableau}\qquad \text{(height: $D-3$ boxes)}
\end{equation} (see \cite{Curtright:1980yk} for early work on mixed symmetry gauge fields). Again, the field and its dual are both symmetric tensors when $D = 4$, and linearized gravity has a continuous $SO(2)$ symmetry \cite{Henneaux:2004jw}.

In this chapter, we first review these facts in arbitrary dimension. A crucial step is the rewriting of the equations of motion as twisted self-duality conditions. We then review the equations of motion for three free bosonic fields in six space-time dimensions. The first is the well-known chiral two-form in six-dimensions; the other two are chiral mixed symmetry fields first considered by Hull in \cite{Hull:2000zn,Hull:2000rr}. Remarkably, the duality rotations for electromagnetism and linearized gravity in four dimensions can be understood geometrically from the dimensional reduction of these fields, as was first noticed in \cite{Verlinde:1995mz} for vector fields and in \cite{Hull:2000rr} for gravity.

The discussion of this chapter is done at the level of equations of motion. The same conclusions also hold at the level of the action, as we will see in the next chapters.

\section{Twisted self-duality of \texorpdfstring{$p$}{p}-forms}

We begin by shortly reviewing the simpler case of $p$-form fields, which follow a similar pattern. As we saw in part \ref{PART:SUGRA}, a $p$-form field $A$ is equivalent to a $(D-p-2)$-form field $\tilde{A}$, defined by
\begin{equation}
\star F = d\tilde{A}, \quad F = dA\, .
\end{equation}
Indeed, the equations of motion for $A$ are $d\star F = 0$, which imply $\star F = d\tilde{A}$ for some $\tilde{A}$ using the Poincaré lemma. If we define $\tilde{F} = d\tilde{A}$, the relation between $F$ and $\tilde{F}$ can be written as the twisted self-duality condition
\begin{equation}\label{eq:twistedpform}
\star \cF = \rho \, \cF, \quad\text{with}\quad \cF = \begin{pmatrix} F \\ \tilde{F} \end{pmatrix}, \quad \rho = \begin{pmatrix}
0 & 1 \\ (-1)^{(p+1)(D-1)+1} & 0
\end{pmatrix} .
\end{equation}
This equation is equivalent to the original equations of motion $d \star F = 0$, but puts $A$ and $\tilde{A}$ on the same footing.

For the case of a vector field in four dimensions ($p = 1$, $D = 4$), these equations have a continuous $SO(2)$ symmetry, acting as
\begin{equation}
\begin{pmatrix}
A \\ \tilde{A}
\end{pmatrix} \to \begin{pmatrix} \cos \theta & \sin \theta \\ - \sin \theta & \cos \theta \end{pmatrix} \begin{pmatrix}
A \\ \tilde{A}
\end{pmatrix}
\end{equation}
on the field $A$ and its dual $\tilde{A}$. This is absent in other dimensions, where a vector field and its dual have different index structures.

\section{Duality in linearized gravity}

Linearized gravity is described by a symmetric tensor $h_{\mu\nu}$ (graviton field), with gauge invariances and invariant curvature
\begin{equation}\label{eq:Rofh}
\delta h_{\mu\nu} = 2 \, \pd_{(\mu} \xi_{\nu)}, \qquad R\indices{^{\mu\nu}_{\rho\sigma}}[h] = \pd^{[\mu} \pd_{[\rho} h\indices{^{\nu]}_{\sigma]}} .
\end{equation}
The tensor $R$ is just, up to a factor, the linearized Riemann tensor for the metric $g_{\mu\nu} = \eta_{\mu\nu} + h_{\mu\nu}$. It satisfies the usual symmetry properties
\begin{align}\label{eq:22riem}
R_{\mu\nu\rho\sigma} = - R_{\nu\mu\rho\sigma} = - R_{\mu\nu\sigma\rho}, \quad
R_{\mu\nu\rho\sigma} + R_{\nu\rho\mu\sigma} + R_{\rho\mu\nu\sigma} = 0
\end{align}
described by the $(2,2)$ Young diagram\footnote{See appendix \ref{app:youngpoincare} for our Young tableau conventions. The convention to keep in mind is that the integers in $[p,q, \dots ]$ denote the heights of the columns, while they are the lengths of the rows in $(a,b,\dots)$.}, $R\sim\tyng{2,2}$. (These imply also $R_{\mu\nu\rho\sigma} = R_{\rho\sigma\mu\nu}$.) It also satisfies the differential Bianchi identity
\begin{equation}\label{eq:bianchiriem}
\pd_{[\mu} R_{\nu\rho]\sigma\tau} = 0 .
\end{equation}
Conversely, any four-index tensor satisfying the properties \eqref{eq:22riem} and \eqref{eq:bianchiriem} can be written as in \eqref{eq:Rofh} for some $h$. This fact follows from the generalized Poincaré lemmas of \cite{Olver_hyper,DuboisViolette:1999rd,DuboisViolette:2001jk,Bekaert:2002dt} for tensors with mixed Young symmetries, reviewed in appendix \ref{app:poincare}. The equations of motion are simply the vanishing of the linearized Ricci tensor,
\begin{equation}\label{eq:ricci}
R\indices{^{\mu\nu}_{\rho\nu}} = 0 .
\end{equation}

\subsection{In four dimensions}

We follow the presentation of references \cite{Henneaux:2004jw,Bunster:2012km,Bunster:2013oaa}.

The dual Riemann tensor is defined by
\begin{equation}
(* R)_{\mu\nu\rho\sigma} = \frac{1}{2}\, \varepsilon_{\mu\nu\lambda\tau} R\indices{^{\lambda\tau}_{\rho\sigma}} \, .
\end{equation}
It is identically traceless because of the cyclic identity satisfied by the Riemann tensor (last of \eqref{eq:22riem}),
\begin{equation}
(* R)\indices{^{\mu\nu}_{\rho\nu}} = 0 \quad \Leftrightarrow \quad R_{[\mu\nu\rho]\sigma} = 0 .
\end{equation}
On-shell, i.e., when \eqref{eq:ricci} is satisfied, it also satisfies the cyclic identity
\begin{equation}
(* R)_{[\mu\nu\rho]\sigma} = 0 \quad \Leftrightarrow \quad R\indices{^{\mu\nu}_{\rho\nu}} = 0 \, .
\end{equation}
Therefore, just like Maxwell's theory, linearized gravity has a symmetry under the exchange of the Riemann tensor with its dual. This exchanges the equations of motion with the first (algebraic) Bianchi identity.

The dual of the Riemann tensor also satisfies the second (differential) Bianchi identity on-shell,
\begin{equation}\label{eq:bianchiriemdual}
\pd_{[\mu} (*R)_{\nu\rho]\sigma\tau} = 0 .
\end{equation}
Therefore, there exists a graviton field $\tilde{h}_{\mu\nu}$ of which $*R$ is the Riemann tensor,
\begin{equation}
(*R)\indices{^{\mu\nu}_{\rho\sigma}}[h] = \pd^{[\mu} \pd_{[\rho} \tilde{h}\indices{^{\nu]}_{\sigma]}} \equiv \tilde{R}\indices{^{\mu\nu}_{\rho\sigma}}
\end{equation}
or, in shorthand notation, $*R = \tilde{R}$. The relation between the Riemann and its dual can be written as the twisted self-duality
\begin{equation}\label{eq:gravduality4}
\cR = \cS * \cR,
\end{equation}
with
\begin{equation}
\cR = \col{R}{\tilde{R}}, \quad \cS = \begin{pmatrix}
0 & -1 \\ 1 & 0
\end{pmatrix} .
\end{equation}
As we have seen, the linearized Einstein equations are equivalent to this equation, which has the advantage of putting the graviton and its dual on the same footing.

Just as in the case of the vector field in four dimensions, these equations have continuous $SO(2)$ symmetry, rotating the field $h$ and its dual $\tilde{h}$ into each other. They induce rotations between $R$ and $* R$.

\subsection{In $D$ dimensions}

In other dimensions, the dual graviton is a mixed symmetry field described by the $[D-3,1]$ Young tableau
\begin{equation}
T_{\mu_1 \dots \mu_{D-3} \nu} \sim \begin{ytableau}
{} & {} \\
{} \\
\none[\vdots] \\
\none \\ 
{} \\
\end{ytableau}\, ,
\end{equation}
i.e.,
\begin{equation}
T_{\mu_1 \dots \mu_{D-3} \nu} = T_{[\mu_1 \dots \mu_{D-3}] \nu}, \quad T_{[\mu_1 \dots \mu_{D-3} \nu]} = 0.
\end{equation}
The gauge invariances of this field are
\begin{equation}
\delta T_{\mu_1 \dots \mu_{D-3} \nu} = \pd_{[\mu_1} \sigma_{\mu_2 \dots \mu_{D-3}] \nu} + \pd_{[\mu_1} \alpha_{\mu_2 \dots \mu_{D-3}] \nu} + (-1)^{D} \pd_\nu \alpha_{\mu_1 \dots \mu_{D-3}} ,
\end{equation}
where $\sigma$ and $\alpha$ have the $[D-4,1]$ and $[D-3]$ symmetry, respectively. The equations of motion for this field are again the vanishing of the trace
\begin{equation}
E\indices{^\nu_{\mu_2 \dots \mu_{D-2} \nu \rho}}[T] = 0,
\end{equation}
where $E$ is the gauge-invariant curvature
\begin{equation}
E\indices{^{\mu_1 \dots \mu_{D-2}}_{\nu \rho}}[T] = \pd^{[\mu_1} \pd_{[\nu} T\indices{^{\mu_2 \dots \mu_{D-2}]}_{\rho]}}\, .
\end{equation}
This curvature is a $[D-2,2]$ tensor that satisfies the differential Bianchi identities
\begin{equation}\label{eq:bianchiE}
\pd^{[\mu_1} E\indices{^{\mu_2 \dots \mu_{D-2}]}_{\nu \rho}} = 0, \quad \pd_{[\sigma} E\indices{^{\mu_1 \dots \mu_{D-2}}_{\nu \rho]}} = 0 .
\end{equation}

The equivalence between this field and the ordinary graviton can be seen at the level of the massless little group $SO(D-2)$ by dualizing one index,
\begin{equation}
h_{ij} = \varepsilon_{ik_1 \dots k_{D-3}} T\indices{^{k_1 \dots k_{D-3}}_j} .
\end{equation}
The symmetry condition on $h$ is equivalent to the tracelessness of $T$, and vice-versa. At the level of the equations of motion, we can follow the same procedure as before: we define the dual of the Riemann as
\begin{equation}
(*R)_{\mu_1 \dots \mu_{D-2} \nu_1 \nu_2} = \frac{1}{2} \varepsilon_{\mu_1 \dots \mu_{D-2} \rho_1 \rho_2} R\indices{^{\rho_1 \rho_2}_{\nu_1 \nu_2}} .
\end{equation}
Again, this dualization exchanges trace and cyclic identities,
\begin{equation}
(* R)\indices{^\nu_{\mu_2 \dots \mu_{D-2} \nu \rho}} = 0 \quad \Leftrightarrow \quad R_{[\mu\nu\rho]\sigma} = 0
\end{equation}
and
\begin{equation}
(* R)_{[\mu_1 \dots \mu_{D-2} \nu_1] \nu_2} = 0 \quad \Leftrightarrow \quad R\indices{^{\mu\nu}_{\rho\nu}} = 0 .
\end{equation}
This also implies that $*R$ satisfies the same differential Bianchi identities \eqref{eq:bianchiE} as $E[T]$. Therefore, again using the relevant Poincaré lemmas, there exists a field $T$ such that, on-shell, we have
\begin{equation}
* R = E[T] .
\end{equation}
The relation between $h$ and its dual $E$ can be summarized in the twisted self-duality equation
\begin{equation}\label{eq:twistgrav}
\cR = \cS * \cR, \qquad \cR = \col{R}{E},
\end{equation}
with $\cS$ as in \eqref{eq:gravduality4}. This is equivalent to the linearized Einstein's equations in any dimension and puts $h$ and $T$ on the same footing. The $SO(2)$ symmetry of the $D = 4$ case is absent here, however, since the field and its dual don't have the same number of indices.

\section{Reduction of chiral fields}
\label{sec:selfdualfields}

We now review the equations of motion and dimensional reduction to five and four dimensions of three different types of chiral bosonic fields in six dimensions. They enable a remarkable geometric realization of the $SO(2)$ duality rotations as a rotation of the two internal coordinates\footnote{There is an important comment to be made here. We are considering the reduction on a fixed torus with flat metric $\delta_{ij}$. Therefore, we find only $SO(2)$ rotations here instead of the full $S$-duality $\SL$ of references \cite{Verlinde:1995mz,Hull:2000rr}. The other generators of $SL(2,\R)$ change the shape of the internal torus; this corresponds to changing the coupling constants of the four-dimensional theory as
\[ \tau \mapsto \frac{a \tau + b}{c \tau + d}\, \qquad \begin{pmatrix}
a & b \\ c & d
\end{pmatrix} \in \SL\, .\]
The group $SO(2) \subset \SL$ we have here is the isotropy subgroup, leaving $\tau$ (and the internal torus) unchanged. The possibility of full gravitational $S$-duality (and, in particular, weak/strong coupling $\tau \mapsto -1/\tau$ duality) is certainly tantalizing and deserves further study. This falls however outside of the scope of this thesis.}.

\subsection{Reduction of the chiral $2$-form}

The field is an antisymmetric tensor $A_{\mu\nu} = - A_{\nu\mu}$ in six dimensions, with gauge transformations $\delta A_{\mu\nu} = 2 \pd_{[\mu} \lambda_{\nu]}$.

\subsubsection{Equations of motion}

The invariant field strength is just $F = dA$ or, in components,
\begin{equation}
F_{\mu\nu\rho} = 3 \, \pd_{[\mu} A_{\nu\rho]} = \pd_\mu A_{\nu\rho} + \pd_\nu A_{\rho\mu} + \pd_\rho A_{\mu\nu} .
\end{equation}
Since $(*)^2 = 1$ in this case, the equation $* F = F$ is a consistent first order equation of motion for $A$, which in components reads
\begin{equation}\label{eq:sd2form}
F_{\alpha_1 \alpha_2 \alpha_3} = \frac{1}{3!} \varepsilon_{\alpha_1 \alpha_2 \alpha_3 \beta_1 \beta_2 \beta_3} F^{\beta_1 \beta_2 \beta_3} .
\end{equation}
It implies the usual equation of motion $d * F = 0$ because of the Bianchi identity $dF = 0$, but is stronger. In fact, the number of degrees of freedom carried by a chiral (self-dual) two-form is three, which is half the number carried by a two-form satisfying $d * F = 0$ (this was already discussed in appendix \ref{app:pforms}).

\subsubsection{Dimensional reduction to five dimensions}

In the reduction to five dimensions, $\hat{A}_{\mu\nu}$ gives a vector and a two-form,
\begin{equation}
A_{\mu} \equiv \hat{A}_{\mu 5}, \quad B_{\mu\nu} \equiv \hat{A}_{\mu\nu}\, .
\end{equation}
The self-duality equation \eqref{eq:sd2form} in six dimensions implies that these two fields are dual to each other: the field strengths are
\begin{align}
F_{\mu\nu}[A] = \hat{F}_{\mu\nu 5}[\hat{A}] \, , \quad F_{\mu\nu\rho}[B] = \hat{F}_{\mu\nu\rho}[\hat{A}]\, .
\end{align}
Condition \eqref{eq:sd2form} then gives
\begin{equation}
F_{\mu\nu}[A] = - \frac{1}{3!} \varepsilon_{\mu\nu\alpha\beta\gamma} F^{\alpha\beta\gamma}[B]\, ,
\end{equation}
which is exactly the equations of motion for a single vector field in five dimensions, written in the twisted self-duality form \eqref{eq:twistedpform}. In this way, dimensional reduction of the chiral $2$-form in six dimensions gives one vector in five dimensions. (Remark that a usual vector field in six dimensions gives, in five dimensions, a vector field \emph{and} a scalar.)

\subsubsection{Dimensional reduction to four dimensions}

Reduction to four dimensions then gives a vector field and a scalar. This is clear since there is only one vector field in five dimensions; however, it is interesting to go directly from six dimensions to four dimensions. The fields in four dimensions are
\begin{align}
B_{\mu\nu} = \hat{A}_{\mu\nu}, \quad A_{\mu\, i} = \hat{A}_{\mu i}, \quad \phi = \varepsilon^{ij} \hat{A}_{ij}\, ,
\end{align}
where the $i, j, \dots$ label the internal (flat) directions. The self-duality condition in six dimensions implies that the various curvatures satisfy
\begin{equation}
F_{\mu\nu}[A_i] = \frac{1}{2} \varepsilon_{\mu\nu\rho\sigma} \varepsilon_{ij} F^{\rho\sigma}[A^j], \qquad \pd_\mu \phi = \varepsilon_{\mu\nu\rho\sigma} F^{\nu\rho\sigma}[B]
\end{equation}
(internal indices are raised with the flat metric $\delta^{ij}$). Therefore, one has indeed one vector field and one scalar field only, with equations of motion written in twisted self-duality form.

Now, rotations in the internal space precisely induce $SO(2)$ duality rotations of the vector field in four dimensions, while $\phi$ is invariant. In this way, duality invariance in four dimensions is understood geometrically from the existence of a self-dual two-form in six dimensions.

\subsection{Reduction of the $(2,2)$-tensor}

This case is described by the $(2,2)$ Young tableau,
\begin{equation}
T_{\alpha_1 \alpha_2 \beta_1 \beta_2} \sim \yng{2,2}\, .
\end{equation}
This means that the gauge field $T_{\alpha_1 \alpha_2 \beta_1 \beta_2}$ satisfies the symmetry properties
\begin{equation}
T_{\alpha_1 \alpha_2 \beta_1 \beta_2} = - T_{\alpha_2 \alpha_1 \beta_1 \beta_2} = - T_{\alpha_1 \alpha_2 \beta_2 \beta_1} \, , \quad T_{[\alpha_1 \alpha_2 \beta_1] \beta_2} =0
\end{equation}
(symmetries of the Riemann tensor).
The gauge symmetries are
\begin{equation}
\delta T_{\alpha_1 \alpha_2 \beta_1 \beta_2} = \mathbb{P}_{(2,2)} \left( \partial_{\alpha_1} \eta_{\beta_1 \beta_2 \alpha_2} \right),
\end{equation}
where $\eta_{ \beta_1 \beta_2 \alpha_2} $ is an arbitrary $(2,1)$-tensor and $\mathbb{P}_{(2,2)}$ is the projector on the $(2,2)$ symmetry. The gauge transformations therefore read explicitly
\begin{equation}
\delta T_{\alpha_1 \alpha_2 \beta_1 \beta_2} = - \frac{1}{2} \left( \eta_{\alpha_1 \alpha_2 [\beta_1, \beta_2]} + \eta_{\beta_1 \beta_2 [\alpha_1, \alpha_2]} \right) \, ,
\end{equation}
where the comma is a derivative.

This field (with self-dual equations of motion written below) reduces in $5$ spacetime dimensions to the standard description of linearized gravity; we will therefore call it the ``exotic graviton" field.

\subsubsection{Equations of motion}

The gauge invariant curvature, or ``Riemann tensor", is then a tensor of type $(2,2,2)$, $R \sim \tyng{2,2,2}\,$. It is defined in components by
\begin{equation}
R_{\alpha_1 \alpha_2 \alpha_3 \beta_1 \beta_2 \beta_3} = \partial_{[\alpha_1}T_{\alpha_2 \alpha_3][ \beta_1 \beta_2, \beta_3]}
\end{equation} and one has the symmetry properties
\begin{equation}
R_{\alpha_1 \alpha_2 \alpha_3 \beta_1 \beta_2 \beta_3}= R_{[\alpha_1 \alpha_2 \alpha_3] \beta_1 \beta_2 \beta_3}= R_{\alpha_1 \alpha_2 \alpha_3 [\beta_1 \beta_2 \beta_3]}
\end{equation}
as well as
\begin{equation}
R_{[\alpha_1 \alpha_2 \alpha_3 \beta_1] \beta_2 \beta_3} = 0. \label{cyclic}
\end{equation}
It also satisfies the differential Bianchi identity
\begin{equation}
\pd_{[\alpha_1} R_{\alpha_2 \alpha_3 \alpha_4] \beta_1 \beta_2 \beta_3} = 0 \, .
\end{equation}

The equations of motion for a general $(2,2)$-tensor express that the corresponding $(2,2)$ ``Ricci tensor", i.e., the trace 
$ R_{\alpha_1 \alpha_2  \beta_1 \beta_2 } \equiv  R_{\alpha_1 \alpha_2 \alpha_3 \beta_1 \beta_2 \beta_3} \eta^{\alpha_3 \beta_3}$ of the Riemann tensor, vanishes,
\begin{equation}
 R_{\alpha_1 \alpha_2  \beta_1 \beta_2 } = 0. \label{EOM0}
\end{equation}
This is a direct generalization of linearized Einstein's equations.

In six spacetime dimensions, the dual $* R$ of the Riemann tensor on, say, the first three indices,
\be
* R_{\alpha_1 \alpha_2 \alpha_3 \beta_1 \beta_2 \beta_3} = \frac{1}{3!} \varepsilon_{\alpha_1 \alpha_2 \alpha_3 \lambda_1 \lambda_2 \lambda_3} R\indices{^{\lambda_1 \lambda_2 \lambda_3}_{\beta_1 \beta_2 \beta_3}},
\ee
is identically traceless because of the cyclic identity \eqref{cyclic}, i.e., 
\be
 *R_{\alpha_1 \alpha_2  \beta_1 \beta_2 } = 0.
\ee
This implies that a $(2,2)$-tensor field $T$ with a self-dual or anti-self-dual Riemann tensor
 \begin{equation}
 R =  * R \label{SD22}
 \end{equation}
 (self-duality) or $R = - * R$ (anti-self-duality)
 is automatically a solution of the equations of motion \eqref{EOM0}).   Note that this implies that $* R$ is also a $(2,2,2)$ tensor on-shell. The condition \eqref{SD22} is consistent because $(*)^2 = + 1$ in this case.

This condition halves the number of degrees of freedom: they go from $10$ in equation \eqref{EOM0} to $5$ upon imposing \eqref{SD22}. Indeed, just as in the case of the chiral $2$-form, any $(2,2)$-tensor can be split into a self-dual and an anti-self-dual part.

\subsubsection{Dimensional reduction to five dimensions}

Upon reduction to five dimensions, $\hat{T}_{\mu\nu\rho\sigma}$ yields three fields,
\begin{equation}
h_{\mu\nu} \equiv \frac{4}{9} \hat{T}_{\mu 5 \nu 5}, \quad T_{\mu\nu\rho} \equiv \frac{2}{3} \hat{T}_{\mu\nu\rho 5}, \quad T_{\mu\nu\rho\sigma} \equiv \hat{T}_{\mu\nu\rho\sigma}
\end{equation}
(we include combinatorial factors for future convenience).
The last of those is pure gauge in five space-time dimensions\footnote{The easiest way to see this is that the corresponding representation of the five-dimensional massless little group $SO(3)$ has no degree of freedom. Indeed, it has $6$ (components of a $(2,2)$-tensor) minus $6$ (tracelessness condition) independent components. The rule of thumb is the following (see e.g. \cite{hamermesh1962group,fulton1991representation,Bekaert:2006py}): in space-time dimension $D$, a massless mixed symmetry tensor has no degree of freedom if the sum of the heights of the first two columns of its Young tableau is strictly greater that $D-2$.}. Therefore, we are left with a symmetric field (the graviton) and a $(2,1)$ field (dual graviton). The self-duality equation \eqref{SD22} in six dimensions precisely implies that those are dual to each other: indeed, their curvatures are
\begin{align}
R\indices{^{\mu\nu}_{\rho\sigma}}[h] = \hat{R}\indices{^{\mu\nu 5}_{\rho\sigma 5}}[\hat{T}]\, , \quad E\indices{^{\mu\nu\rho}_{\sigma\tau}}[T] = \hat{R}\indices{^{\mu\nu\rho}_{\sigma \tau 5}}[\hat{T}]
\end{align}
and the self duality equation then gives
\begin{equation}
R\indices{^{\mu\nu}_{\rho\sigma}}[h] = - \frac{1}{3!} \varepsilon^{\mu\nu\alpha\beta\gamma} E_{\alpha\beta\gamma\rho\sigma}[T]\, .
\end{equation}
Those are exactly linearized Einstein's equations in the twisted self-duality form \eqref{eq:twistgrav}. Therefore, dimensional reduction of the chiral $(2,2)$-tensor in six dimensions gives one graviton in five dimensions, and nothing else.

\subsubsection{Dimensional reduction to four dimensions}

Since the reduction to five dimensions only gave one graviton, reduction to four dimensions yields a graviton, a vector field and a scalar field. As in the case of the chiral two-form, it is interesting to go directly from six dimensions: duality rotations for the graviton and the vector field are then realized geometrically from rotations in the internal coordinates.

Coming from six dimensions, it is shown in \cite{Hull:2000rr} using the equation of motion \eqref{EOM0} for a non-chiral $(2,2)$-tensor that the independent fields in four dimensions can be taken as
\begin{align}
h_{\mu\nu\, ij} \sim \hat{T}_{\mu(ij)\nu}, \quad A_{\mu\, i} \sim \varepsilon^{jk} \hat{T}_{\mu ijk}, \quad \phi \sim \varepsilon^{ij} \varepsilon^{kl} \hat{T}_{ijkl}, \quad B_{\mu\nu} \sim \varepsilon^{ij} \hat{T}_{\mu\nu ij}\, ,
\end{align}
where the $i, j, \dots$ label the internal (flat) directions and $h_{\mu\nu\, ij}$ is trace-free,
\begin{equation}
\delta^{ij} h_{\mu\nu\, ij} = 0 .
\end{equation}
We define the two gravitons $h_{\mu\nu}$ and $\tilde{h}_{\mu\nu}$ by
\begin{equation}
(h_{\mu\nu\,ij}) = \begin{pmatrix}
h_{\mu\nu} & \tilde{h}_{\mu\nu} \\ \tilde{h}_{\mu\nu} & - h_{\mu\nu} 
\end{pmatrix} .
\end{equation}

Now, the self-duality condition in six dimensions further reduces the independent degrees of freedom by a factor of two. It implies that $h_{\mu\nu}$ and $\tilde{h}_{\mu\nu}$ are dual to each other. The same goes for the vector fields $A_\mu \equiv A_{\mu 4}$ and $\tilde{A}_\mu \equiv A_{\mu 5}$, and for the $(\phi, B_{\mu\nu})$ pair. From their internal indices, it is clear that $SO(2)$ rotations of the internal space induce the duality rotations in four dimensions. More precisely, under a rotation of angle $\theta$ in the internal plane, the fields transform as
\begin{equation}\label{eq:4dduality22}
\begin{pmatrix}
h \\ \tilde{h}
\end{pmatrix} \rightarrow \begin{pmatrix} \cos 2 \theta & \sin 2 \theta \\ - \sin 2 \theta & \cos 2 \theta \end{pmatrix} \begin{pmatrix}
h \\ \tilde{h}
\end{pmatrix}\, , \qquad \begin{pmatrix}
A \\ \tilde{A}
\end{pmatrix} \rightarrow \begin{pmatrix} \cos \theta & \sin \theta \\ - \sin \theta & \cos \theta \end{pmatrix} \begin{pmatrix}
A \\ \tilde{A}
\end{pmatrix}\, ,
\end{equation}
while $\phi$ is of course a singlet.

\subsection{Reduction of the $(2,1)$-tensor}

The ``exotic graviton-photon" $\phi_{\alpha_1 \alpha_2 \beta_1}$ is given by the $(2,1)$ Young tableau,
\begin{equation}
    \phi_{\alpha_1 \alpha_2 \beta_1} 
\sim 
\ydiagram{2,1} \,,
\end{equation}
which means that it satisfies
\begin{equation}
\phi_{[\alpha_1 \alpha_2] \beta_1} =  \phi_{\alpha_1\alpha_2  \beta_1}, \quad \phi_{[\alpha_1 \alpha_2 \beta_1]} = 0 \, .
\end{equation}
The gauge symmetries are given by \cite{Curtright:1980yk}
\begin{align} \label{eq:phigauge}
\delta \phi_{\alpha_1 \alpha_2 \beta_1} = \pd_{[\alpha_1}S_{\alpha_2 ]\beta_1} +\pd_{[\alpha_1}A_{\alpha_2 ]\beta_1}- \pd_{\beta_1} A_{\alpha_1 \alpha_2}  \, ,
\end{align}
where $S$ and $A$ are arbitrary symmetric and antisymmetric tensors, respectively.

This field reduces in $5$ spacetime dimensions to the standard Pauli-Fierz field, plus a standard massless vector field.  It provides for that reason what can be regarded as an exotic description of the combined Einstein-Maxwell system. This is to be constrasted with the usual Pauli-Fierz field $h_{\mu \nu}$ in $6$ spacetime dimensions, which gives one graviton, one photon and one massless scalar upon Kaluza-Klein reduction to $5$ spacetime dimensions, and with the exotic graviton of the previous section, which only gives one graviton.

\subsubsection{Equations of motion}

The corresponding gauge invariant curvature (the Riemann tensor) is defined by 
\begin{equation}
 R_{\alpha_1 \alpha_2 \alpha_3 \beta_1 \beta_2} 
\equiv 
\pd_{[\alpha_1}\phi_{\alpha_2 \alpha_3][ \beta_1, \beta_2]}
= \frac{1}{12} \pd_{\alpha_1}\!\pd_{\beta_2}\phi_{\alpha_2 \alpha_3 \beta_1} \pm \cdots \, .
\end{equation}
It is a tensor of Young symmetry type $(2,2,1)$, $R\sim \tyng{2,2,1}\,$, 
which means that it satisfies
\begin{equation}\label{eq:symR}
R_{\alpha_1 \alpha_2 \alpha_3 \beta_1 \beta_2 }= R_{[\alpha_1 \alpha_2 \alpha_3] \beta_1 \beta_2 }= R_{\alpha_1 \alpha_2 \alpha_3 [\beta_1 \beta_2]}\, , \qquad R_{[\alpha_1 \alpha_2 \alpha_3 \beta_1] \beta_2 } = 0 \, .
\end{equation}
Its definition in terms of $\phi$ also implies that it satisfies the differential Bianchi identities
\begin{align}\label{Bianchi}
\pd_{[\mu}  R_{\nu \rho \sigma]\alpha \beta} = 0\, ,  \qquad R_{\mu \nu \rho [\alpha \beta , \gamma]} =0 \, .
\end{align}

The equations of motion for $\phi$ are then given by the self-duality condition on the Riemann tensor on the first group of indices,
\begin{equation}
  R_{\alpha_1 \alpha_2 \alpha_3 \beta_1 \beta_2}
  =
  \frac{1}{3!}\varepsilon_{\alpha_1 \alpha_2 \alpha_3 \gamma_1 \gamma_2 \gamma_3}R\indices{^{\gamma_1 \gamma_2 \gamma_3 }_{\beta_1 \beta_2}} \,.
\label{self-duality}
\end{equation}
Because of the cyclic identity satisfied by $R$ (last of \eqref{eq:symR}), this condition implies the usual equation of motion
\begin{equation}
R\indices{^{\mu\nu\rho}_{\sigma\rho}} = 0 
\end{equation}
for a non-chiral $(2,1)$ field, but is stronger and only half the number of degrees of freedom are propagating ($8$ instead of $16$), as in the previous cases.

\subsubsection{Dimensional reduction to five dimensions}

In five dimensions, we have the four fields
\begin{equation}
T_{\mu\nu\rho} = \hat{\phi}_{\mu\nu\rho}, \quad h_{\mu\nu} = \hat{\phi}_{5(\mu\nu)}, \quad B_{\mu\nu} = \hat{\phi}_{5[\mu\nu]}, \quad A_\mu = \hat{\phi}_{\mu55} \, .
\end{equation}
The components $\hat{\phi}_{\mu\nu 5}$ are not independent from those: they are given by $\hat{\phi}_{\mu\nu 5} = - 2 \hat{\phi}_{5[\mu\nu]}$ because of the cyclic identity satisfied by $\phi$.

It then follows from the self-duality condition in $D=6$ that $T_{\mu\nu\rho}$ is dual to the graviton $h_{\mu\nu}$, and that $B_{\mu\nu}$ is dual to the vector field $A_{\mu}$ \cite{Hull:2000zn}. All in all, this describes one metric and one vector field, as announced.

\subsubsection{Dimensional reduction to four dimensions}

In four dimensions, one gets two scalar fields, two vector fields, and linearized gravity. The way this arises from six dimensions is through the reduction
\begin{equation}
h_{\mu\nu\, i} \sim \hat{\phi}_{i(\mu\nu)}, \quad A_{\mu\, ij} \sim \hat{\phi}_{\mu(ij)}, \quad B_{\mu} \sim \varepsilon^{ij} \hat{\phi}_{\mu ij}, \quad \phi_i \sim \varepsilon^{jk} \hat{\phi}_{jki}, \quad B_{\mu\nu\, i} \sim \hat{\phi}_{\mu\nu i} ,
\end{equation}
with $\delta^{ij} A_{\mu\, ij} = 0$ (by a gauge transformation) and the following duality relations:
\begin{itemize}
\item the two $2$-forms $B_{\mu\nu \,i}$ are dual to the two scalars $\phi_i$;
\item the two gravitons $h_{\mu\nu} = h_{\mu\nu\, 4}$ and $\tilde{h}_{\mu\nu\, 5}$ are dual to each other; and
\item the two remaining vectors contained in $A_{\mu\,ij}$, defined by
\begin{equation}
(A_{\mu\,ij}) = \begin{pmatrix}
A_{\mu} & \tilde{A}_{\mu} \\ \tilde{A}_{\mu} & - A_{\mu} 
\end{pmatrix} \, ,
\end{equation}
are also dual to each other.
\end{itemize}
Then, by rotating the internal space, the two scalars rotate into each other, and the vector $B_\mu$ is a singlet. For the other fields, it gives the duality transformations
\begin{equation}
\begin{pmatrix}
h \\ \tilde{h}
\end{pmatrix} \rightarrow \begin{pmatrix} \cos \theta & \sin \theta \\ - \sin \theta & \cos \theta \end{pmatrix} \begin{pmatrix}
h \\ \tilde{h}
\end{pmatrix}\, , \qquad \begin{pmatrix}
A \\ \tilde{A}
\end{pmatrix} \rightarrow \begin{pmatrix} \cos 2 \theta & \sin 2 \theta \\ - \sin 2 \theta & \cos 2 \theta \end{pmatrix} \begin{pmatrix}
A \\ \tilde{A}
\end{pmatrix}
\end{equation}
between the fields and their duals. Notice that the relative angles are switched with respect to \eqref{eq:4dduality22}: this is because $h$, $\tilde{h}$ form a vector in internal space, while $A$, $\tilde{A}$ parametrize a traceless symmetric matrix. This is reversed with respect to the internal index structure of the previous case.
\chapter{First-order actions for twisted self-duality}
\label{chap:twisted}

In this chapter, we write first-order actions for $p$-forms and linearized gravity in which the field and its dual appear on the same footing. Just like in the case of vector fields reviewed in chapter \ref{chap:emduality}, this is done by going to the Hamiltonian formalism and solving the constraints. This was done in \cite{Bunster:2011qp} for the case of $p$-forms, thereby recovering the actions first written in \cite{Sen:1994eb}. For linearized gravity, this was first done in \cite{Henneaux:2004jw} for dimension four and then in \cite{Bunster:2013oaa} for dimension five, with indications towards the general case. Actions for the self-dual fields in six dimensions considered in section \ref{sec:selfdualfields} will be written in the next chapter using the same techniques; they will give the actions constructed here upon dimensional reduction.

We begin by reviewing the case of $p$-forms, following \cite{Bunster:2011qp}. We then go to the case of linearized gravity in arbitrary dimension \cite{Lekeu:2018kul}, generalizing the results of \cite{Henneaux:2004jw,Bunster:2013oaa}. Solving the constraints requires the introduction of new variables of mixed Young symmetry, called the ``prepotentials"\footnote{Those were called ``superpotentials" in \cite{Henneaux:2004jw}. The terminology has since settled to prepotentials; this is because the usual variables of the variational principle (the potentials) are given by derivatives of the prepotentials.}. A new feature in the case of gravity with respect to the $p$-form case is that the action is invariant under local Weyl transformations of the prepotentials; this requires therefore the systematic construction of invariant tensors. Those are called the Cotton tensors by analogy with three-dimensional gravity. In dimension four, the two prepotentials have the same index structure; the $SO(2)$ symmetry of linearized gravity is then manifest at the level of the action as rotations of those variables \cite{Henneaux:2004jw}.

Finally, we also apply this treatment to the Rarita-Schwinger spin 3/2 field (gravitino) in arbitrary dimension, and write the corresponding supersymmetry transformations of linearized supergravity using these prepotential variables, thereby generalizing the work of \cite{Bunster:2012jp}\footnote{A similar reformulation was already carried out for the four-dimensional linearized ``hypergravity" (a supermultiplet combining the graviton and a spin $5/2$ field), without systematic use of conformal tools, in \cite{Bunster:2014fca}.}.

The original results of this chapter appeared in the preprint \cite{Lekeu:2018kul}, written in collaboration with A. Leonard. I would also like to thank N. Boulanger, M. Henneaux, J. Matulich and S. Prohazka for useful discussions and comments on this work.

\section{Two-potential action for \texorpdfstring{$p$}{p}-forms}

It is appealing to have an action principle in which the field and its dual appear on the same footing. This is not the case for the free actions for a $p$-form or a $(D-p-2)$-form, since only one field appears but not the other. The auxiliary action \eqref{eq:pformaux} of appendix \ref{app:pformduality} does not work either, since the two fields are not on the same footing (see also section \ref{sec:embeddingtensor} for a similar situation). As in the case of vector fields in four dimensions, the way to achieve this goal is to go to the Hamiltonian formalism and solve the constraints (see chapter \ref{chap:emduality}). The resulting action gives the equations of motion in their twisted self-duality form.

For a free $p$-form, the only constraint appearing in the Hamiltonian is the momentum constraint (Gauss' law)
\begin{equation}
\pd_{i_1} \pi^{i_1 i_2 \dots i_p} = 0 .
\end{equation}
It is solved by introducing the potential $Z_{k_1 \dots k_{d-p-1}}$ as
\begin{equation}\label{eq:piofZ}
\pi^{i_1 \dots i_p} = \cB^{i_1 \dots i_p}[Z] \equiv \frac{1}{(d-p-1)!}\varepsilon^{i_1 \dots i_p j k_1 \dots k_{d-p-1}} \pd_j Z_{k_1 \dots k_{d-p-1}} \, ,
\end{equation}
where $d = D-1$ is the space dimension and we introduced the definition of the magnetic field $\cB[Z]$ of a $(d-p-1)$-form. The new field $Z$ is just the spatial components of the dual $(D-p-2)$-form $\tilde{A}$. The expression \eqref{eq:piofZ} can be inserted back in the Hamiltonian action; this gives \cite{Bunster:2011qp}
\begin{equation}\label{eq:twistedaction}
S[A,Z] = \int \dtdx \left( \frac{1}{p!} \,\cB^{i_1 \dots i_p}[Z] \dot{A}_{i_1 \dots i_p} - \cH \right),
\end{equation}
where the Hamiltonian energy density is
\begin{equation}
\cH = \frac{1}{2} \left( \frac{1}{p!} \, \cB^{i_1 \dots i_p}[Z] \cB_{i_1 \dots i_p}[Z] + \frac{1}{(d-p-1)!} \, \cB^{i_1 \dots i_{d-p-1}}[A] \cB_{i_1 \dots i_{d-p-1}}[A] \right)
\end{equation}
and the magnetic field of $A$ is defined as
\begin{equation}
\cB^{i_1 \dots i_{d-p-1}}[A] \equiv \frac{1}{p!} \varepsilon^{i_1 \dots i_{d-p-1} j k_1 \dots k_p} \pd_j A_{k_1 \dots k_p} .
\end{equation}
Up to integration by parts, the kinetic term can be written as
\begin{align}
\frac{1}{p!} \cB^{i_1 \dots i_p}[Z] \dot{A}_{i_1 \dots i_p} &= \frac{(-1)^{(p+1)d+1}}{(d-p-1)!} \, \cB^{i_1 \dots i_{d-p-1}}[A] \dot{Z}_{i_1 \dots i_{d-p-1}} \\
&= \frac{1}{2} \, \rho_{MN} \, \dot{\cA}^M \cdot \cB[\cA^N] \, ,
\end{align}
where the two potentials are collected in $(\cA^M) = (A, Z)$ and the index contraction is implicit (and divided by the factorial of the number of contracted indices). In this way, the ``twist matrix" $\rho$ of \eqref{eq:twistedpform} appears in the action, which then truly treats both fields on the same footing. Note also that doing the same procedure, but starting from the free action for $\tilde{A}$ instead of $A$, yields the same final action. Conversely, one can go back from this action to the free (second order) action for $A$ or $\tilde{A}$ by eliminating one half of the variables.

The equations of motion following from this action are the spatial curl of the twisted self-duality equations \eqref{eq:twistedpform}. Therefore, only the spatial components of the form fields appear (this is also the case in the action \eqref{eq:twistedaction}, of course). Nevertheless, these equations are equivalent to the full set \eqref{eq:twistedpform}: the temporal components $\cA_{0jk\cdots}$ are recovered by using the Poincaré lemma (indeed, the difference between $\dot{\cA}_{ij\cdots}$ and $\cF_{0ijk\cdots}$ is a total (spatial) derivative of the temporal components). This replacement of $\dot{\cA}_{ij\cdots}$ by $\cF_{0ijk\cdots}$ is also allowed at the level of the action, because the magnetic fields are identically transverse.

Finally, we note that in the case $p=1$, $D=4$, the action reduces to
\be
S = \frac{1}{2} \int \dtdx[3] \left(\varepsilon_{MN} \dot{\mathcal{A}}^N_i \mathcal{B}^{Mi} -  \delta_{MN} \mathcal{B}^M_i \mathcal{B}^{Ni}\right) \, ,
\ee
with manifest $SO(2)$ invariance\footnote{There is a sign discrepancy in the kinetic term with respect to the action \eqref{eq:symlag} of chapter \ref{chap:emduality} (without scalars). This is just a redefinition of the potential $Z$ (compare \eqref{eq:piofZ} and \eqref{eq:momentumZ}).}. As discussed above, this continuous symmetry is of course absent when the field and its dual have different index structures.

\section{Conformal geometry}
\label{sec:conformal}

The goal of this section is to analyse the properties of certain conformal fields in $d$ Euclidean dimensions. In addition to their gauge transformations, those fields also enjoy local Weyl symmetries (hence the name ``conformal" by abuse of teminology). The geometric tensors for those symmetries are systematically constructed.

The cases we analyse here are those relevant for solving the constraints appearing in the Hamiltonian formulation of linearized gravity in $D = d+1$ spacetime dimensions. The pattern is well established and follows references \cite{Bunster:2012km,Bunster:2013oaa,Henneaux:2015cda,Henneaux:2016zlu}. The main feature of these cases is that the analogue of the Weyl tensor (the traceless part of the curvature tensors) identically vanishes; one must therefore construct the analogue of the Cotton tensor of three-dimensional gravity. These generalized Cotton tensors satisfy two important properties.
\begin{enumerate}
\item They completely control Weyl invariance: the Cotton tensor is zero if and only if the field can be written as a the sum of a gauge and a Weyl transformation.
\item They are divergenceless and also obey some appropriate trace condition. Conversely, any traceless, divergenceless tensor with the symmetries of the Cotton tensor can be written as the Cotton tensor of some field. (The ambiguities in determining this field are precisely the gauge and Weyl transformations that we consider.)
\end{enumerate}
The first of these properties is equivalent to the fact that any function of the fields that is invariant under gauge and Weyl transformations can be written as a function of the Cotton tensor only. The second property is the one that allows us to solve the constraints. The proofs of those two properties are presented in appendix \ref{app:cotton}. They strongly rely on the generalized Poincaré lemmas of \cite{Olver_hyper,DuboisViolette:1999rd,DuboisViolette:2001jk,Bekaert:2002dt} for tensors of mixed Young symmetry reviewed in appendix \ref{app:youngpoincare}.

\subsection{Bosonic $[d-2,\,1]$-field}
\label{app:confphi}

We first consider a bosonic field $\phi_{i_1 \dots i_{d-2} j }$ with the symmetries of the two-column $[d-2, \,1]$ Young tableau, i.e.,
\begin{equation}
\phi_{i_1 \dots i_{d-2} j } = \phi_{[i_1 \dots i_{d-2}] j }, \quad \phi_{[i_1 \dots i_{d-2} j] } = 0 ,
\end{equation}
with the local gauge and Weyl symmetries
\begin{equation}\label{eq:phigaugeapp}
\delta \phi\indices{^{i_1 \dots i_{d-2}}_{j} } = \partial^{[i_1} M\indices{^{i_2 \dots i_{d-2}]}_j} + \partial_j A^{i_1 \dots i_{d-2}} + \partial^{[i_1} A\indices{_j^{i_2 \dots i_{d-2}]}} + \delta^{[i_1}_j B^{i_2 \dots i_{d-2}]},
\end{equation}
where $M$ has the $[d-3,1]$ mixed symmetry and $A$, $B$ are totally antisymmetric. The tensors $M$ and $A$ are the usual gauge parameters for a mixed symmetry field while $B$ is the parameter for the local Weyl transformations of $\phi$.

\paragraph{Einstein tensor.} The Einstein tensor is obtained by taking a curl of $\phi$ on both groups of indices,
\begin{equation}
G\indices{^{i_1 \dots i_{d-2}}_{j}} [\phi] = \partial^k \partial_m \phi\indices{^{l_1 \dots l_{d-2}}_{n} } \varepsilon^{mni_1 \dots i_{d-2}} \varepsilon_{jkl_1 \dots l_{d-2}} .
\end{equation}
It is invariant under the gauge symmetries parametrized by $M$ and $A$. It is also identically divergenceless (on both groups of indices) and has the $[d-2,1]$ Young symmetry,
\begin{equation}
\pd_{i_1} G\indices{^{i_1 \dots i_{d-2}}_{j}} [\phi] = 0, \quad \pd^j G\indices{^{i_1 \dots i_{d-2}}_{j}} [\phi] = 0, \quad G\indices{_{[i_1 \dots i_{d-2}j]}} [\phi] = 0 \, .
\end{equation}
The converse of these properties is also true and is easily proven using the generalized Poincaré lemma of \cite{Bekaert:2002dt} for mixed symmetry fields.
\begin{itemize}
\item The condition $G[\phi] = 0$ implies that the field is is pure gauge,
\begin{equation}\label{eq:gaugeEinsteinphi}
G\indices{^{i_1 \dots i_{d-2}}_{j}} [\phi] = 0 \quad\Leftrightarrow\quad \phi\indices{^{i_1 \dots i_{d-2}}_{j} } = \partial^{[i_1} M\indices{^{i_2 \dots i_{d-2}]}_j} + \partial_j A^{i_1 \dots i_{d-2}} + \partial^{[i_1} A\indices{_j^{i_2 \dots i_{d-2}]}}
\end{equation}
for some $A$ and $M$ with the appropriate symmetry.
\item Any divergenceless tensor of $[d-2,1]$ symmetry is the Einstein tensor of some $[d-2,1]$ field,
\begin{equation}
\partial_{i_1} T\indices{^{i_1 \dots i_{d-2}}_{j}} = 0 \quad \Leftrightarrow \quad T = G[\phi] \;\text{ for some } \phi.
\end{equation}
(The divergencelessness of $T$ in the $j$ index follows from the cyclic identity $T_{[i_1 \dots i_{d-2} j]} = 0$.)
\end{itemize}

\paragraph{Schouten tensor.}  The Einstein tensor is not invariant under the Weyl symmetries parametrized by $B$, under which it transforms as
\begin{equation}\label{eq:varGphi}
\delta G\indices{^{i_1 \dots i_{d-2}}_{j}} [\phi] = \partial^k \partial_m B^{l_2 \dots l_{d-2} } \varepsilon^{mni_1 \dots i_{d-2}} \varepsilon_{jknl_2 \dots l_{d-2}} = - (d-1)! \partial^k \partial_m B^{l_2 \dots l_{d-2} } \delta^{m i_1 \dots i_{d-2}}_{jkl_2 \dots l_{d-2}} .
\end{equation}
For the trace $G^{i_2 \dots i_{d-2}} [\phi] \equiv G\indices{^{j i_2 \dots i_{d-2}}_{j}} [\phi]$, this implies
\begin{equation}\label{eq:vartraceGphi}
\delta G^{i_2 \dots i_{d-2}} [\phi] = \partial^k \partial_m B^{l_2 \dots l_{d-2} } \varepsilon^{mnj i_2 \dots i_{d-2}} \varepsilon_{jknl_2 \dots l_{d-2}} = 2(d-2)! \partial^k \partial_m B^{l_2 \dots l_{d-2} } \delta^{m i_2 \dots i_{d-2}}_{k l_2 \dots l_{d-2}}
\end{equation}
From the Einstein tensor and its trace, one can then define the Schouten tensor
\begin{equation}\label{eq:SofGphi}
S\indices{^{i_1 \dots i_{d-2}}_{j}} [\phi] = G\indices{^{i_1 \dots i_{d-2}}_{j}} [\phi] - \frac{d-2}{2} \,\delta^{[i_1}_j G^{i_2 \dots i_{d-2}]}[\phi]
\end{equation}
which has the important property of transforming simply as
\begin{equation}\label{eq:transfSchphi}
\delta S\indices{^{i_1 \dots i_{d-2}}_{j}} [\phi] = - (d-2)!\,\partial_j\partial^{[i_1} B^{i_2 \dots i_{d-2}]}
\end{equation}
under a Weyl transformation, as follows from \eqref{eq:varGphi} and \eqref{eq:vartraceGphi}.

The relation between the Einstein and the Schouten can be inverted to
\begin{equation}
G\indices{^{i_1 \dots i_{d-2}}_{j}} [\phi] = S\indices{^{i_1 \dots i_{d-2}}_{j}} [\phi] - (d-2) \,\delta^{[i_1}_j S^{i_2 \dots i_{d-2}]}[\phi],
\end{equation}
where $S^{i_2 \dots i_{d-2}} [\phi] \equiv S\indices{^{j i_2 \dots i_{d-2}}_{j}}$ is the trace of the Schouten tensor.
The divergencelessness of $G$ is then equivalent to
\begin{equation}\label{eq:bianchiSphi}
\partial_{i_1} S\indices{^{i_1 \dots i_{d-2}}_{j}} [\phi] - \partial_j S^{i_2 \dots i_{d-2}} = 0 .
\end{equation}
It also implies the identity
\begin{equation}
\partial^j S\indices{^{i_1 \dots i_{d-2}}_{j}} [\phi] - (d-2) \partial^{[i_1} S^{i_2 \dots i_{d-2}]} = 0
\end{equation}
because $G$ is divergenceless on its last index.

\paragraph{Cotton tensor.} The Cotton tensor is defined as
\begin{equation}\label{eq:cottonphi}
D^{i_1 \dots i_{d-2} \, j_1 \dots j_{d-2}}[\phi] = \varepsilon^{i_1 \dots i_{d-2}kl} \partial_k S\indices{^{j_1 \dots j_{d-2}}_{l}} [\phi] .
\end{equation}
Because of the transformation law \eqref{eq:transfSchphi} of the Schouten tensor, it is invariant under the full gauge and Weyl transformations \eqref{eq:phigaugeapp}. Moreover, $D[\phi] = 0$ implies that the field takes the form \eqref{eq:phigaugeapp} (see appendix \ref{app:cotton}). It also satisfies the following properties:
\begin{itemize}
\item it has the $[d-2,d-2]$ Young symmetry,
\begin{equation}\label{eq:cyclicDphi}
D_{[i_1 \dots i_{d-2} \, j_1] j_2 \dots j_{d-2}}[\phi] = 0\, ;
\end{equation}
\item it is divergenceless,
\begin{equation}
\partial^{i_1}D_{i_1 \dots i_{d-2}\, j_1 \dots j_{d-2}}[\phi] = 0\, ;
\end{equation}
and
\item its complete trace vanishes,
\begin{equation}
D\indices{^{i_1 \dots i_{d-2}}_{i_1 \dots i_{d-2}}}[\phi] = 0 \, .
\end{equation}
\end{itemize}
The first of these properties is equivalent to the identity \eqref{eq:bianchiSphi}. The second is evident from the definition of $D$; divergencelessness on the second group of indices then follows from the cyclic identity \eqref{eq:cyclicDphi}. Finally, the last property follows from the cyclic identity for the Schouten tensor.
Conversely, any tensor that satisfies these three properties must be the Cotton tensor of some $[d-2,1]$ field, as is proved in appendix \ref{app:cotton}.

Note that the transformation property \eqref{eq:transfSchphi} of the Schouten tensor implies that the tensor
\begin{equation}
D'_{ij}[\phi] = \varepsilon_{ikl_1 \dots l_{d-2}} \partial^k S\indices{^{l_1 \dots l_{d-2}}_j}[\phi]
\end{equation}
is also invariant under Weyl transformations of $\phi$. (With respect to definition \eqref{eq:cottonphi}, $D'$ is obtained by taking the curl of $S$ on the other group of indices.) Therefore, it must be a function of the Cotton tensor. Indeed, using the cyclic identity $S_{[i_1 \dots i_{d-2} j]} = 0$, one finds
\begin{equation}\label{eq:dprimephi}
D\indices{^{\prime i}_j}[\phi] = (d-2) D\indices{^{ik_2 \dots k_{d-2}}_{jk_2 \dots k_{d-2}}}[\phi] .
\end{equation}

\subsection{Bosonic $[d-2,\,d-2]$-field}
\label{app:confP}

We consider now a bosonic field $P_{i_1 \dots i_{d-2} \, j_1 \dots j_{d-2} }$ with the symmetries of the $[d-2, \,d-2]$ Young tableau, i.e.,
\begin{equation}
P_{i_1 \dots i_{d-2} \, j_1 \dots j_{d-2} } = P_{[i_1 \dots i_{d-2}] \, j_1 \dots j_{d-2}} = P_{i_1 \dots i_{d-2} \, [j_1 \dots j_{d-2}] }, \quad P_{[i_1 \dots i_{d-2} \, j_1] j_2 \dots j_{d-2} } = 0 .
\end{equation}
This field has the gauge symmetries
\begin{align} \label{eq:Pgaugeapp}
\delta P\indices{^{i_1 \dots i_{d-2}}_{j_1 \dots j_{d-2} }} = \; &\alpha\indices{^{i_1 \dots i_{d-2}}_{[j_1 \dots j_{d-3}, j_{d-2}] }} + \alpha\indices{_{j_1 \dots j_{d-2}}^{[i_1 \dots i_{d-3}, i_{d-2}] }} \nonumber \\
&+ \delta^{i_1 \dots i_{d-2}}_{j_1 \dots j_{d-2}} \, \xi ,
\end{align}
where $\alpha$ has the $[d-2,\,d-3]$ Young symmetry and the comma denotes the derivative, as before. The $\alpha$ transformations are the usual gauge tranformations for a $[d-2, \,d-2]$ mixed symmetry field, while the $\xi$ transformations are the local Weyl symmetries in this case.

\paragraph{Einstein tensor.} The Einstein tensor of $P$ is defined as
\begin{equation}
G_{ij}[P] = \varepsilon_{ikm_1 \dots m_{d-2}} \varepsilon_{jln_1 \dots n_{d-2}} \partial^k \partial^l P^{m_1 \dots m_{d-2} \, n_1 \dots n_{d-2} } .
\end{equation}
It is invariant under the $\alpha$ gauge symmetries. It is also symmetric and divergenceless. As in the previous case, the converse is also true: the condition $G_{ij}[P] = 0$ implies that $P$ is pure gauge, and any symmetric divergenceless tensor is the Einstein tensor of some field $P$ with the $[d-2,d-2]$ symmetry. The proof of these two theorems is direct using the Poincaré lemma of \cite{DuboisViolette:1999rd,DuboisViolette:2001jk} for rectangular Young tableaux.

\paragraph{Schouten tensor.} Under a Weyl transformation, we have
\begin{equation}
\delta G_{ij}[P] = (d-2)! (\delta_{ij} \lap \xi - \partial_i \partial_j \xi )
\end{equation}
and $\delta G[P] = (d-1)! \lap \xi$ for the trace $G[P] \equiv G\indices{^i_i}[P]$. The Schouten tensor is then defined as
\begin{equation}
S_{ij}[P] = G_{ij}[P] -  \frac{1}{d-1} \, \delta_{ij}\, G[P] .
\end{equation}
It transforms in a simple way under Weyl transformations,
\begin{equation}\label{eq:deltaSP}
\delta S_{ij}[P] = -(d-2)! \,\partial_i \partial_j \xi .
\end{equation}
It is also symmetric. One can invert the relation between the Schouten and the Einstein as $G_{ij} = S_{ij} - \delta_{ij} S$. Therefore, the divergencelessness of $G_{ij}$ implies that the Schouten tensor satisfies
\begin{equation}\label{eq:SPdiv}
\partial^i S_{ij} = \partial_j S\, .
\end{equation}

\paragraph{Cotton tensor.} The Cotton tensor is then defined as
\begin{equation}
D_{i_1 \dots i_{d-2} \, j }[P] = \varepsilon_{i_1 \dots i_{d-2} kl } \partial^k S\indices{^l_j}[P] .
\end{equation}
It is invariant under all the gauge symmetries \eqref{eq:Pgaugeapp} as a consequence of \eqref{eq:deltaSP}, and, conversely, the condition $D[P] = 0$ implies that $P$ is of the form \eqref{eq:Pgaugeapp}.
It is a tensor of $[d-2,1]$ mixed symmetry,
\begin{equation}\label{eq:DPsym}
D_{i_1 \dots i_{d-2} \, j } = D_{[i_1 \dots i_{d-2}] \, j }, \quad D_{[i_1 \dots i_{d-2} \, j]} = 0.
\end{equation}
The second of these equations is equivalent to \eqref{eq:SPdiv}. Moreover, because the Schouten is symmmetric, the Cotton is traceless,
\begin{equation}
D\indices{^j_{i_2 \dots i_{d-2} \, j }} = 0 .
\end{equation}
Lastly, the Cotton is also identically divergenceless on the first group of indices,
\begin{equation}
\partial^{i_1} D_{i_1 \dots i_{d-2} \, j } = 0 .
\end{equation}
The divergencelessness on the last index, $\partial^j D_{i_1 \dots i_{d-2} \, j } = 0$, is then a consequence of this and the second equation of \eqref{eq:DPsym}. Conversely, any $[d-2,1]$ field that is traceless and divergenceless is the Cotton tensor of some $P$.

\section{Prepotential action for linearized gravity}
\label{sec:graviton}

The results of the previous sections allow us to solve the constraints appearing in the Hamiltonian formulation of linearized gravity. In doing so, the dynamical variables are expressed in terms of two fields $\phi_{i_1 \dots i_{d-2} j}$ and $P_{i_1 \dots i_{d-2} j_1 \dots j_{d-2}}$ of respective $[d-2,1]$ and $[d-2,d-2]$ Young symmetry, called ``prepotentials". This generalizes to arbitrary dimension the work of \cite{Henneaux:2004jw,Bunster:2012km} and \cite{Bunster:2013oaa}, to which our results reduce in $D = 4$ and $5$ respectively. A further improvement over that previous work is the complete rewriting of the action in terms of the relevant Cotton tensors, which makes its gauge and Weyl invariance manifest\footnote{References \cite{Bunster:2012km,Bunster:2013oaa} already contain this rewriting of the kinetic term in four and five dimensions, but not the rewriting of the Hamiltonian.}.

\subsection{Hamiltonian action}

The Pauli-Fierz Lagrangian for linearized gravity is
\begin{equation} \label{LPF}
\mathcal{L}_\text{PF} = -\frac{1}{4} \partial_\mu h_{\nu\rho} \partial^\mu h^{\nu\rho} + \frac{1}{2} \partial_\mu h^{\mu\nu} \partial^\rho h_{\rho\nu} - \frac{1}{2} \partial_\mu h^{\mu\nu} \partial_\nu h\indices{^\rho_\rho} + \frac{1}{4} \partial_\mu h\indices{^\nu_\nu} \partial^\mu h\indices{^\rho_\rho} .
\end{equation}
It can also be written as
\begin{equation}
\mathcal{L}_\text{PF} = - \frac{3}{2} \delta^{\mu\nu\rho}_{\alpha\beta\gamma} \partial_\mu h\indices{_\nu^\gamma} \partial^\alpha h\indices{^\beta_\rho}
\end{equation}
and is invariant under the gauge transformations $\delta h_{\mu\nu} = 2 \partial_{(\mu} \xi_{\nu)}$.

The conjugate momenta are given by
\begin{align} \label{pi}
\pi_{ij} &= \frac{\partial \mathcal{L}_{PF}}{\partial \dot{h}^{ij}} = \frac{1}{2} \left( \dot{h}^{ij} - \delta_{ij} \dot{h} \right) - \partial_{(i} n_{j)} + \delta_{ij} \partial_k n^k \, ,
\end{align}
while the momenta conjugate to $n_i= h_{0i}$ and $N=h_{00}$ vanish identically. 
From this, the relation \eqref{pi} can be inverted to get
\begin{equation}
\dot{h}_{ij} = 2 \left( \pi_{ij} + \partial_{(i} n_{j)} - \delta_{ij} \frac{\pi}{d-1} \right).
\end{equation}
The canonical Hamiltonian density is then (up to total derivatives)
\begin{equation}
\mathcal{H}^\text{can} = \pi_{ij} \dot{h}^{ij} - \mathcal{L}_\text{PF} = \mathcal{H} + 2 n_i \mathcal{C}^i + \frac{1}{2} N \mathcal{C},
\end{equation}
where the Hamiltonian energy density is
\begin{align}
\mathcal{H} = \mathcal{H}_\pi + \mathcal{H}_h, \quad \mathcal{H}_\pi = \pi_{ij} \pi^{ij} - \frac{\pi^2}{d-1}, \quad \mathcal{H}_h = \frac{3}{2} \delta^{ijk}_{abc} \, \partial_i h\indices{_j^c} \partial^a h\indices{^b_k}
\end{align}
and the constraints are
\begin{align}
\mathcal{C} &= \partial_i \partial_j h^{ij} - \lap h, \\
\mathcal{C}^i &= - \partial_j \pi^{ij} .
\end{align}
Finally, the Hamiltonian action $S_H = \int\! dt \, d^d\! x\, ( \pi_{ij} \dot{h}^{ij} - \mathcal{H}^\text{can} )$ is
\begin{equation} \label{eq:hamgraviton}
S_H[h_{ij},\pi^{ij},n_i,N] = \int \! dt \, d^d\! x \left( \pi_{ij} \dot{h}^{ij} - \mathcal{H} - 2 n_i \mathcal{C}^i - \frac{1}{2} N \mathcal{C} \right).
\end{equation}
The dynamical variables are the space components $h_{ij}$ and their conjugate momenta $\pi^{ij}$. The other components of $h_{\mu\nu}$, namely $n_i = h_{0i}$ and $N = h_{00}$, only appear as Lagrange multipliers for the constraints $\mathcal{C}^i = 0$ (momentum constraint) and $\mathcal{C} = 0$ (Hamiltonian constraint).

\subsection{Hamiltonian constraint}

Up to a gauge transformation $\delta h_{ij} = 2 \partial_{(i} \xi_{j)}$, the constraint $\mathcal{C} = 0$ is solved by
\begin{equation}\label{eq:hprep}
h_{ij} = 2\,\varepsilon_{l_1 \dots l_{d-2} k (i} \partial^k \phi\indices{^{l_1 \dots l_{d-2}}_{j)}},
\end{equation}
where $\phi$ is a mixed symmetry field with the symmetries of the two-column $[d-2,1]$ Young tableau\footnote{\label{footnote:sign} Note that in $D=5$, this definition differs by a sign from \cite{Bunster:2013oaa}. The choice we make here is more convenient in arbitrary dimension.}, as is easily checked by direct substitution. This is the direct generalization of the expressions of \cite{Henneaux:2004jw,Bunster:2012km,Bunster:2013oaa}. The prepotential $\phi$ is determined up to a gauge and Weyl transformation of the form \eqref{eq:phigaugeapp}. Indeed, when it is plugged in equation \eqref{eq:hprep}, the second term of \eqref{eq:phigaugeapp} reproduces the gauge transformation of $h_{ij}$ with gauge parameter
\begin{equation}
\xi_i = \varepsilon_{l_1 \dots l_{d-2} k i} \partial^k A^{l_1 \dots l_{d-2}}
\end{equation}
and the other terms drop out. Remark also that the expression \eqref{eq:hprep} is traceless because of the symmetry of $\phi$ (the trace of the graviton is pure gauge).

Equation \eqref{eq:hprep} implies that the spatial components of the linearized Riemann tensor are given in terms of $\phi$ by
\begin{equation}\label{eq:Rofphi}
R\indices{^{ij}_{kl}}[h[\phi]] = \partial^{[i} \partial_{[k} h\indices{^{j]}_{l]}}[\phi] = - \frac{1}{2} \frac{1}{(d-2)!^2} \,\varepsilon^{ija_1 \dots a_{d-2}} \varepsilon_{klb_1\dots b_{d-2}} D\indices{_{a_1 \dots a_{d-2}}^{b_1 \dots b_{d-2}}}[\phi] .
\end{equation}
Using the properties of the previous section, it is easily proved that formula \eqref{eq:hprep} is unique (up to overall factors and gauge transformations). Let us spell out the proof.
\begin{enumerate}
\item The constraint $\mathcal{C} = 0$ is equivalent to the tracelessness condition
\begin{equation}
R\indices{^{ij}_{ij}} = 0
\end{equation}
on the linearized Riemann tensor of $h$. We can dualize this Riemann tensor on both groups of indices to get a tensor with $[d-2,\, d-2]$ Young symmetry (because $R$ itself has the $[2,2]$ symmetry). The constraint $\mathcal{C} = 0$ is then equivalent to the complete tracelessness of this tensor; it must therefore be equal to the Cotton tensor of some $\phi$. Dualizing back to $R$ (and adjusting factors), this gives equation \eqref{eq:Rofphi}.
\item We have established that the relation \eqref{eq:Rofphi} between the Riemann tensor of $h[\phi]$ and the Cotton tensor of $\phi$ must hold. Because of gauge invariance, any expression $h_{ij} = h_{ij}[\phi]$ for $h_{ij}$ differs from \eqref{eq:hprep} only by a gauge transformation of $h_{ij}$, i.e., a term of the form $\pd_{(i} \xi_{j)}$. Moreover, invariance of the Cotton tensor implies that a gauge and Weyl transformation of $\phi$ must induce a gauge variation of $h_{ij}$.
\end{enumerate}

\subsection{Momentum constraint}

The resolution of the momentum constraint $\mathcal{C}^i = 0$ is straightforward. It gives
\begin{equation}
\pi_{ij} = \varepsilon_{ikm_1 \dots m_{d-2}} \varepsilon_{jln_1 \dots n_{d-2}} \partial^k \partial^l P^{m_1 \dots m_{d-2}  n_1 \dots n_{d-2} } \equiv G_{ij}[P],
\end{equation}
where the $[d-2\, ,d-2]$ field $P$ is determined up to the gauge and Weyl symmetries \eqref{eq:Pgaugeapp}. This follows from the construction of the Einstein tensor of $P$ carried out in the previous section.

\subsection{Prepotential action}
\label{sec:actiongraviton}

We can now plug back these solutions in the Hamiltonian action \eqref{eq:hamgraviton}. The dynamical variables $h_{ij}$ and $\pi^{ij}$ are expressed in terms of their prepotentials $\phi$ and $P$ respectively. The constraints vanish identically, so the Lagrange multipliers $n$ and $N$ disappear from the action.

The kinetic term is thus
\begin{align}
\pi_{ij} \,\dot{h}^{ij} &= 2 \,G_{ij}[P]\, \varepsilon^{l_1 \dots l_{d-2} ki} \partial_k \dot{\phi}\indices{_{l_1 \dots l_{d-2}}^j}\, .
\end{align}
Integrating by parts, this can be written as
\begin{align}
\pi_{ij} \dot{h}^{ij} &= - 2 D_{i_1 \dots i_{d-2}j}[P] \,\dot{\phi}^{i_1 \dots i_{d-2}j} \\
&= 2 D_{i_1 \dots i_{d-2}j_1 \dots j_{d-2}}[\phi] \, \dot{P}^{i_1 \dots i_{d-2}j_1 \dots j_{d-2}} \label{eq:kineticPDphi} \\
&= D_{i_1 \dots i_{d-2}j_1 \dots j_{d-2}}[\phi] \, \dot{P}^{i_1 \dots i_{d-2}j_1 \dots j_{d-2}} - D_{i_1 \dots i_{d-2}j}[P] \,\dot{\phi}^{i_1 \dots i_{d-2}j}  .
\end{align}
For the Hamiltonian terms, we find
\begin{align}
\cH_\pi &= G_{ij}[P] S^{ij}[P] \\
&= P_{i_1 \dots i_{d-2}j_1 \dots j_{d-2}}\, \varepsilon^{i_1 \dots i_{d-2} kl} \partial_k D\indices{^{j_1 \dots j_{d-2}}_l}[P]
\end{align}
and
\begin{align}
\cH_h &= \frac{1}{(d-2)!} G_{i_1 \dots i_{d-2}j}[\phi] \, S^{i_1 \dots i_{d-2}j}[\phi] \\
&= \frac{1}{(d-2)!} \phi_{i_1 \dots i_{d-2}j} \, \varepsilon^{l_1 \dots l_{d-2}jk} \partial_k D\indices{^{i_1 \dots i_{d-2}}_{l_1 \dots l_{d-2}}}[\phi]\, ,
\end{align}
again up to total derivatives.

All in all, the action for linearized gravity becomes
\begin{align}
S[\phi, P] = \int \dtdx \,\Big( &D_{i_1 \dots i_{d-2}j_1 \dots j_{d-2}}[\phi] \, \dot{P}^{i_1 \dots i_{d-2}j_1 \dots j_{d-2}} - D_{i_1 \dots i_{d-2}j}[P] \,\dot{\phi}^{i_1 \dots i_{d-2}j} \nn \\
& - P_{i_1 \dots i_{d-2}j_1 \dots j_{d-2}}\, \varepsilon^{i_1 \dots i_{d-2} kl} \partial_k D\indices{^{j_1 \dots j_{d-2}}_l}[P] \label{eq:actiongravprepot} \\
& - \frac{1}{(d-2)!} \phi_{i_1 \dots i_{d-2}j} \, \varepsilon^{l_1 \dots l_{d-2}jk} \partial_k D\indices{^{i_1 \dots i_{d-2}}_{l_1 \dots l_{d-2}}}[\phi] \Big) . \nn
\end{align}
It is manifestly invariant under gauge and Weyl transformations of the prepotentials.

We obtained this action starting from the Pauli-Fierz action for linearized gravity. One could also start from the action for the dual graviton field $T_{\mu_1 \dots \mu_{D-3} \nu}$: this gives the same action, as shown explicitely for $D = 5$ in \cite{Bunster:2013oaa}. As in the case of $p$-forms, the role of the two prepotentials are then interchanged. Again, one can go back from \eqref{eq:actiongravprepot} to the usual second-order actions for the graviton or its dual by eliminating one-half on the variables.

In four dimensions, the two prepotentials are symmetric tensors, and the action reads
\begin{align}
S[\phi^a] = \int \dtdx[3] \,\left( \varepsilon_{\tta\ttb}\, D_{ij}[\phi^\tta] \, \dot{\phi}^{\ttb\, ij} - \delta_{\tta\ttb}\, \phi^\tta_{ij} \, \varepsilon^{jkl} \partial_k D\indices{^i_l}[\phi^\ttb] \right) \, ,
\end{align}
where we wrote $(\phi_{ij}^\tta) = (\phi_{ij}, P_{ij})$. It has a manifest $SO(2)$ invariance. This realizes the duality rotations of linearized gravity in a local way at the level of the action, as was first discovered in \cite{Henneaux:2004jw}.

\subsection{Equations of motion}

The equations of motion coming from the variation of $P$ are
\begin{equation}\label{eq:Ddotphi}
\dot{D}^{i_1 \dots i_{d-2} j_1 \dots j_{d-2}}[\phi] = - \varepsilon^{i_1 \dots i_{d-2} kl} \partial_k D\indices{^{j_1 \dots j_{d-2}}_l}[P] ,
\end{equation}
while the variation of $\phi$ yields the equation
\begin{equation}\label{eq:DdotP}
\dot{D}_{i_1 \dots i_{d-2} j}[P] = \varepsilon_{l_1 \dots l_{d-2}jk} \partial^k D\indices{_{i_1 \dots i_{d-2}}^{l_1 \dots l_{d-2}}}[\phi] .
\end{equation}
In terms of the original dynamical variables, they are the spatial curl of the spatial components of the twisted self-duality equation \eqref{eq:twistgrav}. They are equivalent to the full covariant equation \eqref{eq:twistgrav}, which is in turn equivalent to the linearized Einstein equations. This is proved using the relevant Poincaré lemmas to reintroduce the missing components. This proof appears for $D=4$ and $5$ in \cite{Bunster:2012km,Bunster:2013oaa}; the generalization to arbitrary $D$ is direct and we will not repeat it here.

\section{Gravitino field and supersymmetry}

In this section, we rewrite the action of the free Rarita-Schwinger field (gravitino) in terms of an antisymmetric rank  $d-2$ tensor-spinor, which we also call the prepotential of the gravitino. This was done in \cite{Lekeu:2018kul} by generalizing the results of \cite{Bunster:2012jp} to arbitrary dimension, with the improvement of writing the action in terms of the appropriate Cotton tensor of section \ref{app:conffermion}. We then write the supersymmetry transformations in the prepotential formalism.

The gravitino field is a vector-spinor $\psi_\mu$. The action is
\begin{equation}\label{eq:gravitinoactioncov}
S = - \int \!d^D\!x\, \bar{\psi}_\mu \gamma^{\mu\nu\rho} \partial_\nu \psi_\rho \, ,
\end{equation}
where the Dirac adjoint is $\bar{\psi}_\mu = i \psi_\mu^\dagger \gamma^0$. It is invariant under the gauge transformation
\begin{equation}\label{eq:gaugegravitino}
\delta \psi_\mu = \partial_\mu \nu
\end{equation}
with $\nu$ an arbitrary spinor (see appendix \ref{app:gammamatrices} for gamma matrix and spinor conventions).

\subsection{Conformal geometry of a fermionic $(d-2)$-form}
\label{app:conffermion}

As a preliminary, we consider the conformal geometry of an antisymmetric tensor-spinor $\chi_{i_1 \dots i_{d-2}}$, with the gauge and Weyl transformations
\begin{equation} \label{eq:chigaugeapp}
\delta \chi_{i_1 \dots i_{d-2}} = (d-2)\, \partial_{[i_1} \eta_{i_2 \dots i_{d-2}]} + \gamma_{i_1 \dots i_{d-2}} \rho\, ,
\end{equation}
where $\eta_{i_1 \dots i_{d-3}}$ and $\rho$ are spinor fields. This is the case relevant for solving the constraint associated to the gauge invariance \eqref{eq:gaugegravitino} of the gravitino field.

\paragraph{Einstein tensor.} The invariant tensor for the $\eta$ transformations is of course the curl
\begin{equation}
G_i[\chi] = \varepsilon_{ijk_1\dots k_{d-2}} \partial^j \chi^{k_1 \dots k_{d-2}} ,
\end{equation}
which we also call ``Einstein tensor" by analogy with the previous cases. It satisfies the following properties, which are easily proved using the usual Poincaré lemma with a spectator spinor index.
\begin{itemize}
\item It is invariant under the $\eta$ gauge transformations. Conversely, $G_i[\chi] = 0$ implies that $\chi_{i_1 \dots i_{d-2}}$ is pure gauge,
\begin{equation}
G_i[\chi] = 0 \quad\Leftrightarrow\quad \chi_{i_1 \dots i_{d-2}} = (d-2) \partial_{[i_1} \eta_{i_2 \dots i_{d-2}]} \;\text{ for some } \eta.
\end{equation}
\item It is identically divergenceless, $\partial^i G_i = 0$. Conversely, any divergenceless vector-spinor is the Einstein tensor of some antisymmetric tensor-spinor $\chi_{i_1 \dots i_{d-2}}$,
\begin{equation}
\partial^i T_i = 0 \quad \Leftrightarrow \quad T_i = G_i[\chi] \;\text{ for some } \chi.
\end{equation}
\end{itemize}

\paragraph{Schouten tensor.} As before, the Einstein tensor is not invariant under Weyl transformations. We have
\begin{equation}
\delta G^i[\chi] = \varepsilon^{ijl_1 \dots l_{d-2}} \gamma_{l_1\dots l_2}\partial_j \rho  = i^{m+1} (d-2)!\, \gamma^{ij} \gamma_0 \hat{\gamma}\, \partial_j \rho\, ,
\end{equation}
where $m = \lfloor (d+1)/2 \rfloor$ and the dimension-dependent matrix $\hat{\gamma}$ is defined in appendix \ref{app:gammamatrices}.

The Schouten tensor is then defined as
\begin{equation}
S_i[\chi] = \frac{1}{d-1} \left( \gamma_{ij} - (d-2) \delta_{ij} \right) G^j[\chi] .
\end{equation}
Using the gamma matrix identity \eqref{eq:gammadelta}, one can see that it transforms simply as a total derivative under Weyl rescalings,
\begin{equation}
\delta S_i [\chi] = \partial_i \nu, \qquad \nu = i^{m+1} (d-2)! \gamma_0 \hat{\gamma}\, \rho .
\end{equation}
The Einstein can be written in terms of the Schouten as
\begin{equation}\label{eq:GofSchi}
G_i[\chi] = \gamma_{ij} S^j[\chi] ,
\end{equation}
which implies that the Schouten identically satisfies
\begin{equation}\label{eq:dSfermionic}
\gamma_{ij} \partial^i S^j[\chi] = 0 .
\end{equation}

\paragraph{Cotton tensor.} The invariant tensor for Weyl transformations is the Cotton tensor
\begin{equation}\label{eq:defcottonchi}
D_{i_1 \dots i_{d-2}}[\chi] = \varepsilon_{i_1\dots i_{d-2}jk} \partial^j S^k[\chi] .
\end{equation}
It is identically divergenceless. Its complete gamma-trace also vanishes,
\begin{equation}\label{eq:cottongammatrace}
\gamma^{i_1 \dots i_{d-2}} D_{i_1 \dots i_{d-2}} = 0,
\end{equation}
as follows from \eqref{eq:dSfermionic} and identity \eqref{eq:gij}. Conversely, $D[\chi] = 0$ implies that $\chi$ takes the form \eqref{eq:chigaugeapp}, and any divergenceless, rank $d-2$ antisymmetric tensor-spinor satisfying the complete gamma-tracelessness condition \eqref{eq:cottongammatrace} is the Cotton tensor of some $\chi$. The proof of those properties is done in appendix \ref{app:cotton}.

\subsection{Hamiltonian action}

The action \eqref{eq:gravitinoactioncov} is already in first-order (Hamiltonian) form. Splitting space and time indices, it is equal to
\begin{equation} \label{eq:hamgravitino}
S_H = \int \!dt\,d^d\!x\, \left( \eta^i \dot{\psi}_i - \mathcal{H} - \psi_0^\dagger \mathcal{D} - \mathcal{D}^\dagger \psi_0 \right)\, ,
\end{equation}
where the conjugate momentum, the Hamiltonian density and the constraint are
\begin{align}
\eta^i &= - i \psi^\dagger_j \gamma^{ji}\, , \\
\mathcal{H} &= \bar{\psi}_i \gamma^{ijk} \partial_j \psi_k\, , \\
\mathcal{D} &= -i \gamma^{ij} \partial_i \psi_j \, .
\end{align}
The momentum conjugate to the temporal component $\psi_0$ identically vanishes, and $\psi_0$ only appears as a Lagrange multiplier for the constraint $\mathcal{D} = 0$.

\subsection{Solving the constraint}

The constraint $\mathcal{D} = 0$ is equivalent to
\begin{equation}
\partial_i \left( \gamma^{ij} \psi_j \right) = 0 .
\end{equation}
Application of the standard Poincaré lemma (with a spectator spinor index) then gives
\begin{equation}
\gamma^{ij} \psi_j = \varepsilon^{ik l_1 \dots l_{d-2}} \partial_k \chi_{l_1 \dots l_{d-2}}
\end{equation}
for some fermionic field $\chi_{l_1 \dots l_{d-2}}$ with $d-2$ antisymmetric indices.
Using the gamma matrix identity \eqref{eq:gammadelta}, we then get $\psi_i$ as
\begin{equation}\label{eq:psichi}
\psi_i = \frac{1}{d-1} \left( \gamma_{ij} - (d-2) \delta_{ij} \right) \varepsilon^{jk l_1 \dots l_{d-2}} \partial_k \chi_{l_1 \dots l_{d-2}} \equiv S_i[\chi] \, ,
\end{equation}
where we recognize the Schouten tensor of $\chi$ as defined in section \ref{app:conffermion}. This expression reproduces the result of \cite{Bunster:2012jp} for $d=3$ (we also remark that this result was already obtained for $d=4$, albeit in a different context and in Lorentzian rather than Euclidean dimension, in the early work \cite{Deser:1980hy}). Again, $\chi$ is called the ``prepotential" of the gravitino field. It is determined up to the local gauge and Weyl transformations \eqref{eq:chigaugeapp}. Indeed, as discussed in section \ref{app:conffermion}, the transformation properties of the Schouten are exactly such that a gauge and Weyl transformation of $\chi$ produces a gauge transformation of $\psi$.

\subsection{Prepotential action}

We can now plug back the prepotential expression $\psi_i = S_i[\chi]$ into the action \eqref{eq:hamgravitino}. The kinetic term is
\begin{equation}
-i S^\dagger_i[\chi] \gamma^{ij} \dot{S}_j[\chi] = - i \chi^\dagger_{i_1 \dots i_{d-2}} \dot{D}^{i_1 \dots i_{d-2}}[\chi],
\end{equation}
where $D_{i_1 \dots i_{d-2}}[\chi]$ is the Cotton tensor of $\chi$ defined in \eqref{eq:defcottonchi} and equality holds up to a total derivative. The Hamiltonian is
\begin{align}
\bar{S}_i [\chi] \gamma^{ijk} \partial_j S_k[\chi] &= -i \, \frac{(-i)^{m+1}}{(d-3)!} G_i^\dagger[\chi]\, \gamma_{l_1 \dots l_{d-3}} \hat{\gamma} \,D^{il_1\dots l_{d-3}}[\chi] \\
&= -i \, \frac{(-i)^{m+1}}{(d-3)!} \, \chi_{i_1 \dots i_{d-2}}^\dagger \varepsilon^{i_1 \dots i_{d-2} jk} \gamma^{l_1 \dots l_{d-3}} \hat{\gamma}\, \partial_j D_{k l_1 \dots l_{d-3}}[\chi] ,
\end{align}
where $m = \lfloor D/2 \rfloor = \lfloor (d+1)/2 \rfloor$ and we used identity \eqref{eq:gijk} and integration by parts. The constraint $\mathcal{D}$ and its Lagrange multiplier $\psi_0$ disappear, since now $\mathcal{D} = 0$ identically. Putting things together, the action is
\begin{equation}\label{eq:actionchi}
S[\chi] = -i \int \!dt\,d^d\!x\, \chi_{i_1 \dots i_{d-2}}^\dagger \left( \dot{D}^{i_1 \dots i_{d-2}}[\chi] - \frac{(-i)^{m+1}}{(d-3)!} \varepsilon^{i_1 \dots i_{d-2} jk} \gamma^{l_1 \dots l_{d-3}} \hat{\gamma}\, \partial_j D_{k l_1 \dots l_{d-3}}[\chi] \right) .
\end{equation}
It can also be written in a slightly more aesthetic way as
\begin{equation}\label{eq:actiongravitinotilde}
S[\chi] = -i \int \!dt\,d^d\!x\, \chi_{i_1 \dots i_{d-2}}^\dagger \left( \dot{D}^{i_1 \dots i_{d-2}}[\chi] - \varepsilon^{i_1 \dots i_{d-2} jk} \partial_j \tilde{D}_{k}[\chi] \right) ,
\end{equation}
where we define $\tilde{D}_i[\chi]$ as the contraction
\begin{equation}
\tilde{D}_i[\chi] = \frac{(-i)^{m+1}}{(d-3)!} \,\gamma^{l_1 \dots l_{d-3}} \hat{\gamma}\, D_{i l_1 \dots l_{d-3}}[\chi],
\end{equation}
which fulfills $\gamma^i \tilde{D}_i = 0$ and $\partial^i \tilde{D}_i = 0$.


\subsection{Equations of motion}

The equations of motion following from the action \eqref{eq:actiongravitinotilde} are
\begin{equation}\label{eq:eomchi}
\dot{D}^{i_1 \dots i_{d-2}}[\chi] = \varepsilon^{i_1 \dots i_{d-2} jk} \partial_j \tilde{D}_{k}[\chi] .
\end{equation}
It is interesting to show explicitly the equivalence between this equation and the Rarita-Schwinger equation, even though it was already proven in the previous sections at the level of the action.

The Rarita-Schwinger equation is
\begin{equation} \label{eomgravitino}
\gamma^{\mu\nu\rho} F_{\nu\rho} = 0 ,
\end{equation}
where $F_{\mu\nu} = 2 \partial_{[\mu} \psi_{\nu]}$ is the field strength of the gravitino.
Taking $\mu = 0$ and $\mu = i$, it is equivalent to the two equations
\begin{equation} \label{RSsplit}
\gamma^{ij} F_{ij} = 0, \quad \gamma^{ijk} F_{jk} - 2 \gamma^0 \gamma^{ij} F_{0j} = 0 .
\end{equation}
The first of those is the constraint $\mathcal{D} = 0$ and the second is the dynamical equation.

On the other hand, it follows from definition \eqref{eq:psichi} that the Cotton tensor is related to the field strength $F_{ij}$ as
\begin{equation}\label{eq:DofF}
D_{i_1 \dots i_{d-2}}[\chi] = \frac{1}{2}\varepsilon_{i_1 \dots i_{d-2} j k} F^{jk}, \quad\tilde{D}_i = \frac{1}{2} \gamma^0 \gamma_{ijk} F^{jk} .
\end{equation}
Using these relations, one finds that \eqref{eq:eomchi} is equivalent to the equations
\begin{equation}\label{eq:eomchiprime}
\dot{F}_{ij} = \gamma^0 \partial_{[i} \gamma_{j]kl} F^{kl}, \quad \gamma^{ij} F_{ij} = 0.
\end{equation}
One must keep in mind that $F_{ij}$ is expressed in terms of the prepotential $\chi$ in \eqref{eq:DofF}: this is equivalent to the constraint $\gamma^{ij} F_{ij} = 0$, which therefore must supplement the first equation of \eqref{eq:eomchiprime} if we write it in terms of $\psi_\mu$.

Instead of proving the equivalence between \eqref{eq:eomchi} and \eqref{eomgravitino} directly, we will prove that the sets \eqref{RSsplit} and \eqref{eq:eomchiprime} are equivalent. Notice that \eqref{RSsplit} contains the time components $\psi_0$ of the field $\psi_\mu$, while \eqref{eq:eomchiprime} does not but has one more derivative. This is not a problem but the usual feature of the prepotential formalism: the extra components $\psi_0$ come from \eqref{eq:eomchiprime} using the Poincaré lemma.

\begin{itemize}
\item \eqref{RSsplit} $\Rightarrow$ \eqref{eq:eomchiprime}:

Contracting the second equation with $\gamma_i$ and using the constraint, the Rarita-Schwinger equation also implies
\begin{equation}\label{eq:electricconstraint}
\gamma^i F_{0i} = 0 .
\end{equation}
(Another, equivalent way to show this is by contracting the covariant equation \eqref{eomgravitino} with $\gamma_\mu$: this gives $0 = \gamma^{\nu\rho} F_{\nu\rho} = \gamma^{ij} F_{ij} + 2 \gamma^0 \gamma^i F_{0i}$, which implies \eqref{eq:electricconstraint} using the constraint.)
This then gives the identity
\begin{equation}
\gamma^{ij} F_{0j} = ( \gamma^i \gamma^j - \delta^{ij} ) F_{0j} = - F\indices{_0^i} .
\end{equation}
Using this in the dynamical equation and multiplying by $\gamma^0$, we get
\begin{equation}
\gamma^0\gamma^{ijk} F_{jk} = 2 F\indices{_0^i}.
\end{equation}
This equation still contains the time component $\psi_0$, while \eqref{eq:eomchi} does not; we can get rid of them by taking an extra curl,
\begin{equation}
\gamma^0 \partial^{[i} \gamma^{j]kl} F_{kl} = 2 \partial^{[i} F\indices{_0^{j]}} .
\end{equation}
The right-hand side is equal to $2 \partial^{[i} \dot{\psi}^{j]} = \dot{F}^{ij}$, which then proves \eqref{eq:eomchiprime}.

\item \eqref{eq:eomchiprime} $\Rightarrow$ \eqref{RSsplit}:

Using $\dot{F}^{ij} = 2 \partial^{[i} \dot{\psi}^{j]}$, the first equation of \eqref{eq:eomchiprime} becomes
\begin{equation}
\partial_{[i} ( 2 \dot{\psi}_{j]} ) = \partial_{[i} (\gamma^0 \gamma_{j]kl} F^{kl} ).
\end{equation}
The usual Poincaré lemma then implies that $2 \dot{\psi}_{j} = \gamma^0 \gamma_{jkl} F^{kl} + 2 \partial_j \lambda$ for some spinor $\lambda$ that we are free to call $\psi_0$. This gives
\begin{equation}\label{eq:nearlyRS}
2F_{0j} = \gamma^0 \gamma_{jkl} F^{kl} .
\end{equation}
Now, contracting this equation with $\gamma^j$ and using the constraint, we get $\gamma^j F_{0j} = 0$; this in turn implies
\begin{equation}
F_{0j} = ( \delta_{ij} - \gamma_j\gamma_i )F\indices{_0^i} = \gamma_{ij} F\indices{_0^i} .
\end{equation}
We then recover \eqref{RSsplit} by putting this back in \eqref{eq:nearlyRS} and multiplying by $\gamma^0$.
\end{itemize}

\subsection{Supersymmetry transformations}
\label{sec:susylingrav}

The sum of the actions \eqref{LPF} and \eqref{eq:gravitinoactioncov} is invariant under the rigid supersymmetry transformations
\begin{align}
\delta h_{\mu\nu} &= \frac{1}{2}\, \bar{\epsilon}\,\gamma_{(\mu} \psi_{\nu )} + \text{h.c.} = \frac{1}{2}\, \bar{\epsilon}\,\gamma_{(\mu} \psi_{\nu )} - 
\frac{1}{2}\, \bar{\psi}_{(\mu }\gamma_{\nu )}\epsilon,
\\
\delta \psi_{\mu} &= \frac{1}{4}\, \partial_{\rho} h_{ \mu\nu}\, \gamma^{\nu\rho}\epsilon ,
\end{align}
where $\epsilon$ is a constant spinor parameter\footnote{We call these ``supersymmetry transformations" because they involve a fermionic parameter. Of course, they do not satisfy the supersymmetry algebra unless $D = 4$ because of the mismatch of the number of degrees of freedom.}.
In this section, we prove that the corresponding variations of the prepotentials are
\begin{align}
\delta \phi\indices{_{l_1 \dots l_{d-2} j}} &=
\frac{1}{4}\,\mathbb{P}_{[d-2,1]} \left( \bar{\epsilon} \gamma_j \chi_{l_1 \dots l_{d-2}} \right) + \hc \label{eq:deltaSUSYphi} \\
\delta P_{i_1 \dots i_{d-2} j_1 \dots j_{d-2}} &= - \frac{i^{m+1}}{4(d-2)!} \, \mathbb{P}_{[d-2, d-2]} \left( \bar{\epsilon} \, \hat{\gamma} \,\gamma_{j_{d-2} \dots j_1} \chi_{i_1 \dots i_{d-2}} \right) + \hc \label{eq:deltaSUSYP}
\end{align}
for the bosonic prepotentials and $\delta \chi = \delta_\phi \chi + \delta_P \chi$ with
\begin{align}
\delta_\phi \chi_{i_1 \dots i_{d-2}} = - \frac{(-i)^{m+1}}{2(d-2)!} \bigg[ &\varepsilon_{jkl_1 \dots l_{d-2}} \pd^j \phi\indices{_{i_1 \dots i_{d-2}}^k} \gamma^{l_1 \dots l_{d-2}} \label{eq:deltaSUSYphichi} \\
&+ \frac{(d-2)(d-3)}{2}\varepsilon_{[i_1|jk_1 \dots k_{d-2}} \pd^j \phi\indices{^{k_1 \dots k_{d-2}}_{|i_2}} \gamma_{i_3 \dots i_{d-2}]} \bigg] \hat{\gamma} \gamma^0 \epsilon \nn
\end{align}
and
\begin{equation}
\delta_P \chi_{i_1 \dots i_{d-2}} = - \frac{1}{2} \partial^p P\indices{_{i_1 \dots i_{d-2}}^{q_1 \dots q_{d-2}}}\varepsilon_{jpq_1 \dots q_{d-2}} \gamma^j \gamma^0 \epsilon \label{eq:deltaSUSYPchi}
\end{equation}
for the fermionic one. This is the generalization of the transformations found in \cite{Bunster:2012jp} for $d=3$ to arbitrary dimension. (Note that the second line of \eqref{eq:deltaSUSYphichi} is absent in $d=3$.)

\subsubsection{Variation of the first bosonic prepotential $\phi$}

The spatial components of the covariant expression give
\begin{equation}\label{eq:deltahij}
\delta h_{ij} = \frac{1}{2} \bar{\epsilon}\gamma_{(i} \psi_{j )} + \hc.
\end{equation}
By expressing $\psi_i$ in terms of its prepotential, one gets
\begin{align}
\delta h_{ij} &= \frac{1}{2} \bar{\epsilon}\gamma_{(i} \psi_{j)} + \hc \\
&= \frac{1}{2(d-1)} \bar{\epsilon}\gamma_{(i} \left( \gamma_{j)k} - (d-2) \delta_{j)k} \right) \varepsilon^{km l_1 \dots l_{d-2}} \partial_m \chi_{l_1 \dots l_{d-2}} + \hc \\
&= \frac{1}{2(d-1)} \bar{\epsilon} \left( \delta_{ij} \gamma_{k} - (d-1) \delta_{k(j}\gamma_{i)} \right) \varepsilon^{km l_1 \dots l_{d-2}} \partial_m \chi_{l_1 \dots l_{d-2}} + \hc \\
&= \frac{1}{2(d-1)} \bar{\epsilon} \partial_m \left( \delta_{ij} \gamma_{k}\varepsilon^{km l_1 \dots l_{d-2}} \chi_{l_1 \dots l_{d-2}} 
- (d-1) \gamma_{(i}\varepsilon_{j)}^{\phantom{j)}m l_1 \dots l_{d-2}} \chi_{l_1 \dots l_{d-2}}  \right) + \hc \, .
\end{align}
By noting that a full antisymmetrization over all the indices of the $\varepsilon$ together with the $j$ index of the $\delta$ vanishes identically, we can rewrite the first term as
\begin{align}
\bar{\epsilon} \partial^m  \delta_{ij} \gamma^{k}\varepsilon_{km l_1 \dots l_{d-2}} \chi^{l_1 \dots l_{d-2}} 
&= \bar{\epsilon} \partial^m  \delta_{ik} \gamma^{k}\varepsilon_{jm l_1 \dots l_{d-2}} \chi^{l_1 \dots l_{d-2}} \nn \\
&\quad + \bar{\epsilon} \partial^m  \delta_{im} \gamma^{k}\varepsilon_{kj l_1 \dots l_{d-2}} \chi^{l_1 \dots l_{d-2}} \nn \\
&\quad + \left(d-2\right)\bar{\epsilon} \partial^m  \delta_{il_1} \gamma^{k}\varepsilon_{km j l_2 \dots l_{d-2}} \chi^{l_1 \dots l_{d-2}} \\ &= \bar{\epsilon} \partial^m  \gamma_{i}\varepsilon_{jm l_1 \dots l_{d-2}} \chi^{l_1 \dots l_{d-2}} \nn \\
&\quad + \bar{\epsilon} \partial_i \gamma^{k}\varepsilon_{kj l_1 \dots l_{d-2}} \chi^{l_1 \dots l_{d-2}} \nn \\
&\quad + \left(d-2\right)\bar{\epsilon} \partial^m \gamma^{k}\varepsilon_{km j l_2 \dots l_{d-2}} \chi_i^{\ l_2 \dots l_{d-2}} .
\end{align}
Since this expression is already symmetric in $ij$, we can explicitly symmetrize it again. The second term then appears as a gauge variation of $h_{ij}$ and can be discarded. We thus obtain, up to a gauge transformation,
\begin{equation} \label{eq:varhij}
\delta h_{ij} = \frac{d-2}{2(d-1)} \bar{\epsilon} \partial_m \left( 
 \gamma_{k}\varepsilon_{(i}^{\phantom{(i} km l_2 \dots l_{d-2}} \chi_{j) l_2 \dots l_{d-2}}
-  \gamma_{(i}\varepsilon_{j)}^{\phantom{j)}m l_1 \dots l_{d-2}} \chi_{l_1 \dots l_{d-2}} \right) + \hc\, .
\end{equation}
On the other hand, carrying out the projection in \eqref{eq:deltaSUSYphi},
\begin{align}
\delta \phi\indices{_{l_1 \dots l_{d-2} j}}
&= \frac{1}{4}\,\mathbb{P}_{[d-2,1]} \left( \bar{\epsilon} \gamma_j \chi_{l_1 \dots l_{d-2}} \right) + \hc \nn \\
&= \frac{1}{4} \left[\bar{\epsilon} \gamma_j \chi_{l_1 \dots l_{d-2}} - \bar{\epsilon} \gamma_{[j} \chi_{l_1 \dots l_{d-2}]}\right] + \hc \nn \\
&= \frac{d-2}{4(d-1)}\left[\bar{\epsilon} \gamma_j \chi_{l_1 \dots l_{d-2}} + \bar{\epsilon}\gamma_{[l_1\vert} \chi_{ j \vert l_2 \dots l_{d-2} ] }\right] + \hc\, ,
\end{align}
the ansatz gives
\begin{align}
\delta h_{ij} &= 2\,\varepsilon^{l_1 \dots l_{d-2}}_{\phantom{l_1 \dots l_{d-2}} k(i\vert} \, \partial^k \delta \phi_{l_1 \dots l_{d-2}\vert j)} \nn \\
&= \frac{d-2}{2(d-1)} \bar{\epsilon} \partial_m \left[\varepsilon^{l_1 \dots l_{d-2}m}_{\phantom{l_1 \dots l_{d-2}m} (i\vert} \,  \gamma_{\vert j)} \chi_{l_1 \dots l_{d-2}}
+ \varepsilon^{k l_2 \dots l_{d-2}m}_{\phantom{kl_1 \dots l_{d-2}m} (i\vert} \, \gamma_{k} \chi_{\vert j) l_2 \dots l_{d-2} } \right] ,
\end{align}
which is the same expression as \eqref{eq:varhij}.

\subsubsection{Variations of the fermionic prepotential}

Splitting time and space, one has $\delta \psi_i = \delta_h \psi_i + \delta_\pi \psi_i$ for the spatial components of the gravitino, with
\begin{align}
\delta_h \psi_i &= \frac{1}{4} \partial_k h_{ij} \gamma^{jk} \epsilon, \\
\delta_\pi \psi_i &= \frac{1}{2} \left( \pi_{ij} - \delta_{ij} \frac{\pi}{d-1} \right) \gamma^j \gamma^0 \epsilon .
\end{align}
Expressing this in terms of prepotentials, we see that the variation $\delta \chi = \delta_\phi \chi + \delta_P \chi$ of the fermionic prepotential must be such that
\begin{equation}\label{eq:schoutensusy}
S_i[\delta_\phi \chi] = \frac{1}{4} \partial_k h_{ij}[\phi] \gamma^{jk} \epsilon, \quad S_i[\delta_P \chi] = \frac{1}{2} S_{ij}[P] \gamma^j \gamma^0 \epsilon .
\end{equation}

\paragraph{Variation $\delta_\phi \chi$ containing the first prepotential.}

First, it follows from \eqref{eq:schoutensusy} that the variation of the Cotton tensor of $\chi$ is
\begin{align}
D_{i_1 \dots i_{d-2}}[\delta_\phi \chi] &= - \frac{1}{4} \varepsilon_{i_1 \dots i_{d-2}kl} R\indices{^{kl}_{pq}}[h[\phi]] \gamma^{pq} \epsilon \\
&= \frac{1}{4(d-2)!} D\indices{_{i_1 \dots i_{d-2}}^{j_1 \dots j_{d-2}}}[\phi] \varepsilon_{j_1 \dots j_{d-2} pq} \gamma^{pq} \epsilon, \label{eq:deltachiD}
\end{align}
where we used the relation \eqref{eq:Rofphi} between the Riemann of $h[\phi]$ and the Cotton of $\phi$. Writing both sides as the curl of the respective Schouten tensors, this gives
\begin{equation}
S_i[\delta_\phi \chi] = \frac{1}{4(d-2)!} S\indices{^{j_1 \dots j_{d-2}}_i} [\phi] \varepsilon_{j_1 \dots j_{d-2} kl} \gamma^{kl} \epsilon
\end{equation}
up to a total derivative that can cancelled by adding the appropriate Weyl transformation to $\delta_\phi \chi$. (This is more convenient than \eqref{eq:schoutensusy} because it involves directly the Schouten tensors on both sides.) We now write this in terms of the Einstein tensors: this gives
\begin{equation}
G_i[\delta_\phi \chi] = - \frac{(-i)^{m+1}}{2(d-2)!}\left( G_{j_1 \dots j_{d-2} i}[\phi] \gamma^{j_1 \dots j_{d-2}}  + \frac{(d-2)(d-3)}{2} G\indices{_{k i j_3 \dots j_{d-2}}^k} [\phi] \gamma^{j_3 \dots j_{d-2}} \right) \hat{\gamma} \gamma^0 \epsilon .
\end{equation}
The computation uses relations \eqref{eq:GofSchi}, \eqref{eq:SofGphi} and the gamma matrix identity \eqref{eq:g2gd-2}. From there, the variation $\delta_\phi \chi$ is found by writing the right-hand side as a curl: this yields expression \eqref{eq:deltaSUSYphichi}.

\paragraph{Variation $\delta_P \chi$ containing the second prepotential.}

Using the identity $\gamma_{ij} S^j[\chi] = G_i[\chi]$, we must have for the Einstein tensor
\begin{equation}
G_i[\delta_P \chi] = \frac{1}{2} S\indices{^j_k}[P] \gamma_{ij} \gamma^k \gamma^0 \epsilon .
\end{equation}
The variations $\delta_P \chi$ can the be identified up to a gauge transformation by writing the right-hand sides as a curl. Using the identity $\gamma_{ij} \gamma^k = \gamma\indices{_{ij}^k} + \gamma_i \delta^k_j - \gamma_j \delta^k_i$ and the fact that $S_{ij}[P]$ is symmetric, we get
\begin{align}
\frac{1}{2} S\indices{^j_k}[P] \gamma_{ij} \gamma^k \gamma^0 \epsilon &= - \frac{1}{2} (S_{ij}[P] - \delta_{ij} S\indices{^k_k}[P] ) \gamma^j \gamma^0 \epsilon = - \frac{1}{2} G_{ij}[P] \gamma^j \gamma^0 \epsilon \\
&= \varepsilon_{ikl_1\dots k_{d-2}} \partial^k \left( - \frac{1}{2}\varepsilon_{jpq_1 \dots q_{d-2}} \partial^p P^{l_1 \dots l_{d-2} q_1 \dots q_{d-2}} \gamma^j \gamma^0 \epsilon \right)
\end{align}
from which we deduce the expression \eqref{eq:deltaSUSYPchi}.

\subsubsection{Variation of the second bosonic prepotential $P$}

Finally, we determine the variation of $P$ from the invariance of the kinetic term of the prepotential action. Equation \eqref{eq:deltachiD} implies for the hermitian conjugate
\begin{equation}
(D_{i_1 \dots i_{d-2}}[\delta_\phi \chi])^\dagger = - \frac{i^m}{2(d-2)!} \bar{\epsilon} \, \hat{\gamma} \gamma_{j_{d-2} \dots j_1} D\indices{_{i_1 \dots i_{d-2}}^{j_1 \dots j_{d-2}}}[\phi] ,
\end{equation}
where we used the gamma matrix identity \eqref{eq:gij}. The variation of the kinetic term of $\chi$ is then readily computed. It cancels with the variation of the bosonic kinetic term (more conveniently written in the form ``$P \, \dot{D}[\phi]$'', see \eqref{eq:kineticPDphi}) provided the variation of $P$ is given by \eqref{eq:deltaSUSYP}.

Checking directly that \eqref{eq:deltaSUSYP} reproduces the supersymmetry variation of the momentum $\pi_{ij}$ is cumbersome because of the many terms contained in the projection. However, it is not necessary since, if the variations of the other fields are known, the variation of $\pi_{ij}$ is uniquely determined by the invariance of the kinetic term in the Hamiltonian action.

\chapter{Self-dual actions}
\label{chap:selfdual}

In this chapter, we write actions for the six-dimensional self-dual fields described in chapter \ref{chap:gravduality}. The construction of an action principle for chiral fields is notoriously subtle \cite{Marcus:1982yu}. Nevertheless, one can formulate an action principle which is covariant but not manifestly so, as was first achieved for chiral $p$-forms in \cite{Henneaux:1988gg} by generalizing the work of \cite{Floreanini:1987as} for chiral bosons in two space-time dimensions.

The way the action can be constructed is by first rewriting the equations of motion in a manner that contains only the purely spatial components of the fields. The resulting equations still contain, in the cases of the $(2,2)$ and $(2,1)$-tensors, a constraint on these components that can be solved by the means of appropriate prepotentials and conformal geometry. The resulting equations then derive directly form an action principle.

Equivalently (following \cite{Deser:1997se,Bekaert:1998yp}), this action can be obtained by starting from the Hamiltonian action for a non-chiral field and solving the constraints, as in the previous chapter. A change of variables in the resulting action produces the sum of the actions for the chiral and anti-chiral components (which have the same structure but with the opposite sign for the kinetic term). Since these actions are decoupled, one can consistently drop the anti-chiral part to get the action for a chiral field. As in the first approach, this also gets rid of the temporal components of the fields, since they only appear as Lagrange multipliers for the constraints that have been solved.

Although not manifestly so, these actions are Poincaré-invariant, just like for the Hamiltonian actions of usual relativistic field theories. An important difference, however, is that they are \emph{intrinsically Hamiltonian}, i.e., first-order in the time derivatives. Indeed, there is no natural split of the variables into $q$'s and $p$'s; one cannot therefore naturally eliminate half of the variables (``conjugate momenta'') to go to a conventional second order formulation (in order to be able to do so, one would need to keep the anti-chiral part). The situation, thus, is not that there is no action, but that there is no natural second-order action. This is in line with the fact that the (independent) self-duality conditions are after all of the first order in the time derivatives, as we will see.

It should be noted that extra fields and non-polynomial terms (even in the free case) can be added to the action of a chiral $p$-form to restore manifest Lorentz invariance \cite{Pasti:1995ii,Pasti:1995tn,Pasti:1996vs}. It would be interesting to see if this could also be achieved here. Nevertheless, space-time covariance can also be controlled with Hamiltonian tools \cite{Dirac:1962aa,Schwinger:1963zza} (see also \cite{Bunster:2012hm} for the complementarity between explicit space-time covariance and duality).

Concerning the exotic graviton, the original contributions of this chapter were published in the paper \cite{Henneaux:2016opm} written in collaboration with M. Henneaux and A. Leonard. For the $(2,1)$-tensor, they appeared in the recent preprint \cite{Henneaux:2018rub}, written in collaboration with M. Henneaux, J. Matulich and S. Prohazka. Finally, for the exotic gravitino introduced in the last section, they are part of \cite{Henneaux:2017xsb}, also written with M. Henneaux and A. Leonard. 

\section{The chiral \texorpdfstring{$2$}{2}-form}

We begin by reviewing the action for the chiral $2$-form, following \cite{Henneaux:1988gg} and \cite{Deser:1997se,Bekaert:1998yp}, since the other fields follow the same pattern.

\subsection{Equations of motion}

Splitting its components into time an space, the self-duality equation of motion $ F = * F$ reads
\begin{equation}\label{eq:selfdual2formsplit}
F_{0ij} = \frac{1}{3!} \varepsilon_{ijklm} F^{klm} \, .
\end{equation}
Defining the electric and magnetic fields as usual by
\begin{align}
\cE_{ij} &= F_{0ij} = \dot{A}_{ij} + 2 \pd_{[i} A_{j]0}\, , \label{eq:electric2form}\\
\cB_{ij} &= \frac{1}{3!} \varepsilon_{ijklm} F^{klm} = \frac{1}{2} \varepsilon_{ijklm} \, \pd^k A^{lm} \, ,
\end{align}
it states that the electric field is equal to the magnetic field,
\begin{equation}\label{eq:E=Btwoform}
\cE_{ij} = \cB_{ij} \, .
\end{equation}
This equation still contains the temporal components $A_{0i}$ of the two-form, but they appear only through a total derivative (see \eqref{eq:electric2form}). To get rid of those components, it is therefore sufficient to take a further curl, giving
\begin{equation}\label{eq:curlEB2form}
\varepsilon_{ijklm} \, \pd^k \left( \cE^{lm} - \cB^{lm} \right) = \varepsilon_{ijklm} \, \pd^k \left( \dot{A}^{lm} - \cB^{lm} \right) = 0\, .
\end{equation}
This equation now only contains the spatial components $A_{ij}$ of the field, and is of course implied by the original equation $F = * F$. In fact, one can also go back from \eqref{eq:curlEB2form} to \eqref{eq:selfdual2formsplit}: indeed, \eqref{eq:curlEB2form} implies that
\begin{equation}
\dot{A}^{lm} - \cB^{lm} = \pd^{[l} \lambda^{m]}
\end{equation}
for some $\lambda^m$, using the (usual) Poincaré lemma. We are free to set $\lambda_m \equiv 2 A_{0m}$, which gives the original equation \eqref{eq:selfdual2formsplit}.

\subsection{Chiral action}

The equations of motion \eqref{eq:curlEB2form} follow simply from the variation of the action
\begin{equation}\label{eq:actiontwoform}
S[A] = - \frac{1}{4} \int \dtdx[5] \, A^{ij} \,\varepsilon_{ijklm} \, \pd^k \left( \dot{A}^{lm} - \cB^{lm} \right) \, ,
\end{equation}
as is easily checked\footnote{Note that the extra curl is needed: the naive action
\[
S_\text{wrong}[A_{ij},A_{0i}] = \int \dtdx[5] \, A^{ij} \left( \cE_{ij} - \cB_{ij} \right)
\]
contains no time derivative (they appear only through a total derivative) and can therefore not give the correct equations of motion. Moreover, it is not gauge invariant (since the electric field is not identically divergenceless) and would give the wrong canonical dimension to the field.}. It can also be written as
\begin{align}
S[A] &= - \frac{1}{2} \int \dtdx[5] \, A^{ij} \left( \dot{\cB}_{ij} - \frac{1}{2} \varepsilon_{ijklm} \, \pd^k \cB^{lm} \right) \\
&= \frac{1}{2} \int \dtdx[5] \, \cB^{ij} \left( \cE_{ij} - \cB^{ij} \right)\, ,
\end{align}
where we used an integration by parts in the second line. We can also replace $\dot{A}_{ij}$ by the electric field $\cE_{ij}$ everywhere, since the temporal components $A_{0i}$ drop out anyway. This last expression is the original one of \cite{Henneaux:1988gg} and makes the gauge invariance of the action manifest.

As we discussed above, this action yields the non-covariant equations of motion \eqref{eq:curlEB2form}, which are nevertheless physically equivalent to the original (covariant) equations $F = * F$. It is expressed purely in terms of spatial objects.

\subsection{Split of the non-chiral Hamiltonian action}

There is another way to get the action \eqref{eq:actiontwoform} of the previous section: it is by splitting the Hamiltonian action for a generic two-form in self-dual and anti-self dual components \cite{Deser:1997se,Bekaert:1998yp}. This makes the link between chiral and non-chiral actions very explicit.

We start with the action for a non-chiral action written in its two-potential formulation,
\begin{equation}
S[A,Z] = \frac{1}{4} \int\dtdx[5] \left( \cB_{ij}[Z] \dot{A}^{ij} + \cB_{ij}[A] \dot{Z}^{ij} - \cB_{ij}[A] \cB^{ij}[A] - \cB_{ij}[Z] \cB^{ij}[Z] \right)\, .
\end{equation}
This is the action \eqref{eq:twistedaction} of the previous chapter (i.e., the Hamiltonian action with constraints solved) in the case $p = 2$, $D = 6$. Now, doing the change of variables
\begin{equation}
A^\pm = A \pm Z
\end{equation}
splits the action into two non-interacting parts,
\begin{equation}
S[A^+, A^-] = \frac{1}{2} \int \dtdx[5] \, \cB^{ij}[A^+] \left( \dot{A}^+_{ij} - \cB_{ij}[A^+] \right) + \frac{1}{2} \int \dtdx[5] \, \cB^{ij}[A^-] \left( \dot{A}^-_{ij} + \cB_{ij}[A^-] \right) \, .
\end{equation}
The first part is the action for a chiral tensor and the second is the action for an anti-chiral tensor. Since there are no cross-terms, it is consistent to only keep one or the other\footnote{Another way to understand this is to realize that setting $A^- = 0$ is the same as imposing $A = Z$: this is exactly setting the two-form and its dual to be equal (recall from the previous chapter that the $Z$'s are exactly the spatial components of the dual form).}. This gives the action written above from the Hamiltonian point of view.

\section{The \texorpdfstring{$(2,2)$}{(2,2)}-tensor}

We now follow the same procedure for the chiral $(2,2)$-tensor (exotic graviton), which was first presented in \cite{Henneaux:2016opm}. First, we prove that the self-duality condition is equivalent to a set of equations containing only the spatial components $T_{ijkl}$: schematically,
\begin{equation}
  R = * R
  \quad\Leftrightarrow\quad
  \mathcal{E} = \mathcal{B}
  \quad\Leftrightarrow\quad
  \curl(\mathcal{E} - \mathcal{B})=0 \, \text{ and } \,\bar{\cE}=0 \, .
\end{equation}
where $\mathcal{E}$ and $\mathcal{B}$ denote the electric and magnetic fields of $T_{\mu\nu\rho\sigma}$, respectively, and where the bar stands for the trace (see section \ref{sec:electr-magn-field22} for details). In addition to the dynamical equation, the last set contains a constraint on the spatial $T_{ijkl}$ variables (tracelessness of the electric field).  Second, we show that this constraint can be solved by expressing the field in terms of a new variable $Z$, called the prepotential. In terms of $Z$, the equation $\curl(\mathcal{E} - \mathcal{B})=0$ becomes
\begin{equation}
\dot D[Z] = \curl \, D[Z]\, ,
\end{equation}
where $D[Z]$ is the Cotton tensor of $Z$, defined in section \ref{sec:chiralaction22}. These final equations have the advantage that they allow for the remarkably simple variation principle
\begin{align} 
S[Z] &= \frac12 \int \dtdx[5]\,  Z_{mnrs} \left(\dot{D}^{mn rs} [Z] - \frac12\varepsilon^{mnijk}\partial_{k}D\indices{_{ij}^{rs}}[Z] \right)  \, .
\end{align}

\subsection{Electric and magnetic fields}
\label{sec:electr-magn-field22}

\paragraph{Definitions.}

To identify a subset of the equations $R = * R$ derivable from a variational principle, we introduce the electric and magnetic fields.  The electric field contains the components of the curvature tensor with the maximum number of indices equal to the time direction $0$, namely two, ${\mathcal E}^{ijkl}  \sim R^{0ij0kl} $, or equivalently on-shell, the components of the curvature with no index equal to zero. Since, in 5 dimensions, the curvature tensor $R_{pijqkl}$ of $T_{ijkl}$ is completely determined by the Einstein tensor
\begin{equation}
G\indices{^{ij}_{kl}} = \frac{1}{(3!)^2} R\indices{^{abc def}} \varepsilon\indices{_{abc}^{ij}} \varepsilon_{defkl} = R\indices{^{ij}_{kl}} - 2 \delta^{[i}_{[k} R\indices{^{j]}_{l]}} + \frac{1}{3} \delta^i_{[k} \delta^j_{l]} R
\end{equation}
(the analogue of the Weyl tensor identically vanishes), one defines explicitly the electric field as
\be {\mathcal E}^{ijkl} \equiv G^{ijkl}. \ee
Here, $R\indices{^{ij}_{kl}} = R\indices{^{mij}_{mkl}} $, $R\indices{^{j}_{l}} = R\indices{^{mij}_{mil}}$ and $R = R\indices{^{mij}_{mij}} $ are the successive traces of the Riemann. (Similar conventions will be adopted below for the traces of the tensors that appear.)
 The electric field has the $(2,2)$ Young symmetry and is identically transverse, $ \partial_i {\mathcal E}^{ijkl} = 0$.  

On the other hand, the magnetic field contains the components of the curvature tensor with only one index equal to $0$,
 \be
 {\mathcal B}_{ijkl}  = \frac{1}{3!} R\indices{_{0ij}^{abc}} \varepsilon_{abckl} .
 \ee
It is identically traceless, ${\mathcal B}^{jl} \equiv {\mathcal B}^{ijkl} \delta_{ik} = 0$, and transverse on the second pair of indices, $ \partial_k {\mathcal B}^{ijkl} = 0$.
 
\paragraph{First step.} The self-duality equation \eqref{SD22} implies
\be
{\mathcal E}^{ijrs} - {\mathcal B}^{ijrs} = 0 . \label{E=B22}
\ee
Conversely, the equation \eqref{E=B22} implies all the components of the self-duality equation \eqref{SD22}. This is verified in appendix \ref{app:eom} by repeating the argument of \cite{Bunster:2012km} given there for a $(2)$-tensor, which is easily adapted to a $(2,2)$-tensor.

This equation implies that the electric field is also traceless on-shell,
 \be
 {\mathcal E}^{ik} \equiv {\mathcal E}^{ijkl}\delta_{jl} = 0 \, , \label{trace0}
 \ee
as follows by taking the trace of \eqref{E=B22} since the magnetic field is identically traceless. Similarly, the magnetic field has the $(2,2)$ Young symmetry on-shell.

The trace condition \eqref{trace0} appears as a constraint on the initial conditions on the spatial variables $T_{ijrs}$ because it does not involve time derivatives.  There is no analogous constraint in the $p$-form case; this explains why the introduction of prepotentials is necessary here.


\paragraph{Second step.} To get  equations that involve only the spatial components $T_{ijrs}$, we proceed as in the $2$-form case and take the curl of \eqref{E=B22}, i.e.,
\be
\varepsilon^{mnijk}\partial_{k}\left( \cE\indices{_{ij}^{rs}} - \cB\indices{_{ij}^{rs}} \right)=0 , \label{E2=B2}
\ee
eliminating thereby  the gauge components $T_{0jrs}$. We also retain the equation \eqref{trace0}, which is a consequence of \eqref{E=B22} involving only the electric field.  There is no loss of physical information in going from \eqref{E=B22} to the system \eqref{trace0}, \eqref{E2=B2}. Indeed, as shown in appendix \ref{app:eom}, if \eqref{trace0} and \eqref{E2=B2} are fulfilled, one recovers \eqref{E=B22} up to a term that can be absorbed in a redefinition of $T_{0jrs}$.  The use of \eqref{trace0} is crucial in this argument.  It is in the form \eqref{trace0}, \eqref{E2=B2} that the self-duality equations $R = * R$ can be derived from a variational principle.

\subsection{Prepotential and action principle}
\label{sec:chiralaction22}

To achieve the goal of constructing the action for the chiral tensor, we first solve the constraint (\ref{trace0}) by introducing a prepotential $Z_{ijrs}$ for $T_{ijrs}$, as in the previous chapter.

Explicitly, the prepotential $Z_{ijrs}$ provides a parametrization of the most general $(2,2)$ tensor field $T_{ijrs}$ that solves the constraint \eqref{trace0}. One has
\be 
T_{ijrs} = {\mathbb P}_{(2,2)} \left( \frac{1}{3!} \varepsilon\indices{_{ij}^{k mn}} \partial_k Z_{mn rs} \right) + \text{(gauge transf.)}, \label{TZ}
\ee
which is a direct generalisation of the formula given in \cite{Henneaux:2004jw} for a symmetric tensor. This prepotential is determined up to the gauge symmetries
\be
\delta Z_{ijrs}  = {\mathbb P}_{(2,2)} \left(\partial_{i} \xi_{rs j} + \lambda_{ir}\delta_{js}\right) \label{Weyl}
\ee
where $\xi_{rs j}$ is a $(2,1)$-tensor parametrizing the ``spin-$(2,2)$ gauge transformations" of the prepotential and $\lambda_{ir}$ a symmetric tensor parametrizing its ``linearized spin-$(2,2)$ Weyl rescalings".

Because the Weyl tensor of a $(2,2)$-tensor identically vanishes, the relevant tensor that controls Weyl invariance is the Cotton tensor, defined as
\be D_{ij kl} = \frac{1}{3!} \varepsilon_{ijabc} \partial^{a} S\indices{^{bc}_{kl}},\label{eq:cotton22}
\ee
where
\be
S\indices{^{ij}_{kl}} = G\indices{^{ij}_{kl}} - 2 \delta^{[i}_{[k} G\indices{^{j]}_{l]}} + \frac{1}{3} \delta^i_{[k} \delta^j_{l]} G
\ee
is the Schouten tensor which has the key property of transforming simply as $\delta S\indices{^{ij}_{kl}} = -\frac{4}{27} \partial^{[j}\partial_{[k} \lambda\indices{^{i]}_{l]}}$ under Weyl rescalings. The Cotton tensor satisfies the following two properties.
\begin{itemize}
\item A necessary and sufficient condition for $Z_{ijrs}$ to be pure gauge is that its Cotton tensor vanishes.
\item The Cotton tensor $D_{ij kl}$  is a $(2,2)$-tensor which is gauge invariant under (\ref{Weyl}), as well as identically transverse and traceless, $\partial_i D^{ij rs} = 0 = D^{ij rs}  \delta_{js}$. Moreover, the converse is true: any field satisfying these three properties is the Cotton tensor of some prepotential $Z_{ijrs}$ determined up to the transformations \eqref{Weyl} (conformal Poincaré lemma).
\end{itemize}
The proof of those properties follows the exact same steps as in the case of gravity (see appendix \ref{app:cotton}) or higher-spin fields \cite{Henneaux:2015cda}. We do not reproduce them here.

The relation \eqref{TZ} implies that 
\be
\mathcal{E}^{ij rs} [T[Z]] \equiv G^{ij rs} [T[Z]] = D^{ij rs} [Z]
\ee
and gives the most general solution for $T_{ij rs}$ subject to the constraint that $\mathcal{E}^{ij rs}$ is traceless (this is proved in \cite{Henneaux:2015cda} for general higher spins described by completely symmetric tensors and in the previous chapter for gravity in arbitrary dimension, and is easily extended to tensors with mixed Young symmetry).

It also follows from \eqref{TZ} that 
\be
\frac12 \varepsilon^{mnijk}\partial_{k}\cB\indices{_{ij}^{rs}} = \dot{D}^{mn rs} [Z]
\ee
and therefore, in terms of the prepotential $Z_{ijrs}$, the self-duality condition \eqref{E2=B2} reads
\be
\frac12 \varepsilon^{mnijk}\partial_{k}D\indices{_{ij}^{rs}}[Z]  - \dot{D}^{mn rs} [Z]=0,
\ee
an equation that we can rewrite as
\be L^{mnrs \vert ijpq} Z_{ijpq} = 0 \, ,\label{E3=B3}
\ee
where the differential operator $L^{mnrs \vert ijpq}$ contains four derivatives and can easily be read off from \eqref{E3=B3}. The operator $L^{mnrs \vert ijpq}$ is {\em symmetric}, so that one can form the action 
\begin{align} 
S[Z] &= \frac12 \int \dtdx[5]\,  Z_{mnrs} \left(L^{mnrs \vert ijpq} Z_{ijpq}\right) \nn \\
&= \frac12 \int \dtdx[5]\,  Z_{mnrs} \left(\dot{D}^{mn rs} [Z] - \frac12\varepsilon^{mnijk}\partial_{k}D\indices{_{ij}^{rs}}[Z] \right) \label{Action}
\end{align}
which yields \eqref{E3=B3} as equations of motion. This action is the central result of this section. 

\subsection{Split of the non-chiral Hamiltonian action}

As in the case of chiral forms, the action \eqref{Action} can be obtained from the Hamiltonian action of a non-chiral $(2,2)$-tensor by splitting it in chiral and anti-chiral components.

The Lagrangian for a non-chiral $(2,2)$-tensor $T_{\mu\nu\rho\sigma}$ is given by \cite{Boulanger:2004rx}
\begin{equation}
\mathcal{L} = - \frac{5}{2} \,\delta^{\mu_1 \dots \mu_5}_{\nu_1 \dots \nu_5} \, M\indices{^{\nu_1\nu_2\nu_3}_{\mu_1\mu_2}} \, M\indices{_{\mu_3\mu_4\mu_5}^{\nu_4\nu_5}},
\end{equation}
where $M_{\mu\nu\rho \sigma\tau} = \partial_{[\mu}T_{\nu\rho]\sigma\tau}$ and $\delta^{\mu_1 \dots \mu_5}_{\nu_1 \dots \nu_5} = \delta^{\mu_1}_{[\nu_1} \dotsm \delta^{\mu_5}_{\nu_5]}$.
Its associated Hamiltonian action is
\begin{equation}\label{eq:hamaction22}
S_H = \int \! dt \, d^5\! x \left( \pi_{ijkl} \dot{T}^{ijkl} - \mathcal{H} - n_{ijk} \,\mathcal{C}^{ijk} - n_{ij}\, \mathcal{C}^{ij} \right),
\end{equation}
where the Hamiltonian density is $\mathcal{H} = \mathcal{H}_\pi + \mathcal{H}_T$ with
\begin{align}
\mathcal{H}_\pi &= 3 \left( \pi^{ijkl}\pi_{ijkl} - 2 \pi^{ij}\pi_{ij} + \frac{1}{3} \pi^2\right)\, , \\
\mathcal{H}_T &= \frac{5}{2}\, \delta^{i_1 \dots i_5}_{j_1 \dots j_5} \, M\indices{^{j_1 j_2 j_3}_{i_1 i_2}} \, M\indices{_{i_3 i_4 i_5}^{j_4 j_5}} \, .
\end{align}
In \eqref{eq:hamaction22}, the components $n_{ijk} = - 4 T_{ij0k}$ and $n_{ij} = 6 T_{0i0j}$ of $T$ with some indices equal to zero are the Lagrange multipliers for the constraints
\begin{align}
\mathcal{C}^{ijk} &\equiv \partial_l \pi^{ijlk} = 0\, ,\\
\mathcal{C}^{ij} &\equiv \mathcal{E}\indices{^{ikj}_k}[T] = 0\, .
\end{align}
Those constraints are solved by introducing two prepotentials $Z^{(1)}_{ijkl}$ and $Z^{(2)}_{ijkl}$ through
\begin{align}
\pi^{ijkl} &= G^{ijkl}[Z^{(1)}] \, ,\\
T_{ijkl} &= \frac{1}{3} \,{\mathbb P}_{(2,2)} \left( \varepsilon\indices{_{ij}^{abc}} \partial_a Z^{(2)}_{bckl} \right)\, .
\end{align}
In terms of prepotentials, we then have up to a total derivative
\begin{align}
\pi_{ijkl} \dot{T}^{ijkl} &= 2\, Z^{(1)}_{ijkl} \dot{D}^{ijkl}[Z^{(2)}] \, ,\\
\mathcal{H}_\pi &= 3\, G_{ijkl}[Z^{(1)}]S^{ijkl}[Z^{(1)}]\, , \\ \mathcal{H}_T &= 3\, G_{ijkl}[Z^{(2)}]S^{ijkl}[Z^{(2)}]\, .
\end{align}
Again up to a total derivative, one has
\begin{equation}
G_{ijkl}[Z]S^{ijkl}[Z\up{a}] = \frac{1}{3!} Z\up{a}_{ijkl} \varepsilon^{ijabc}\partial_a D\indices{_{bc}^{kl}}[Z\up{a}] \, .
\end{equation}
Therefore, defining the prepotentials $Z^{\pm}_{ijkl} = Z^{(1)}_{ijkl} \pm Z^{(2)}_{ijkl}$, the action splits into two parts,
\be
S[Z^+, Z^-] = S^+[Z^+] - S^-[Z^-] \, .
\ee
The action $S^+[Z^+]$ is exactly the action \eqref{Action} for a chiral tensor, while $S^-[Z^-]$ is the analog action for an anti-chiral tensor (which differs from \eqref{Action} only by the sign of the second term).

\section{The \texorpdfstring{$(2,1)$}{(2,1)}-tensor}

We now write the action for the chiral $(2,1)$-tensor (exotic graviton-photon). The procedure is the same as in the $(2,2)$ case.

\subsection{Electric and magnetic fields}
\label{sec:electr-magn-field21}

In this subsection, we show how the equation $R = * R$ can be replaced by equations involving the spatial components $\phi_{ijk}$ only.

\paragraph{Definitions.}

We define the electric field of $\phi$ by
 \begin{equation}
 \mathcal{E}^{ijklm} \equiv \frac{1}{2! 3!} R_{pqrab} \varepsilon^{pqrlm}\varepsilon^{abijk}\, .
 \end{equation}
 It contains only the purely spatial components $\phi_{ijk}$ and has Young symmetry {\tiny $\ydiagram{2,2,1}$}. It is also identically transverse in both groups of indices, i.e., $\pd_i{\mathcal E}^{ijklm}=0$ and $\pd_l{\mathcal E}^{ijklm}=0$.

The magnetic field is defined by the components of the curvature tensor with only one $0$,
 \begin{equation}
 {\mathcal B}_{ijklm} \equiv \frac{1}{2!} R\indices{_{0lm}^{ab}} \varepsilon_{abijk}\, .
 \end{equation}
It is identically double-traceless, $\bbar{\cB}_{i}\equiv {\mathcal B}\indices{_{ilm}^{lm}}=0$, and transverse on the first group of indices, $\pd^{i}{\mathcal B}_{ijklm}=0$.

\paragraph{First step.}

The self-duality equation \eqref{self-duality} implies $\mathcal{E} = \mathcal{B}$,
\begin{equation}
{\mathcal E}^{ijklm} - {\mathcal B}^{ijklm}=0\, .
\label{E=B21}
\end{equation}
Conversely, even though it does not contain the $\phi_{i00}$ variables, \eqref{E=B21} implies all the components of the self-duality equation \eqref{self-duality} (see Appendix \ref{app:dem1}).
As a consequence of the double-tracelessness of the magnetic field, equation \eqref{E=B21} implies that the electric
field is double-traceless on-shell, i.e.,
\begin{equation}
\bbar{\mathcal E}^{i} \equiv{\mathcal E}\indices{^{ilm}_{lm}}=0 \, .
\label{dtraceE}
\end{equation}
Similarly, the magnetic field has the $(2,2,1)$ symmetry on-shell.

\paragraph{Second step.}

The equation \eqref{E=B21} still contains the $\phi_{0jk}$ components. To get rid of those, we take the curl on the second group of indices,
\begin{equation}
\curl_2(\mathcal{E} - \mathcal{B}) \equiv \varepsilon_{abcpq}\pd^{a}({\mathcal E}\indices{_{ijk}^{bc}}-{\mathcal B}\indices{_{ijk}^{bc}})=0 \, .
\label{curlE=B21}
\end{equation}
This equation needs to be supplemented by \eqref{dtraceE}
which is a consequence of \eqref{E=B21} 
that contains only $\phi_{ijk}$ components.

As we show in Appendix \ref{app:dem2}, there is no loss of information in going from \eqref{E=B21} to the system \eqref{dtraceE}, \eqref{curlE=B21}. This system is therefore equivalent to the original equations $R = * R$.

\subsection{Prepotential and action principle}
\label{sec:chiralaction}

In order to construct an action for the chiral tensor, we solve the constraint \eqref{dtraceE} by introducing a prepotential $Z_{abcij}$ for $\phi_{ijk}$, as before.
Explicitly, the field can be written as
\begin{align}
\phi_{ijk} &= {\frac{1}{12}}{\mathbb P}_{(2,1)} \left( \partial^a Z\indices{^{bcd}_{ij}} \varepsilon_{kabcd} \right) + \text{(gauge)} \nonumber \\
&= \frac{1}{18} \left( \partial^a Z\indices{^{bcd}_{ij}} \varepsilon_{kabcd} - \partial^a Z\indices{^{bcd}_{k[i}} \varepsilon_{j]abcd} \right) + \text{(gauge)} \, .
\label{phiofZ}
\end{align}
The prepotential $Z_{abcij}$ has the $(2,2,1)$ Young symmetry, $Z \sim \tyng{2,2,1}\,$, i.e.,
\begin{equation}
Z_{abcij} = Z_{[abc]ij} = Z_{abc[ij]}\, , \quad Z_{[abci]j} = 0 \, .
\end{equation}
The fact that this expression solves the constraint is easily checked. Moreover, if $\phi$ is determined up to the gauge transformations \eqref{eq:phigauge}, the prepotential is determined up to the gauge and Weyl symmetries
\begin{equation}
\delta Z\indices{^{abc}_{ij}} = \delta_{g}Z\indices{^{abc}_{ij}} + \delta_{w}Z\indices{^{abc}_{ij}}\, ,
\label{invZ}
\end{equation}
where
\begin{align}
\delta_{g}Z\indices{^{abc}_{ij}} &= \pd^{[a}\alpha\indices{^{bc]}_{ij}}+ \pd^{[a}\beta\indices{^{bc]}_{[ij]}}- \frac{2}{3}\beta\indices{^{abc}_{[i,j]}} = {\mathbb P}_{(2,2,1)}\left( \frac{4}{3} \pd^{a}\alpha\indices{^{bc}_{ij}} + \pd_i \beta\indices{^{abc}_{j}}\right)\, , \label{eq:gaugeZ} \\
\delta_{w}Z\indices{^{abc}_{ij}} &= \rho^{[a}\delta^{bc]}_{ij}  = {\mathbb P}_{(2,2,1)}\left(\frac{4}{3} \rho^{a}\delta^{bc}_{ij}\right)\, , \label{eq:WeylZ}
\end{align}
as is again easily checked by direct substitution. Here, $\alpha\indices{^{ab}_{cd}}$, $\beta\indices{^{abc}_{d}}$ are $(2,2)$ and $(2,1,1)$ tensors, respectively, and the vector $\rho^{a}$ parametrizes Weyl rescalings.

The tensor that is invariant under the gauge transformations \eqref{eq:gaugeZ} is the Einstein tensor
\begin{equation}
    G\indices{_{de}^{l}}[Z] \equiv \frac{2}{3!4!}\varepsilon\indices{^{l}_{spqr}}\varepsilon_{deijk}\pd^{s}\pd^{i}Z^{pqrjk}\, .
\end{equation}
As usual, it is not invariant under the Weyl transformations \eqref{eq:WeylZ}. The invariant tensor controlling this Weyl invariance is the Cotton tensor
\begin{equation}\label{eq:defDtext}
    D_{abcde}[Z] \equiv \frac{1}{2}\varepsilon_{abclm}\pd^{m} S\indices{_{de}^{l}}[Z]\, ,
\end{equation}
where $S\indices{_{de}^{l}}[Z]$ is the Schouten tensor, which is defined from the Einstein by
\begin{equation}\label{eq:SofGtext}
S\indices{_{de}^{l}}[Z] \equiv G\indices{_{de}^{l}}[Z] + \frac{2}{3} \delta^{l}_{[d}G\indices{_{e]p}^p}[Z]
\end{equation}
and transforms as
\begin{equation}
\delta S\indices{_{de}^{l}}=\frac{1}{18}\pd_{[d}\pd^{l}\rho_{e]}
\end{equation}
under Weyl rescalings.

The Cotton tensor satisfies the three properties of the electric field:
\begin{itemize}
    \item it is of Young symmetry type $(2,2,1)$;
    \item it is transverse in both groups of indices; and
    \item it is double-traceless.
\end{itemize}
Therefore, the conformal Poincaré lemma states that there must exist a field $Z\indices{^{pqrjk}}$ such that
\begin{equation}\label{eq:E=DZ}
\mathcal{E}\indices{_{abc}^{ij}} = D\indices{_{abc}^{ij}}[Z]\, .
\end{equation}
This is proven as in the previous cases (see \cite{Henneaux:2018rub} for details). Equation \eqref{phiofZ} is such that \eqref{eq:E=DZ} is satisfied; moreover, it is uniquely determined from this condition up to gauge and Weyl transformations. Equation \eqref{phiofZ} is therefore the general solution to the constraint.

In addition, this implies for the curl of the magnetic field
\begin{equation}
    \frac{1}{2}\varepsilon_{abcpq}\pd^{a}{\mathcal B}\indices{_{ijk}^{bc}}[\phi[Z]]= \dot{D}_{ijkpq}[Z] \,.
\end{equation}
In terms of the prepotential $Z$, equation \eqref{curlE=B21} therefore reads
\begin{equation}
\frac{1}{2}\varepsilon_{abcpq}\pd^{a}D\indices{_{ijk}^{bc}}\left[Z\right]-\dot{D}_{ijkpq}\left[Z\right]=0\, .
\label{eomZ}
\end{equation}
This equation follows from the variation of the action
\begin{equation}
    S[Z] = \frac{1}{2}\int \!dt\,d^5\!x\, Z_{ijkpq}\left(\dot{D}^{ijkpq}[Z] - \frac{1}{2} \varepsilon^{abcpq}\pd_{a}D\indices{^{ijk}_{bc}}\left[Z\right]  \right) \, .
    \label{actionZ}
\end{equation}
The action \eqref{actionZ} for the chiral field of $(2,1)$ Young symmetry type constitutes the central result of this section.

\subsection{Split of the non-chiral Hamiltonian action}
\label{app:ham21}

We now show that the action \eqref{actionZ} for the chiral $(2,1)$-tensor can also be obtained by splitting the first-order (Hamiltonian) action for a non-chiral tensor into its chiral and anti-chiral parts.

The Lagrangian for a non-chiral $(2,1)$-tensor $\phi_{\mu\nu\rho}$ is given by \cite{Curtright:1980yk}
\begin{equation}
\mathcal{L} = - 12\, \delta^{\mu_1 \mu_2 \mu_3 \mu_4}_{\nu_1 \nu_2 \nu_3 \nu_4} M\indices{^{\nu_1 \nu_2 \nu_3}_{\mu_4}} M\indices{_{\mu_1 \mu_2 \mu_3}^{\nu_4}}\, ,
\end{equation}
with $M_{\mu\nu\rho\sigma} = \partial_{[\mu} \phi_{\nu\rho]\sigma}$.
The associated Hamiltonian action is
\begin{equation}
S_H = \int \! dt \, d^5\! x \, \left( \pi_{ijk} \dot{\phi}^{ijk} - \mathcal{H} - n_{ij} \,\mathcal{C}^{ij} - N_{i}\, \mathcal{C}^{i} \right) \, ,
\end{equation}
where the Hamiltonian density $\mathcal{H} = \mathcal{H}_\pi + \mathcal{H}_\phi$ is
\begin{align}
\mathcal{H}_\pi &= \frac{1}{4} \left( \pi^{ijk}\pi_{ijk} + \frac{2}{3} \pi\indices{^{ij}_j} \pi\indices{_{ik}^k} \right) \, , \\
\mathcal{H}_\phi &= 12\, \delta^{i_1 \dots i_4}_{j_1 \dots j_4} \, M\indices{^{j_1 j_2 j_3}_{i_4}} \, M\indices{_{i_1 i_2 i_3}^{j_4}} \, .
\end{align}
The temporal components $n_{ij} = \phi_{0ij}$ and $N_{i} = \phi_{i00}$ of $\phi_{\mu\nu\rho}$ are Lagrange multipliers enforcing the constraints
\begin{align}
\mathcal{C}^{ij} &\equiv 2 \,\partial_k \left( \pi^{ijk} - \pi^{kij} \right) = 0\, ,\\
\mathcal{C}_i &\equiv 12 \,\delta^{abc}_{ijk} \, \partial^j \partial_a \phi\indices{_{bc}^k} = 0 \, .
\end{align}
The first (momentum) constraint $\mathcal{C}^{ij} = 0$ is equivalent to
\begin{equation}\label{mom2}
\partial_k \pi^{kij} = 0 \qquad ( \Rightarrow \; \partial_k \pi^{ijk} = 0)
\end{equation}
because of the cyclic identity $\pi^{[ijk]} = 0$, and the second (Hamiltonian) constraint $\mathcal{C}_i = 0$ is equivalent to the double tracelessness of the electric field defined in section \ref{sec:electr-magn-field21},
\begin{equation}
\mathcal{E}\indices{_{ijk}^{jk}}[\phi] = 0 \, .
\end{equation}

The momentum constraint is solved by introducing a first prepotential $Z^{(1)}_{ijk lm}$ of $(2,2,1)$ symmetry, in terms of which the momentum reads
\begin{equation}
\pi_{ijk} = 6 \, G_{ijk}[Z^{(1)}]
\end{equation}
(we include an extra factor for convenience). The Hamiltonian constraint is solved by introducing a second prepotential $Z^{(2)}_{ijk lm}$, as in formula \eqref{phiofZ}. The various terms in the Hamiltonian action are then, up to a total derivative,
\begin{align}
\pi_{ijk} \dot{\phi}^{ijk} &= 2\, Z^{(1)}_{abcij} \dot{D}^{abcij}[Z^{(2)}] \, , \\
\mathcal{H}_\pi &= 18\, G_{ijk}[Z^{(1)}] S^{ijk}[Z^{(1)}] = \frac{1}{2} Z^{(1)\, abc pq} \varepsilon_{pqijk} \partial^{k} D\indices{_{abc}^{ij}}[Z^{(1)}]\, , \\
\mathcal{H}_\phi &= 18\, G_{ijk}[Z^{(2)}] S^{ijk}[Z^{(2)}] = \frac{1}{2} Z^{(2) abc pq} \varepsilon_{pqijk} \partial^{k} D\indices{_{abc}^{ij}}[Z^{(2)}]\, .
\end{align}
Since the constraints are identically satisfied, the Lagrange multipliers $n_{ij}$ and $N_i$ disappear from the action.

We now do the change of variables
\begin{equation}
Z^\pm = Z^{(1)} \pm Z^{(2)} \; \Leftrightarrow \; Z^{(1)} = \frac{1}{2} \left( Z^+ + Z^- \right)\, , \; Z^{(2)} = \frac{1}{2} \left( Z^+ - Z^- \right)\, .
\end{equation}
This splits the action into two parts, $S[Z^+, Z^-] = S^+[Z^+] - S^-[Z^-]$, where $S^+$ is exactly the action \eqref{actionZ} obtained in section \ref{sec:chiralaction}. The action $S^{-}$ is the same, but with the sign of the second term flipped: it is the action for an anti-chiral $(2,1)$-tensor.

\section{The exotic gravitino}

As we will see in the next chapter, the spectrum of the the maximally supersymmetric $\cN = (4,0)$ and $\cN = (3,1)$ theories also contain left-handed fermionic two-forms $\Psi_{\mu\nu}$ (``exotic gravitinos"). We now describe this field in detail.

\subsection{Fermionic two-form}

The action for a fermionic two-form $\Psi_{\mu\nu}$ is given by a straightforward generalization of the Rarita-Schwinger action, 
\begin{equation}
S = \int \dx[6]\, \bar{\Psi}_{\mu\nu} \Gamma^{\mu\nu\rho\sigma\tau} \partial_\rho \Psi_{\sigma\tau} \label{RS2F6}
\end{equation}
(with $\bar{\Psi}_{\mu\nu} \equiv i \Psi_{\mu\nu}^\dagger \Gamma^0$), where we have taken the spacetime dimension to be six from the outset although the action \eqref{RS2F6} makes sense in any number of spacetime dimensions. It is invariant under the gauge transformations
\begin{equation}
\delta \Psi_{\mu\nu} = \partial_\mu \lambda_\nu - \partial_\nu \lambda_\mu = 2 \partial_{[\mu} \lambda_{\nu]} \, ,
\end{equation}
for which the invariant field strength is
\begin{equation}
H_{\mu\nu\rho} = \partial_\mu \Psi_{\nu\rho} + \partial_\nu \Psi_{\rho\mu} + \partial_\rho \Psi_{\mu\nu} = 3 \partial_{[\mu} \Psi_{\nu\rho]} .
\end{equation}
The equation of motion is the generalized Rarita-Schwinger equation
\begin{equation} \label{eom2form}
\Gamma^{\mu\nu\alpha\beta\gamma} H_{\alpha\beta\gamma} = 0 .
\end{equation}
In six spacetime dimensions, one can also impose a positive or negative chirality condition (Weyl spinors)  and we shall assume from now on that 
\be
\Gamma_7 \Psi_{\lambda \mu} = \Psi_{\lambda \mu} \label{eq:ChiralPsi}
\ee
(our conventions on $\Gamma$-matrices in six spacetime dimensions are collected in appendix \ref{app:gammamatrices}).  This implies that 
\be
\Gamma_7 H_{\mu\nu\rho} = H_{\mu\nu\rho}
\ee
and that the gauge parameter $\lambda_\nu$ must also be taken to have positive chirality, 
\be
\Gamma_7 \lambda_\nu = \lambda_\nu.
\ee

The equations of motion (\ref{eom2form}) can be split into space and time as follows,
\begin{equation}
\Gamma^{iabc} H_{abc} = 0, \quad \Gamma^{ijabc} H_{abc} + 3 \Gamma^0 \Gamma^{ijab} H_{0ab} = 0 . \label{eq:eomBis}
\end{equation}
The first equation is a constraint on $\Psi_{mn}$ and its spatial gradients, and the second involves the time derivatives of $\Psi_{mn}$ and is dynamical.  They are of first order.  In fact, the action itself is already in Hamiltonian form and can be decomposed as
\begin{equation} \label{eq:2formactionham}
S = \int \!dt \,d^5\!x\, \left( \eta^{ij} \dot{\Psi}_{ij} - \mathcal{H} + \Psi_{0i}^\dagger \mathcal{C}^i + \mathcal{C}^{i\dagger} \Psi_{0i} \right)\, ,
\end{equation}
where the conjugate momentum, the Hamiltonian density and the constraint are
\begin{align}
\eta^{ij} &= -i \Psi^\dagger_{kl} \Gamma^{klij} \\
\mathcal{H} &= -\bar{\Psi}_{ij} \Gamma^{ijklm} \partial_k \Psi_{lm} \\
\mathcal{C}^i &= 2i \Gamma^{ijkl} \partial_j \Psi_{kl} .
\end{align}
The momentum conjugate to $\Psi_{0i}$ identically vanishes, and $\Psi_{0i}$ only appears in the action as a Lagrange multiplier for the constraint $\mathcal{C}^i = 0$, which is of course just $\Gamma^{iabc} H_{abc} = 0$.

\subsection{Self-duality condition}

The equations of motion for the chiral spinorial two-form are equivalent  to the set formed by the self-duality condition on its gauge-invariant curvature $H$ and the constraint,
\begin{equation} \label{twisted2form}
H =  \star H  , \quad  \Gamma^{iabc} H_{abc} = 0 \, .
\end{equation}
The goal of this subsection is to establish this fact.

The  self-duality condition is consistent because $(\star)^2 = 1$, just as for ordinary two-forms in six spacetime dimensions.  If the spinorial two-form were not chiral, the self-duality condition would involve $\Gamma_7$ and would read
\be
 H = \Gamma_7 \star H,
 \ee
 involving both self-dual and anti-self-dual components, which are separated by diagonalizing $\Gamma_7$.
In components, the self-duality condition reads
\begin{equation}
H_{0ij} = \frac{1}{3!} \varepsilon_{ijabc}  H^{abc} \; \Leftrightarrow\; H_{abc} = \frac{1}{2} \varepsilon_{abcij}  H\indices{_0^{ij}} .
\end{equation}
The proof of the equivalence of \eqref{eom2form} and \eqref{twisted2form} goes as follows.
\begin{itemize}
\item \eqref{eom2form} $\Rightarrow$ \eqref{twisted2form}: Contracting the dynamical equation with $\Gamma_i$, the constraint implies also
\begin{equation}
\Gamma^{ijk} H_{0jk} = 0, \quad \Gamma^{jk} H_{0jk} = 0 .
\end{equation}
(The second equation follows from the first by contracting with $\Gamma_i$.) Then, contracting the standard identities on the $\Gamma$-matrices
$\Gamma^i \Gamma^{jab} = \Gamma^{ijab} + 3 \delta^{i[j} \Gamma^{ab]}$ and 
$\Gamma^{ij} \Gamma^{ab} = \Gamma^{ijab} + 4 \delta^{[i[a} \Gamma^{b]j]} -2 \delta^{[i[a} \delta^{b]j]} $
with $H_{0ab}$, we get
\begin{align}
\Gamma^{ijab} H_{0ab} = 2 \delta^{a[i} \Gamma^{j]b} H_{0ab}\, , \quad \Gamma^{ijab} H_{0ab} = 4 \delta^{a[i} \Gamma^{j]b} H_{0ab} +2 H\indices{_0^{ij}}\, ,
\end{align}
which together imply
\begin{equation}
\Gamma^{ijab} H_{0ab} = -2 H\indices{_0^{ij}} .
\end{equation}
With this relation, the dynamical equation becomes
\begin{equation}
\Gamma^{ijabc} H_{abc} - 6 \Gamma^0 H\indices{_0^{ij}} = 0 .
\end{equation}
Now, using the identity $\Gamma^{ijabc} = - \varepsilon^{ijabc} \Gamma_0 \Gamma_7$,  multiplying by $\Gamma_0$ and using the chirality condition, we derive
\begin{equation}
\varepsilon^{ijabc}  H_{abc} - 6 H\indices{_0^{ij}} = 0,
\end{equation}
which is exactly the self-duality condition in components.
\item \eqref{twisted2form} $\Rightarrow$ \eqref{eom2form}: Contracting the $abc$ component of the self-duality equation with $\Gamma^{iabc}$ and using the constraint, we get
\begin{equation}
\Gamma^i H_{0ij} = 0, \quad \Gamma^{ij} H_{0ij} = 0 .
\end{equation}
(The second one follows from the first upon contraction with $\Gamma^j$.)
We now use the identity
$\Gamma^{ija} \Gamma^b = \Gamma^{ijab} + 3 \Gamma^{[ij} \delta^{a]b}$,
along with  $\Gamma^{ij} \Gamma^{ab} = \Gamma^{ijab} + 4 \delta^{[i[a} \Gamma^{b]j]} -2 \delta^{[i[a} \delta^{b]j]}$. Contracting them with $H_{0ab}$, we find again $\Gamma^{ijab} H_{0ab} = -2 H\indices{_0^{ij}}$. Using this in the self-duality equation, we obtain the dynamical equation of motion for $\Psi_{\lambda \mu}$ (second of equation (\ref{eq:eomBis})).
\end{itemize}
We thus conclude that the spinorial two-form in six dimensions possesses the remarkable property that its field strength is self-dual, just as the field strengths of its $\tyng{2,2}$, $\tyng{2,1}$ and $\tyng{1,1}$ bosonic counterparts.  The self-duality condition does not provide a complete description in the fermionic case, however, since it must be supplemented by the condition $\Gamma^{iabc} H_{abc} = 0$.

\subsection{Conformal geometry for a spinorial $2$-form in five spatial dimensions}
\label{app:cotton5}

We now want to solve the constraint $\Gamma^{iabc} H_{abc} = 0$ in order to have an action for the spinorial two-form that parallels that of the bosonic chiral fields. In this case, the prepotential is an antisymmetric tensor-spinor $\chi_{ij}$, with gauge symmetries
\begin{equation} \label{eq:gaugeweyl6d}
\delta \chi_{ij} = \partial_{[i} \eta_{j]} + \Gamma_{[i} \rho_{j]} .
\end{equation}
The goal of this section is to construct its ``geometry'', i.e., the invariants that can be built out of $\chi_{ij}$ and its derivatives.  We also derive the main properties of these invariants.  We shall develop the formalism without imposing the chirality condition $\Gamma_7  \chi_{ij} =  \chi_{ij}$.  It can of course be imposed; in that case, $\Gamma_7$ should be replaced in the formulas below by the identity when acting on spinors.

\paragraph{Einstein tensor.}

First, we construct tensors that are invariant under generalized diffeomorphisms.  Of course, it is just enough to take the exterior derivative of $\chi_{ij}$.  By dualizing, one gets the Einstein tensor
\begin{equation}
G_{ij}[\chi] = \varepsilon_{ijklm} \partial^k \chi^{lm}, \quad G_{ij}[\chi] = G_{[ij]}[\chi]\, ,
\end{equation}
which is not only invariant under the $\eta$ transformations but also identically divergenceless,
\be
\partial_i G^{ij}[\chi] = 0 \, .
\ee
We have furthermore the properties
\begin{itemize}
\item $\eta$ triviality criterion:
\begin{equation}
G_{ij}[\chi] = 0 \quad\Leftrightarrow\quad \chi_{ij} = \partial_{[i} \eta_{j]} \;\text{ for some } \eta_i\, ;
\end{equation}
\item divergence:
\begin{equation}
\partial^i T_{ij} = 0 \quad \Leftrightarrow \quad T_{ij} = G_{ij}[\chi] \;\text{ for some } \chi_{ij} .
\end{equation}
\end{itemize}
These are just two applications of the usual Poincar\'e lemma with a spectator spinor index in form degrees $2$ and $3$. The first property implies that the most general invariant under the $\eta$ transformation is a function of the Einstein tensor and its derivatives. 

Under generalized Weyl transformations, we have
\begin{equation}
\delta G_{ij} = \varepsilon_{ijklm} \Gamma^l \partial^k \rho^m = - \Gamma_{ijkm} \Gamma_7 \Gamma_0 \partial^k \rho^m\, ,
\end{equation}
which gives for the gamma-traces
\begin{align}
\delta ( \Gamma^j G_{ij} ) &= 2 \Gamma_{ikm} \Gamma_7 \Gamma_0 \partial^k \rho^m \;\Rightarrow\; \delta ( \Gamma_{[i} \Gamma^k G_{j]k} ) = 2 \Gamma_{[i} \Gamma_{j]km} \Gamma_7 \Gamma_0 \partial^k \rho^m \, ,\\
\delta ( \Gamma^{ij} G_{ij} ) &= 6 \Gamma_{km} \Gamma_7 \Gamma_0 \partial^k \rho^m \;\Rightarrow\; \delta ( \Gamma_{ij} \Gamma^{kl} G_{kl} ) = 6 \Gamma_{ij} \Gamma_{km} \Gamma_7 \Gamma_0 \partial^k \rho^m\, .
\end{align}

\paragraph{Schouten tensor.}
\label{app:subSchouten}
We define the Schouten tensor as
\begin{align}
S_{ij} &= G_{ij} + \Gamma_{[i} \Gamma^k G_{j]k} - \frac{1}{6} \Gamma_{ij} \Gamma^{kl} G_{kl}\, .
\end{align}
It is again an antisymmetric tensor, which transforms as
\begin{align}
\delta S_{ij} = \left( - \Gamma\indices{_{ij}^{km}} + 2 \Gamma_{[i} \Gamma\indices{_{j]}^{km}} - \Gamma_{ij} \Gamma^{km}  \right) \Gamma_7 \Gamma_0 \partial_k \rho_m 
\end{align}
under generalized Weyl transformations.
Using the identites $\Gamma_{[m} \Gamma\indices{_{n]}^{rs}} = \Gamma\indices{_{mn}^{rs}} + 2 \delta^{[r}_{[m} \Gamma\indices{^{s]}_{n]}}$  and $\Gamma_{mn} \Gamma^{rs} = \Gamma\indices{_{mn}^{rs}} + 4 \delta_{[m}^{[r} \Gamma\indices{^{s]}_{n]}} -2 \delta^{rs}_{mn}$, it can be seen that the combination in brackets reduces to $2 \delta^{km}_{ij}$, so that the Schouten simply transforms as
\begin{align}
\delta S_{ij} = \partial_{[i} \nu_{j]}, \quad \nu_j = 2 \Gamma_7 \Gamma_0 \rho_j .
\end{align}
It is this simple transformation law of the Schouten tensor that motivates its definition.

Using $\Gamma_i \Gamma^k = \delta^k_i + \Gamma\indices{_i^k}$, the Schouten tensor can be rewritten as
\begin{equation} \label{eq:defschouten2}
S_{ij} = - \left( \delta^{[k}_{[i} \Gamma\indices{_{j]}^{l]}} + \frac{1}{6} \Gamma_{ij} \Gamma^{kl} \right) G_{kl}
\end{equation}
and then, using the identity $\delta^{rs}_{mn} = \frac12 \left(  \delta^{[p}_{[m} \Gamma\indices{_{n]}^{q]}} + \frac16 \Gamma_{mn} \Gamma^{pq} \right) \Gamma\indices{_{pq}^{rs}}$, we can write the Einstein tensor in terms of the Schouten tensor as
\begin{equation}
G_{ij} = - \frac{1}{2}\Gamma_{ijkl} S^{kl} . 
\end{equation}
Therefore, the Schouten satisfies
\begin{equation}
\Gamma_{ijkl} \partial^j S^{kl} = 0
\end{equation}
as a consequence of $\partial^j G_{ij} = 0$.
We have also the following direct properties.
\begin{itemize}
\item Pure gauge property:
\begin{equation}
S_{ij}[\chi] = \partial_{[i} \nu_{j]} \quad\Leftrightarrow\quad \chi_{ij} = \partial_{[i} \eta_{j]} + \Gamma_{[i} \rho_{j]} \;\text{ for some } \eta_i \quad (\rho_i = - \frac{1}{2} \Gamma_0 \Gamma_7 \nu_i)\, .
\end{equation}
\item Constraint property:
\begin{equation}
\Gamma_{ijkl} \partial^j T^{kl} = 0 \quad\Leftrightarrow\quad T_{ij} = S_{ij}[\chi] \;\text{ for some } \chi_{ij}\, .
\end{equation}
This is the property that underlies the introduction of the prepotential for the chiral spinorial two-form $\Psi_{ij}$.
\end{itemize}
Finally, using the identities $\Gamma_{mn} \Gamma^{rs} = \Gamma\indices{_{mn}^{rs}} + 4 \delta_{[m}^{[r} \Gamma\indices{^{s]}_{n]}} -2 \delta^{rs}_{mn}$ and $\Gamma^{ijkl} = - \varepsilon^{ijklm} \Gamma_m \Gamma_0 \Gamma_7$, we can also rewrite the Schouten tensor as
\begin{equation} \label{a49}
S_{ij} = \frac{1}{3} G_{ij} + \frac{1}{3} \Gamma\indices{_{[i}^k} G_{j]k} + \frac{1}{6} \varepsilon_{ijklm} \Gamma^k \Gamma_0 \Gamma_7 G^{lm} \, ,
\end{equation}
a form which is useful for some of the computations of the next chapter.

\paragraph{Cotton tensor.}
The Cotton tensor is defined by taking the curl of the Schouten tensor
\begin{equation}\label{eq:relDS}
D_{ij}[\chi] = G_{ij} [S[\chi]] = \varepsilon_{ijklm} \partial^k S^{lm}[\chi] .
\end{equation}
It is invariant by construction under both generalized diffeomorphisms and Weyl transformations \eqref{eq:gaugeweyl6d}. It is also divergenceless, $\partial_i D^{ij} = 0$, and also gamma-traceless, $\Gamma^i D_{ij} = 0$, because of the identity $\Gamma_{ijkl} \partial^j S^{kl} = 0$.
Its key properties are the following.
\begin{itemize}
\item Pure gauge condition: the prepotential $\chi_{ij}$ is pure gauge if and only if its Cotton tensor vanishes,
\begin{equation}
D_{ij}[\chi] = 0 \quad\Leftrightarrow\quad \chi_{ij} = \partial_{[i} \eta_{j]} + \Gamma_{[i} \rho_{j]} \;\text{ for some } \eta_i \text{ and } \rho_i.
\end{equation}
This also means that any invariant under \eqref{eq:gaugeweyl6d} is a function of the Cotton tensor and its derivatives.
\item Tracelessness and divergencelessness conditions: if a spinorial antisymmetric tensor is both divergenceless and $\Gamma$-traceless, then it is equal to the Cotton tensor of some antisymmetric spinorial prepotential,
\begin{equation}
\partial_i T^{ij} = 0, \quad \Gamma^i T_{ij} = 0 \quad\Leftrightarrow\quad T_{ij} = D_{ij}[\chi] \;\text{ for some } \chi_{ij} .
\end{equation}
\end{itemize}
Both results directly follow from the Poincar\'e lemma and the above definitions.

Using formula \eqref{a49} for the Schouten tensor, we can express the Cotton tensor in terms of the Einstein tensor as
\begin{equation}
D_{ij} = \frac{1}{3} \left( \varepsilon_{ijklm} \partial^{k} G^{lm} + \varepsilon_{ijklm} \partial^k \Gamma^{lp} G\indices{^m_p} + 2 \partial^k \Gamma_k \Gamma_0 \Gamma_7 G_{ij} \right) .
\end{equation}
The second term can be written as $\frac{1}{3} \Gamma^{pq} \varepsilon_{pqkl[i} \partial^k G\indices{^l_{j]}}$ using identity \eqref{eq:g2epsilon}, which in this case $(d=5)$ reads
\begin{equation}
\varepsilon_{abcdi} \Gamma^{ij} = \frac{1}{2} \left( \delta^j_a \varepsilon_{bcdkl} - \delta^j_b \varepsilon_{acdkl} - \delta^j_c \varepsilon_{badkl} - \delta^j_d \varepsilon_{abckl} \right) \Gamma^{kl}.
\end{equation}
Note also that the identities
\begin{align}
\Gamma^{a_1 \dots a_k i} D_{ij} &= - k \Gamma^{[a_1 \dots a_{k-1} } D\indices{^{a_k]}_j}\, , \\
\Gamma^{a_1 \dots a_k i j} D_{ij} &= - k (k-1) \Gamma^{[a_1 \dots a_{k-2}} D^{a_{k-1} a_k]}
\end{align}
follow from the gamma-tracelessness of the Cotton tensor.

\subsection{Prepotential and action principle}

As we have seen, the Lagrangian formulations of the bosonic chiral fields have the following important properties: 1) they involve only spatial tensors, and, in particular, the temporal components of the fields (which are pure gauge) are absent; 2) the variational equations of motion are not the self-duality conditions ``on the nose'' but an equivalent differential version of them in which the temporal components have been eliminated by taking the appropriate curls.

A similar formulation for the chiral spinorial two-form exists and is in fact necessary if one wants to exhibit supersymmetry.  It is obtained by solving the constraint $\Gamma^{iabc} H_{abc} = 0$ in terms of a ``prepotential'', getting rid thereby of its Lagrange multiplier $\Psi_{0 i}$. Once this is done, the only relevant equation that is left is the self-duality condition on the curvature, or rather its differential version
\be
\varepsilon^{rsmij} \partial_m\left(H_{0ij} - \frac{1}{3!} \varepsilon_{ijabc}  H^{abc} \right) = 0 \; \Leftrightarrow\; \varepsilon^{rsmij} \partial_m\left(\partial_0 \Psi_{ij}  \right)  - 2 \, \partial_m H^{mrs}= 0 \label{eq:SDPSI}
\ee
that does not contain $\Psi_{0 i}$.

As shown in the previous section, the general solution of the constraint $\Gamma^{iabc} H_{abc} = 0$ can be written in terms of a chiral antisymmetric tensor-spinor $\chi_{ij}$ ($\Gamma_7 \chi_{ij} = \chi_{ij}$, $\chi_{ij} = - \chi_{ji}$) as
\begin{equation}
\Psi^{ab}  = S^{ab}[\chi], \label{eq:RelPsiChi}
\end{equation}
where $S^{ab}[\chi]$ is the Schouten tensor of $\chi_{ij}$ defined in equation \eqref{eq:defschouten2} above.
We call $\chi_{ij}$ the prepotential of $\Psi_{ij}$. In other words, the chiral two-form is the Schouten tensor of the prepotential, because of the ``constraint property of the Schouten tensor'' established in section \ref{app:subSchouten}.

Furthermore, if $\Psi_{ij}$ is determined up to a gauge transformation $\delta\Psi_{ij} = 2 \partial_{[i} \lambda_{j]}$, then $\chi_{ij}$ is determined up to
\begin{equation}
\delta \chi_{ij} = \partial_{[i} \eta_{j]} + \Gamma_{[i} \rho_{j]}, \label{eq:GaugeChi}
\end{equation}
where $\lambda_i = \Gamma_0 \rho_i$. This is because the Einstein tensor of $\Psi_{ij}$ is equal to the Cotton tensor of the prepotential, and follows then directly from the discussion in the previous section. The first term in (\ref{eq:GaugeChi}) is the standard gauge transformation of a spinorial two-form, the second term is a generalized Weyl transformation.  When inserted into the Schouten tensor, the first term drops out, while the generalized Weyl transformation induces precisely the gauge transformation of $\Psi_{ij}$.

We now use formula \eqref{eq:RelPsiChi} in the action \eqref{eq:2formactionham}. Since the constraint $\mathcal{C}^i = 0$ is automatically satisfied, the Lagrange multipliers $\Psi_{0i}$ disappear. 
The kinetic term becomes
\begin{align}
-i \Psi^\dagger_{ij} \Gamma^{ijkl} \dot{\Psi}_{kl} = -i S_{ij}^\dagger[\chi] \gamma^{ijkl} \dot{S}_{kl}[\chi] = -2 i \chi^\dagger_{ij} \dot{D}^{ij}[\chi]\, ,
\end{align}
where $D^{ij}[\chi]$ is the Cotton tensor of $\chi$ defined above, and the Hamiltonian is
\begin{align}
\mathcal{H} &= -\bar{\Psi}_{ij} \Gamma^{ijklm} \partial_k \Psi_{lm} =  i \Psi^\dagger_{ij} \varepsilon^{ijklm} \partial_k \Psi_{lm} \\
&= i S^\dagger_{ij}[\chi] D^{ij}[\chi] = -i \chi^\dagger_{ij} \varepsilon^{ijklm} \partial_k D_{lm}[\chi]\, .
\end{align}
The action is therefore
\begin{align}
S[\chi] = -2i \int \!dt \,d^5\!x\, \chi^\dagger_{ij} \left( \dot{D}^{ij}[\chi] - \frac{1}{2} \varepsilon^{ijklm} \partial_k D_{lm}[\chi] \right). \label{eq:exoticgravitino}
\end{align}
The equation of motion obtained by varying the prepotential is
\be
\dot{D}^{ij}[\chi] - \frac{1}{2} \varepsilon^{ijklm} \partial_k D_{lm}[\chi] = 0\, ,
\ee
which is nothing but the self-duality condition in the form \eqref{eq:SDPSI}, as follows from definitions \eqref{eq:RelPsiChi} and \eqref{eq:relDS}.

\subsection{Dimensional reduction}
\label{sec:DimRed}

We now perform the dimensional reduction of the spinorial two-form $\Psi_{\lambda \mu}$ and show that it reduces to a single gravitino in five dimensions.

An explicit realization of the six-dimensional gamma matrices is given by the block form
\begin{align}
\Gamma_\mu &= \sigma_1 \otimes \gamma_\mu = \begin{pmatrix}
0 & \gamma_\mu \\
\gamma_\mu & 0
\end{pmatrix} \quad(\mu = 0, \dots, 4), \nn \\
\Gamma_5 &= \sigma_2 \otimes I= \begin{pmatrix}
0 & - iI \\ iI & 0
\end{pmatrix}, \label{eq:gammasixd}
\end{align}
each block being $4 \times 4$. The first five are given in terms of the five-dimensional gamma matrices (in particular  $\gamma_4 = i \gamma_0 \gamma_1 \gamma_2 \gamma_3$ is what is usually called $\gamma_5$). In this representation, the $\Gamma_7$ matrix is diagonal,
\begin{equation}
\Gamma_7 = \begin{pmatrix}
I & 0 \\ 0 & -I
\end{pmatrix} .
\end{equation}
Therefore, a left-handed $2$-form $\Psi_{\mu\nu} = P_L \Psi_{\mu\nu}$ takes the simple form
\begin{equation}
\Psi_{\mu\nu} = \begin{pmatrix}
\hat{\psi}_{\mu\nu} \\ 0
\end{pmatrix}
\end{equation}
and its Dirac conjugate is $\bar{\Psi}_{\mu\nu} = \begin{pmatrix}
0 & \bar{\hat{\psi}}_{\mu\nu}
\end{pmatrix}$ .
The six-dimensional field $\hat{\psi}_{\mu\nu}$ ($\mu,\nu = 0, \dots 5$) then reduces to two fields $\psi_{\mu\nu} = \hat{\psi}_{\mu\nu}$ and $\psi_\mu = \hat{\psi}_{\mu5}$ ($\mu,\nu = 0, \dots 4$) in five dimensions.

Using $\Gamma^{\mu_1 \dots \mu_5} = - \varepsilon^{\mu_1 \dots \mu_5 \nu} \Gamma_\nu \Gamma_7$ and $\Gamma_7 \Psi_{\mu\nu} = \Psi_{\mu\nu}$, the Lagrangian for the spinorial $2$-form in six dimensions can be rewritten as
\begin{equation}
\mathcal{L} = - \bar{\Psi}_{\mu\nu} \varepsilon^{\mu\nu\rho\sigma\tau\lambda} \Gamma_\lambda \partial_\rho \Psi_{\sigma\tau} .
\end{equation}
Using the above decomposition, it becomes
\begin{align}
\mathcal{L} = - i \bar{\psi}_{\mu\nu} \varepsilon^{\mu\nu\rho\sigma\tau} \partial_\rho \psi_{\sigma\tau} + 2 \bar{\psi}_{\mu\nu} \varepsilon^{\mu\nu\rho\sigma\tau} \gamma_\rho \partial_\sigma \psi_{\tau} + 2 \bar{\psi}_{\mu} \varepsilon^{\mu\nu\rho\sigma\tau} \gamma_\nu \partial_\rho \psi_{\sigma\tau} .
\end{align}
Now, one can eliminate the five-dimensional spinorial two-form $\psi_{\mu\nu}$ using its own equation of motion.  Indeed, varying the action with respect to $\psi_{\mu\nu}$ yields \begin{equation}
\varepsilon^{\mu\nu\rho\sigma\tau} \partial_\rho \left( i \psi_{\sigma\tau} + 2 \gamma_{\sigma} \psi_\tau \right) = 0\, ,
\end{equation}
from which one derives
\begin{equation}
\psi_{\mu\nu} = 2i\gamma_{[\mu} \psi_{\nu]} + \partial_{[\mu} \Lambda_{\nu]}
\end{equation} for some $\Lambda_\nu$. Inserting this expression in the Lagrangian, one gets
\begin{equation}
\mathcal{L} = - 4 i \bar{\psi}_{\mu } \varepsilon^{\mu\nu\rho\sigma\tau} \gamma_{\sigma\tau} \partial_\nu \psi_{\rho }  = - 8 \bar{\psi}_{\mu} \gamma^{\mu\nu\rho} \partial_\nu \psi_{\rho}\, ,
\end{equation}
which is exactly the Rarita-Schwinger action for $\psi_\mu$  (after rescaling $\psi_\mu \rightarrow \psi_\mu / 2 \sqrt{2}$).  We can conclude that the dimensional reduction of the chiral spinorial two-form in six dimensions gives indeed a single Rarita-Schwinger field in five dimensions.

It is interesting to point out that if instead of the standard description of gravity based on a symmetric tensor,   one uses the description involving the dual graviton given by a $\tyng{2,1}$-tensor in five dimensions, and keeps the two-form $\psi_{\mu \nu}$ instead of the Rarita-Schwinger field $\psi_\mu$, one gets the dual description of five-dimensional linearized supergravity  alluded to in \cite{Curtright:1980yk}.

\chapter{Maximal supersymmetry in six dimensions}
\label{chap:6dsusy}

In $D=6$ spacetime dimensions, three different maximal supersymmetry algebras exist: the $\cN = (4,0)$ and $\cN = (3,1)$ chiral algebras, and the $\cN = (2,2)$ non-chiral algebra \cite{Nahm:1977tg,Strathdee:1986jr}.  While the theory realizing the $\cN = (2,2)$ supersymmetry algebra is well known and just the toroidal dimensional reduction of maximal supergravity in $11$ dimensions, the theories realizing the other two superalgebras (if they exist) are more mysterious.  This is because they would involve, in place of the standard spin-2 field describing gravity, the chiral tensor fields with mixed Young symmetries of the previous chapters. In view of the subtleties of writing an action principle for chiral bosonic fields \cite{Marcus:1982yu}, and the various no-go theorems preventing the interactions of tensor fields with mixed Young symmetries \cite{Bekaert:2002uh,Boulanger:2004rx,Bekaert:2004dz,Ciobirca:2004yu,Bizdadea:2009zc} or involving chirality conditions \cite{Bekaert:1999dp,Bekaert:1999sq,Bekaert:2000qx}, the maximal chiral supersymmetry in six dimensions algebras were largely ignored in the literature.

A notable exception is the work  \cite{Hull:2000zn,Hull:2000rr}, in which it was argued that the $\cN = (4,0)$ theory could emerge as the strong coupling limit of maximal supergravity in five spacetime dimensions. The covariant field content and free equations of motion for the $\cN = (4,0)$ and the $\cN = (3,1)$ multiplets were also written down.

The goal of this chapter is to write the free actions and supersymmetry transformations of the $\cN = (4,0)$ and $\cN = (3,1)$ theories. The explicit check that both theories reduce correctly to linearized maximal supergravity in five dimensions is done in appendix \ref{app:redlinsugra}. The inclusion of (presumably non-local) consistent interactions, if they exist, remains of course an important challenge.

The results of this chapter were presented in the paper \cite{Henneaux:2017xsb} for the $\cN = (4,0)$ theory, written with M. Henneaux and A. Leonard, and in \cite{Henneaux:2018rub} for the $\cN = (3,1)$ theory, written with M. Henneaux, J. Matulich and S. Prohazka.

\section{The \texorpdfstring{$\cN = (4,0)$}{N = (4,0)} theory}

The bosonic field content of the $\cN = (4,0)$ theory is given by
a chiral tensor of mixed Young symmetry $\tyng{2,2}$ (``exotic graviton''),  27 chiral two-forms and 42 scalars, making 128 physical degrees of freedom. This can be written as
\be (5,1;1) \oplus (3,1; 27)  \oplus (1,1;42)\, , \ee
where the first two integers label the representation of the little algebra $so(4) \simeq su(2) \oplus su(2)$, and the last number refers to the representation of the $R$-symmetry algebra, which is $usp(8)$ in this case.

The fermionic field content is given by 8 fermionic chiral $2$-forms (``exotic gravitini'') and  48 spin-$1/2$ fields,
\be
(4,1; 8) \oplus (2,1; 48)\, ,
\ee
which matches the number 128 of bosonic degrees of freedom.  The fields fit into a unitary supermultiplet  of the six-dimensional superconformal group $OSp(8^* \vert 8)$ \cite{Hull:2000zn,Chiodaroli:2011pp}. Moreover, the theory is expected to have $E_{6(6)}$ symmetry (like maximal supergravity in 5 dimensions), the chiral $2$-forms being in the  $\mathbf{27}$ and the scalars parametrizing the coset $E_{6(6)}/USp(8)$, which has the correct dimension $78-36 = 42$.

\subsection{Explicit form of the action}

The action is a sum of five terms, one for each set of fields in the supermultiplet,
\be \label{eq:action40}
S = S_{\tyng{2,2}} +  S_{\tyng{1,1}_F} + S_{\tyng{1,1}_B} + S_{\frac12} + S_{0} .
\ee
We  describe each of these contributions in turn.

\paragraph{The exotic graviton.}

This is a real tensor field of mixed $(2,2)$ Young symmetry with self-dual field strength. The action for such a chiral tensor was derived in the previous chapter; the variables of the variational principle are the components of the prepotential $Z_{ijkl}$. It is real, $(Z_{ijkl})^* = Z_{ijkl}$, and a singlet under $USp(8)$.
The action reads
\begin{equation}
S_{\tyng{2,2}}[Z] = \frac{1}{2} \int \! d^6\! x \, Z_{mnrs} \left( \dot{D}^{mnrs} - \frac{1}{2} \varepsilon^{mnijk} \partial_k D\indices{_{ij}^{rs}} \right)\, . \label{eq:actionZ}
\end{equation}
where $D_{ij kl}$ is the Cotton tensor of $Z^{mnrs}$ written in \eqref{eq:cotton22}.

The Cotton tensor contains three derivatives of the prepotential and the action \eqref{eq:actionZ} is therefore of fourth order in derivatives (but of first order in the time derivative).

\paragraph{The exotic gravitini.}

These have been discussed in the previous chapter. As shown there, the exotic gravitini  are described by fermionic $2$-form prepotentials $\chi^A_{ij}$ which are chiral,
\begin{equation}
\Gamma_7 \chi^A_{ij} = + \chi^A_{ij}.
\end{equation}
The exotic gravitini transform in the ${\mathbf 8}$ of the $R$-symmetry group is $USp(8)$.  So there are 8 gravitini, labelled by one $USp(8)$ index $A$ ($A = 1, \cdots, 8$). The reality conditions are given by the symplectic Majorana conditions,
\begin{equation}
 \chi^*_{A\,ij} = \Omega_{AB} \mathcal{B} \chi^B_{ij} .
\end{equation}
These can be consistently imposed, as discussed in Appendix \ref{app:usp8}. The action reads
\begin{equation}
S_{\tyng{1,1}_F}[\chi] = -2i \int \!d^6\!x\, \chi^{\dagger}_{A\, ij} \left( \dot{D}^{A\,ij} - \frac{1}{2} \varepsilon^{ijklm} \partial_k D^A_{lm} \right)\, ,
\end{equation}
where the $D^{A\,ij} \equiv D^{ij}[\chi^A]$ are the Cotton tensors. This action is of third order in derivatives (but again of first order in the time derivative).

\paragraph{The chiral $2$-forms.}

The chiral two-forms are in the $\mathbf{27}$ of $USp(8)$.  There are thus 27 of them, labelled by the antisymmetric pair $[AB]$ with the constraint
\begin{equation}
A^{AB}_{ij} \Omega_{AB} = 0 . \label{eq:Cons2F}
\end{equation}
The reality condition $(A^{AB}_{ij})^* = A^{AB}_{ij}$ is not compatible with $USp(8)$ covariance, as discussed in Appendix \ref{app:usp8}. Instead, we impose
\begin{equation}\label{eq:Areality}
A^*_{AB\, ij} = \Omega_{AA'} \Omega_{BB'} A^{A'B'}_{ij} .
\end{equation}
The consistency condition $(A^{AB}_{ij})^{**} = A^{AB}_{ij}$ is satisfied because there is an even number of $\Omega$ matrices in equation \eqref{eq:Areality}.

Their action reads
\begin{equation}
S_{\tyng{1,1}_B}[A] = -\frac{1}{2} \int \! d^6\!x \, A^*_{AB\,ij} \left( \dot{\mathcal{B}}^{AB\,ij} - \frac{1}{2}\varepsilon^{ijklm} \partial_k \mathcal{B}^{AB}_{lm} \right) \, ,
\end{equation}
where the magnetic fields $\mathcal{B}^{AB\, ij}$ are given by
\be
\mathcal{B}^{AB\, ij} = \frac{1}{2} \varepsilon^{ijklm} \partial_k A^{AB}_{lm} .
\ee
This action is of second order in derivatives (and of first order in the time derivative).

\paragraph{The spin-$1/2$ fields.}

These are Dirac fermions $\psi^{ABC}$ subject to the chirality condition
\begin{equation}
\Gamma_7 \psi^{ABC} = + \psi^{ABC}
\end{equation}
and transforming in the $\mathbf{48}$ of $USp(8)$, i.e., they are labelled by a completely antisymmetric triplet of indices $[ABC]$ with the constraint
\begin{equation}
\psi^{ABC} \Omega_{AB} = 0 . \label{eq:Cons12F}
\end{equation}
The reality conditions are
\begin{equation}
\psi^*_{ABC} = \Omega_{AA'} \Omega_{BB'} \Omega_{CC'} \mathcal{B} \psi^{A'B'C'} .
\end{equation}
There are no gauge transformations and the action is just a sum of Dirac actions,
\begin{align}
S_{\frac12}[\psi] &= - \int \! d^6\! x \, \bar{\psi}_{ABC} \Gamma^\mu \partial_\mu \psi^{ABC} \nonumber \\
&= \int \! d^6\! x \,i \psi_{ABC}^{\dagger} \left( \dot{\psi}^{ABC} - \Gamma^0 \Gamma^i \partial_i \psi^{ABC} \right) .
\end{align}
It is of first order in derivatives.

\paragraph{The scalar fields.}

They transform in the $\mathbf{42}$ of $USp(8)$ and are therefore labelled by a completely antisymmetric quadruplet of indices $[ABCD]$ with the constraint
\begin{equation}
\phi^{ABCD} \Omega_{AB} = 0 .\label{eq:ConsScal}
\end{equation}
The reality conditions are
\begin{equation}
\phi^*_{ABCD} = \Omega_{AA'} \Omega_{BB'} \Omega_{CC'} \Omega_{DD'} \phi^{A'B'C'D'} .
\end{equation}
The momenta $\pi^{ABCD}$ satisfy the same conditions.
There are no gauge transformations and the action in Hamiltonian form reads
\begin{equation}
S_{0}[\phi,\pi] = \int \! d^6\! x \, \left( \pi^*_{ABCD} \dot{\phi}^{ABCD} - \frac{1}{2} \pi^*_{ABCD} \pi^{ABCD} - \frac{1}{2} \partial_i \phi^*_{ABCD} \partial^i \phi^{ABCD}\right)  .
\end{equation}

\subsection{Supersymmetry transformations}
\label{sec:susy40trsf}

The action \eqref{eq:action40} is invariant under $\cN = (4,0)$ supersymmetry, as we will check now.

The supersymmetry variations only mix fields that have one more or one less $USp(8)$ index, i.e.,
\begin{equation}
Z_{ijkl} \,\longleftrightarrow\, \chi^A_{ij} \,\longleftrightarrow\, A^{AB}_{ij} \,\longleftrightarrow\, \psi^{ABC} \,\longleftrightarrow\, \pi^{ABCD},\, \phi^{ABCD} .
\end{equation}
In terms of representations of the little algebra $so(4) \simeq su(2) \oplus su(2)$, this corresponds to fields of neighbouring ``spin'': $(5,1)$, $(4,1)$, $(3,1)$, $(2,1)$ and $(1,1)$. We take the supersymmetry parameters to be symplectic Majorana-Weyl spinors of negative chirality,
\begin{equation}
\Gamma_7 \epsilon^A = - \epsilon^A, \quad \epsilon^*_A = \Omega_{AB} \mathcal{B} \epsilon^B ,
\end{equation}
and so $\Gamma_7 \Gamma_0 \epsilon^A = + \Gamma_0 \epsilon^A$.
The canonical dimensions of the various objects (fields and supersymmetry parameter) appearing in the supersymmetry variations are gathered in table \ref{tab:candim}, where we have also listed for completeness the canonical dimension of $\partial_\mu$.  
\begin{table}
\centering
\begin{tabular}{cccccccc}
$Z_{ijkl}$ & $\chi^A_{ij}$ & $A^{AB}_{ij}$ & $\psi^{ABC}$ & $\pi^{ABCD}$ & $\phi^{ABCD}$ & $\epsilon^A$ & $\partial_\mu$ \\ \midrule
$1$ & $3/2$ & $2$ & $5/2$ & $3$ & $2$ & $-1/2$ & $1$
\end{tabular}
\caption[Canonical dimension of fields of the $\cN = (4,0)$ theory]{Canonical dimension of the various objects appearing in the supersymmetry transformations of the $\cN = (4,0)$ theory.}
\label{tab:candim}
\end{table}

The variations containing fields with one index more are easy to guess: the $USp(8)$ index on the supersymmetry parameter $\epsilon^A$ must be contracted, and not many possibilities remain with the correct dimension, spatial index structure, $USp(8)$ covariance, chirality, and reality conditions. From those variations, we get the others by requiring the invariance of the kinetic terms and projecting on the appropriate $USp(8)$ representation.
The end result is
\begin{align}
\delta_{\epsilon} Z_{ijkl} &= \alpha_1 \mathbb{P}_{(2,2)} \left( \bar{\epsilon}_A \Gamma_{ij} \chi^A_{kl} \right)\, ,  \\
\delta_{\epsilon} \chi^A_{ij} &= - \frac{\alpha_1}{4.3!^3} \partial^r Z\indices{_{ij}^{kl}} \varepsilon_{pqrkl} \Gamma^{pq} \Gamma^0 \epsilon^A + \frac{\alpha_2}{2} A_{ij}^{AB} \Omega_{BC} \Gamma^0 \epsilon^C\, , \\
\delta_{\epsilon} A^{AB}_{ij} &= \alpha_2 \left( 4 \bar{\epsilon}_C S^{[A}_{ij} \Omega^{B]C} + \frac{1}{2} \Omega^{AB} \bar{\epsilon}_{C} S^C_{ij} \right) + \alpha_3 \bar{\epsilon}_{C} \Gamma_{ij} \psi^{ABC}\, , \\
\delta_{\epsilon} \psi^{ABC} &= - \frac{\alpha_3}{2} \Gamma^{ij} \Gamma^0 \left( \mathcal{B}^{[AB}_{ij} \epsilon^{C]} - \frac{1}{3} \Omega^{[AB} \mathcal{B}^{C]D}_{ij} \Omega_{DE} \epsilon^E \right)\, , \\
&\quad + \alpha_4 \left( \pi^{ABCD} \Omega_{DE} \Gamma^0 \epsilon^E + \partial_i \phi^{ABCD} \Omega_{DE} \Gamma^i \epsilon^E \right) \nonumber \\
\delta_{\epsilon} \phi^{ABCD} &= \alpha_4 \left( 2 \bar{\epsilon}_{E} \psi^{[ABC} \Omega^{D]E} + \frac{3}{2} \bar{\epsilon}_E \Omega^{[AB} \psi^{CD]E} \right)\, , \\
\delta_{\epsilon} \pi^{ABCD} &= \alpha_4 \left( 2 \bar{\epsilon}_{E} \Gamma^0 \Gamma^i \partial_i \psi^{[ABC} \Omega^{D]E} + \frac{3}{2} \bar{\epsilon}_E \Gamma^0 \Gamma^i \Omega^{[AB} \partial_i \psi^{CD]E} \right)\, ,
\end{align}
where $ \mathbb{P}_{(2,2)}$ is the projector on the $(2,2)$-Young symmetry, which takes the explicit form
\begin{equation}
\mathbb{P}_{(2,2)} \left( \bar{\epsilon}_A \Gamma_{ij} \chi^A_{kl} \right) = \frac{1}{3} \left( \bar{\epsilon}_A \Gamma_{ij} \chi^A_{kl} + \bar{\epsilon}_A \Gamma_{kl} \chi^A_{ij} -2 \bar{\epsilon}_A \Gamma_{[i[k} \chi^A_{l]j]} \right)
\end{equation}
in this case, and where $S_{ij}^A$ is the Schouten tensor of $\chi^A_{ij}$.  Here,  the matrix $\Omega^{AB}$ (with indices up) is defined through $\Omega^{AB}\Omega_{CB} = \delta^A_C$  and is numerically equal to $ \Omega_{AB}$ (see appendix \ref{app:gammamatrices}).   Having $\epsilon^A$ to be of negative chirality while the fields have positive chirality makes the variations of the bosonic fields not identically zero and gives the correct chirality to those of the fermionic fields. 

These transformations leave not only the kinetic term invariant but one  also verifies that they leave the Hamiltonian invariant. The real constants $\alpha_1$ to $\alpha_4$ are free at this stage since the action is invariant for any values of them. They will be fixed in equation \eqref{eq:RelAlpha} below (up to an overall factor) through the requirement that the supersymmetry transformations close according to the standard supersymmetry algebra.

For later purposes, it is convenient to compute the supersymmetry transformations of the gauge-invariant tensors.  They are
\begin{align}
\delta_{\epsilon} D_{ijkl}[Z] &= \frac{\alpha_1}{(3!)^3} \mathbb{P}_{(2,2)} \left( \bar{\epsilon}_A \varepsilon_{pqrij} \Gamma^{pq} \partial^r D^A_{kl} \right)\, , \\
\delta_{\epsilon} D_{ij}[\chi^A] &= - \frac{\alpha_1}{4} D_{ijkl}[Z] \Gamma^{kl} \Gamma^0 \epsilon^A \\
&\quad +\frac{\alpha_2}{3} \left( \varepsilon_{ijklm} \partial^k \mathcal{B}^{AB\,lm} + \varepsilon_{ijklm} \partial^k \Gamma\indices{^l_p}\mathcal{B}^{AB\,mp} + 2 \Gamma_k \Gamma_0 \partial^k \mathcal{B}^{AB}_{ij} \right) \Omega_{BC} \Gamma^0 \epsilon^C\, , \nonumber \\
\delta_{\epsilon} \mathcal{B}^{AB}_{ij} &= \alpha_2 \left( 2 \bar{\epsilon}_C D^{[A}_{ij} \Omega^{B]C} + \frac{1}{4} \Omega^{AB} \bar{\epsilon}_{C} D^C_{ij} \right) + \frac{\alpha_3}{2} \bar{\epsilon}_C \varepsilon_{ijklm} \Gamma^{lm} \partial^k \psi^{ABC}\, .
\end{align}
These variations involve only the gauge invariant objects, as they should.

\subsection{Supersymmetry algebra}
\label{sec:susy40alg}

As we mentioned above, the action of the $\cN = (4,0)$ theory is Poincar\'e invariant, although not manifestly so. The Poincar\'e generators $P_\mu$ and $M_{\mu \nu}$  close therefore according to the Poincar\'e algebra.  

In order to establish the $\cN = (4,0)$ supersymmetry algebra, one needs to verify that the anticommutator of two supersymmetries gives a space-time translation, as well as the other commutation relations involving the supercharges.  This is the question on which we focus in this section. Before proceeding with the computation, we stress that in the standard Hamiltonian formalism, the variation  $v^0 P_0 F$ of a dynamical variable $F$ under a time translation is given by its Poisson bracket $ [F,H]$ with the Hamiltonian times the parameter $v^0$ of the time translation.  It is a function of the phase space variables only and not of their time derivatives, but when one uses the equations of motion, $v^0 P_0 F$ becomes of course equal to $v^0 \partial_0 F$.  The same feature holds for the above ``Hamiltonian-like''  first-order action of the $\cN = (4,0)$ theory.
Similarly,  the supersymmetry transformations do not involve time derivatives of the variables.  For this reason, their algebra  cannot contain time derivatives either.  This is again a well known feature of the Hamiltonian formalism,  when transformations are written in terms of phase space variables, which occurs in the Hamiltonian formulation of  non-exotic supersymmetric theories  as well. To compare with the familiar form of the supersymmetry algebra acting on the fields where time derivatives appear, one must use the equations of motion.  

We first compute the anticommutator of two supersymmetry transformations.  We carry this task for the gauge-invariant curvatures, for which the computation is simpler.  For non gauge-invariant fields,  the commutator of supersymmetries  may indeed give additional gauge or Weyl transformations terms.  

On the Cotton tensor of $Z$, we find
\begin{align}
[\delta_{\epsilon_1}, \delta_{\epsilon_2} ] D\indices{^{ij}_{kl}}[Z] 
&= \frac{\alpha_1^2}{(3!)^3} \partial_r D\indices{^{ij}_{kl}}[Z] \left( \bar{\epsilon}_{2A} \Gamma^r \epsilon_1^A\right) \nonumber \\
&\quad - \frac{\alpha_1^2}{(3!)^3} \mathbb{P}_{(2,2)} \partial_r D_{klpn} \varepsilon^{ijpqr} \bar{\epsilon}_{2A} \Gamma\indices{^n_q} \Gamma^0 \epsilon_1^A \nonumber \\
&\quad + \frac{\alpha_1^2}{2(3!)^3} \mathbb{P}_{(2,2)}\varepsilon^{pqrij} \partial_p D_{qrkl} \bar{\epsilon}_{2A} \Gamma^0
\epsilon_1^A \nonumber \\
&\quad - (1\leftrightarrow2) + (\text{terms containing } A^{AB}_{ij}) .
\end{align}
The first term is a spatial translation. The second term vanishes: it follows from the symplectic Majorana reality conditions that $\bar{\epsilon}_{2A} \Gamma\indices{^n_q} \Gamma^0 \epsilon_1^A$ is symmetric in $1$, $2$, see appendix \ref{app:gammamatrices}. (The other terms are antisymmetric under the exchange of $1$ and $2$.) 
Using the equation of motion for $Z$, the curl appearing in the third term becomes a time derivative. Finally, the extra terms containing the bosonic two-forms $A^{AB}_{ij}$ can be shown to vanish. We therefore find indeed a space-time translation,
\begin{equation}
[\delta_{\epsilon_1}, \delta_{\epsilon_2} ] D\indices{^{ij}_{kl}}[Z] = v^\mu \partial_\mu D\indices{^{ij}_{kl}}[Z], \quad v^\mu = - \frac{2 \alpha_1^2}{(3!)^3} (\bar{\epsilon}_{1A} \Gamma^\mu \epsilon_2^A).
\end{equation}
We proceed in a similar fashion for the other fields. We collect here a few identities useful for this computation.
\begin{enumerate}
\item For the commutator on fermionic fields, one needs the following Fierz rearrangement identity, valid for two spinors $\epsilon_1$, $\epsilon_2$ of negative chirality and a spinor $\eta$ of positive chirality:
\begin{equation}
(\bar{\epsilon}_1 \eta) \epsilon_2 = \frac{1}{4}\left[ (\bar{\epsilon}_1 \Gamma^0 \epsilon_2) \Gamma^0 \eta - (\bar{\epsilon}_1 \Gamma^i \epsilon_2) \Gamma_i \eta + \frac{1}{2} (\bar{\epsilon}_1 \Gamma_{0ij} \epsilon_2) \Gamma^{0ij} \eta\right] .
\end{equation}
This follows from the completeness relations of gamma matrices, and from the duality relations between rank $r$ and rank $6-r$ antisymmetric products of gamma matrices \cite{freedman_vanproeyen_2012}. It is independent from any Majorana condition on the spinors and is therefore also valid when $\epsilon_1$, $\epsilon_2$ and $\eta$ carry free symplectic indices.
\item The Cotton tensor of $\chi_{ij}$ satisfies the massless Dirac equation
\begin{equation}
\Gamma^0 \Gamma^k \partial_k D_{ij} = \dot{D}_{ij}.
\end{equation}
It is a consequence of its equation of motion \eqref{eq:eomchi} and of the $\Gamma$-tracelessness identity $\Gamma^i D_{ij} = 0$.
\item The scalar field satisfies
\begin{equation}
2\phi^{[ABC|E|} \Omega^{D]N} \Omega_{EM} + 3 \Omega^{[AB} \phi^{CD]NE} \Omega_{EM} + 2 \phi^{[ABC|N|} \delta^{D]}_M = \frac{1}{2} \phi^{ABCD} \delta^N_M .
\end{equation}
This identity follows from
\begin{equation}\label{eq:9indices}
\Omega^{[AA'}\Omega^{BB'}\Omega^{CC'}\Omega^{DD'}\delta^{N]}_M = 0
\end{equation}
by contracting with $\phi^{PQRS}\Omega_{PA'}\Omega_{QB'}\Omega_{RC'}\Omega_{SD'}$. Equation \eqref{eq:9indices} holds because the indices go from $1$ to $8$ and the antisymmetrization is performed on nine indices.
\end{enumerate}
After this is done, we find \begin{equation}
[\delta_{\epsilon_1}, \delta_{\epsilon_2} ] \Phi = - \kappa^2\, (\bar{\epsilon}_{1A} \Gamma^\mu \epsilon_2^A) \, \partial_\mu \Phi
\end{equation}
on any (gauge-invariant) field $\Phi$, provided the following relations hold between the constants $\alpha_i$:
\begin{equation}
\frac{2\alpha_1^2}{(3!)^3} = \alpha_2^2 = \frac{2\alpha_3^2}{3} = \frac{\alpha_4^2}{2} \equiv \kappa^2 . \label{eq:RelAlpha}
\end{equation}
The supersymmetry algebra dictates  the equality of the numbers of bosonic and fermionic degrees of freedom through the anticommutation relation $\{Q,Q\} \sim P$.  This equality holds only for the full spectrum, and not for any of the individual actions obtained by dropping from the theory the fields that do not transform under the individual fermionic symmetries with one single non-vanishing $\alpha_i$ ($i = 1, \cdots, 4$). Consequently, the fact that the constants $\alpha_i$ are related by the supersymmetry algebra does not come as a surprise.

Writing the generator of supersymmetry transformations as $\bar{\epsilon}_A Q^A$, we therefore get the algebra
\begin{equation}
\{ Q^A_\alpha, Q^B_\beta \} = \kappa^2 \Omega^{AB} \left( P_L \Gamma^\mu C^{-1} \right)_{\alpha\beta} P_\mu . 
\end{equation}
The remaining relations involving the supercharge are easy to derive. First, one notes that the supersymmetry transformations commute with $\partial_\mu$ since they do not depend explicitly on the spacetime coordinates. So one has  $[Q^A_\alpha, P_\mu] = 0$. Then, one observes that the Poincar\'e invariance of the action forces the supercharges to transform in a definite representation of the Lorentz group.  Knowing the transformation properties of the supercharges under spatial rotations (which we do since $SO(5)$ covariance is manifest) determines the commutators $[Q^A_\alpha, M_{ij}] $ of the supercharges with the spatial rotation generators $M_{ij}$. The ambiguity as to which representation of the Lorentz group ($(2,1)$ or $(1,2)$) actually occurs is resolved by considering the manifestly Lorentz invariant transformation rules of the scalars, which shows that the supercharges transform in the $(2,1)$.  Taking into account the $USp(8)$ transformation properties of the supercharges, one thus gets $Q_{\frac12} \sim (2,1; 8)$.

Finally, we note that there exist other supersymmetric multiplets containing the exotic graviton, with lower supersymmetry ($\cN = (1,0)$, $(2,0)$ or $(3,0)$). These are listed in \cite{deWit:2002vz}. The corresponding actions are easy to write down by appropriate truncations of the $\cN = (4,0)$-theory; we refer to \cite{Henneaux:2017xsb} for details. Upon dimensional reduction, they yield the (linearized) versions of the $\cN=2$, $\cN=4$ and $\cN=6$ theories in 5 dimensions with $USp(\cN)$ $R$-symmetry \cite{Cremmer:1980gs,Chamseddine:1980sp,Awada:1985ep,Gunaydin:1985cu}.

\section{The \texorpdfstring{$\cN = (3,1)$}{N = (3,1)} theory}
\label{sec:Action}

We also have all the necessary tools to write down the action of the free $\cN = (3,1)$ theory. The bosonic field content of this theory is
\begin{equation}
(4,2;1,1) \oplus (3,1;6,2) \oplus (2,2;14,1) \oplus (1,1;14',2)\, , 
\end{equation}
where again the labels refer to the representation of the little algebra $su(2) \oplus su(2)$ and under the $R$-symmetry $usp(6) \oplus usp(2)$. This is one exotic chiral field of Young symmetry type $(2,1)$, $12$ chiral $2$-forms, $14$ vectors and $28$ scalars. The fermionic spectrum is
\begin{equation}
(3,2;6,1) \oplus (4,1;1,2) \oplus (2,1; 14, 2) \oplus (1,2;14',1)\, ,
\end{equation}
i.e., $2$ exotic gravitini, $6$ standard chiral gravitini (Rarita-Schwinger fields), as well as $28$ left-handed and $14$ right-handed spin-$\frac12$-fields. Those fields are displayed in Table \ref{Table1}, where their transformation properties both under the little algebra and the $R$-symmetry are given, as well as reality properties and canonical dimensions. At the interacting level, the theory is expected to have $F_{4(4)}$ symmetry, with the scalars parametrizing the coset $F_{4(4)}/USp(6) \times USp(2)$, the vectors and chiral $2$-forms being in the fundamental $\mathbf{26}$ representation (more details in section \ref{sec:commentF4}).

\begin{table}
\centering
$
\arraycolsep=5pt
\begin{array}{l l | l  c c c}
 \multicolumn{2}{c|}{su(2) \oplus su(2)}         & \multicolumn{1}{c}{\multirow{2}*{\text{Reality}}} & \multicolumn{1}{c}{\multirow{2}*{$usp(6)$\text{ irr.}}}                                                                                                                        & \multicolumn{1}{c}{\multirow{2}*{\text{Chirality}}}                                                                                                                                                   & \multicolumn{1}{c}{\multirow{2}*{\text{Dim.}}} \\ 
 \multicolumn{2}{c|}{\oplus\, usp(6)\oplus usp(2)} &                                                                                                                                                                                                                                              &                                                  &                                                                                                                                                                                                       & \\ \otoprule %
  (4,2;1,1)                                     & {\tyng{2,1}}                            & (Z\indices{^{ijk}_{pq}})^{*}=Z\indices{^{ijk}_{pq}}                                                                                                                            & -                                                                     &                   -                          & 1                                           \\ \midrule
  (3,1;6,2)                                     & {\tyng{1,1}}                           & A^{*}_{\tta \alpha ij} =    \Omega_{\tta \ttb} \varepsilon_{\alpha\beta}  A^{\ttb \beta}_{ij}                                                                                                         & -                                                                                &                                          -    & 2                                           \\ \midrule
  \multirow{2}*{$(2,2;14,1)$}                   & \multirow{2}*{{\tyng{1}}}                           & V_{\tta \ttb i}^{*} =  \Omega_{\tta \ttc} \Omega_{\ttb \ttd} V^{\ttc \ttd}_{i}                                                                                                                              & \Omega_{\tta \ttb} V^{\tta \ttb}=0                                                                        &                                           -   & \multirow{2}*{2} \\ 
                                                &                                                              & W_{\tta \ttb ijk}^{*} =  \Omega_{\tta \ttc} \Omega_{\ttb \ttd} W^{\ttc \ttd}_{ijk}                                                                                                                          & \Omega_{\tta \ttb} W^{\tta \ttb}_{ijk}=0                       & - \\ \midrule
 \multirow{2}*{$(1,1;14^{\prime} ,2)$}          & \multirow{2}*{$\bullet$}                                     & \phi^{*}_{\tta \ttb \ttc \alpha} =  \Omega_{\tta \tta^{\prime}} \Omega_{\ttb \ttb^{\prime}}\Omega_{\ttc \ttc^{\prime}} \varepsilon_{\alpha \alpha^{\prime}}\phi^{\tta^\prime \ttb^\prime \ttc^\prime \alpha^{\prime}} & \Omega_{\tta \ttb} \phi^{\tta \ttb \ttc \alpha}=0                   &          -        & 2 \\
                                                &                                                              & \pi^{*}_{\tta \ttb \ttc \alpha} =  \Omega_{\tta \tta^{\prime}} \Omega_{\ttb \ttb^{\prime}}\Omega_{\ttc \ttc^{\prime}} \varepsilon_{\alpha \alpha^{\prime}} \pi^{\tta^\prime \ttb^\prime \ttc^\prime \alpha^{\prime}}  & \Omega_{\tta \ttb} \pi^{\tta \ttb \ttc \alpha}=0                                                                          &                                          -    & 3 \\ \otoprule
  (3,2;6,1)                                     & {\tyng{1}}_{\text{F}}                               & \theta^{*}_{\tta ijk} = \Omega_{\tta \ttb} \cB \theta^{\ttb}_{ijk}                                                                                                                                   &  -                                                                      & \Gamma_{7}\theta=+\theta                      & \frac{3}{2} \\ \midrule
  (4,1;1,2)                                     & {\tyng{1,1}}_{\text{F}}                          & \chi^{*}_{\alpha ij} =\varepsilon_{\alpha\beta} \cB \chi_{ij}^{\beta}                                                                                                            &     -                                               & \Gamma_{7}\chi=+\chi                          & \frac{3}{2} \\ \midrule
  (2,1;14,2)                                    & \bullet_{\text{F}}                                           & \psi^{*}_{\tta \ttb \alpha} =  \Omega_{\tta \tta^{\prime}} \Omega_{\ttb \ttb^{\prime}} \varepsilon_{\alpha \alpha^{\prime}} \cB \psi^{\tta^\prime \ttb^\prime  \alpha^{\prime}}                    &  \Omega_{\tta \ttb} \psi^{\tta \ttb \alpha}=0                                                                                                  & \Gamma_{7}\psi=+\psi                          & \frac{5}{2} \\ \midrule
  (1,2;14^{\prime},1)                           & \bullet_{\text{F}}                                           & \widetilde{\psi}^{*}_{\tta \ttb \ttc} =  \Omega_{\tta \tta^{\prime}} \Omega_{\ttb \ttb^{\prime}}\Omega_{\ttc \ttc^{\prime}}  \cB \widetilde{\psi}^{\tta^\prime \ttb^\prime \ttc^\prime}                                & \Omega_{\tta \ttb} \widetilde{\psi}^{\tta \ttb \ttc}=0                                                  & \Gamma_{7}\widetilde{\psi}=-\widetilde{\psi} & \frac{5}{2}
\end{array}
$
\caption[Fields of the $\cN = (3,1)$ theory]{The fields of the $\cN = (3,1)$ theory (in the prepotential formulation) and their transformation properties. We have indicated the space-time transformation properties by the corresponding Young diagram (with an extra $F$ index in the case of fermions). Indices $\tta, \ttb = 1,..,6$ and $\alpha, \beta = 1, 2$ label the fundamental representations of $usp(6)$ and $usp(2)$, respectively. Quantities with multiple indices transform
in the corresponding tensor product.  We also write the reality, irreducibility and (in the case of fermions) chirality conditions they satisfy. Canonical dimension is indicated in the last column.}
\label{Table1}
\end{table}

The action is a sum of eight terms, one for each type of field,
\begin{equation}
    S= S_{{\tyng{2,1}}}+ S_{{\tyng{1,1}}}+S_{1} +S_{0}+ S^{L}_{{\tyng{1,1}}_{\text{F}}}+S^{L}_{3/2}+S^{L}_{1/2} + S^{R}_{1/2} \, .
\end{equation}

\subsection{Bosonic fields}

\paragraph{Chiral $(2,1)$-tensor.}

This is the real tensor field of mixed Young symmetry $(2, 1)$ type with self-dual field strength described in the previous chapters. The action is written in terms of the prepotential $Z_{ijklm}$; it was derived in subsection \ref{sec:chiralaction} and reads
\begin{equation}
    S_{\tyng{2,1}}[Z] = \frac{1}{2}\int \!d^6 \!x\, Z_{ijklm}\left(\dot{D}^{ijklm}[Z] - \frac{1}{2} \varepsilon^{abclm}\pd_{a}D\indices{^{ijk}_{bc}}\left[Z\right]  \right)        \, , 
\end{equation}
where the Cotton tensor $D^{ijklm}[Z]$ is explicitly given in equation \eqref{eq:defDtext}.

\paragraph{Chiral $2$-forms.}

The theory contains $12$ chiral 2-forms $A^{\tta\alpha}_{ij}$, described by the action
\begin{equation}
S_{\tyng{1,1}}[A] = - \frac{1}{2}\int \!d^6 \!x\, A_{\tta\alpha ij}^{*} \left( \dot \cB^{\tta\alpha ij}[A]- \frac{1}{2} \varepsilon^{ijklm} \pd_{k} \cB^{\tta\alpha}_{lm}[A] \right)   \, ,
\end{equation}
where the magnetic fields $\cB^{\tta \alpha}_{ij}$ are given by
\begin{equation}
    \cB^{\tta \alpha i j}[A] = \frac{1}{2}\varepsilon^{ijklm}\pd_{k}A^{\tta \alpha}_{lm} \, .
\end{equation}

\paragraph{Vector fields.}

The theory possesses 14 vector fields $V_\mu^{\tta\ttb}$.  We write the action in its first order, twisted self-duality form
\begin{align}
S_1[V,W] =- \frac{1}{2} \int \!d^6 \!x\, \Big[ &V^{*}_{\tta \ttb i} \left( \dot \cB^{\tta \ttb i}[W] + \frac{1}{3!} \varepsilon^{ijklm} \pd_{j} \cB^{\tta \ttb}_{klm}[V] \right) \nonumber \\
& \left.+\frac{1}{3!} W^{*}_{\tta\ttb ijk} \left(-\dot{\cB}^{\tta \ttb ijk}[V] +\epsilon^{ijklm}\pd_{l} \cB^{\tta \ttb}_{m}[W] \right)\right]\, ,
\end{align}
where the magnetic fields are
\begin{align}
    \cB^{\tta \ttb ijk}[V] = \varepsilon^{ijklm}\pd_{l} V^{\tta \ttb}_{m} \, , \quad \cB^{\tta \ttb i}[W] = \frac{1}{3!}\varepsilon^{ijklm}\pd_{j} W^{\tta \ttb}_{klm} \, . 
\end{align}

\paragraph{Scalar fields.}

The action for the 28 scalar fields $\phi^{\tta \ttb \ttc \alpha}$ can be written in Hamiltonian form in the following way:
\begin{equation}
    S_0[\phi,\pi] = \frac{1}{2} \int \!d^6 \!x\, \left(2 \,\pi^{*}_{\tta \ttb \ttc \alpha}  \dot{\phi}^{\tta \ttb \ttc \alpha} -\pi_{\tta \ttb \ttc \alpha}^{*} \pi^{\tta \ttb \ttc \alpha} - \pd_{i} \phi_{\tta \ttb \ttc \alpha}^{*} \pd^{i} \phi^{\tta \ttb \ttc \alpha} \right)  \, .
\end{equation}

\subsection{Fermionic fields}

\paragraph{Exotic gravitini.}

The $2$ left-handed exotic gravitini are described by fermionic $2$-form prepotentials $\chi^{\alpha}_{ij}$. Their action in terms of prepotentials was derived in the previous chapter and reads
\begin{equation}
S^L_{{\tyng{1,1}}_{\text{F}}}[\chi] = -2 i \int \!d^6 \!x\, \chi^{\dagger}_{\alpha ij} \left(\dot D^{\alpha ij}[\chi] - \frac{1}{2}\varepsilon^{ijklm} \pd_{k} D^{\alpha}_{lm}[\chi] \right). 
\end{equation}

\paragraph{Gravitini.}

The theory also contains $6$ left-handed gravitini, described by the prepotentials $\theta^\tta_{ijk}$. The action is
\begin{equation}\label{eq:actiongravitini}
    S^L_{3/2}[\theta] = -i \int \!d^6 \!x\, \theta^{\dagger}_{\tta ijk} \left(\dot D^{\tta ijk}[\theta] - \varepsilon^{ijklm} \pd_{j} \widetilde{D}^{\tta}_{m}[\theta] \right) \, ,
\end{equation}
where the Cotton tensor is defined by
\begin{equation}
D^{\tta ijk}[\theta] = \varepsilon^{ijklm} \pd_{l}S^{\tta}_{m}[\theta] \, ,
\end{equation}
and the Schouten by
\begin{equation}
    S^{\tta}_{i}[\theta]=\frac{1}{4}\left( \Gamma_{ij} - 3 \delta_{ij}\right) \varepsilon^{jklmn}\pd_{k}\theta^{\tta}_{lmn}
\end{equation}
as in chapter \ref{chap:twisted}. In \eqref{eq:actiongravitini}, the quantity $\widetilde{D}^\tta_m$ is the gamma-matrix contraction
   \begin{equation}
           \widetilde{D}^\tta_{m} = \frac{1}{2} \Gamma^{ab} D^\tta_{mab} \, .
   \end{equation}

\paragraph{Spin $1/2$ fields.}

Finally, the theory possesses $28$ and $14$ left-handed and right-handed spin $1/2$ fields, which we write as $\psi^{\tta \ttb \alpha}$ and $\widetilde{\psi}^{\tta \ttb \ttc}$, respectively. Their action is simply
\begin{equation}
    S^{L}_{1/2}[\psi] = i \int \!d^6 \!x\, \psi^{\dagger}_{\tta \ttb \alpha} \left(\dot \psi^{\tta \ttb \alpha} - \Gamma^{0} \Gamma^{i} \pd_{i} \psi^{\tta \ttb \alpha} \right)
\end{equation}
and
\begin{equation}
    S^{R}_{1/2}[\widetilde{\psi}] = i \int \!d^6 \!x\, \widetilde{\psi}^{\dagger}_{\tta \ttb \ttc} \left(\dot{\widetilde{\psi}}{}^{\tta \ttb \ttc} - \Gamma^{0} \Gamma^{i} \pd_{i} \widetilde{\psi}^{\tta \ttb \ttc} \right) \, .
\end{equation}

\subsection{Comments on the \texorpdfstring{$F_{4(4)}$}{F 4(4)} symmetry}
\label{sec:commentF4}

As was noticed by Hull, ${USp(6) \times USp(2)}$ is the maximal compact subgroup of the exceptional group $F_{4(4)}$ and the number of scalar fields is the same as the dimension of the coset space
\begin{equation}\label{eq:F4coset}
\frac{F_{4(4)}}{USp(6) \times USp(2)} \, .
\end{equation}
By analogy with the well-known situation in extended supergravities, it is natural to conjecture that, in the interacting theory, the scalar fields indeed take values in \eqref{eq:F4coset} \cite{Hull:2000zn}.

The $F_{4(4)}$ symmetry should then also extend as a symmetry of the full bosonic sector. It is interesting to notice that the lowest-dimensional irreducible representation of $F_{4(4)}$ is exactly of dimension $26$: therefore, we conjecture that the $12$ chiral $2$-forms combine with the $14$ vector fields into an irreducible multiplet, while the chiral tensor is a singlet under $F_{4(4)}$. This is supported by the branching rules of representations for the embeddings
\begin{equation}
USp(6) \times USp(2) \,\subset\, F_{4(4)} \,\subset\, E_{6(6)}
\end{equation}
(see for example \cite{Yamatsu:2015npn} for tables).
\begin{enumerate}
\item Under its maximal compact subgroup $USp(6) \times USp(2)$, the fundamental representation of $F_{4(4)}$ decomposes as
\begin{equation}
\mathbf{26} = (6,2) \oplus (14,1),
\end{equation}
which are exactly the $USp(6) \times USp(2)$ transformation rules of the chiral forms and vector fields.
\item Under its $F_{4(4)}$ subgroup, the fundamental representation of $E_{6(6)}$ decomposes as
\begin{equation}
\mathbf{27} = \mathbf{26} \oplus \mathbf{1} .
\end{equation}
It is well known that the $27$ vector fields of maximal supergravity in five dimensions fall into an irreducible $E_{6(6)}$ multiplet \cite{Cremmer:1979uq}. From the dimensional reduction of the
$(3,1)$-theory, $26 = 12 + 14$ of those vector fields arise from the six-dimensional chiral forms and vector fields, and one comes from the reduction of the chiral $(2,1)$ tensor. Therefore, the well-known  $E_{6(6)}$ symmetry of five-dimensional maximal supergravity seems to support this conjecture.
\end{enumerate}
Of course, it is impossible to prove this in the absence of a consistent interacting theory. Moreover, this symmetry would be very peculiar since it would mix fields with different spacetime indices (see also \cite{Ananth:2017nbi} for a similar situation). Nevertheless, we feel that these remarks are intriguing and could have useful implications for the interacting theory, if it exists.

\subsection{Supersymmetry}
\label{sec:susy31}

We now establish that the action of the previous sections is invariant under $\cN = (3,1)$ supersymmetry. The supersymmetry parameters are $\epsilon^\tta$ and $\et^\alpha$. As their indices suggest, they transform in the fundamental of $usp(6)$ and $usp(2)$, respectively, and are inert under the other factor. They satisfy symplectic Majorana-Weyl conditions gathered in table \ref{tab:susy} and their canonical dimension is $-1/2$.
\begin{table}
    \centering
    $
\begin{array}{l | c c c}

             & \multicolumn{1}{c}{\text{Reality}}                         & \multicolumn{1}{c}{\text{Chirality}}     & \multicolumn{1}{c}{\text{Dimension}} \\ \midrule
  \epsilon^{\tta} \; & \;  \epsilon_{\tta}^{*}= \Omega_{\tta\ttb} \,\cB\, \epsilon^{\ttb}         \; & \; \Gamma_{7} \epsilon^{\tta} = -\epsilon^{\tta} & -1/2                         \\
 \et^{\alpha} \; & \; \et_{\alpha}^{*} = \varepsilon_{\alpha\beta} \,\cB\, \et^{\beta} \; & \; \Gamma_{7}\et^{\alpha} = + \et^{\alpha} & -1/2 
\end{array}
$
    \caption{Supersymmetry parameters of the $\cN = (3,1)$ theory.}
    \label{tab:susy}
\end{table}

The supersymmetry transformations on the various fields are then
\begin{align}
\delta Z\indices{^{ijk}_{lm}} &= \mathbb{P}_{(2,2,1)}\left( \beta_1\bar{\epsilon}_\tta \,\Gamma_{lm} \theta^{\tta ijk}  + \beta_{2} \bar{\epsilon}_{\alpha}\left(\Gamma^{ijk} \chi_{lm}^{\alpha}+12 \Gamma^{[i}\delta^{j}_{[l}\chi\indices{^{k]\alpha}_{m]}} \right)\right)\, , \\
\delta\theta^{\tta}_{ijk} &= -\frac{\beta_1}{8 \cdot 3!^2} \left(\pd^r Z\indices{_{ijk}^{ab}}\varepsilon_{pqrab}\Gamma^{pq} \Gamma^{0}\epsilon^{\tta} + \frac{2}{3} \pd^{a} Z\indices{^{bcd}_{[ij}}\varepsilon_{k]abcd}\Gamma^{0}\epsilon^{\tta}\right)\nonumber \\
&\quad + \beta_{3}\left(\varepsilon_{\alpha \beta} A^{\tta \alpha}_{[ij}\Gamma_{k]}\Gamma^{0}\tilde{\epsilon}^{\beta}\right)+ \beta_{4}\left( V\indices{^{\tta\ttb}_{[i}}\Omega_{\ttb \ttc} \Gamma_{jk]} \Gamma^0 \epsilon^\ttc +\frac{1}{3} W^{\tta\ttb}_{ijk} \Omega_{\ttb\ttc} \Gamma^{0} \epsilon^\ttc\right) \, , \\
\delta\chi^{\alpha }_{ij} &= \frac{\beta_{2}}{4\cdot 4!}\left(\varepsilon^{abcde}\pd_{e}Z_{abc ij} \Gamma_{d}\Gamma^{0}\tilde{\epsilon}^{\alpha} \right)+ \beta_{5}\left( \frac{1}{2} A^{\tta \alpha}_{ij}\Omega_{\tta\ttb}\Gamma^{0}\epsilon^{\ttb}\right)\, , \\
\delta A^{\tta\alpha}_{ij} &= \beta_{3}\left(-4 \varepsilon^{\alpha \beta}\bar{\tilde{\epsilon}}_{\beta } \Gamma_{[i} S^{\tta}_{j]}[\theta]\right)+\beta_{5} \left(4 \Omega^{\tta\ttb} \bar{\epsilon}_\ttb S^{\alpha}_{ij}[\chi] \right)+\beta_{6} \left(-2 \bar{\epsilon}_{\ttb} \Gamma_{ij} \psi^{\tta\ttb \alpha} \right)\, , \\
\delta V^{\tta\ttb}_{i} &= - \beta_{4} \left(4  \Omega^{\ttc[\tta}\bar{\epsilon}_\ttc S^{\ttb]}_{i}[\theta]+\frac{2}{3} \Omega^{\tta\ttb} \bar{\epsilon_\ttc} S^\ttc_{i}[\theta]\right)  + \beta_{7} \left( 2 \bar{\tilde{\epsilon}}_\alpha \Gamma_{i} \psi^{\tta\ttb\alpha}\right) + \beta_{8}\left( 2 \bar{\epsilon}_\ttc  \Gamma_{i} \tilde{\psi}^{\tta\ttb\ttc}\right)\, , \\
\delta W^{\tta\ttb}_{ijk} &= - \beta_{4} \left(12  \Omega^{\ttc[\tta} \bar{\epsilon}_\ttc \Gamma_{[ij} S^{\ttb]}_{k]}[\theta]+2\Omega^{\tta\ttb} \bar{\epsilon}_\ttc \Gamma_{[ij} S^{\ttc}_{k]}[\theta] \right) \nonumber \\
&\quad + \beta_{7} \left( 2 \bar{\tilde{\epsilon}}_\alpha \Gamma_{ijk} \psi^{\tta\ttb\alpha}\right) + \beta_{8}\left(2 \bar{\epsilon}_\ttc \Gamma_{ijk} \tilde{\psi}^{\tta\ttb\ttc}\right) \, ,
\end{align}
\begin{align}
\delta \psi^{\tta\ttb\alpha} &= \beta_{7} \left( \cB^{\tta\ttb}_{i}[W] \Gamma^i +\frac{1}{3!} \cB^{\tta\ttb}_{ijk}[V] \Gamma^{ijk} \right) \Gamma^{0} \tilde{\epsilon}^{\alpha} \nonumber \\
&\quad +\beta_{9}\left( \pi^{\tta\ttb\ttc\alpha} \Omega_{\ttc\ttd}\Gamma^0 \epsilon^{\ttd} +  \pd_i \phi^{\tta\ttb\ttc\alpha} \Omega_{\ttc\ttd} \Gamma^{i}\epsilon^{\ttd} \right) \nonumber \\
&\quad +\beta_{6} \Gamma^{ij} \Gamma^{0} \left(\cB^{[\tta|\alpha}_{ij}[A]  \epsilon^{|\ttb]}-\frac{1}{6}\Omega^{\tta \ttb} \cB^{\ttc \alpha}_{ij}[A] \Omega_{\ttc\ttd} \epsilon^{\ttd}\right)\, , \\
\delta \tilde{\psi}^{\tta\ttb\ttc} &= \beta_{10}\left( \pi^{\tta\ttb\ttc\alpha} \varepsilon_{\alpha \beta} \Gamma^{0} \tilde{\epsilon}^{\beta} +  \pd_i \phi^{\tta\ttb\ttc\alpha} \Gamma^{i}\varepsilon_{\alpha \beta}\tilde{\epsilon}^{\beta}\right)\nonumber \\
&\quad +\beta_{8} \Gamma^i \Gamma^{0} \left( \cB^{[\tta\ttb}_{i}[W] \epsilon^{\ttc]} -\frac{1}{2}\Omega^{[\tta \ttb} \cB^{\ttc]\ttd}_{i}[W] \Omega_{\ttd \tte}\epsilon^{\tte} \right) \nonumber \\
&\quad  +\frac{\beta_8}{3!} \Gamma^{ijk}\Gamma^{0}\left( \cB^{[\tta\ttb}_{ijk}[V] \epsilon^{\ttc]}-\frac{1}{2} \Omega^{[\tta \ttb}\cB^{\ttc]\ttd}_{ijk}[V] \Omega_{\ttd \tte} \epsilon^{\tte}\right) \, , \\
\delta \pi^{\tta\ttb\ttc\alpha} &= \beta_{10} \left( -2\varepsilon^{\alpha \beta } \bar{\tilde{\epsilon}}_{\beta} \Gamma^{i} \Gamma^{0} \pd_i \tilde{\psi}^{\tta\ttb\ttc}\right)+ \beta_{9} \left(- 2 \bar{\epsilon}_{\ttd} \Gamma^{i} \Gamma^{0} \pd_i \psi^{[\tta\ttb| \alpha} \Omega^{\ttc]\ttd} - \bar{\epsilon}_{\ttd} \Gamma^{i} \Gamma^{0}  \Omega^{[\tta\ttb}\pd_i \psi^{\ttc]\ttd\alpha} \right)\, , \\
\delta \phi^{\tta\ttb\ttc\alpha} &= \beta_{10} \left( 2 \varepsilon^{\alpha \beta} \bar{\tilde{{\epsilon}}}_\beta \tilde{\psi}^{\tta\ttb\ttc}\right) + \beta_{9} \left(2 \bar{\epsilon}_{\ttd} \psi^{[\tta\ttb |\alpha}\Omega^{\ttc]\ttd} + \bar{\epsilon}_\ttd \Omega^{[\tta\ttb} \psi^{\ttc]\ttd\alpha}  \right) \, . 
\end{align}
They were found from the following requirements:
\begin{enumerate}[a)]
\item they leave the action invariant;
\item symplectic indices, reality and chirality conditions must match; and
\item a gauge and Weyl transformation of the right hand-side must induce a gauge and Weyl transformation of the field of the left-hand side.
\end{enumerate}
Conditions b) and c) are actually sufficient to fix nearly all the variations. The remaining transformations (and relative factors) are then found from condition a). A consequence of condition c) is that the supersymmetry transformations of the invariant curvatures (Cotton tensors and magnetic fields) can be expressed purely with invariant objects. This is indeed what we find (see \cite{Henneaux:2018rub} for explicit formulas).
The action is invariant for any values of the (real) constants $\beta_1$ to $\beta_{10}$, which are fixed by the supersymmetry algebra (as in the case of the $\cN = (4,0)$ theory): they must satisfy
\begin{equation}\label{eq:relbetas}
\frac{\beta_1^2}{36} = \frac{\beta_2^2}{8} = 4 \beta_3^2 = 4 \beta_4^2 = 2 \beta_5^2 = 4 \beta_6^2 = 4 \beta_7^2 = \frac{4 \beta_8^2}{3} = \frac{2 \beta_9^2}{3} = 2 \beta_{10}^2 \equiv \kappa^2 .
\end{equation}
The proof is a bit technical and presented in Appendix \ref{app:redlinsugra}. The method we follow is dimensional reduction and comparison with linearized maximal supergravity in five dimensions \cite{Cremmer:1979uq}: this allows us to bypass the somewhat cumbersome direct computation of the $\cN = (3,1)$ commutators.

\bookmarksetup{startatroot}
\nnchapter{Conclusions and perspectives}

In this thesis, we have studied several aspects of electric-magnetic dualities and their gravitational generalizations, with a focus on maximal supergravity. Along the way, we have obtained a variety of original results, which we gather here one last time:
\begin{itemize}
\item an explicit proof that the hidden $E_{8(8)}$ symmetry of maximal supergravity in three dimensions appears if and only if the eleven-dimensional Chern-Simons coupling takes the specific value predicted by supersymmetry \cite{Henneaux:2015opa};
\item a manifestly $E_{7(7)}$-invariant formula for the entropy (horizon area) of non-extremal black holes in $D = 4$, $\cN = 8$ supergravity \cite{Compere:2015roa};
\item a complete characterization of the BRST cohomology of non-minimally coupled scalar-vector models, putting the embedding tensor formalism on a strong mathematical foundation \cite{Henneaux:2017kbx,Barnich:2017nty};
\item a formulation of the linearized supergravity mutiplet (spin $2$ and spin $3/2$) that puts the graviton and its dual on the same footing in all dimensions \cite{Lekeu:2018kul}; and
\item free actions for the exotic $\cN = (4,0)$ and $\cN = (3,1)$ theories, including the action for self-dual fields, the supersymmetry tranformations and the dimensional reduction to maximal supergravity in five dimensions \cite{Henneaux:2016opm,Henneaux:2017xsb,Henneaux:2018rub}.
\end{itemize}
There remain of course a lot of intriguing open questions and perspectives for future work. Let us now try to give a broad (and, of course, non-exhaustive) overview of those as conclusion.

\subsubsection*{Infinite-dimensional algebras}

There has been considerable speculation as to whether the infinite-dimensional algebras $E_{10}$ or $E_{11}$ could be the fundamental symmetry structures of maximal supergravity or $M$-theory \cite{West:2001as,Damour:2002cu}. However, the dynamical realisation of these symmetries is still far from clear (see for example \cite{Bossard:2017wxl} for recents developments).
In particular, the interpretation of the fields above the level of the dual graviton is not straightforward. It could be hoped that our action of chapter \ref{chap:twisted} in the case $D = 11$ can help in unravelling these intriguing symmetries. It may also be that one actually needs a more general algebraic structure, see for example \cite{HenryLabordere:2002dk,Henneaux:2010ys,Henneaux:2015gya,Bossard:2017wxl}. On a more positive note, the gaugings of maximal supergravity (through the embedding tensor) have also found an interpretation within $E_{11}$, see \cite{Riccioni:2007au,Bergshoeff:2007qi}.

Finally, it is interesting to remark that the ten-dimensional maximal supergravities (type IIA and IIB) fit perfectly in the $E_{11}$ and $E_{10}$ structures\cite{Schnakenburg:2001he,Schnakenburg:2002xx,West:2004st,Kleinschmidt:2004dy,Kleinschmidt:2004rg,Kleinschmidt:2006tm,Riccioni:2007au,Henneaux:2008nr}. In particular, the self-duality condition in the type IIB case is properly incorporated in the infinite-dimensional algebra\footnote{See also \cite{Henneaux:2015gya} for a similar case of $\cN = (1,0)$ supergravity with a chiral $2$-form, where the group $F_{4}$ plays a role. There also, the infinite-dimensional algebra properly incorporates the self-duality condition.}. It would certainly be interesting if the chiral $\cN = (4,0)$ and $\cN = (3,1)$ theories in six dimensions could also find their place within this framework.

\subsubsection*{Fermions and supersymmetry}

In chapter \ref{chap:cscouplings}, we saw a situation where supersymmetry and the requirement of exceptional symmetry give the same prediction. This has also been noticed in numerous other situations, for example in \cite{deBuyl:2005zy,deBuyl:2005sch,Damour:2005zs,Steele:2010tk} in the context of infinite-dimensional algebras, or in \cite{Hohm:2013vpa,Hohm:2013uia,Hohm:2014fxa} for exceptional field theory.
Still, it is fair to say that the precise link between supersymmetry and hidden symmetries (a question which was already asked in \cite{Julia:1980gr}) is not yet clear and deserves further study.

\subsubsection*{BRST-BV deformations}

We proved in part \ref{PART:GAUGINGS} that the most general local deformation of ungauged supergravity Lagrangians corresponds to the gauging of rigid electric symmetries of the ungauged theory. In particular, there seems to be no way to gauge the electric-magnetic duality rotations. There are two ways (that we can see) to avoid this theorem and construct new deformations of supergravity. The first one would be to work with non-localities in a controlled way. The second would be to gauge symmetries that are not of the electric type (in the sense of chapter \ref{chap:vsgaugings}). If one does not impose Poincaré invariance, those certainly exist: see for example \cite{Brandt:2001hs} (and section 6 of \cite{Barnich:2017nty}) for the gauging of the conformal symmetries of electromagnetism in four dimensions. However, when Poincaré invariance is imposed (for example by the possibility of coupling to gravity, see \cite{Barnich:1995ap}), only electric type symmetries are expected to remain. This is proved in \cite{Torre:1994kb} in the absence of scalars and for some particular examples with scalars in \cite{Barnich:2017nty}. It would certainly be satisfying if one could prove that all Poincaré-invariant symmetries are indeed of the electric type.

Inclusion of fermions would also be interesting within this framework. Although this is not expected to give new constraints on deformations \cite{Trigiante:2016mnt}, the BRST method could provide some insight on the uniqueness properties of extended supergravity theories, along the lines of \cite{Boulanger:2001wq,Boulanger:2018fei}.

\subsubsection*{Dual (super)gravity}

It would be of interest to extend the formulation of chapter \ref{chap:twisted} to fully interacting gravity. However, much like for the case of Yang-Mills theory, there is no clear path to the resolution of the non linear constraints and duality invariance in four dimensions is excluded for fully interacting gravitation \cite{Deser:2005sz}. Moreover, no-go theorems exist for local self-interactions of the dual graviton (Curtright field), see \cite{Bekaert:2002uh,Bekaert:2004dz}. Notwithstanding these difficulties, see \cite{Barnich:2008ts} for an introduction of sources and \cite{Julia:2005,Julia:2005ze,Leigh:2007wf,Hortner:2016omi} for the inclusion of a cosmological constant. 

It could be that the situation is the same as in Yang-Mills theory, which is not duality-invariant classically, but where there are strong indications that its supersymmetric extensions  have this duality at the quantum level \cite{Montonen:1977sn,Seiberg:1994rs,Vafa:1994tf,Seiberg:1994aj}. The analogue of this would have very intriguing consequences for supergravity \cite{Hull:2000rr,Hull:2000zn,Hull:2001iu}. Another logical possibility is of course that the duality is just absent in the interacting theory and is an artifact of the free limit.

Let us also mention that duality was clarified for the massless, massive and ``partially-massless" gravitons in the recent preprint \cite{Boulanger:2018shp}, including arbitrary dimension and cosmological constant. The actions they use include both the graviton and dual graviton fields in a manifestly Lorentz-invariant manner, but they are not on the same footing (see also \cite{Boulanger:2003vs,Boulanger:2008nd}): it would be interesting to understand how they relate to the actions of the chapter \ref{chap:twisted}. These methods were also used in \cite{Boulanger:2012df,Boulanger:2012mq,Boulanger:2015mka,Bergshoeff:2016ncb} to investigate the ``double dual graviton" and other exotic dualizations of $p$-forms and mixed symmetry gauge fields. Again, how this is realized in the first-order, unconstrained Hamiltonian formalism remains to be explored.

Finally, the prepotential formalism for the gravitino field developed in chapter \ref{chap:twisted} should provide an avenue for exploring the dual formulations of supergravity alluded to in \cite{Curtright:1980yk}, where the gravitino is replaced by a fermionic $(D-3)$-form.

\subsubsection*{Higher-spin fields}

Just as for spin $1$ (abelian vector) and $2$ (linearized gravity), \emph{all} free fields of higher spin $s > 2$ in fact also have electric-magnetic symmetries \cite{Hull:2001iu,Bekaert:2003az,Boulanger:2003vs,Deser:2004xt}. In the bosonic case, a manifestly duality-invariant rewriting of the free theory in four dimensions has been achieved in \cite{Henneaux:2015cda,Henneaux:2016zlu}, again in terms of prepotentials enjoying a linearized Weyl invariance. A similar reformulation is still missing in the fermionic case.

Generalizing the ideas of part \ref{PART:6D}, this duality could be understood geometrically from the reduction of a mixed symmetry field in six dimensions of the form
\[ \ydiagram{3,3} \cdots \ydiagram{2,2} \qquad \text{($s$ columns)} \]
satisfying a self-duality equation of motion for its curvature. An action principle giving these equations of motion should be constructed in terms of prepotentials along the lines of chapter \ref{chap:selfdual}, which would reduce to the manifestly duality-invariant actions of \cite{Henneaux:2015cda,Henneaux:2016zlu}. This would require the identification of the appropriate conformal geometry and Cotton tensors.
The analysis of these questions would of course be greatly simplified by a general construction of the conformal geometry and invariant tensors for mixed symmetry fields in the spirit of \cite{Olver_hyper,DuboisViolette:1999rd,DuboisViolette:2001jk,Bekaert:2002dt}, generalizing the scattered results of references \cite{Bunster:2012km,Bunster:2013oaa,Henneaux:2015cda,Henneaux:2016zlu,Henneaux:2016opm,Henneaux:2017xsb,Henneaux:2018rub}.

Let us also note that the status of higher-spin fields within the $E_{10}$ framework is still unclear, even though the tensionless limit of string theory contains an infinite tower of massless higher-spin gauge fields (see \cite{deBuyl:2004ps,Henneaux:2011mm,Kleinschmidt:2013eka}).

\subsubsection*{Chiral fields and exotic supergravities}

For the chiral fields of part \ref{PART:6D}, the natural question is again that of interactions. By analogy with the case of chiral forms, where local self-interactions are excluded \cite{Bekaert:1999dp,Bekaert:1999sq,Bekaert:2000qx}, these interactions are most likely non-local. A strong constraint for the interacting $\cN = (4,0)$ and $\cN = (3,1)$ theories is that they reduce to maximal supergravity in five dimensions, with $E_{6(6)}$ symmetry and well-known interactions. The situation is similar to the problem of finding the interacting $\cN = (2,0)$ theory with tensor multiplets that reduces to five-dimensional super-Yang-Mills. Contrary to that case, however, there is no string theory construction to convince us that the interacting theories actually exist.

Nevertheless, some avenues of exploration can be imagined. First of all, the $\cN = (4,0)$ theory has been suggested to arise as the strong-coupling limit of $\cN = 8$ supergravity, in analogy with the relation between the $\cN = (2,0)$ theory and super-Yang-Mills. The theory should be conformal in this limit; therefore, understanding the conformal symmetry at the level of the prepotentials is an important point. The $\cN = (4,0)$ exotic supergravity was further argued in \cite{Anastasiou:2013hba,Borsten:2017jpt,Anastasiou:2017taf} to find a natural place in a ``conformal magic pyramid'', and to be obtained by squaring two maximal tensor multiplets of same chirality. Finding the terms yielding the cosmological constant for linearized gravity in the prepotential formulation \cite{Julia:2005,Julia:2005ze,Leigh:2007wf,Hortner:2016omi} upon dimensional reduction could also be a useful first step. Finally, we note that non-linear Cotton tensors for higher-spin fields in three dimensions were constructed in \cite{Linander:2016brv}. A generalization of those to the present case could be useful for the construction of interactions in the prepotential formalism.

\appendix
\chapter{Appendices to part \ref{PART:SUGRA}}
\label{chap:app1}

\section{Conventions for part \ref{PART:SUGRA}}
\label{app:diffgeo}

The flat space metric $\eta_{\mu\nu}$ is of ``mostly plus'' signature,
\begin{equation}
\eta = \diag(-1, +1, \dots, +1) .
\end{equation}
The epsilon tensor $\varepsilon_{\mu_1 \dots \mu_D}$ is totally antisymmetric with
\begin{equation}
\varepsilon_{01\dots (D-1)} = \sqrt{|\det g|}, \quad \varepsilon^{01\dots (D-1)} = \frac{(-1)^t}{\sqrt{|\det g|}},
\end{equation}
where $t$ is the number of negative eigenvalues of $g_{\mu\nu}$.
When $g_{\mu\nu} = \eta_{\mu\nu}$, this is
\begin{equation}
\varepsilon_{012\dots (D-1)} = +1, \quad \varepsilon^{012\dots (D-1)} = -1 .
\end{equation}
We have the identity
\begin{equation}
\varepsilon_{\mu_1\dots\mu_p\nu_{p+1}\dots\nu_D}\varepsilon^{\mu_1\dots\mu_p\rho_{p+1}\dots\rho_D} = (-1)^t\,p!(D-p)!\, \delta_{[\nu_{p+1}}^{\rho_1}\delta_{\nu_2}^{\rho_2}\dotsm\delta_{\nu_D]}^{\rho_D}.
\end{equation}
Symmetrization and antisymmetrization are done with weight one, i.e.
\begin{equation}
T_{(\mu_1 \dots \mu_k)} = \frac{1}{k!} \sum_{\sigma \in S_k} T_{\mu_{\sigma(1)} \dots \mu_{\sigma(k)}}, \quad T_{[\mu_1 \dots \mu_k]} = \frac{1}{k!} \sum_{\sigma \in S_k} (-1)^{|\sigma|} T_{\mu_{\sigma(1)} \dots \mu_{\sigma(k)}},
\end{equation}
where the sum runs over all permutations of $k$ elements and $|\sigma|$ is the signature of the permutation $\sigma$. With this normalization, they are projectors, $T_{((\mu_1 \dots \mu_k))} = T_{(\mu_1 \dots \mu_k)}$ and $T_{[[\mu_1 \dots \mu_k]]} = T_{[\mu_1 \dots \mu_k]}$. We use the notation
\begin{equation}
\delta^{\mu_1 \dots \mu_k}_{\nu_1 \dots \nu_k} = \delta^{\mu_1}_{[\nu_1} \dots \delta^{\mu_k}_{\nu_k]}
\end{equation}
for antisymmetrized products of Kronecker deltas.

The components of a $p$-form are
\begin{equation}
\omega = \frac{1}{p!} \omega_{\mu_1\mu_2 \dots \mu_p}dx^{\mu_1} \wedge dx^{\mu_2} \wedge \dots \wedge dx^{\mu_p} .
\end{equation}
The exterior derivative is defined by
\begin{equation}
d \omega = \frac{1}{p!} \partial_\nu \omega_{\mu_1 \dots \mu_p}dx^{\nu}\wedge dx^{\mu_1} \wedge \dots \wedge dx^{\mu_p},
\end{equation}
which in components gives
\begin{equation}
(d\omega)_{\mu_1 \dots \mu_{p+1}} = (p+1) \partial_{[\mu_1} \omega_{\mu_2 \dots \mu_{p+1}]} .
\end{equation}
The Hodge dual is defined on the basis of $p$-forms by
\begin{equation}
\star (dx^{\mu_1} \wedge \dots \wedge dx^{\mu_p}) = \frac{1}{(D-p)!} \varepsilon\indices{_{\nu_1 \dots \nu_{D-p}}^{\mu_1 \dots \mu_p}} dx^{\nu_1} \wedge \dots \wedge dx^{\nu_{D-p}} . 
\end{equation}
It takes a $p$-form to a $D-p$ form, where $D$ is the dimension of the manifold. In components, we have
\begin{equation}
(\star \omega)_{\nu_1 \dots \nu_{D-p}} = \frac{1}{p!} \varepsilon\indices{_{\nu_1 \dots \nu_{D-p}}^{\mu_1 \dots \mu_p}} \omega_{\mu_1 \dots \mu_p} .
\end{equation}
The Hodge dual squares to $\pm 1$ according to
\begin{equation}
(\star)^2 = (-1)^{p(D-p) + t}.
\end{equation}
Writing $\1$ for the function that is equal to one everywhere (a $0$-form), we get the invariant volume form (a $D$-form) as
\begin{align}
\star \1 &= \frac{1}{D!} \varepsilon_{\mu_1 \dots \mu_D} dx^{\mu_1} \wedge \dots \wedge dx^{\mu_D} \nonumber \\
&= \sqrt{|\det g|} \,d^D\!x,
\end{align}
where we defined $d^D\!x = dx^0 \wedge dx^1 \wedge \dots \wedge dx^{D-1}$.
Another useful identity is
\begin{equation}
\star \omega \wedge \eta = \star \eta \wedge \omega = \frac{1}{p!} \omega_{\mu_1 \dots \mu_p} \eta^{\mu_1 \dots \mu_p} \,\sqrt{|\det g|} \,d^D\!x,
\end{equation}
where $\omega$ and $\eta$ are both $p$-forms.

\section{Duality of \texorpdfstring{$p$}{p}-form fields}
\label{app:pforms}

This appendix is devoted to a brief introduction to the dynamics of $p$-form gauge fields and their duality properties. We follow the presentation of the book \cite{freedman_vanproeyen_2012}.

\subsection{Preliminaries}

The theory of a free $p$-form field $A\dwn{p}$ is invariant under the gauge symmetries
\begin{equation}
\delta A\dwn{p} = d \gL\dwn{p-1},
\end{equation}
where $\gL\dwn{p-1}$ is an arbitrary $(p-1)$-form. The invariant curvature for this gauge transformation is
\begin{equation}
F\dwn{p+1} = d A\dwn{p} .
\end{equation}
In addition to the Bianchi identity $dF\dwn{p+1} = 0$, which is a consequence of $d^2 = 0$, it satisfies the free equation of motion
\begin{equation}\label{eq:pformeom}
d \star F\dwn{p+1} = 0
\end{equation}
which follows from the variation of the action
\begin{equation}\label{eq:pformaction}
S_p = - \frac{1}{2} \int \star F\dwn{p+1} \wedge F \dwn{p+1} = - \frac{1}{2\cdot (p+1)!} \int \dx \sqrt{-g}\, F_{\mu_1 \dots \mu_{p+1}} F^{\mu_1 \dots \mu_{p+1}} .
\end{equation}
The number of degrees of freedom carried by such a field is
\begin{equation}\label{eq:dofpform}
\# \text{ d.o.f. } = \begin{pmatrix}
D-2 \\ p
\end{pmatrix},
\end{equation}
which is equal to the number of independent components of a rank $p$ antisymmetric tensor of the massless little group $SO(D-2)$, as it should. Note that this is zero for $p > D-2$.

\subsection{Duality between $p$-forms and $(D-p-2)$-forms}
\label{app:pformduality}

From the equation of motion \eqref{eq:pformeom}, the Poincaré lemma implies that there exists a $(D-p-2)$-form $\tilde{A}\dwn{D-p-2}$ such that
\begin{equation}\label{eq:D-p-2form}
\star F\dwn{p+1} = d \tilde{A}\dwn{D-p-2} .
\end{equation}
(The Hodge dual $\star$ takes a $q$-form to a $(D-q)$-form; therefore, $\star F$ is a $(D-p-1)$-form and $\tilde{A}$ must indeed be a $(D-p-2)$-form.) We ignore topological obstructions here.

The field $\tilde{A}$ satisfies all the equations of a free $(D-p-2)$-form. Indeed, equation \eqref{eq:D-p-2form} determines $\tilde{A}$ only up to an exterior derivative, which means that $\tilde{A}$ has the gauge transformation
\begin{equation}
\delta \tilde{A}\dwn{D-p-2} = d \tilde{\gL}\dwn{D-p-3}
\end{equation}
where $\tilde{\gL}$ is an arbitrary $(D-p-3)$-form. Moreover, by taking the Hodge dual of \eqref{eq:D-p-2form}, one gets $\star \tilde{F} = \pm F$, where
\begin{equation}
\tilde{F}\dwn{D-p-1} = d \tilde{A}\dwn{D-p-2}
\end{equation}
is the curvature of $\tilde{A}$. The Bianchi identity for $F$ then implies the usual equations of motion for $\tilde{F}$,
\begin{equation}
d \star \tilde{F}\dwn{D-p-1} = 0 .
\end{equation}
In other words, a $p$-form and a $(D-p-2)$-form are physically equivalent in $D$ dimensions. The counting of degrees of freedom also matches, since
\begin{equation}
\begin{pmatrix}
D-2 \\ p
\end{pmatrix} = \begin{pmatrix}
D-2 \\ D-p-2
\end{pmatrix} .
\end{equation}

This can be implemented a the level of the action \eqref{eq:pformaction} by going through the following procedure:
\begin{enumerate}
\item First, one introduces $\tilde{A}$ as a Lagrange multiplier for the Bianchi identity of $F$,
\begin{equation}\label{eq:pformaux}
S[F,\tilde{A}] = \int \left( - \frac{1}{2} \star F\dwn{p+1} \wedge F\dwn{p+1} + (-1)^{D-p-1} \tilde{A}\dwn{D-p-2} \wedge d F\dwn{p+1}  \right)
\end{equation}
(with a sign included for convenience). In this action, the fundamental variables are $F$ and $\tilde{A}$. The equations of motion are $\star F = d \tilde{A}$ (from the variation of $F$) and $dF = 0$ (from the variation of $\tilde{A}$). Those equations are equivalent to those coming from the action \eqref{eq:pformaction}: indeed, $dF = 0$ implies the existence of some $A$ with $F = dA$, and $\star F = d\tilde{A}$ then implies the equation of motion $d\star F = 0$.
\item We can then choose to keep either $F$ or $\tilde{A}$ in the action \eqref{eq:pformaux}, by solving the equations of motion coming from the variation of the other field and reinserting the solution back into \eqref{eq:pformaux}. Keeping $F$ (i.e., writing $F = dA$) removes the second term of \eqref{eq:pformaux}, giving back the original action \eqref{eq:pformaction}. On the other hand, keeping $\tilde{A}$ (i.e., writing $F = (-1)^{(p+1)(D-1)+1}\, \star \tilde{F}$) gives the action
\begin{equation}
S_{D-p-2} = - \frac{1}{2} \int \star \tilde{F}\dwn{D-p-1} \wedge \tilde{F}\dwn{D-p-1},
\end{equation}
which is indeed the action for a free $(D-p-2)$-form.
\end{enumerate}
This procedure works as long as the equations of motion of the original $p$-form can be written as $dG = 0$ for some $G$, which allows us to introduce $\tilde{A}$ as $G = d\tilde{A}$. This is the case, for example, for eleven-dimensional supergravity \eqref{eq:11lag} and the class of scalar-vector Lagrangians \eqref{eq:lag4dvectors} considered in chapter \ref{chap:emduality}. Most notably, this is \emph{not} the case for Yang-Mills theory.

\subsection{Self-dual forms}

In even dimensions, the curvature of a $(D/2-1)$-form has the same form degree as its Hodge dual. We can then wonder if the first-order equation of motion
\begin{equation}\label{eq:selfdualform}
F\dwn{D/2} = \star F\dwn{D/2}
\end{equation}
can be imposed. By taking the Hodge dual of this equation, one sees that it can be consistent if and only if the Hodge dual squares to the identity, $\star^2 = 1$. This is true for dimensions of the form $D = 4n +2$, i.e. $D = 2$, $6$, $10$, and so on.

When this is the case, equation \eqref{eq:selfdualform} is a consistent, first order equation of motion for a form field. It implies the second-order equation of motion \eqref{eq:pformeom}, as can be seen by taking an exterior derivative of \eqref{eq:selfdualform} and using the Bianchi identity, but is stronger. It propagates half the number of degrees of freedom as \eqref{eq:pformeom}.
An easy way to see this is to notice that any curvature can be written as
\begin{equation}
F = F^+ + F^-, \quad F^\pm = \frac{1}{2} \left( F \pm \star F \right), \quad \star F^\pm = \pm F^\pm.
\end{equation}
It can therefore always be split into a self-dual and anti-self-dual part. In terms of the fields $A$ and $\tilde{A}$, this corresponds to setting $A = \pm \tilde{A}$.

There is no quadratic, manifest Lorentz-invariant action principle that gives \eqref{eq:selfdualform} as equations of motion \cite{Marcus:1982yu}. Non-manifestly Lorentz-invariant \cite{Henneaux:1988gg,Deser:1997se,Bekaert:1998yp} or non-quadratic \cite{Pasti:1995ii,Pasti:1995tn,Pasti:1996vs} actions nevertheless exist (see part \ref{PART:6D}).

\section{Dimensional reduction on a circle}
\label{app:dimred}

We start by reviewing the necessary concepts and formulas of dimensional reduction on a circle. We follow the notations and presentation of \cite{Cremmer:1997ct} and of the lectures \cite{PopeKKLectures}. We keep only the massless modes in the reduction, i.e., all fields are independent of the $S^1$ coordinate $z$.

\subsection{Reduction of the metric}

A $(D+1)$-dimensional metric $\hat{g}$ splits into a metric $g_{\mu\nu}$, a vector field $A_\mu$ and a scalar field $\phi$ according to which of the indices are in the internal $z$ direction,
\begin{equation}
g_{\mu\nu} \sim \hat{g}_{\mu\nu}, \quad \cA_{\mu} \sim \hat{g}_{\mu z}, \quad \phi \sim \hat{g}_{zz} .
\end{equation}
More precisely, the usual Kaluza-Klein ansatz is to take the $(D+1)$-dimensional metric as
\begin{equation}\label{eq:KKansatz}
ds_{D+1}^2 = e^{2\alpha \phi} g_{\mu\nu} dx^\mu dx^\nu + e^{2 \beta \phi} (\cA_\mu dx^\mu + dz)^2,
\end{equation}
where the coefficients $\alpha$, $\beta$ are
\begin{equation}
\alpha^2 = \frac{1}{2(D-2)(D-1)}, \quad \beta = - \alpha (D-2).
\end{equation}
This choice of the constants gives canonically normalized kinetic terms for the Einstein-Hilbert and scalar Lagrangians on $D$ dimensions.
The reduced Lagrangian then takes the form
\begin{equation}
\cL\up{D} = R \star \1 - \frac{1}{2} \star d \phi \wedge d \phi - \frac{1}{2} e^{-2(D-1) \alpha \phi} \star \cF \wedge \cF ,
\end{equation}
with $\cF = d\cA$. It is invariant under the gauge transformations $\delta \cA =  d\lambda$, which comes from a reparametrization $\delta z = \lambda(x)$ of the $S^1$ coordinate. It is also invariant under a constant shift of the dilaton field accompanied by a rescaling of the vector field,
\begin{equation}
\phi \rightarrow \phi + c, \quad \cA \rightarrow e^{(D-1)\alpha\, c} \cA .
\end{equation}
As discussed in \cite{Cremmer:1997ct,PopeKKLectures}, this stems from the invariance of higher-dimensional equations of motion under a rescaling of the whole metric.

\subsection{Reduction of $p$-forms}

Upon dimensional reduction from $D+1$ to $D$ dimensions, a $p$-form field $\hat{A}\dwn{p}$ splits into two parts: a $p$-form $A\dwn{p}$ where all the indices are $D$-dimensional, and a $(p-1)$-form $B\dwn{p-1}$ where one index is in the extra $z$ dimension,
\begin{equation}
\hat{A}\dwn{p} = A\dwn{p} + B\dwn{p-1} \wedge dz . 
\end{equation}
(Because $\hat{A}_{\mu_1 \dots \mu_p}$ is an antisymmetric tensor, it cannot have two $z$ indices.)

The $(D+1)$-dimensional curvature $\hat{F}\dwn{p+1} = d \hat{A}\dwn{p}$ similarly splits in two parts,
\begin{align}
\hat{F}\dwn{p+1} &= d A\dwn{p} + d B\dwn{p-1} \wedge dz \\
&= dA\dwn{p} - dB\dwn{p-1} \wedge \cA + dB\dwn{p-1} \wedge \left( \cA + dz \right) \\
&\equiv F\dwn{p+1} + F\dwn{p} \wedge \left( \cA + dz \right), \label{eq:fieldstrengthreduc}
\end{align}
where it is convenient to make the vector field $\cA$ coming from the reduction \eqref{eq:KKansatz} of the metric appear. This defines the lower-dimensional field strengths
\begin{equation}
F\dwn{p+1} = dA\dwn{p} - dB\dwn{p-1} \wedge \cA\, , \quad F\dwn{p} = dB\dwn{p-1}\, ,
\end{equation}
which are invariant under the gauge transformations
\begin{equation}
\delta A\dwn{p} = d\Lambda\dwn{p-1} + B\dwn{p-1} \wedge d\lambda, \quad \delta B\dwn{p-1} = d\Lambda\dwn{p-2} .
\end{equation}
The second term in $\delta A\dwn{p}$ compensates the variation of $\cA$ which appears in the field strength.

With this notation, the reduction of the $p$-form Lagrangian \eqref{eq:pformaction} is simply
\begin{equation}
-\frac{1}{2} \hat{\star} \hat{F}\dwn{p+1} \wedge \hat{F} \dwn{p+1} = -\frac{1}{2} e^{-2p\alpha \phi} \star F\dwn{p+1} \wedge F\dwn{p+1} - \frac{1}{2} e^{2(D-p-1)\alpha \phi} \star F\dwn{p} \wedge F\dwn{p}.
\end{equation}
The dilaton shift symmetry $\phi \rightarrow \phi + c$ is also a symmetry of this Lagrangian, provided the $p$-forms scale as
\begin{equation}
A\dwn{p} \rightarrow e^{p \alpha\, c} A\dwn{p}, \quad B\dwn{p-1} \rightarrow e^{-(D-p-1)\alpha\, c} B\dwn{p-1} .
\end{equation}
Notice that the scalings of $B\dwn{p-1}$ and $\cA$ conspire in such a way that the complete field strength $F\dwn{p+1}$ scales in the same way as $A\dwn{p}$.

\subsection{Three dimensions and the Ehlers group}

When reducing gravity from four to three dimensions, one gets (as discussed above) one metric, one vector field and one scalar field, with Lagrangian
\begin{equation}\label{eq:3Dlag}
\cL\up{3} = R \star \1 - \frac{1}{2} \star d \phi \wedge d \phi - \frac{1}{2} e^{-2\phi} \star \cF \wedge \cF .
\end{equation}
The equation of motion for the vector field is $d\left( e^{-2\phi} \star \cF \right) = 0$.
In three dimensions, the vector field can therefore be dualized to a scalar field $\chi$ defined by
\begin{equation}\label{eq:defchi}
e^{-2\phi} \star \cF = d\chi,
\end{equation}
resulting in a theory with one metric and two scalar fields. This is done at the level of the Lagrangian by following the procedure of section \ref{app:pformduality}:
\begin{enumerate}
\item We add the scalar field $\chi$ as a Lagrange multiplier for the Bianchi identity $d\cF = 0$ of the vector field,
\begin{equation}\label{eq:3Dlagaux}
\cL\up{3} = R \star \1 - \frac{1}{2} \star d \phi \wedge d \phi - \frac{1}{2} e^{-2\phi} \star \cF \wedge \cF - \chi\, d\cF .
\end{equation}
The equation of motion obtained by varying $\cF$ is $e^{-2\phi} \star \cF = d\chi$, which can be solved algebraically for $\cF$ as
\begin{equation}
\cF = - e^{2\phi} \star d\chi
\end{equation}
using the property $\star^2 = - 1$ in this case.
\item Writing $\cF = - e^{2\phi} \star d\chi$ in \eqref{eq:3Dlagaux} and integrating by parts, we get the action
\begin{align}
\cL\up{3} &= R \star \1 - \frac{1}{2} \star d \phi \wedge d \phi - \frac{1}{2} e^{2\phi} \star d \chi \wedge d \chi \\
&= \sqrt{-g} \,d^3\!x \left( R - \frac{1}{2} \partial_\mu \phi\partial^\mu \phi - \frac{1}{2} e^{2\phi} \partial_\mu\chi\partial^\mu \chi \right) \label{eq:ehlerslag}
\end{align}
which is equivalent to \eqref{eq:3Dlag}.
\end{enumerate}
Remarkably, this action has an $SL(2,\R)$ global symmetry acting on the scalars. The easiest way to see this is to define the complex scalar field
\begin{equation}
\tau = \chi + i e^{-\phi},
\end{equation}
in terms of which we have
\begin{equation}\label{eq:taulag}
-\frac{1}{2} \partial_\mu \phi\partial^\mu \phi - \frac{1}{2} e^{2\phi} \partial_\mu\chi\partial^\mu \chi = -\frac{\partial_\mu \tau \,\partial^\mu \tau^*}{2 \Im(\tau)^2} .
\end{equation}
This Lagrangian is invariant under the fractional linear transformations
\begin{equation}\label{eq:tautransf}
\tau \rightarrow \frac{a \tau + b}{c \tau + d},
\end{equation}
where $a$, $b$, $c$, $d$ are real parameters satisfying $ad - bc = 1$. Indeed, from \eqref{eq:tautransf}, one gets
\begin{equation}
\partial_\mu \tau \rightarrow \frac{\partial_\mu \tau}{(c\tau + d)^2}, \quad \partial_\mu \tau^* \rightarrow \frac{\partial_\mu \tau^*}{(c\tau^* + d)^2}, \quad \Im(\tau) \rightarrow \frac{\Im(\tau)}{(c\tau + d)(c \tau^* + d)},
\end{equation}
from which the invariance of \eqref{eq:taulag} follows.
This group of transformation is isomorphic to $SL(2,\R)$: to the transformation \eqref{eq:tautransf}, we can associate the real matrix of determinant one
\begin{equation}
\gL = \begin{pmatrix}
a & b \\ c & d
\end{pmatrix} \in SL(2,\R),
\end{equation}
and the composition of transformations \eqref{eq:tautransf} corresponds to multiplication of those matrices.
The transformations \eqref{eq:tautransf} can be written in terms of $\phi$ and $\chi$: they are
\begin{align}
e^{\phi'} &= (c \chi + d)^2 e^\phi + c^2 e^{-\phi} \label{eq:transfphi} \\
\chi' e^{\phi'} &= (c \chi + d)(a \chi + b) e^\phi + ac\, e^{-\phi} . \label{eq:transfchi}
\end{align}
Infinitesimally, the three generators of $SL(2,\R)$ act as follows:
\begin{equation} \label{eq:sl2infscalars}
\begin{array}{lll}
h: \qquad & \delta \phi = 2 \varepsilon, & \delta \chi = - 2 \varepsilon \chi, \\
e: \qquad & \delta \phi = 0, & \delta \chi = \varepsilon, \\
f: \qquad & \delta \phi = 2 \varepsilon \chi, & \delta \chi = \varepsilon ( e^{-2\phi} - \chi^2) .
\end{array}
\end{equation}
The first two symmetries are easily understood. The $h$ is the dilaton shift symmetry that is always present in the 
$S^1$ reduction, as discussed above. The $e$ is a constant shift of $\chi$: it must be present because $\chi$ is defined only through its derivative, see equation \eqref{eq:defchi}. These two symmetries generate the standard Borel subgroup $B^+$ of $SL(2,\R)$.

The third ($f$) symmetry, however, is unexpected and enhances the symmetry group from $B^+$ to the full $\SL$. It is often referred to as the ``hidden symmetry" of Einstein gravity and was first discovered by Ehlers \cite{Ehlers}. Its appearance is analogous to the appearance of the exceptional $E$-series groups in maximal supergravity.

\section{Simple Lie algebras}
\label{app:liealg}

We collect here some important definitions and formulas of the theory of simple Lie algebras. Useful references include \cite{fuchs2003symmetries,fulton1991representation,Henneaux:2007ej}.

A finite-dimensional, complex, simple Lie algebra of rank $r$ is completely specified by its Cartan matrix $A$ as the Lie algebra generated by the $3r$ generators
\begin{equation}
\{\, h_i, e_i, f_i \;|\; i = 1, 2, \dots, r \,\}
\end{equation}
subject to the Chevalley-Serre relations
\begin{align}\label{eq:chevalleyserre}
[h_i,h_j] &= 0 \\
[e_i,f_j] &= \delta_{ij}h_i \\
[h_i,e_j] &= A_{ij} e_j &\qquad [h_i,f_j] &= -A_{ij}f_j \\
\ad_{e_i}^{1-A_{ij}}(e_j) &=0 &\qquad \ad_{f_i}^{1-A_{ij}}(f_j) &= 0
\end{align}
(no sum over repeated indices).
For a simple, finite-dimensional Lie algebra, the Cartan matrix is a square $r\times r$ matrix that satisfies the following five properties:
\begin{enumerate}[a)]
\item $A_{ii}=2$ (no sum on $i$),
\item $A_{ij} = 0 \iff A_{ji} = 0$,
\item $A_{ij} \in \Z_{\leq 0}$ for $i\neq j$,
\item $\det A > 0$,
\item $A$ cannot be put in block-diagonal form by reordering the lines and columns.
\end{enumerate}
Conditions a) to c) should hold for any $i, j = 1, \dots r$. In property c), $\Z_{\leq 0}$ is the set of non-positive integers. Property d) ensures the finite-dimensionality of the algebra; it can be dropped to define general Kac-Moody algebras (see for example \cite{Henneaux:2007ej} for a review). If property e) is dropped, one simply gets a semi-simple Lie algebra, whose simple factors have Cartan matrices given by the diagonal blocks.

A Cartan matrix can be neatly encoded in a Dynkin diagram, defined as follows:
\begin{enumerate}[a)]
\item There is one node in the diagram for each $i= 1, \dots, r$,
\item The nodes $i$ and $j$ are connected by $\max\left( |A_{ij}|, |A_{ji}| \right)$ lines,
\item There is an arrow from $i$ to $j$ if $|A_{ji}| > |A_{ij}|$.
\end{enumerate}
Note that the last property is convention-dependent; it is switched if one uses $A^T$ instead of $A$ as Cartan matrix. (We follow \cite{Henneaux:2007ej}; the books \cite{fuchs2003symmetries,fulton1991representation} have the opposite convention.) What is independent of convention, however, is the fact that the arrow points from the long to the short root, as we will see below.

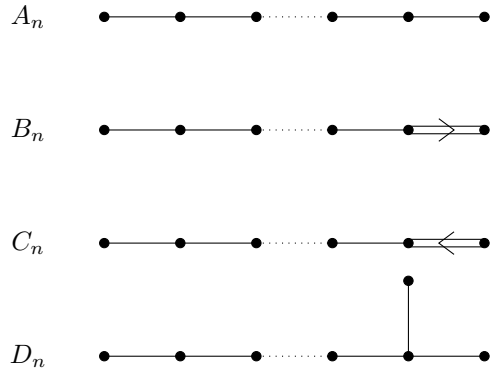
\begin{figure}
\centering
\begin{tikzpicture}
\tikzset{v/.style={circle,fill,inner sep=0pt,minimum size=3.5pt,draw}}

\node at (-1,0) {$A_n$};
\draw (0,0) node[v]{} -- (1,0) node[v]{} -- (2,0) node[v]{};
\draw[dotted] (2,0) -- (3,0);
\draw (3,0) node[v]{} -- (4,0) node[v]{} -- (5,0) node[v]{};

\begin{scope}[shift={(0,-1.5)}]
\node at (-1,0) {$B_n$};
\draw (0,0) node[v]{} -- (1,0) node[v]{} -- (2,0) node[v]{};
\draw[dotted] (2,0) -- (3,0);
\draw (3,0) node[v]{} -- (4,0) node[v]{};
\begin{scope}[shift={(4,0)}]
\draw (0, 0.05) -- (1, 0.05);
\draw (0, -0.05) -- (1, -0.05);
\draw (0.4, 0.15) -- (0.6, 0) -- (0.4, -0.15);
\end{scope}
\draw (5,0) node[v]{};
\end{scope}

\begin{scope}[shift={(0,-3)}]
\node at (-1,0) {$C_n$};
\draw (0,0) node[v]{} -- (1,0) node[v]{} -- (2,0) node[v]{};
\draw[dotted] (2,0) -- (3,0);
\draw (3,0) node[v]{} -- (4,0) node[v]{};
\draw (5,0) node[v]{};
\begin{scope}[shift={(5,0)},scale=-1]
\draw (0, 0.05) -- (1, 0.05);
\draw (0, -0.05) -- (1, -0.05);
\draw (0.4, 0.15) -- (0.6, 0) -- (0.4, -0.15);
\end{scope}
\end{scope}

\begin{scope}[shift={(0,-4.5)}]
\node at (-1,0) {$D_n$};
\draw (0,0) node[v]{} -- (1,0) node[v]{} -- (2,0) node[v]{};
\draw[dotted] (2,0) -- (3,0);
\draw (3,0) node[v]{} -- (4,0) node[v]{} -- (5,0) node[v]{};
\draw (4,0) -- (4,1) node[v]{};
\end{scope}

\end{tikzpicture}
\caption[Dynkin diagrams of the simple Lie algebras $A_n$, $B_n$, $C_n$ and $D_n$]{The four infinite families of finite-dimensional complex simple Lie algebras. Each diagram contains $n$ nodes. They are the algebras of the classical matrix groups over the complex numbers: $A_n \simeq \mathfrak{sl}(n,\C)$, $B_n \simeq \mathfrak{so}(2n+1,\C)$, $C_n \simeq \mathfrak{sp}(2n, \C)$ and $D_n \simeq \mathfrak{so}(2n,\C)$.}
\label{fig:cartaninfinite}
\end{figure}

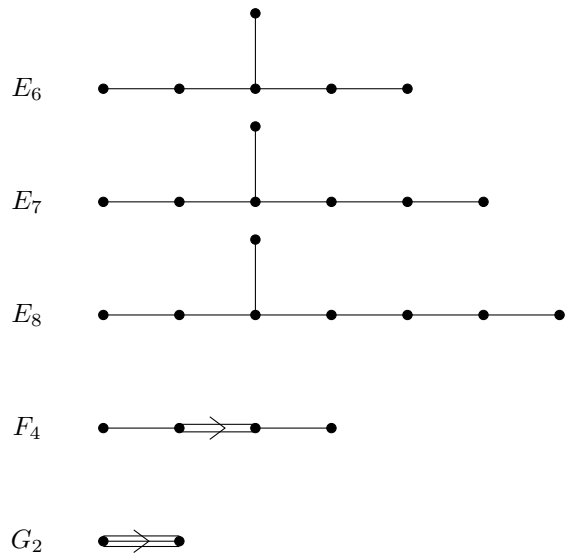
\begin{figure}
\centering
\begin{tikzpicture}
\tikzset{v/.style={circle,fill,inner sep=0pt,minimum size=3.5pt,draw}}

\node at (-1,0) {$E_{6}$};
\draw (0,0) node[v]{} -- (1,0) node[v]{} -- (2,0) node[v]{} -- (3,0) node[v]{} -- (4,0) node[v]{};
\draw (2,0) -- (2,1) node[v]{};

\begin{scope}[shift={(0,-1.5)}]
\node at (-1,0) {$E_{7}$};
\draw (0,0) node[v]{} -- (1,0) node[v]{} -- (2,0) node[v]{} -- (3,0) node[v]{} -- (4,0) node[v]{} -- (5,0) node[v]{};
\draw (2,0) -- (2,1) node[v]{};
\end{scope}

\begin{scope}[shift={(0,-3)}]
\node at (-1,0) {$E_{8}$};
\draw (0,0) node[v]{} -- (1,0) node[v]{} -- (2,0) node[v]{} -- (3,0) node[v]{} -- (4,0) node[v]{} -- (5,0) node[v]{} -- (6,0) node[v]{};
\draw (2,0) -- (2,1) node[v]{};
\end{scope}

\begin{scope}[shift={(0,-4.5)}]
\node at (-1,0) {$F_{4}$};
\draw (0,0) node[v]{} -- (1,0) node[v]{};
\draw (2,0) node[v]{} -- (3,0) node[v]{};
\begin{scope}[shift={(1,0)}]
\draw (0, 0.05) -- (1, 0.05);
\draw (0, -0.05) -- (1, -0.05);
\draw (0.4, 0.15) -- (0.6, 0) -- (0.4, -0.15);
\end{scope}
\end{scope}

\begin{scope}[shift={(0,-6)}]
\node at (-1,0) {$G_{2}$};
\draw (0,0) node[v]{} -- (1,0) node[v]{};
\begin{scope}[shift={(0,0)}]
\draw (0, 0.07) -- (1, 0.07);
\draw (0, -0.07) -- (1, -0.07);
\draw (0.4, 0.15) -- (0.6, 0) -- (0.4, -0.15);
\end{scope}
\end{scope}
\end{tikzpicture}
\caption[Dynkin diagrams of the exceptional simple Lie algebras $E_6$, $E_7$, $E_8$, $F_4$ and $G_2$]{The Dynkin diagrams of the five exceptional simple Lie algebras $E_6$, $E_7$, $E_8$, $F_4$ and $G_2$.}
\label{fig:cartanexceptional}
\end{figure}

\subsection*{Important subalgebras}

One can split those algebras as
\begin{equation}\label{eq:triangulardecomposition}
\mathfrak{g} = \mathfrak{n}_- \oplus \mathfrak{h} \oplus \mathfrak{n}_+,
\end{equation}
where $\mathfrak{n}_-$ and $\mathfrak{n}_+$ are the subalgebras spanned by the nested commutators $[ f_{i_1}, [f_{i_2}, \dots [f_{i_{k-1}},f_{i_k} ]\dots]]$, and $[ e_{i_1}, [e_{i_2}, \dots [e_{i_{k-1}},e_{i_k} ]\dots]]$ respectively, and $\mathfrak{h}$ is the subalgebra spanned by the $h_i$. The subalgebra $\mathfrak{h}$ is the Cartan subalgebra of $\mathfrak{g}$; it is abelian and of dimension $r$. Note that the sum in the triangular decomposition \eqref{eq:triangulardecomposition} is a direct sum of vector spaces, not of Lie algebras, since the factors do not commute.

The Borel subalgebras $\mathfrak{b}_+$ and $\mathfrak{b}_-$ are defined by
\begin{equation}
\mathfrak{b}_\pm = \mathfrak{h} \oplus \mathfrak{n}_\pm .
\end{equation}

A parabolic subalgebra of $\mathfrak{g}$ is any subalgebra that contains the positive Borel subalgebra $\mathfrak{b}_+$. They are in one-to-one correspondence with the subsets of nodes in the Dynkin diagram: namely, for each $\Sigma \subset \{\, 1, 2,  \dots, r \,\}$ we have the parabolic subalgebra
\begin{equation}
\mathfrak{p}(\Sigma) = \braket{f_i \,|\, i \in \Sigma} \oplus \mathfrak{b}_+,
\end{equation}
where by $\braket{f_i \,|\, i \in \Sigma}$ we mean the subalgebra of $\mathfrak{n}_-$ spanned by the $f_i$ with $i \in \Sigma$ and their nested commmutators.

The Chevalley involution $\tau$ exchanges the positive and negative algebras; it is defined by
\begin{equation}
\tau(h_i) = - h_i \qquad \tau(e_i) = -f_i \qquad \tau(f_i) = - e_i.
\end{equation}
It is an Lie algebra automorphism of $\mathfrak{g}$.
The set of elements fixed under $\tau$ is the maximal compact subalgebra $\mathfrak{k}$ of $\mathfrak{g}$,
\begin{equation}
\mathfrak{k} = \{\, x \in \mathfrak{g} \,|\, \tau(x) = x \,\} .
\end{equation}
It is the subalgebra spanned by the $k_i = e_i - f_i$ and their commutators. The generalized transpose $\#$ is the opposite of the Chevalley involution,
\begin{equation}
x^\# = - \tau(x) ,
\end{equation}
so that the elements of $\mathfrak{k}$ are the ``antisymmetric" elements, $x \in \mathfrak{k} \Leftrightarrow x^\# = - x$. Since $\tau$ is an involution, the Lie algebra $\mathfrak{g}$ splits as the direct sum of vector spaces
\begin{equation}
\mathfrak{g} = \mathfrak{k} \oplus \mathfrak{p} ,
\end{equation}
where $\mathfrak{p} = \{\, x \in \mathfrak{g} \,|\, \tau(x) = -x \,\} = \{\, x \in \mathfrak{g} \,|\, x^\# = x \,\}$ (not to be confused with the parabolic subalgebras introduced above). The commutator structure is then
\begin{equation}
[\mathfrak{k}, \mathfrak{k}] \subset \mathfrak{k}, \quad [\mathfrak{k}, \mathfrak{p}] \subset \mathfrak{p}, \quad
[\mathfrak{p}, \mathfrak{p}] \subset \mathfrak{k}
\end{equation}
(note that $\mathfrak{p}$ is \emph{not} a subalgebra of $\mathfrak{g}$).

\subsection*{Roots}

The action of $\mathfrak{h}$ on $\mathfrak{n}_\pm$ is diagonal. The roots of $\mathfrak{g}$ are defined as the eigenvalues of this action, i.e. the linear forms $\alpha$ defined on $\mathfrak{h}$ for which the equation $[h, x] = \alpha(h) x$ has non-zero solutions $x \in \mathfrak{g}$ for all $h \in \mathfrak{h}$. For example, one has from \eqref{eq:chevalleyserre}
\begin{equation}
[h, e_i] = \alpha_i(h) e_i
\end{equation}
for all $h \in \mathfrak{h}$ and every generator $e_i$, where
$\alpha_i$ is the linear form defined on $\mathfrak{h}$ by
\begin{equation}
\alpha_i(h_j) = A_{ji}
\end{equation}
and extend by linearity to the whole of $\mathfrak{h}$. The $\alpha_i$ are called the simple roots of the Lie algebra $\mathfrak{g}$. On nested commutators, we have
\begin{equation}
[h, \,[ e_{i_1}, \dots [e_{i_{k-1}},e_{i_k} ]\dots]] = (\alpha_{i_1} + \dots + \alpha_{i_k})(h)\, [ e_{i_1}, \dots [e_{i_{k-1}},e_{i_k} ]\dots]\, .
\end{equation}
Therefore, $\alpha_{i_1} + \dots + \alpha_{i_k}$ is a root as long as $[ e_{i_1}, \dots [e_{i_{k-1}},e_{i_k} ]\dots]]$ is non-zero. Then, its opposite is also a root,
\begin{equation}
[h, \,[ f_{i_1}, \dots [f_{i_{k-1}},f_{i_k} ]\dots]] = -(\alpha_{i_1} + \dots + \alpha_{i_k})(h)\, [ f_{i_1}, \dots [f_{i_{k-1}},f_{i_k} ]\dots]\, .
\end{equation}
In particular, the $-\alpha_i$ are always roots. Moreover, roots are always either positive, i.e. linear combinations of the simple roots $\alpha_i$ with non-negative coefficients, or negative (linear combinations of the $\alpha_i$ with non-positive coefficients).

Since $\mathfrak{g}$ is simple, its Killing form is non-degenerate. It therefore induces a non-degenerate (in fact, Euclidean) scalar product $(\cdot\,|\,\cdot)$ on the root space. In terms of this scalar product, the Cartan matrix is expressed with the simple roots as
\begin{equation}\label{eq:Ascalarproduct}
A_{ij} = 2 \frac{(\alpha_i|\alpha_j)}{(\alpha_i|\alpha_i)} .
\end{equation}
Therefore, rule c) in the definition of the Dynkin diagram becomes
\begin{enumerate}
\item[c')] There is an arrow from $i$ to $j$ if $(\alpha_i|\alpha_i) > (\alpha_j|\alpha_j)$
\end{enumerate}
so that the arrow points from long to short roots.
This scalar product is uniquely determined from equation \eqref{eq:Ascalarproduct} and a choice of normalization (usually $(\alpha_i|\alpha_i) = 2$) for the longest simple roots.

\section{Scalar coset Lagrangians}
\label{app:cosetlagrangians}

The scalar fields appearing in maximal supergravity parametrize homogeneous spaces of the form $G/K$, where $G$ is a semi-simple Lie group and $K$ its maximal compact subgroup (see table \ref{table:scalarcosets}). We review their construction here, following the presentation of references \cite{PopeKKLectures,Henneaux:2007ej} to which we refer for further details.

\subsection{Non-linear sigma models}

A scalar sigma model describes scalar fields as a function
\begin{equation}
\phi : M \rightarrow \Sigma
\end{equation}
from a manifold $M$ endowed with a metric $\gamma$, to some \emph{target space} $\Sigma$ with metric $g$. The dynamics of the scalar fields is specified by the action
\begin{equation}\label{eq:sigmamodel}
S[\phi] = \int_M \dx \,\sqrt{|\det \gamma|} \gamma^{\mu\nu} \,\pd_\mu \phi^i \pd_\nu \phi^j \,g_{ij}(\phi)  \qquad (D = \dim M).
\end{equation}
There are as many scalar fields $\phi^i$ as the dimension of $\Sigma$. If $g_{ij}(\phi) = \delta_{ij}$ for all $\phi$, this is just $\dim(\Sigma)$ free scalar fields on $M$; however, in general $\Sigma$ is a non-trivial curved manifold and its metric $g_{ij}(\phi)$ induces couplings between the $\phi^i$.

Note also that isometries of the target space $\Sigma$ induce global symmetries of the action \eqref{eq:sigmamodel}. Indeed, if $\xi^i$ is a Killing vector field of the metric $g_{ij}$, i.e.
\begin{equation}
\xi^k \pd_k g_{ij} + g_{ik} \,\pd_j \xi^k + g_{jk} \,\pd_i \xi^k = 0, \qquad \pd_i \equiv \frac{\pd}{\pd \phi^i}\, ,
\end{equation}
then the infinitesimal variation $\delta \phi^i = \xi^i(\phi)$ of the scalar fields preserves \eqref{eq:sigmamodel}.

\subsection{Lagrangian and symmetries}

In the cases relevant for extended supergravities, $M$ is space-time, $\Sigma$ is the coset space $G/K$, and $g_{ij}(\phi)$ is the metric induced on $G/K$ by the Killing metric of $G$.

More explicitely, we can parametrize the (left) coset space
\begin{equation}
G/K = \{ [\cV] = K\, \cV \,|\, \cV \in G \}
\end{equation}
by a $x$-dependent element $\cV$ of $G$. However, two elements $\cV$ and $\cV' = k \cV$ of $G$ that differ by left-multiplication by an element $k$ of the denominator correspond to the same coset, $[\cV] = [\cV']$. This is implemented by the requirement that the transformation
\begin{equation}\label{eq:cosetgauge}
\cV \rightarrow k \cV \qquad (k \in K),
\end{equation}
for any (possibly $x$-dependent) element $k$ of $K$, is a gauge symmetry of the action. Moreover, we have the natural action $[\cV] \rightarrow [\cV g]$ of $G$ on $G/K$ by right-multiplication, which on the coset representative $\cV$ reads
\begin{equation}
\cV \rightarrow \cV g \qquad (g \in G) .
\end{equation}
This will be a global symmetry of the action. All in all, the $G/K$ coset Lagrangian has the symmetry
\begin{equation}
K_\text{local} \times G_\text{global},
\end{equation}
which is indeed a direct product since left and right multiplication commute.

To construct the action, we first consider the Maurer-Cartan one-from $d\cV \cV^{-1}$, which is a one-form on $M$ with values in the Lie algebra $\mathfrak{g}$ of $G$. We then use the fact that $K$ is in our case the maximal compact subgroup of $G$, i.e., the exponential of the maximal compact subalgebra $\mathfrak{k}$ of $\mathfrak{g}$. As is recalled in appendix \ref{app:liealg}, $\mathfrak{k}$ is the set of elements $x \in \mathfrak{g}$ that satisfy $x^\# = - x$, where $\#$ is the generalized transpose. This generalized transpose allows to split $d\cV \cV^{-1}$ as
\begin{equation}
d\cV \cV^{-1} = P + Q,
\end{equation}
with
\begin{equation}
P = \frac{1}{2} \left( d\cV \cV^{-1} + (d\cV \cV^{-1})^\# \right), \qquad Q = \frac{1}{2} \left( d\cV \cV^{-1} - (d\cV \cV^{-1})^\# \right).
\end{equation}
The action for the $G/H$ coset Lagrangian is then simply
\begin{equation}\label{eq:cosetlagrangianP}
S = - \int_M \dx \,\sqrt{|\det \gamma|} \gamma^{\mu\nu} \left( P_\mu(x) | P_\nu(x) \right),
\end{equation}
where $(\cdot |\cdot)$ is the Killing metric on the Lie algebra $\mathfrak{g}$.

It has the global $G$ symmetry since $d\cV \cV^{-1}$, and hence $P$, is invariant under $\cV \rightarrow \cV g$. A short computation also shows that it is invariant under local $K$ transformations \cite{Henneaux:2007ej}. Intuitively, this is because the $Q$-component of $d\cV \cV^{-1}$, which is ``along the gauge direction", does not appear in the action.

\subsection{Borel gauge and the Iwasawa decomposition}

We now introduce an explicit parametrization of the coset space $G/K$ by $n = \dim G - \dim K$ scalar fields. To do so, we take the coset representative in the form
\begin{equation}\label{eq:cosetBorel}
\cV(x) = g_H \, g_N, \quad g_H = \exp\left( \frac{1}{2} \sum_{i=1}^r \phi^i(x)\, h_i \right), \quad g_N = \exp\left( \sum_{\alpha > 0} \chi^\alpha(x)\, e_\alpha \right) .
\end{equation}
Here, $g_H$ is in the exponential of the Cartan subalgebra $\mathfrak{h}$ and defines the $r = \text{rank}(G)$ scalar fields $\phi^i$, which are the dilatons in the supergravity context. The second factor $g_N$ is the exponential of a general element of the upper triangular subalgebra $\mathfrak{n}_+$: the sum runs over all positive roots $\alpha$, with $e_\alpha$ the corresponding raising operator (such that $[h, e_\alpha] = \alpha(h) e_\alpha$ for all $h \in \mathfrak{h}$). It defines the scalar fields $\chi^\alpha$, which are the axion fields of supergravity. This choice of representative is called the \emph{Borel gauge}, since $\cV$ in then an element of the positive Borel subgroup $B^+$ of $G$.

The number of scalar fields is correct: indeed, if $n_+$ ($= n_-$) denotes the number of positive (negative) roots of $\mathfrak{g}$, we have the formulas $\dim(G) = r + n_+ + n_- = r + 2 n_+$ and $\dim(K) = n_+$, from which
\begin{align}
\text{number of scalar fields} = r + n_+ = (r + 2 n_+) - n_+ = \dim(G) - \dim(K)
\end{align}
follows. To show that we can always choose to go to the Borel gauge, we use the Iwasawa decomposition, which asserts that any element $g \in G$ can be written as the product
\begin{equation}
g = g_K \, g_H \, g_N,
\end{equation}
where $g_K$, $g_H$, $g_N$ are elements of the compact, Cartan and upper triangular subgroups respectively. Therefore, writing $\cV$ in this form and doing a gauge transformation \eqref{eq:cosetgauge} with parameter $k = g_K^{-1}$ brings us to the Borel gauge \eqref{eq:cosetBorel}. (This decomposition is valid when the group $G$ is in its maximally non-compact (or \emph{split}) form, which is always the case here. See \cite{PopeKKLectures} for details on the general case.)

This completly gauge-fixed parametrization is convenient, since it contains exactly the physical number of fields, instead of having $\dim(K)$ extra ``pure gauge" fields. However, it introduces a subtlety for the global $G$ symmetry: the $G$-transformed coset representative $\cV g$ is in general not in the Borel gauge anymore. However, since $\cV g$ is still an element of $G$, the Iwasawa decomposition again assures us that a compensating gauge transformation $k \in K$ (which in general depends on $\cV$ and $g$) exists such that $\cV$ is again in the Borel gauge. The global $G$ symmetry is thus
\begin{equation}
\cV \,\rightarrow \, k(\cV, g) \,\cV\, g, \qquad g \in G, \quad k(\cV, g) \in K,
\end{equation}
where the ``compensator" $k(\cV, g)$ is uniquely determined from the fact that the right-hand side is in the Borel gauge.

There is a way to forget about this compensator, which is to introduce the group element $\cM$ defined from $\cV$ by
\begin{equation}
\cM = \cV^\# \cV .
\end{equation}
Since the elements of $K$ satisfy $k^\# k = I$ by definition, the compensator drops out of the transformation law and the global symmetry $G$ acts simply as
\begin{equation}
\cM \rightarrow g^\# \cM g .
\end{equation}
In terms of this matrix $\cM$, the Lagrangian \eqref{eq:cosetlagrangianP} becomes
\begin{equation}\label{eq:cosetlagrangianM}
S = - \frac{1}{4} \int_M \dx \,\sqrt{|\det \gamma|} \gamma^{\mu\nu} \,(\mathcal{M}^{-1} \partial_\mu \,\mathcal{M}\,|\,\mathcal{M}^{-1} \partial_\nu \mathcal{M}).
\end{equation}
We finish this section with a few comments on notation:
\begin{itemize}
\item We have an extra minus sign in \eqref{eq:cosetlagrangianP} and \eqref{eq:cosetlagrangianM} with respect to \cite{Henneaux:2007ej}. This gives the usual sign of the scalar kinetic terms in our applications where $\gamma$ is the space-time metric.
\item It is often convenient to split the upper-triangular element $g_N$ in \eqref{eq:cosetBorel} as a product of several exponentials, see for example the coset representatives \eqref{eq:G2coset} or \eqref{eq:N8coset}. This is a harmless field redefinition of the axions $\chi^\alpha$.
\item In a suitable matrix representation of $G$ in which the Killing metric is simply given by the trace, one may write
\begin{equation}
-(\mathcal{M}^{-1} \partial_\mu \,\mathcal{M}\,|\,\mathcal{M}^{-1} \partial_\nu \mathcal{M}) = - \Tr(\mathcal{M}^{-1} (\partial_\mu \,\mathcal{M}\,) \mathcal{M}^{-1} \partial_\nu \mathcal{M}) = \Tr(\partial_\mu \cM^{-1}\, \partial_\nu \mathcal{M}),
\end{equation}
which is the form of the action \eqref{eq:cosetlagrangianM} used in \cite{Cremmer:1997ct,PopeKKLectures}.
\end{itemize}

\subsection{Example: $SL(2,\R)/SO(2)$}
\label{app:SL2SO2}

A basis of the $\mathfrak{sl}(2,\R)$ algebra is given by the matrices
\begin{equation}
h=\begin{pmatrix} 1 & 0 \\ 0 & -1 \end{pmatrix},\qquad e=\begin{pmatrix} 0 & 1 \\ 0 & 0 \end{pmatrix},\qquad f=\begin{pmatrix} 0 & 0 \\ 1 & 0 \end{pmatrix},
\end{equation}
with commutation relations
\begin{equation}
[h,e]=2e, \qquad [h,f]=-2f, \qquad [e,f]=h .
\end{equation}
In this basis, the Killing form $(x|y)=\tr(\ad_x \circ \ad_y)=\tr(xy)$ is given by
\begin{equation}
(h|h)=2, \qquad (e|f)=1, \qquad \text{others}=0.
\end{equation}
It is invariant under the adjoint action, i.e. $(S x S^{-1} | S y S^{-1})=(x|y)$ for all $S \in SL(2,\R)$ or, equivalently, $([s,x]|y) + (x|[s,y]) = 0$ for all $s \in \mathfrak{sl}(2,\R)$.
The generalized transpose $\#$ is just the usual matrix transpose. The maximal compact subalgebra $\mathfrak{so}(2)$ is generated by
\begin{equation}
k = e - f = \begin{pmatrix}
0 & 1 \\ -1 & 0
\end{pmatrix} .
\end{equation}

We take the coset representative $\mathcal V$ in the Borel gauge,
\begin{equation}
\mathcal{V} =  e^{\phi h /2} e^{\chi e} = \begin{pmatrix} e^{\phi/2} & \chi e^{\phi/2} \\ 0 & e^{-\phi/2} \end{pmatrix} .
\end{equation}
The Maurer-Cartan one-form is then
\begin{equation}
d \cV \cV^{-1} = \frac{1}{2} d\phi \,h + e^{\phi} d\chi \,e.
\end{equation}
This can be computed either directly using the matrix representation above, or from the exponential form of $\cV$ using the BCH-like formulae
\begin{align}
de^{A} \, e^{-A} &= dA + \frac{1}{2} [A, dA] + \frac{1}{3!} [A, [A, dA]] + \dots \\
e^A B e^{-A} &= B + [A, B] + \frac{1}{2} [A, [A, B]] + \dots = e^{\ad_A} B .
\end{align}
Therefore, $P_\mu = \frac{1}{2} \pd_\mu \phi \, h + \frac{1}{2} e^\phi \pd_\mu \chi \, (e + f)$ and the Lagrangian is
\begin{equation}
\cL_S = (P_\mu | P^\mu) = -\frac{1}{2} \partial_\mu \phi \, \partial^\mu \phi - \frac{1}{2} e^{2\phi} \partial_\mu \chi \, \partial^\mu \chi .
\end{equation}
This is exactly the scalar Lagrangian \eqref{eq:ehlerslag} appearing in the dimensional reduction of Einstein gravity down to three dimensions. This explains nicely why it has a $SL(2,\R)$ symmetry. Equivalently, we can define the matrix
\begin{equation}\label{eq:cosetSL2SO2M}
\mathcal{M} = \mathcal{V}^t \mathcal{V} = \begin{pmatrix} e^\phi & \chi e^\phi \\ \chi e^\phi & e^{-\phi} + \chi^2 e^\phi \end{pmatrix} .
\end{equation}
The Lagrangian is then given by $\cL_S = \Tr(\pd_\mu \cM^{-1} \pd^\mu \cM )/4$.

The global $SL(2,\R)$ symmetry acts on the scalar fields as
\begin{equation}
\mathcal{V} \rightarrow R \mathcal{V} S,
\end{equation}
where $S$ is in $SL(2,\R)$ and $R \in SO(2)$ is the local compensator required to keep $\mathcal{V}$ in the Borel gauge (its explicit form can be found in \cite{PopeKKLectures,Henneaux:2007ej}).
Infinitesimally, this is
\begin{equation}
\delta \mathcal{V} = r \mathcal{V} + \mathcal{V} s .
\end{equation}
On the matrix $\mathcal M$, the symmetry is
\begin{equation}
\mathcal{M} \rightarrow S^t \mathcal{M} S, \qquad \delta \mathcal{M} = s^t \mathcal{M} + \mathcal{M} s .
\end{equation}
Writing
\begin{equation}
S = \begin{pmatrix}
d & b \\ c & a
\end{pmatrix}, \qquad (ad-bc = 1),
\end{equation}
the transformations of $\phi$ and $\chi$ are exactly the transformations \eqref{eq:transfphi} and \eqref{eq:transfchi}.
Notice, however, that $S \neq \gL$: the matrix $\Lambda$ there is related to $S$ as $S = h \Lambda^{-1} h$.

\section{More on \texorpdfstring{$\cN = 8$}{N = 8} supergravity and \texorpdfstring{$E_{7(7)}$}{E7(7)}}
\label{app:N8E7}

In this appendix, we review some explicit formulas and properties of $\cN = 8$ supergravity and its $E_{7(7)}$ symmetry. It is based on \cite{Lu:1995yn,Cremmer:1997ct,Cremmer:1979up,Compere:2015roa}.

\subsection{Field content and reduction ansatz}
\label{app:fieldsN=8}

We follow the notations of \cite{Lu:1995yn,Cremmer:1997ct}. The fields of $D=4$, $\cN = 8$ supergravity are defined through the ansatze
\begin{align}
ds^2_{11} &= e^{\frac{1}{3} \vec{g} \cdot \vec{\phi}} ds_4^2 + \sum_{i=1}^{7} e^{2\vec{\gamma}_i \cdot \vec{\phi}} (\omega^i)^2, \label{eq:ansatzmetric}\\
\hat{A}\dwn{3} &= A_{(3)} + A_{(2)i} \wedge dz^i - \frac{1}{2} A_{(1)ij} \wedge dz^i \wedge dz^j - \frac{1}{6} A_{(0)ijk} \,dz^i \wedge dz^j \wedge dz^k \label{eq:ansatz3form}
\end{align}
for the eleven-dimensional metric and three-form, where all indices $i, j, \dots$ run from one to seven and correspond to the internal dimensions.

The vector $\vec{\phi}$ contains the seven dilatons $\phi_i$. The $7$-component vectors appearing in the reduction of the metric are
\begin{equation}\label{eq:defgg}
\vec{g} = 3 (s_1, s_2, \dotsc, s_7) ,\quad \vec{\gamma}_i = \frac{1}{6} \vec{g} - \frac{1}{2} \vec{f}_i
\end{equation}
with
\begin{equation}\label{eq:deffs}
\vec{f}_i = (\,\underbrace{0, \,\dotsc, \,0}_{i-1}\, ,\, (10-i) s_i, s_{i+1}, \dotsc, s_7) ,\quad s_i = \sqrt{\frac{2}{(10-i)(9-i)}}.
\end{equation}
The one-forms $\omega^i$ appearing in the metric are
\begin{equation}
\omega^i = dz^i + \cA\dwn{1}^i + \cA\indices{_{(0)}^i_j} dz^j .
\end{equation}
They define the seven Kaluza-Klein vectors $\cA\dwn{1}^i$ and the $21$ axion scalar fields $\cA\indices{_{(0)}^i_j}$ (defined only for $j > i$).

The three-form ansatz defines the four-dimensional three-form $A_{(3)}$, the seven two-forms $A_{(2)i}$, the $21$ vector fields $A_{(1)ij}$ and the $35$ scalar fields $A_{(0)ijk}$. As was mentioned before, the three form carries no degree of freedom and the two-forms $A_{(2)i}$ can be dualized to seven scalars $\chi^i$. All in all, this agrees with the $D=4$ entries of tables \ref{table:fieldcontent} (before dualization) and \ref{table:fieldcontentdual} (after dualization).

\subsection{$E_{7(7)}/SU(8)$ scalar coset}

In the notations of \cite{Cremmer:1997ct}, the Borel subalgebra of $E_{7(7)}$ contains 7 Cartan generators, packed in a vector $\vec{H}$, and 63 "raising" operators, which are
\begin{itemize}
\item the raising operators $E\indices{_i^j}$ of the $SL(7,\R)$ subalgebra (defined for $i < j$);
\item an antisymmetric generator $E^{ijk}$;
\item a one-index generator $D_i$.
\end{itemize}
All indices take values from one to eight. The $SL(7,\R)$ subalgebra coresponds to the nodes numbered $2$ to $7$ in figure \ref{fig:E8dynkin}, while $E^{ijk}$ corresponds to the exceptional node $1$. Commutation relations can be found in section 4 of \cite{Cremmer:1997ct}.

The scalar fields of $D=4$, $\cN = 8$ supergravity are then packaged in the $E_{7(7)}/SU(8)$ coset representative
\begin{equation} \label{eq:N8coset}
\cV = \cV_1 \cV_2 \cV_3 \cV_4
\end{equation}
with
\begin{align}
\cV_1 &= \exp\left[\frac{1}{2} \vec{\phi} \cdot \vec{H}\right]\, , \\
\cV_2 &= \prod_{i<j} \exp\left[\cA\indices{_{(0)}^i_j} E\indices{_i^j}\right]\, , \\
\cV_3 &= \exp\left[\sum_{i<j<k}A_{(0)ijk} E^{ijk}\right]\, , \\
\cV_4 &= \exp\left[\sum_i \chi^i D_i\right]\, .
\end{align}
The scalar Lagrangian of $D=4$, $\cN=8$ supergravity is exactly equal to the coset Lagrangian \eqref{eq:cosetlagrangianP} (or \eqref{eq:cosetlagrangianM}) of the previous section for the coset $E_{7(7)}/SU(8)$ parametrized in this way.

\subsection{Action of $E_{7(7)}$ on the vector fields}

The $E_{7(7)}$ symmetry of the scalar sector extends to the vector part as follows: the group $E_{7(7)}$ has a faithful $56$-dimensional representation, which acts linearly on the $28$ vectors $A_{(1)ij}$, $\cA\dwn{1}^i$ and their duals. More precisely:
\begin{itemize}
\item One first dualizes the vectors $A_{(1)ij}$ coming from the three-form to get $21$ vectors $B\dwn{1}^{ij}$;
\item Those vectors, along with the seven vectors $\cA\dwn{1}^i$ coming from the metric, form the $28$-dimensional antisymmetric representation of $SL(8,\R)$. Indeed, defining the tensor $A^{ab}$ by
\begin{equation}
A^{ij} = B\dwn{1}^{ij}, \quad A^{i8} = - A^{8i} = \cA\dwn{1}^i,
\end{equation}
the action is invariant under $SL(8,\R)$ transformations
\begin{equation}
A^{ab} \rightarrow A^{\prime ab} = S\indices{^a_c} S\indices{^b_d} A^{cd}
\end{equation}
with $S \in SL(8,\R)$. Here, the indices $a,b$ run from one to eight and can be split as $a = (i, 8)$.
\item The vectors $A^{ab}$ and their duals $\tilde{A}_{ab}$ together form the fundamental, $56$-dimensional representation of $E_{7(7)}$. In that representation, an element of the algebra $\mathfrak{e}_{7(7)}$ is parametrized by a traceless $8\times 8$ matrix $\Lambda\indices{^a_b}$ and an antisymmetric tensor $\Sigma_{abcd}$. It acts on the pair $(A^{ab}, \tilde{A}_{ab})$ as \cite{Cremmer:1979up}
\begin{align}
\delta A^{ab} &= \Lambda\indices{^a_c} A^{cb} + \Lambda\indices{^b_c} A^{ac} + \star\Sigma^{abcd}\tilde{A}_{cd} ,\nn \\
\delta \tilde{A}_{ab} &= - \tilde{A}_{cb} \Lambda\indices{^c_a} - \tilde{A}_{ac} \Lambda\indices{^c_b} + \Sigma_{abcd} A^{cd} \label{eq:E7transf}
\end{align}
where we use the notation $\star\Sigma^{abcd}=\frac{1}{4!}\,\varepsilon^{abcdefgh}\Sigma_{efgh}$.
\end{itemize}
The $SL(8,\R)$ subalgebra of $E_{7(7)}$, obtained in \eqref{eq:E7transf} by setting $\Sigma_{abcd} = 0$, does not mix the vector fields and their duals. This is why it can act locally at the level of the action. A general element of $E_{7(7)}$, however, mixes $A^{ab}$ and $\tilde{A}_{ab}$ and is therefore a symmetry of the action wich acts in a non-local manner. (For more on this phenomenon, see chapter \ref{chap:emduality}.)

\subsection{Some properties of $E_{7(7)}$}
\label{app:E77properties}

We conclude this chapter with some interesting properties of the $E_{7(7)}$ algebra and its fundamental representation \eqref{eq:E7transf}, which will be useful in chapter \ref{chap:blackholes}. For compatibility with the notations of \cite{Cremmer:1979up} (which were also used in \cite{Compere:2015roa}), we drop the distinction between $a,b$ and $i,j$ indices: all indices run from $1$ to $8$ for the remainder of this section.

First of all, the action \eqref{eq:E7transf} of $E_{7(7)}$ can be written in matrix form as
\begin{equation} \label{eq:E7matrices}
\delta \begin{pmatrix} X^{ij} \\ X_{ij} \end{pmatrix} =
\begin{pmatrix}
2\Lambda\indices{^{[i}_{[k}}\delta^{j]}_{l]} & \star\Sigma^{ijkl} \\
\Sigma_{ijkl} & -2\Lambda\indices{^{[i}_{[k}}\delta^{j]}_{l]}
\end{pmatrix}
\begin{pmatrix} X^{kl} \\ X_{kl} \end{pmatrix},
\end{equation}
which gives an explicit representation of $\mathfrak{e}_{7(7)}$ by $56\times 56$ matrices that is useful for explicit (computer-assisted) computations. With the obvious notations, this can be written in the compact form $\delta X = g X$. An important property is that this matrix is \emph{symplectic},
\begin{equation}
g^T \Omega g = \Omega, \qquad \Omega = \begin{pmatrix} 0 & I_{28\times 28} \\ -I_{28\times 28} & 0 \end{pmatrix} ,
\end{equation}
which shows that $E_{7(7)}$ is a subgroup of $Sp(56,\R)$. Moreover, $E_{7(7)}$ leaves the quartic form
\begin{align}\label{eq:I4E7}
\mathcal{I}_4 (X) &= X^{ij}X_{jk}X^{kl}X_{li} -\frac{1}{4}(X^{ij}X_{ij})^2 \nn \\
&\quad +\frac{1}{96} \varepsilon^{ijklmnpq} X_{ij}X_{kl}X_{mn}X_{pq} +\frac{1}{96} \varepsilon_{ijklmnpq} X^{ij}X^{kl}X^{mn}X^{pq} .
\end{align}
invariant. (In fact, $E_{7(7)}$ can be uniquely defined as the largest subgroup of $Sp(56,\R)$ that leaves $\cI_4$ invariant \cite{Cremmer:1979up}.) This quartic invariant is crucial for the entropy of black holes in $\cN = 8$ supergravity (see \cite{Kallosh:1996uy,Compere:2015roa} and chapter \ref{chap:blackholes}).

The commutator $[(\Lambda_1,\Sigma_1), (\Lambda_2,\Sigma_2)] = (\Lambda_3,\Sigma_3)$ of two $\mathfrak{e}_{7(7)}$ elements is parametrized by
\begin{align}
\Lambda\indices{_3^i_j} &= \Lambda\indices{_1^i_k}\Lambda\indices{_2^k_j} - \frac{1}{3} \star\Sigma_1^{iklm} \Sigma_{2klmj} - (1\leftrightarrow 2), \\  
\Sigma_{3ijkl} &= 4 \Lambda\indices{_1^m_{[i} }\Sigma_{2jkl]m} - (1\leftrightarrow 2) .
\end{align}
Writing a basis of $\mathfrak{e}_{7(7)}$ as
\begin{align}
G\indices{_a^b} &= (\Lambda\indices{_a^b},0), & (\Lambda\indices{_a^b})\indices{^i_j}&=\delta^i_a\delta^b_j-\frac{1}{8}\delta^a_b\delta^i_j ,\\
G^{abcd}&=(0,\Sigma^{abcd}), & (\Sigma^{abcd})_{ijkl} &= \delta^{[a}_i\delta^b_j\delta^c_k\delta^{d]}_l,
\end{align}
we get the commutation relations
\begin{align} \label{e7comm}
[ G\indices{_a^b}, G\indices{_c^d}] &= \delta^b_c G\indices{_a^d}-\delta^d_a G\indices{_c^b}, \\
[ G\indices{_a^b}, G^{cdef}] &= 4 \delta^{[c}_a G^{def]b} + \frac{1}{2}\delta^a_b G^{cdef}, \\
[ G^{abcd}, G^{efgh} ] &=\frac{1}{72} \left( G\indices{_k^{[a}} \varepsilon^{bcd]efghk} - G\indices{_k^{[e}}\varepsilon^{fgh]abcdk}\right) .
\end{align}
The link with the notation of \cite{Cremmer:1997ct} for the Borel subalgebra is then
\begin{equation} \label{eq:E7dict}
E\indices{_i^j} = G\indices{_i^j}, \quad E^{ijk} = - 12 \,G^{ijk8}, \quad D_i = G\indices{_i^8}, \quad \vec{H} = \sum_{j=1}^7 \left( -\vec{f}_j+\vec{g} \right) G\indices{_j^j}, 
\end{equation}
where the $7$-component vectors $\vec{f}_j$ and $\vec{g}$ were defined in \eqref{eq:deffs} and \eqref{eq:defgg}.
It can be checked using \eqref{e7comm} that these identifications correctly reproduce the commutation relations of \cite{Cremmer:1997ct}.

The maximal compact subalgebra $\mathfrak{su}(8)$ of $\mathfrak{e}_{7(7)}$ is generated by transformations with $\Lambda\indices{^i_j}=-\Lambda\indices{^j_i}$ and $\Sigma_{ijkl}=-\star\Sigma^{ijkl}$. It can be made manifest via the change of basis
\begin{equation}\label{eq:sl8su8basis}
X_{AB} = \frac{1}{4\sqrt 2} \left( X^{ij} + i X_{ij} \right) (\Gamma^{ij})_{AB},
\end{equation}
where the $8\times 8$ matrices $\Gamma^{ij}$ are $\mathfrak{so}(8)$ generators described below.
In this basis, the $\mathfrak{e}_{7(7)}$ transformations are given by 
\begin{equation} \label{eq:E7transfSU8}
\delta X_{AB} = \Lambda\indices{_A^C} X_{CB} + \Lambda\indices{_B^C} X_{AC} + \Sigma_{ABCD}\bar{X}^{CD},
\end{equation}
where $\bar{X}^{AB} =(X_{AB})^*$, $\Lambda\indices{_A^B}$ is an element of $\mathfrak{su}(8)$ (a traceless antihermitian matrix), and $\Sigma_{ABCD}$ is a complex antisymmetric tensor satisfying the self-duality condition
\begin{equation}\label{eq:E7SL8SU8}
\Sigma_{ABCD} = \frac{1}{24} \varepsilon_{ABCDEFGH} \bar{\Sigma}^{EFGH}, \qquad \bar{\Sigma}^{ABCD} = (\Sigma_{ABCD})^* .
\end{equation}
The link between these parameters and $(\Lambda^i_{\;\; j},\Sigma_{ijkl})$ is given in \cite{Cremmer:1979up}. In this basis, the $\mathfrak{su}(8)$ subalgebra is obtained by simply setting $\Sigma_{ABCD} = 0$. The matrices $\Gamma^{ij}$ used in \eqref{eq:sl8su8basis} are given by
\begin{align}
\Gamma^{ij} &= \gamma^{[i} \gamma^{j]} ,\nn \\
\Gamma^{i 8} &= - \gamma^i \qquad (\Gamma^{8 i} = \gamma^i) ,
\end{align}
(with $i,j=1, \dotsc, 7$ here), where the $\gamma^i$ are $\mathfrak{so}(7)$ gamma matrices satisfying
\begin{equation}
\{ \gamma^i, \gamma^j \} = - 2 \delta^{ij} I_{8\times 8} .
\end{equation}
It follows from this definition that the $\Sigma^{ij}=-\frac{1}{2}\Gamma^{ij}$ satisfy the $\mathfrak{so}(8)$ commutation relations
\begin{equation}
\left[ \Sigma^{ij},\Sigma^{kl}\right] = \delta^{il} \Sigma^{jk} + \delta^{jk} \Sigma^{il} - \delta^{ik} \Sigma^{jl} - \delta^{jl} \Sigma^{ik}.
\end{equation}
For the explicit computations of chapter \ref{chap:blackholes}, we used the real antisymmetric representation
\begin{align}
\gamma^1 &= -i \sigma_3 \otimes \sigma_2 \otimes \sigma_1, \nn \\
\gamma^2 &= i \sigma_3 \otimes \sigma_2 \otimes \sigma_3, \nn \\
\gamma^3 &= -i \sigma_3 \otimes I_{2\times 2} \otimes \sigma_2 ,\nn \\
\gamma^4 &= -i \sigma_1 \otimes \sigma_1 \otimes \sigma_2, \nn \\
\gamma^5 &= -i \sigma_1 \otimes \sigma_2 \otimes I_{2\times 2} ,\nn \\
\gamma^6 &= i \sigma_1 \otimes \sigma_3 \otimes \sigma_2 ,\nn \\
\gamma^7 &= i \sigma_2 \otimes I_{2\times 2} \otimes I_{2\times 2}
\end{align}
where the $\sigma_i$ are the standard Pauli matrices.


\chapter{Appendices to part \ref{PART:GAUGINGS}}

\section{Conventions for part \ref{PART:GAUGINGS}}
\label{app:conv2}

The conventions for differential forms of this part of the thesis follow the review \cite{Barnich:2000zw} and were used in the paper \cite{Barnich:2017nty}. Unfortunately, they are not the same as in part \ref{PART:SUGRA}. We collect the differences here.

The space-time dimension is $n$. The symbol $\epsilon_{\mu_1\dots\mu_n}$ denotes the completely 
antisymmetric Levi-Civita density with the convention that 
$\epsilon^{01\dots n-1}=1\,$ so that $\epsilon_{01\dots n-1}=-1\,$.
For $p$-forms, we sometimes use the notation 
\begin{equation}
(d^{ n - p }x)_{\mu_{n-p+1}\dots\mu_n} \equiv -\frac{1}{p!(n -
  p)!}dx^{\mu_{1}}\dots dx^{\mu_{n-p}}\epsilon_{\mu_1\dots\mu_n}
\end{equation}
for $1\leq p \leq n$, and $d^nx:=dx^0\dots dx^{n-1}\,$.
The Hodge dual of a differential $p$-form
\begin{equation}
\omega^p \equiv \frac{1}{p!}\,dx^{\mu_1}\dots
dx^{\mu_p}\omega_{\mu_1\dots\mu_p}
\end{equation}
is the $(n-p)$-form given, in our convention, by 
\begin{align}
\star\,\omega^p &= \frac{1}{p!(n-p)!}\,dx^{\nu_1}\dots
dx^{\nu_{n-p}}\epsilon_{\nu_1\dots\nu_{n-p}\mu_{n-p+1}\ldots\mu_n}
\omega^{\mu_{n-p+1}\dots\mu_n} \\
&= - (d^{ n - p }x)_{\mu_{n-p+1}\dots\mu_n}
\omega^{\mu_{n-p+1}\dots\mu_n}\;. 
\nonumber
\end{align}
As a consequence, the exterior differential of the dual of a 
$p\,$-form reads 
\begin{align}
d \star \omega^p =-(-)^{n-p} (d^{n-p+1}x)_{\nu_1\dots\nu_{p-1}}
\,\partial_{\mu}\omega^{\mu\nu_1\dots\nu_{p-1}}\;.
\end{align}
Note also that in these formulas, the $p$-form components are written to the \emph{right} of the basis elements $dx^\mu$. This is crucial when the components are anticommuting variables (for example, for the antifields $A^*_{I\mu}$ associated with the vector fields, we write $A^*_I = dx^\mu A^*_{I\mu} = - A^*_{I\mu} dx^\mu$).

\section{Pure gauge and auxiliary fields}
\label{app:auxfields}

As is discussed in section \ref{sec:embeddingtensor}, the extra variables appearing in the embedding tensor formalism are of two types:
\begin{itemize}
\item either they are of pure gauge type and drop out from the Lagrangian (or, equivalently, invariant under arbitrary shifts);
\item or they are auxiliary fields appearing quadratically and undifferentiated in the Lagrangian (they can therefore be eliminated algebraically through their own equations of motion).
\end{itemize}
In this appendix, we show that these variables appearing in addition to the standard variables of the Lagrangian \eqref{eq:lag} do not modify the space of local deformations $H^0(s \vert d)$.

We collectively denote by $W^\alpha$ the fields of pure gauge type.  So, in section \ref{sec:embeddingtensor}, the $W^\alpha$'s stand for the $n_v$ vector potentials $\tilde{A}'_{\hat{U}}$, $A'^{\hat{I}}$ and the (dim$\,G-r$) two-forms $B'_m$.  The auxiliary fields are the $r$ two-forms $\Delta_i$.    For convenience, we absorb in \eqref{eq:lbarB} the linear term in the auxiliary fields by a redefinition $\Delta_i \rightarrow \Delta_i' = \Delta_i + b_i$ where the term $b_i$ is $\Delta_i$-independent.  Once this is done, the dependence of the Lagrangian on the auxiliary fields simply reads $\frac12 \kappa^{ij} \Delta_i' \Delta_j'$ with an invertible quadratic form $\kappa^{ij}$.

The BRST differential acting in the sector of the first type of variables read
\be
s W^\alpha = C^\alpha \, ,\; \; \;   s C^\alpha = 0 \, ,\; \; \; s C^*_\alpha = W^*_\alpha \, ,\; \; \; s W^*_\alpha =0
\ee
where $C^\alpha$ are the ghosts of the shift symmetry and $W^*_\alpha$, $C^*_\alpha$ the corresponding antifields, so that $(C^\alpha,  W^\alpha)$ and $(W^*_\alpha, C^*_\alpha)$  form ``contractible pairs" \cite{Henneaux:1992ig}.  Similarly, the BRST differential acting in the sector of the second type of variables read
\be
s \Delta_i' = 0, \quad s \Delta^{\prime *i} = \kappa^{ij} \Delta_j'
\ee
where $\Delta^{\prime *i}$ are the antifields conjugate to $\Delta_i'$, showing that the auxiliary fields and their antifields form also contractible pairs since $\kappa^{ij}$ is non degenerate.

Now, the above BRST transformations involve no spacetime derivatives so that one can construct a ``contracting homotopy" \cite{Henneaux:1992ig} in the sector of the extra variables $(W^\alpha,  \Delta_i')$, their ghosts, and their antifields, that commutes with the derivative operator $\partial_\mu$.  The algebraic setting
is in fact the same as for the variables of the ``non minimal sector" of \cite{Batalin:1981jr}.  This implies that the extra variables neither contribute to $H^k(s)$ nor to $H^k(s \vert d)$  \cite{Barnich:1994db}.

To be more specific, let us focus on the contractible pair $(C^\alpha, W^\alpha)$.  The analysis proceeds in exactly the same way for the other pairs.  One can write the BRST differential  in that sector as
\be
s = \sum_{\{ \mu \}} \partial_{\mu_1 \cdots \mu_s} C^\alpha \frac{\partial}{\partial _{\mu_1 \cdots \mu_s} W^\alpha}
\ee
where the sum is over all derivatives of the fields. The ``contracting homotopy" $\rho$ is defined as
\be
\rho = \sum_{\{ \mu \}} \partial_{\mu_1 \cdots \mu_s} W^\alpha \frac{\partial}{\partial _{\mu_1 \cdots \mu_s} C^\alpha}  .
\ee
One has by construction 
\be
[ \rho, \partial_\mu] = 0,
\ee
just as 
\be
[ s, \partial_\mu] = 0.
\ee
This is equivalent to $\rho d + d \rho = 0$.  Furthermore, the counting operator $N$ defined by
\be
N = s \rho + \rho s
\ee
is explicitly given by
\be
N = \sum_{\{ \mu \}} \partial_{\mu_1 \cdots \mu_s} C^\alpha \frac{\partial}{\partial _{\mu_1 \cdots \mu_s} C^\alpha} + \sum_{\{ \mu \}} \partial_{\mu_1 \cdots \mu_s} W^\alpha \frac{\partial}{\partial _{\mu_1 \cdots \mu_s} W^\alpha}
\ee
and commutes with the BRST differential,
\be
[N, s] = 0.
\ee
The operator $N$ gives the homogeneity degree in $W^\alpha$, $C^\alpha$ and their derivatives.  So, a polynomial $a$ is such that $Na = k a $, with $k$ a non negative integer, if and only if  it is of degree $k$ in $W^\alpha$, $C^\alpha$ and their derivatives (by Euler's theorem for homogeneous functions). 

Because $N$ commutes with $s$, we can analyze the cocycle condition
\be
s a + d b = 0 \label{eq:cocycleapp}
\ee
at definite polynomial degree, i.e., assume that $N a = k a$, $Nb = kb$ where $k$ is a non-negative integer.  Our goal is to prove that the solutions of \eqref{eq:cocycleapp} are trivial when $k \not=0$, i.e., of the form $a = se + df$, so that one can find, in any cohomological class of $H(s \vert d)$, a representative that does not depend on $W^\alpha$ or $C^\alpha$ -- that is, $W^\alpha$ and $C^\alpha$ ``drop from the cohomology".

To that end, we start from $ a = N \left(\frac{a}{k} \right)$ ($k \not=0$), which we rewrite as
$a = s \rho \left(\frac{a}{k} \right)+ \rho s  \left(\frac{a}{k} \right)$ using the definition of $N$.  The first term is equal to $s \left(\rho \left(\frac{a}{k} \right)\right)$ and hence is BRST-exact.   The second term is equal to $\rho \left(-d \left(\frac{b}{k} \right)\right)$ (using \eqref{eq:cocycleapp}), which is the same as $d \left(\rho \left(\frac{b}{k} \right)\right)$ since $\rho$ and $d$ anticommute.  So it is $d$-exact. We thus have shown that 
\be
a = se + df
\ee
with $e = \rho \left(\frac{a}{k} \right)$ and $f = \rho \left(\frac{b}{k} \right)$.   This is what we wanted to prove.

 We can thus conclude that the space of local deformations $H^0(s \vert d)$ -- and in fact also $H^k(s \vert d)$ for any ghost number --  are the same whether or not one includes the extra fields $(W^\alpha, \Delta_i)$.  This is what we wanted to prove: the BRST cohomologies computed from the Lagrangians \eqref{lagtwoforms} and \eqref{eq:lag} are the same\footnote{See also \cite{Coomans:2010xd}, where a relation between the embedding tensor formalism and the BRST-BV antifield formalism has been considered with a different purpose.}.

\section{Antibracket maps and descents}
\label{sec:antibr-maps-desc}

As discussed in section \ref{sec:antibracket}, the first
obstruction to extending infinitesimal deformations to finite ones is controlled by the antibracket map. We show here how the antibracket
map behaves with respect to the length of shortest non trivial
descent, i.e., the ``depth'' defined in section \ref{sec:lengthdepth}.

{\bf Proposition:} {\em The depth of an image of the antibracket map is less or equal to the depth of
  its most shallow argument.}

{\bf Proof:} Consider
$[\omega^{g_1,n}_{l_1}],[\omega^{g_2,n}_{l_2}]\in H^{*,n}(s|d)$, where
we can assume without loss of generality that $l_1\geqslant l_2$. For the
antibracket, let us not choose the expression with Euler-Lagrange
derivatives on the left and right that is graded antisymmetric without boundary terms, but rather the expression
\begin{equation}
  \label{eq:B2}
  (\omega^{g,n},\cdot)_{\rm
    alt}=\d_{(\nu)}\vddr{(-\star\omega^{g,n})}{\phi^A}\ddll{\cdot}{\d_{(\nu)}\phi^*_A}
    -(\phi^A\leftrightarrow \phi^*_A), 
\end{equation}
which can be obtained by integrations by parts. It satisfies a graded Leibniz
rule on the right and the following version of the graded Jacobi identity without
boundary terms,
\begin{equation}
  \label{eq:B3}
  (\omega^{g_1,n},(\omega^{g_2,n},\cdot)_{\rm alt})_{\rm
    alt}=((\omega^{g_1,n},\omega^{g_2,n})_{\rm alt},\cdot)_{\rm
    alt}+(-)^{(g_1+1)(g_2+1)}(\omega^{g_2,n},(\omega^{g_1,n},\cdot)_{\rm alt})_{\rm
    alt}
\end{equation}
(see appendix B of
\cite{Barnich:1996mr} for details and a proof). 
Furthermore, 
\begin{equation}
  (\omega^{g,n},d (\cdot))_{\rm
    alt}=(-)^{g+1}d((\omega^{g,n}, \cdot)_{\rm
    alt}),\quad (d\omega^{g+1,n-1},\cdot)_{\rm alt}=0. \label{eq:B5}
\end{equation}
Let $S=\int (-\star \cL)$ be the BV master action. We have
$s\cdot=(-\star \cL,\cdot)_{\rm
  alt}$. 
Using these properties, we get 
  \begin{equation}
    \label{eq:B4}
    s(\omega^{g_1,n}_{l_1},\omega^{g_2,n}_{l_2})_{\rm
      alt}+d((\omega^{g_1,n}_{l_1},\omega^{g_2+1,n-1}_{l_2})_{\rm
      alt})=0, \; 
\dots, \; s (\omega^{g_1,n}_{l_1},\omega^{g_2+l_2,n-l_2}_{l_2})_{\rm
      alt}=0, 
  \end{equation}
which proves the proposition.

\section{Derivation of Equation \eqref{EQ:UTRASF}}
\label{app:derivation}

In this appendix, we derive formula \eqref{EQ:UTRASF} for the variation $\delta_u G_I = - (U_u, G_I)$ of the two-form $G_I$ under a $U$-type symmetry, in the case where the Lagrangian (or, equivalently, $G_I$) does not depend on the derivatives of $F^I_{\mu\nu}$. This is done in two steps:
\begin{enumerate}
\item First, we show that
\begin{equation}\label{eq:step1}
\delta_u G_I + (f_u)\indices{^J_I} G_J \approx c_{IJ} F^J + d(\text{invariant})
\end{equation}
for some constants $c_{IJ}$.
\item Then, we prove that the $c_{IJ}$ take the form
\begin{equation}\label{eq:step2}
c_{IJ} = - 2 (h_u)_{IJ} + \lambda^w_u (h_w)_{IJ},
\end{equation}
where the constants $(h_u)_{IJ}$ and $(h_w)_{IJ}$ are those appearing in the currents associated with $U_u$ and $W_w$ respectively.
\end{enumerate}

\subsubsection*{A lemma}

The proof of the above
 steps uses the following result on the $W$-type cohomology classes (with $g=-1$):
\begin{equation}\label{eq:Wlemma}
t_{IJ} F^I F^J \approx d(\text{invariant}) \;\Rightarrow\; t_{IJ} = \sum_{w} \lambda^w (h_w)_{IJ} \,\text{ for some } \lambda^w .
\end{equation}
This is proven as follows: $t_{IJ} F^I F^J \approx d(\text{invariant})$ implies that
\begin{equation}\label{eq:appt}
t_{IJ} F^I F^J + dI + \delta k = 0
\end{equation}
for some gauge invariant $I$ and some $k$ of antifield number $1$, where $\delta$ is here the Koszul-Tate differential. Now, it is proven in \cite{Barnich:2000zw} that $k$ must be gauge invariant; hence, it can be written as
\begin{equation}\label{eq:appk}
k = \hat{K} + d R, \quad \hat{K} = d^4x [ A^{*\mu}_I g^I_\mu + \phi^*_i \Phi^i ]
\end{equation}
for some gauge invariant $R$, $g^I_\mu$ and $\Phi^i$. Indeed, derivatives acting on the antifields contained in $k$ are pushed to the term $dR$ by integration by parts, leaving the form \eqref{eq:appk} where $\hat{K}$ contains only the undifferentiated antifields. Putting this back in \eqref{eq:appt} and using the fact that $\delta \hat{K} = s \hat{K}$ because $\hat{K}$ is gauge invariant, we get
\begin{equation}
s \hat{K} + d\left( t_{IJ} A^I F^J + J \right) = 0
\end{equation}
for some gauge invariant $J = I - \delta R$. This shows that $\hat{K}$ is a $W$-type cohomology class: we can therefore expand $\hat{K}$ in the $W_w$ basis as $\hat{K} = \sum \lambda^w W_w$. In particular, this implies that $t_{IJ} = \sum \lambda^w (h_w)_{IJ}$, which proves the lemma.

\subsubsection*{First step}

We start from the chain of descent equations involving $G_I$,
\begin{equation}\label{eq:descCstar}
s\, d^4x\, C^*_I+d \star A^*_I=0,\quad s \star A^*_I+d
G_I=0,\quad sG_I=0.
\end{equation}
Applying $(U_u, \cdot)_\text{alt}$ to this chain, we get
\begin{align}
s \left[ \,d^4x\, (f_{u})\indices{^J_I} C^*_J \right] + d \left[ (f_{u})\indices{^J_I} \star A^*_J +
\frac{\delta K_u}{\delta A^I} \right] &= 0, \\
s \left[ (f_{u})\indices{^J_I} \star A^*_J +
\frac{\delta K_u}{\delta A^I} \right] + d \left[ - \delta_u G_I \right] &= 0, \\
s \left[ - \delta_u G_I \right] &= 0,
\end{align}
which can be simplified to
\begin{align}
d \left( \frac{\delta K_u}{\delta A^I} \right) &= 0, \label{eq:dK}\\
s \left( \frac{\delta K_u}{\delta A^I} \right) + d \left( - \delta_u G_I - (f_{u})\indices{^J_I} G_J \right) &= 0, \label{eq:sK}\\
s \left( - \delta_u G_I \right) &= 0,
\end{align}
using equations \eqref{eq:descCstar} again.
Equation \eqref{eq:dK} implies that
\begin{equation}
\frac{\delta K_u}{\delta A^I} = d \eta^{-1,2}
\end{equation}
for some $\eta^{-1,2}$ of ghost number $-1$ and form degree $2$. Because the left-hand side is gauge invariant and $\eta^{-1,2}$ is of form degree two, $\eta^{-1,2}$ must also be gauge invariant. This follows from theorems on the invariant cohomology of $d$ in form degree $2$ \cite{Brandt:1989gy,DuboisViolette:1992ye}. Equation \eqref{eq:sK} implies then
\begin{equation}
d\left( \delta_u G_I + (f_{u})\indices{^J_I} G_J + s\eta^{-1,2} \right) = 0 ,
\end{equation}
i.e.
\begin{equation}\label{eq:dGdeta}
\delta_u G_I + (f_{u})\indices{^J_I} G_J + s\eta^{-1,2} = d \eta^{0,1}
\end{equation}
for some $\eta^{0,1}$ of ghost number $0$ and form degree $1$. Again, the left-hand side of this equation is gauge invariant: results on the invariant cohomology of $d$ in form degree $1$ \cite{Brandt:1989gy,DuboisViolette:1992ye} now imply that the non-gauge invariant part of $\eta^{0,1}$ can only be a linear combination of the one-forms $A^I$,
\begin{equation}
\eta^{0,1} = c_{IJ} A^J + \text{(gauge invariant)} .
\end{equation}
Plugging this back in equation \eqref{eq:dGdeta} and using the fact that $s\eta^{-1,2} \approx 0$ (since $\eta^{-1,2}$ is gauge invariant), we recover equation \eqref{eq:step1}. This concludes the first step of the proof.

\subsubsection*{Second step}

For the second step, we introduce
\begin{equation}
N = - \int \!d^4x\,( C^*_I C^I + A^{*\mu}_I A^I_\mu), \quad \hat{N} = (N, \cdot)_\text{alt} .
\end{equation}
The operator $\hat{N}$ counts the number of $A^I$'s and $C^I$'s minus the number of $A^*_I$'s and $C^*_I$'s. Because it carries ghost number $-1$, it commutes with the exterior derivative, $\hat{N} d = d \hat{N}$.
Applying this operator to the equation
\begin{equation}
s U_u + d \left[ (f_u)\indices{^I_J} (\star A^*_I C^J + G_I A^J) + (h_u)_{IJ} F^I A^J  + J_u \right] = 0
\end{equation}
gives
\begin{equation}\label{eq:NsU}
(\int\! G_I F^I, U_u)_\text{alt} + d\left[ (f_u)\indices{^I_J} (\hat{N} + 1)(G_I) A^J + 2 (h_u)_{IJ} F^I A^J  + \hat{N}(J_u) \right] \approx 0 .
\end{equation}
The second term is evident. The first term is
\begin{align}
\hat{N}(sU_u) = (N, (S,U_u)_\text{alt})_\text{alt} &= ( (N, S) , U_u)_\text{alt} + (S, (N,U_u)_\text{alt})_\text{alt}
\end{align}
according to the graded Jacobi identity.
The counting operator $\hat{N}$ kills the $A^{*\mu}_I \partial_\mu C^I$ term in the master action $S$, which implies
\begin{equation}
(N, S) = \int\!d^4x\, A^I_\mu \frac{\delta \mathcal{L}_V}{\delta A^I_\mu} = \int\!d^4x\, A^I_\mu \partial_\nu(\star G_I)^{\mu\nu} = \int\! G_I F^I .
\end{equation}
Similarly, $\hat{N}$ kills the first two terms of $U_u$, leaving $\hat{N} U_u = \hat{N} K_u$ which is gauge invariant. This implies $(S, (N,U_u)_\text{alt})_\text{alt} = s(\hat{N} U_u) \approx 0$.
Therefore, we have indeed
\begin{equation}
\hat{N}(sU_u) \approx (\int G_I F^I, U_u)_\text{alt}
\end{equation}
which proves equation \eqref{eq:NsU}.

We now compute $(\int G_I F^I, U_u)_\text{alt}$ using the result of the first step. We have
\begin{equation}
(\int G_I F^I, U_u)_\text{alt} = \frac{\delta (G_K F^K)}{\delta A^I_\mu} \, \delta_u A^I_\mu + \frac{\delta (G_K F^K)}{\delta \phi^i} \, \delta_u \phi^i .
\end{equation}
This looks like the $U$-variation $\delta_u (G_I F^I)$, but it is not because there are Euler-Lagrange derivatives. For a top form $\omega$, the general rule is \cite{Andersonbook}
\begin{equation}
\delta_Q \omega = Q^a \frac{\delta \omega}{\delta z^a} + d \rho, \quad \rho = \partial_{(\nu)} \left[ Q^a \frac{\delta}{\delta z^a_{(\nu)\rho}} \frac{\partial \omega}{\partial dx^\rho} \right] .
\end{equation}
In our case, this becomes
\begin{align}
\delta_u (G_I F^I) &= \frac{\delta (G_K F^K)}{\delta A^I_\mu} \, \delta_u A^I_\mu + \frac{\delta (G_K F^K)}{\delta \phi^i} \, \delta_u \phi^i + d\rho_A + d\text{(inv)} ,\\
\rho_A &= \partial_{(\nu)} \left( (f_u)\indices{^I_J} A^J_\mu \frac{\delta}{\delta A^I_{\mu, (\nu)\rho}} \frac{\partial (G_K F^K)}{\partial dx^\rho} \right) .
\end{align}
Using property \eqref{eq:step1} and putting together the terms of the form $d\text{(invariant)}$, we get then from \eqref{eq:NsU}
\begin{equation}
(c_{IJ} + 2 (h_u)_{IJ} ) F^I F^J + d \left[ (f_u)\indices{^I_J} A^J (\hat{N} + 1)(G_I) - \rho_A \right] + d \text{(inv)} \approx 0 .
\end{equation}
Now, it is sufficient to prove that
\begin{equation}\label{eq:dxy}
 d \left[ (f_u)\indices{^I_J} A^J (\hat{N} + 1)(G_I) - \rho_A \right] \approx d \text{(inv)}.
\end{equation}
Indeed, this implies $(c_{IJ} + 2 (h_u)_{IJ} ) F^I F^J \approx d \text{(inv)}$, which in turn gives
\begin{equation}
c_{IJ} = - 2 (h_u)_{IJ} + \lambda^w_u (h_w)_{IJ}
\end{equation}
for some constants $\lambda^w_u$ using property \eqref{eq:Wlemma} of the $W$-type cohomology classes.

\subsubsection*{Proof of \eqref{eq:dxy}}

We will actually prove the stronger equation
\begin{equation}\label{eq:xy}
\rho_A = (f_u)\indices{^I_J} A^J (\hat{N} + 1)(G_I)
\end{equation}
in the case where $G_I$ depends on $F$ but not on its derivatives.

To do this, we can assume that $G_I$ a homogeneous function of degree $n$ in $A^I$, i.e. $\hat{N}(G_I) = n G_I$. If it is not, we can separate it into a sum of homogenous parts; the result then still holds because equation \eqref{eq:xy} is linear in $G_I$.

In components, equation \eqref{eq:xy} is
\begin{equation}
\frac{1}{2}\partial_{(\nu)} \left( (f_u)\indices{^I_J} A^J_\mu \frac{\delta}{\delta A^I_{\mu, (\nu)\rho}} G_{K\sigma\tau} F^K_{\lambda\gamma} \varepsilon^{\sigma\tau\lambda\gamma} \right) = (n+1) (f_u)\indices{^I_J} A^J_\lambda G_{I\sigma\tau} \varepsilon^{\rho\lambda\sigma\tau} .
\end{equation}
Under the homogeneity assumption $\hat{N}(G_I) = n G_I$, we have
\begin{equation}
G_{K\sigma\tau} F^K_{\lambda\gamma} \varepsilon^{\sigma\tau\lambda\gamma} = 4 (n+1) \mathcal{L}_V .
\end{equation}
Equation \eqref{eq:xy} now becomes
\begin{equation}\label{eq:xyL}
\frac{1}{2}\partial_{(\nu)} \left( (f_u)\indices{^I_J} A^J_\mu \frac{\delta \mathcal{L}_V}{\delta A^I_{\mu, (\nu)\rho}} \right) = \frac{1}{4} (f_u)\indices{^I_J} A^J_\lambda G_{I\sigma\tau} \varepsilon^{\rho\lambda\sigma\tau} .
\end{equation}
We now use the fact that $G_I$ does not depend on derivatives of $F$, which implies that the higher order derivatives $\partial_{(\nu)}$ are not present and that the Euler-Lagrange derivatives are only partial derivatives. We then have
\begin{equation}
\frac{1}{2} \frac{\delta \mathcal{L}_V}{\delta A^I_{\mu,\rho}} = \frac{\delta \mathcal{L}_V}{\delta F^I_{\rho\mu}} = \frac{1}{4} \varepsilon^{\rho\mu\sigma\tau} G_{I\sigma\tau}
\end{equation}
(see \eqref{eq:47}), which proves \eqref{eq:xyL} in this case.


\chapter{Appendices to part \ref{PART:6D}}

\section{Young tableaux and generalized Poincaré lemmas}
\label{app:youngpoincare}

In this appendix, we fix our conventions and notations for tensors of mixed Young symmetry (i.e., neither totally symmetric nor totally antisymmetric), and review some important formulas and combinatorics. Useful references are \cite{fulton1991representation,Bekaert:2006py,hamermesh1962group,fuchs2003symmetries}.

We also review the Poincaré lemmas for tensors of mixed Young symmetry of \cite{Olver_hyper,DuboisViolette:1999rd,DuboisViolette:2001jk,Bekaert:2002dt} in the particular case of two-column Young diagrams relevant to this part of the thesis. We refer to those references for the general case.

\subsection{Definitions}

A Young diagram is a (finite) collection of $n$ boxes arranged in columns of non-increasing size, for example
\begin{equation}\label{eq:exdia}
Y = \ydiagram{5,4,3,1}\, .
\end{equation}
We use two notations for such a diagram: either as $[h_1, h_2, \dots, h_c]$, where $c$ is the number of columns and each $h_i$ is the height of column $i$, or as $(l_1, l_2, \dots, l_r)$, where $r$ is the number of rows and each $l_i$ is the length of row $i$. For example, the diagram \eqref{eq:exdia} can be written as $Y=[4,3,3,2,1]$ or $Y=(5,4,3,1)$. In both cases, the numbers are not increasing and their sum is equal to the number of boxes,
\begin{align}
h_1 \geq h_2 \geq \dots \geq h_c, \quad l_1 \geq l_2 \geq \dots \geq l_r, \quad \sum_{i=1}^c h_i = \sum_{i=1}^r l_i = n\, .
\end{align}
The number of Young diagrams with $n$ boxes is of course equal to the number of partitions of $n$ as a sum of integers.

A standard Young tableau is a Young diagram filled with the numbers $1$ to $n$ in such a way that the numbers always increase when going to the right or to the bottom. For example, for the diagram $(3,2)=[2,2,1]$, there are five standard tableaux:
\begin{equation}\label{eq:sixtableaux}
\ydiagram{3,2} \quad\rightarrow\quad \ytableaushort{123,45}\, ,\quad \ytableaushort{124,35}\, ,\quad \ytableaushort{125,34}\, ,\quad \ytableaushort{134,25}\, ,\quad \ytableaushort{135,24}\, .
\end{equation}
There is a simple formula for the number of standard tableaux associated to a given diagram. To write it, we first define the hook length $h_{ij}$ of the box at row $i$ and column $j$ by
\begin{equation}
h_{ij} = 1 + l_i + h_j - i -j \, ,
\end{equation}
where $l_i$ is the length of row $i$ and $h_j$ is the height of column $j$, as before. This corresponds to counting the number of boxes below and to the right of $ij$, including the box $ij$ once. In the example of the Young diagram $(5,4,3,1)$, the first diagram below illustrates $h_{22} = 4$ and the second labels all boxes by their hook lengths:
\begin{equation}\label{eq:young13}
\ytableaushort{{}{}{}{}{},{}\bullet\bullet\bullet,{}\bullet{},{}}\, ,\quad \ytableaushort{86531,6431,421,1}\, .
\end{equation}
Now, the number of standard tableaux associated to a given diagram is simply given by
\begin{equation}\label{eq:standardtableaux}
\# \text{ standard tableaux } = \frac{n!}{H}\, ,
\end{equation}
where $H$ is the product of all the hook lengths of the diagram. For the example \eqref{eq:sixtableaux}, we get indeed $5!/(4\cdot 3 \cdot 2) = 5$. It is a classic result that all irreducible finite-dimensional representations of the finite group $S_n$ (permutations of $n$ elements) are labelled by a Young diagram $Y$ with $n$ boxes; their dimension is equal to the number \eqref{eq:standardtableaux} of standard tableaux associated to $Y$.

Given a vector space $V$ of dimension $d$, one can form the space $V^{\otimes n}$ of tensors with $n$ indices. The permutation group $S_n$ acts naturally on this space of tensors by permuting the indices,
\begin{equation}
(\sigma T)^{i_1 i_2 \dots i_n} = T^{i_{\sigma(1)} i_{\sigma(2)} \dots i_{\sigma(n)}}\, , \quad \sigma \in S_n \, ,\, T \in V^{\otimes n} \, .
\end{equation}
To each standard tableau $\lambda$ is then associated a subspace $V_\lambda$ of $V^{\otimes n}$, which can be characterized in two equivalent ways.
\begin{enumerate}
\item It can be seen as the image of the Young projector $\mathbf{Y}_{\!\lambda}$ acting the whole space of tensors $V^{\otimes n}$,
\begin{equation}
V_\lambda = \mathbf{Y}_{\!\lambda}\left( V^{\otimes n} \right) .
\end{equation}
This projector is defined by first symmetrizing over the indices corresponding to the indices in the rows, then antisymmetrizing in the indices of the columns,
\begin{equation}
\mathbf{Y}_{\!\lambda} = a_\lambda \circ s_\lambda\, , \quad
s_\lambda = \sum_{\sigma \in P} \sigma \, , \quad a_\lambda = \sum_{\sigma \in Q} (-1)^{|\sigma|}\sigma,
\end{equation}
where $P$ and $Q$ are the subgroups of $S_n$ preserving the content of the rows and columns, respectively. For example, for the standard Young tableau
\begin{equation}\label{eq:littlehook}
\lambda = \ytableaushort{13,2}
\end{equation}
associated to the Young diagram $Y = (2,1)$, we have
\begin{equation}
P = \{ e, \, (13) \}\, , \quad Q = \{e,\, (12)\}\, , \quad s_\lambda = e + (13)\, , \quad a_\lambda = e - (12)\,
\end{equation}
($e$ is the identity permutation) and therefore
\begin{equation}
\mathbf{Y}_{\!\lambda} = \left( e - (12) \right) \circ \left( e + (13)\right) = e - (12) + (13) - (132) \, .
\end{equation}
Explicitly, its action on a generic three-index tensor $T^{i_1 i_2 i_3}$ is
\begin{equation}\label{eq:Yp21}
(\mathbf{Y}_{\!\lambda} T)^{i_1 i_2 i_3} = T^{i_1 i_2 i_3} - T^{i_2 i_1 i_3} + T^{i_3 i_2 i_1} - T^{i_3 i_2 i_1}\, .
\end{equation}
\item Alternatively, it can be characterized as the subspace of $V^{\otimes n}$ satisfying the two properties
\begin{enumerate}
\item the tensors are totally antisymmetric in the indices corresponding to a column of the Young tableau; and
\item any antisymmetrization over all the indices of a column, plus one index belonging to any other column to its right, vanishes.
\end{enumerate}
In the example of the Young tableau \eqref{eq:littlehook} considered above, this is the space of tensors $T^{i_1i_2i_3}$ with three indices such that
\begin{equation}
T^{i_1 i_2 i_3} = - T^{i_2 i_1 i_3}\, , \quad T^{[i_1 i_2 i_3]} = 0 \, .
\end{equation}
It is easy to see that formula \eqref{eq:Yp21} satisfies those two properties.
\end{enumerate}
For example, the Young diagrams $[n]$ and $(n)$ (which have only one standard tableau) correspond to the space of totally antisymmetric and totally symmetric tensors, respectively.

\subsection{Useful properties}

The Young projectors satisfy the completeness and orthogonality relations
\begin{equation}\label{eq:comport}
e = \sum_{\lambda} \frac{1}{H}\mathbf{Y}_{\!\lambda}\, ,\quad \mathbf{Y}_{\!\lambda} \circ \mathbf{Y}_{\!\mu} = 0 \;\text{ if }\; \lambda \neq \mu\, ,
\end{equation}
but they are not normalized,
\begin{equation}\label{eq:norm}
\mathbf{Y}_{\!\lambda} \circ \mathbf{Y}_{\!\lambda} = H \, \mathbf{Y}_{\!\lambda}\, .
\end{equation}
In these formulas, $H$ is the product of the hook lengths of the diagram, as before. The first of \eqref{eq:comport} gives the decomposition
\begin{equation}\label{eq:tensordecomp1}
V^{\otimes n} = \bigoplus_\lambda V_\lambda
\end{equation}
of the space of all tensors, which generalizes the familiar decomposition of a rank $2$ tensor into symmetric and antisymmetric parts.

The spaces $V_\lambda$ provide irreducible finite-dimensional representations of the linear group $GL(V) = GL(d, \C)$. The representations associated to standard tableaux associated to the same Young diagram are isomorphic. Because of this isomorphism, we will only precise the diagram and not the tableau associated to a given tensor. The decomposition above becomes
\begin{equation}
V^{\otimes n} \simeq \bigoplus_Y \left(V_\lambda\right)^{n!/H} \, ,
\end{equation}
which should be understood as a decomposition of $V^{\otimes n}$ into irreducible representations of the linear group. The sum now runs over different Young diagrams $Y$, and $\lambda$ is any standard tableau associated with $Y$. 

The dimension of the spaces $V_\lambda$ is
\begin{equation}\label{eq:dimvl}
\dim V_\lambda = \prod_{(ij)} \frac{d - i + j}{h_{ij}} \, ,
\end{equation}
where the product goes over all boxes of the diagram, where $i$ is the row and $j$ is the column (this formula does indeed only depend on the diagram and not on the specific tableau). Let us take the example of a $[2,2]$-tensor in dimension $d = 4$. The numerators and denominators of formula \eqref{eq:dimvl} are given by
\begin{equation}
\ytableaushort{45,34}\, , \quad \ytableaushort{32,21} \, .
\end{equation}
The dimension of this space is then $(4^2 \cdot 5 \cdot 3)/(3\cdot 2^2) = 20$, which is indeed the number of independent components of the Riemann tensor in four dimensions. In arbitrary dimension, one recovers the well-known formula $d^2 (d^2-1) / 12$ from the same reasoning. Similarly, it reproduces
\begin{align}
\dim \odot^n(V) = \frac{(d+n-1)!}{n!(d-1)!}\, , \quad \dim \wedge^n(V) = \frac{d!}{n!(d-n)!}
\end{align}
for the number of independent components of symmetric or antisymmetric tensors with $n$ indices.

Let us also give two comments on conventions.
\begin{itemize}
\item In the text, we use the canonical, normalized projection operators
\begin{equation}\label{eq:defP}
\mathbb{P}_Y \equiv \frac{1}{H} \, \mathbf{Y}_{\!\lambda}
\end{equation}
instead of the Young projectors $\mathbf{Y}_{\!\lambda}$ themselves. They are specified only by the Young diagram and not by the standard tableau; the $\lambda$ we use in \eqref{eq:defP} is the canonical one with numbers $1$ to $h_1$ in the first column, $h_1 + 1$ to $h_1 + h_2$ in the second column, and so on. For example, we take
\begin{equation}
\lambda = \ytableaushort{13,24}
\end{equation}
to define $\mathbb{P}_{(2,2)}$ (other choices yield equivalent representations). Furthermore, the division by $H$ ensures that they are idempotent (see \eqref{eq:norm}),
\begin{equation}
\mathbb{P}_Y \circ \mathbb{P}_Y = \mathbb{P}_Y \, .
\end{equation}

\item We are using the convention where the antisymmetrization is done last, so that the antisymmetry in the indices of the columns is manifest. One could also take the opposite convention, with manifest antisymmetry, with equivalent results. In the first definition of $V_\lambda$, this corresponds to writing
\begin{equation}
\mathbf{Y}_{\!\lambda} = s_\lambda \circ a_\lambda .
\end{equation}
In the second definition, the criteria become
\begin{enumerate}[(a)]
\item the tensors are totally symmetric in the indices corresponding to a row of the Young tableau; and
\item any symmetrization over all the indices of a row, plus one index belonging to another row below it, vanishes.
\end{enumerate}
The only place where we use the symmetric convention is in section \ref{sec:parametrization}.
\end{itemize}

\subsection{Generalized Poincaré lemmas}
\label{app:poincare}

We now review the Poincaré lemmas of \cite{Olver_hyper,DuboisViolette:1999rd,DuboisViolette:2001jk,Bekaert:2002dt} in the case of two-column Young diagrams.

\subsubsection{Rectangular case}

We first consider the space of tensors where the height of the two columns differ at most by one. A differential $d$ is introduced on this space by putting exterior derivatives ``column by column", for example
\begin{equation}
\ydiagram{2,1} \overset{d}{\longrightarrow} \ytableaushort{{}{},{} \pd} \overset{d}{\longrightarrow} \ytableaushort{{}{},{} \pd,\pd} \overset{d}{\longrightarrow} \ytableaushort{{}{},{} \pd, \pd \pd} \overset{d}{\longrightarrow} \dots
\end{equation}
when starting with a $(2,1)$-tensor. In components, this corresponds to taking an antisymmetrized derivative on the indices of the corresponding column, then acting with the appropriate Young projector. Explicitely, this gives
\begin{equation}
(d R)_{\mu_1 \dots \mu_{p+1} \nu_1 \dots \nu_p} = \pd_{[\mu_1} R_{\mu_2 \dots \mu_{p+1}] \nu_1 \dots \nu_p}\, ,
\end{equation}
on a rectangular $[p,p]$-tensor (with columns of equal height), and
\begin{equation}
(d T)_{\mu_1 \dots \mu_p \nu_1 \dots \nu_p} = \frac{1}{2} \left( T_{\mu_1 \dots \mu_{p} [\nu_1 \dots \nu_{p-1}, \nu_p]} + T_{\nu_1 \dots \nu_{p} [\mu_1 \dots \mu_{p-1}, \mu_p]} \right)
\end{equation}
on a $[p, p-1]$-tensor, where the comma is the derivative.

This differential does not square to zero but \emph{cubes} to zero instead,
\begin{equation}
d^3 = 0 \, .
\end{equation}
This is because putting two derivatives in the same column gives automatically zero, since indices in the same column are always antisymmetrized. The generalized Poincaré lemma now states that the generalized cohomology associated to $d$ is trivial in the space of rectangular Young tableaux only, i.e., the implications
\begin{align}
dR \quad &\Rightarrow \quad \exists\, h \text{ such that } R = d^2 h, \label{eq:poincare} \\
d^2 R \quad &\Rightarrow \quad \exists\, T \text{ such that } R = d T \label{eq:poincaresquare}
\end{align}
are valid when $R$ has columns of equal height. If $R$ is a $[p,p]$-tensor, then $h$ and $T$ are $[p-1,p-1]$ and $[p,p-1]$-tensors respectively.

Let us give two examples in the context of linearized gravity. Theorem \eqref{eq:poincare} exactly states that any $[2,2]$-tensor $R_{\mu\nu\rho\sigma}$ satisfying the differential Bianchi identity is the linearized Riemann tensor of some metric,
\begin{equation}
\pd_{[\mu} R_{\nu\rho]\sigma\tau} = 0 \quad \Rightarrow \quad R\indices{^{\mu\nu}_{\rho\sigma}} = \pd^{[\mu} \pd_{[\rho} h\indices{^{\nu]}_{\sigma]}}
\end{equation}
for some symmetric tensor field $h_{\mu\nu}$.
Theorem \eqref{eq:poincaresquare} states that, if the linearized Riemann tensor vanishes, then the linearized metric $h_{\mu\nu}$ is pure gauge,
\begin{equation}
\pd^{[\mu} \pd_{[\rho} h\indices{^{\nu]}_{\sigma]}} = 0 \quad \Rightarrow \quad h_{\mu\nu} = \pd_{(\mu} \xi_{\nu)}
\end{equation}
for some vector field $\xi_\mu$.

\subsubsection{Non-rectangular case}

We now indicate the case of non-rectangular Young diagrams. There are now two differentials $d_1$ and $d_2$ acting separately on the two columns, for example
\begin{equation}
\ydiagram{2,1,1} \overset{d_1}{\longrightarrow} \ytableaushort{{}{},{},{},\pd} \overset{d_1}{\longrightarrow} \ytableaushort{{}{},{} ,{},\pd,\pd} \overset{d_1}{\longrightarrow} \dots \quad\text{ and }\quad
\ydiagram{2,1,1} \overset{d_2}{\longrightarrow} \ytableaushort{{}{},{} \pd,{}} \overset{d_2}{\longrightarrow} \ytableaushort{{}{},{} \pd,{} \pd}
\end{equation}
on a $[3,1]$ Young tableau ($d_2$ is not defined on tensors with columns of equal height). As before, these are defined by taking the curl on the appropriate group of indices and then acting with a Young projector. Now, each of these differentials square to zero,
\begin{equation}
(d_1)^2 = 0 = (d_2)^2 .
\end{equation}
The relevant Poincaré lemmas are, in this case,
\begin{align}
d_1 d_2 T = 0 \quad &\Rightarrow \quad \exists\, \alpha, \, \beta \text{ such that } T = d_1 \alpha + d_2 \beta , \\
d_1 T = 0, \, d_2 T = 0 \quad &\Rightarrow \quad \exists\, \rho \text{ such that } T = d_1 d_2 \rho ,
\end{align}
where the tensor fields of the right-hand side have the appropriate symmetry: if $T$ is a $[p, q]$ tensor, then $\alpha$, $\beta$, $\rho$ are $[p-1,q]$, $[p,q-1]$ and $[p-1,q-1]$ tensors respectively.

As an application, these theorems make the gauge theory of a $(2,1)$-tensor very clear:
\begin{enumerate}
\item the curvature tensor is zero if and only if the field is pure gauge,
\begin{equation}
\pd^{[\mu_1} \pd_{[\nu_1} T\indices{^{\mu_2 \mu_3]}_{\nu_2]}} = 0 \quad \Leftrightarrow \quad T_{\mu\nu\rho} = \pd_{[\mu} \alpha_{\nu]\rho} + \pd_{[\mu} \beta_{\nu]\rho} - \pd_\rho \beta_{\mu\nu}\, ,
\end{equation}
where $\alpha$, $\beta$ are symmetric and antisymmetric respectively;
\item any $[3,2]$-tensor $E$ satisfying the differential Bianchi identities of the curvature\footnote{A technical comment is necessary here: the second identity, $E_{\mu_1\mu_2\mu_3[\nu_1\nu_2,\nu_3]} = 0$, is not exactly the equation $d_2 E = 0$ since the left-hand side does not have the $[3,3]$ symmetry (an extra Young projection would be needed). Nevertheless, it has the $[3,3]$ symmetry once the first Bianchi identity $\pd_{[\mu_1} E_{\mu_2\mu_3\mu_4]\nu_1\nu_2} = 0$ is used. Therefore, $d_1 E = 0$ and $d_2 E = 0$ are indeed equivalent to the simple curls appearing in \eqref{eq:bianchi21app}, with no extra Young projection. (The tensor $\pd_{[\mu_1} E_{\mu_2\mu_3\mu_4]\nu_1\nu_2}$ has the $[4,2]$ symmetry, so the first differential Bianchi identity is really $d_1 E = 0$.)} is really the curvature of some $[2,1]$-tensor $T$,
\begin{equation}\label{eq:bianchi21app}
\pd_{[\mu_1} E_{\mu_2\mu_3\mu_4]\nu_1\nu_2} = 0, \; E_{\mu_1\mu_2\mu_3[\nu_1\nu_2,\nu_3]} = 0 \quad \Leftrightarrow \quad E\indices{^{\mu_1\mu_2\mu_3}_{\nu_1\nu_2}} = \pd^{[\mu_1} \pd_{[\nu_1} T\indices{^{\mu_2 \mu_3]}_{\nu_2]}} .
\end{equation}
\end{enumerate}

\section{Cotton tensors of linearized supergravity}
\label{app:cotton}

In this appendix, we prove the two important properties of the Cotton tensors for the graviton and gravitino fields presented in chapter \ref{chap:twisted}. The proofs of those properties for the analogue Cotton tensors of chapter \ref{chap:selfdual} follows exactly the same pattern; we will therefore not repeat the proofs in those cases.

\subsection{Gauge and Weyl invariance}

The first property is that the Cotton tensor completely captures gauge and Weyl invariance.

\paragraph{Bosonic $[d-2,1]$ field.}

We want to prove the implication
\begin{equation}
D_{i_1 \dots i_{d-2} j_1 \dots j_{d-2}}[\phi] = 0 \;\Rightarrow\; \phi\indices{^{i_1 \dots i_{d-2}}_{j} } = \text{(gauge)} + \delta^{[i_1}_j B^{i_2 \dots i_{d-2}]} \;\text{ for some } B.
\end{equation}
(The opposite implication is true by construction.) The proof goes in two steps:
\begin{enumerate}
\item First, $D = 0$ implies that the curl of the Schouten tensor on both groups of indices vanishes,
\begin{equation}
\partial^{[i_1} S\indices{^{i_2 \dots i_{d-1}]}_j} = 0, \qquad \partial_{[j} S\indices{^{i_1 \dots i_{d-2}}_{k]}} = 0 .
\end{equation}
Indeed, the second of those equations follows from the definition of the Cotton tensor, and the first from the relation \eqref{eq:dprimephi} between this curl and the trace of the Cotton. The generalized Poincaré lemma of \cite{Bekaert:2002dt} then implies that $S$ is itself a double curl, i.e.
\begin{equation}
S\indices{^{i_1 \dots i_{d-2}}_{j}} [\phi] = - (d-2)!\,\partial_j\partial^{[i_1} B^{i_2 \dots i_{d-2}]}
\end{equation}
for some antisymmetric $B$. This is precisely the variation of the Schouten induced by a Weyl transformation of $\phi$.
\item Then, using the inversion relations between the Einstein and Cotton tensors, this implies that
\begin{equation}
G\indices{^{i_1 \dots i_{d-2}}_{j}} [\phi - \delta B] = \partial^k \partial_m \left( \phi\indices{^{l_1 \dots l_{d-2}}_{n} } - \delta^{[i_1}_j B^{i_2 \dots i_{d-2}]} \right) \varepsilon^{mni_1 \dots i_{d-2}} \varepsilon_{jkl_1 \dots l_{d-2}} = 0
\end{equation}
where $\delta B$ stands for the tensor $\delta^{[i_1}_j B^{i_2 \dots i_{d-2}]}$. The Poincaré lemma of \cite{Bekaert:2002dt} (property \eqref{eq:gaugeEinsteinphi} of the Einstein tensor) now implies that $\phi - \delta B$ is pure gauge, which is what we wanted to prove.
\end{enumerate}

\paragraph{Bosonic $[d-2,d-2]$ field.} The implication to be proved is now
\begin{equation}
D\indices{^{i_1 \dots i_{d-2}}_{j}}[P] = 0 \;\Rightarrow\; P\indices{^{i_1 \dots i_{d-2}}_{j_1 \dots j_{d-2}} } = \text{(gauge)} + \delta^{i_1 \dots i_{d-2}}_{j_1 \dots j_{d-2}} \,\xi \;\text{ for some } \xi.
\end{equation}
The proof goes as before:
\begin{enumerate}
\item First, $D=0$ implies that the curl of $S_{ij}[P]$ vanishes,
\begin{equation}
\partial_{[i} S_{j]k} = 0.
\end{equation}
The Poincaré lemma of \cite{DuboisViolette:1999rd,DuboisViolette:2001jk} applied to symmetric tensors implies then that
\begin{equation}
S_{ij}[P] = -(d-2)! \pd_i \pd_j \xi
\end{equation}
for some $\xi$, which is again the form induced by a Weyl transformation of $\xi$.
\item Then, this implies that the Einstein tensor of $P - \delta^{d-2} \xi$ vanishes, where $\delta^{d-2} \xi$ stands for $\delta^{i_1 \dots i_{d-2}}_{j_1 \dots j_{d-2}} \,\xi$. The combination $P - \delta^{d-2} \xi$ is therefore pure gauge, which proves the proposition.
\end{enumerate}

\paragraph{Fermionic $[d-2]$ field.} We now should prove
\begin{equation}
D_{i_1 \dots i_{d-2}}[\chi] = 0 \;\Rightarrow\; \chi_{i_1 \dots i_{d-2}} =  \text{(gauge)} + \gamma_{i_1 \dots i_{d-2}} \rho \;\text{ for some } \rho.
\end{equation}
We take the same steps:
\begin{enumerate}
\item First, $D=0$ is equivalent to $\partial_{[i} S_{j]}[\chi] = 0$, which by the usual Poincaré lemma with a spectator spinorial index implies
\begin{equation}
S_i[\chi] = \pd_i \nu
\end{equation}
for some $\nu$ that can always be written as $\nu = i^{m+1} (d-2)! \gamma_0 \hat{\gamma}\, \rho$. This is the form of the Schouten induced by a Weyl transformation of $\chi$.
\item This implies that the Einstein tensor (curl) of $\chi_{i_1 \dots i_{d-2}} - \gamma_{i_1 \dots i_{d-2}} \rho$ is zero. That combination is therefore pure gauge, which is what had to be proven.
\end{enumerate}

\subsection{Conformal Poincaré lemmas}

The second property is a ``conformal version" of the Poincaré lemma: any divergenceless tensor field with the same symmetry and trace properties as the Cotton tensor can always be written as the Cotton tensor of the appropriate field.

\paragraph{Bosonic $[d-2,1]$ field.}

We need to prove that, for any $[d-2, d-2]$ tensor $T_{i_1 \dots i_{d-2} j_1 \dots j_{d-2}}$ that is divergenceless and completely traceless, there exists a $[d-2,1]$ field $\phi\indices{^{i_1 \dots i_{d-2}}_{j} }$ such that $T = D[\phi]$. Moreover, $\phi$ is determined from this condition up to gauge and Weyl transformations of the form \eqref{eq:phigaugeapp}.
\begin{enumerate}

\item First, the divergencelessness of $T$ and the usual Poincaré lemma implies that $T$ can be written as the curl of some other tensor $U$,
\begin{equation}\label{eq:TofU}
T^{i_1 \dots i_{d-2}  j_1 \dots j_{d-2}} = \varepsilon^{i_1 \dots i_{d-2}kl} \partial_k U\indices{^{j_1 \dots j_{d-2}}_{l}} .
\end{equation}
The tensor $U$ is determined up to $U\indices{^{j_1 \dots j_{d-2}}_{l}} \rightarrow U\indices{^{j_1 \dots j_{d-2}}_{l}} + \pd_l V^{j_1 \dots j_{d-2}}$, where $V$ is totally antisymmetric.

\item At this stage, $U$ is not of irreducible Young symmetry $[d-2,1]$, but could have a completely antisymmetric component. However, the condition that $T$ is completely traceless implies that this component satisfies $\pd_{[k} U_{j_1 \dots j_{d-2} l]} = 0$, i.e., is of the form $U_{[j_1 \dots j_{d-2} l]} = \pd_{[l} W_{j_1 \dots j_{d-2}]}$ for some antisymmetric $W$. Now, the ambiguity in $U$ described above can be used to precisely cancel this contribution. We can therefore assume the $U$ has the $[d-2,1]$ symmetry from now on.

\item We now use the fact that $T$ has the $[d-2,d-2]$ symmetry. This implies that
\begin{equation}
\varepsilon_{i_1 \dots i_{d-2} j_1 m} T^{i_1 \dots i_{d-2}  j_1 \dots j_{d-2}} = 0
\end{equation}
which, using the expression \eqref{eq:TofU}, is equivalent to the differential identity
\begin{equation}\label{eq:divU}
\partial_{i_1} U\indices{^{i_1 \dots i_{d-2}}_{j}} [\phi] - \partial_j U^{i_2 \dots i_{d-2}} = 0
\end{equation}
on $U$. We now define another tensor $X$ of $[d-2,1]$ symmetry by the equation
\begin{equation}
X\indices{^{i_1 \dots i_{d-2}}_{j}} = U\indices{^{i_1 \dots i_{d-2}}_{j}} - (d-2) \,\delta^{[i_1}_j U\indices{^{i_2 \dots i_{d-2}]k}_k} .
\end{equation}
The identity \eqref{eq:divU} is then equivalent to the fact that $X$ is divergenceless.

\item We are now ready to conclude. Because $X$ is a divergenceless $[d-2,1]$ tensor, it must be the Einstein tensor of some $[d-2,1]$ field $\phi$, $X = G[\phi]$. The relation between $U$ and $X$ then implies that $U$ is the Schouten tensor of $\phi$, $U = S[\phi]$, and equation \eqref{eq:TofU} gives at last $T = D[\phi]$, which was to be proven.

\item The ambiguity in $\phi$ is easily determined from the first property of the Cotton tensor. Indeed, by linearity, the condition $D[\phi'] = D[\phi]$ is equivalent to $D[\phi'-\phi] = 0$. The first property of the Cotton then implies that $\phi' - \phi$ takes the form ``Weyl $+$ gauge transformation".
\end{enumerate}

\paragraph{Bosonic $[d-2,d-2]$ field.}

For the $[d-2,d-2]$ field, we now prove that for any divergenceless and traceless $[d-2, 1]$ tensor $T_{i_1 \dots i_{d-2} j}$, there exists a $[d-2,d-2]$ field $P_{i_1 \dots i_{d-2} j_1 \dots j_{d-2}}$ such that $T = D[P]$. This field $P$ is determined from this condition up to a gauge and Weyl transformation of the form \eqref{eq:Pgaugeapp}. The proof goes as before:

\begin{enumerate}

\item First, the divergencelessness of $T$ implies that it can be written as a curl,
\begin{equation}\label{eq:TofUP}
T^{i_1 \dots i_{d-2}  j} = \varepsilon^{i_1 \dots i_{d-2}kl} \partial_k U\indices{^{j}_{l}} .
\end{equation}
The tensor $U$ is determined up to $U\indices{^{j}_{l}} \rightarrow U\indices{^{j}_{l}} + \pd_l V^{j}$.

\item The tensor $U$ is not necessarily symmetric. However, the condition that $T$ is traceless implies that the antisymmetric component satisfies $\pd_{[k} U_{jl]} = 0$, i.e., is of the form $U_{[jl]} = \pd_{[l} W_{j]}$ for some vector $W$. By using $V$, we can precisely cancel this contribution and we can assume that $U$ is symmetric.

\item We now use the fact that $T$ satisfies $T_{[i_1 \dots i_{d-2}j]}=0$. This is equivalent to
\begin{equation}
\pd^i U_{ij} - \pd_j U\indices{_k^k} = 0.
\end{equation}
Defining the symmetric tensor $X$ by
\begin{equation}
X_{ij} = U_{ij} - \delta_{ij} U\indices{_k^k},
\end{equation}
that identity is equivalent to the fact that $X$ is divergenceless.

\item We can now conclude: because $X$ is a symmetric divergenceless tensor, it must be the Einstein tensor of some $(d-2,d-2)$ field $P$, $X = G[P]$. The relation between $U$ and $X$ then implies that $U$ is the Schouten tensor of $P$, $U = S[P]$, and equation \eqref{eq:TofUP} gives $T = D[\phi]$.

\item The ambiguity in $P$ is given by the first property of the Cotton tensor. Indeed, the condition $D[P'] = D[P]$ is equivalent to $D[P'-P] = 0$, which shows that $P'$ and $P$ differ by a gauge and Weyl transformation.
\end{enumerate}

\paragraph{Fermionic $(d-2)$-form.}

We finish this appendix by proving the analogous property for the fermionic antisymmetric field, namely: For any antisymmetric, divergenceless and completely gamma-traceless rank $d-2$ tensor-spinor $T_{i_1 \dots i_{d-2}}$, there exists an antisymmetric tensor-spinor $\chi_{i_1 \dots i_{d-2}}$ such that $T = D[\chi]$. Moreover, $\chi$ is determined from this condition up to gauge and Weyl transformations of the form \eqref{eq:chigaugeapp}. We follow the same steps as before, but without the complications of Young symmetries:
\begin{enumerate}
\item The divergencelessness of $T$ implies that it can be written as
\begin{equation}\label{eq:TofUchi}
T^{i_1 \dots i_{d-2}} = \varepsilon^{i_1 \dots i_{d-2}kl} \partial_k U_l,
\end{equation}
where $U$ is determined up to a total derivative $\pd_l V$ for some spinor field $V$.

\item Using the gamma matrix identity \eqref{eq:gij}, the complete gamma-tracelessness
\begin{equation}
\gamma^{i_1 \dots i_{d-2}} T_{i_1 \dots i_{d-2}} = 0    
\end{equation}
of $T$ is equivalent to the differential equation $\gamma_{ij} \pd^i U^j = 0$, which is in turn equivalent to $\pd^i X_i = 0$ if we define $X_i = \gamma_{ij} S^j$.

\item Because $X$ is a divergenceless vector-spinor, it is the Einstein tensor of some antisymmetric tensor-spinor $\chi$, $X = G[\chi]$. As before, the relation between $U$ and $X$ then implies that $U$ is the Schouten tensor of $\chi$, $U = S[P]$, and the relation between $T$ and $U$ then gives $T = D[\phi]$.

\item The ambiguity in $\chi$ is given by the first property of the Cotton tensor: $D[\chi'] = D[\chi]$ is equivalent to $D[\chi'-\chi] = 0$, which shows that $\chi'$ and $\chi$ differ by a transformation of the form \eqref{eq:chigaugeapp}.
\end{enumerate}

\section{Equations of motion of chiral tensors}

In this appendix, we prove the equivalences between the different forms of the self-duality equations for the $(2,2)$ and $(2,1)$ tensors given in the main text. They rely on the generalized Poincaré lemmas reviewed in appendix \ref{app:poincare}.

\subsection{The $(2,2)$-tensor}
\label{app:eom}

\subsubsection{First step: $R = * R \Leftrightarrow \cE = \cB$}

In components, the self-duality equation $R = * R$ reads
\begin{align}
R_{0ijklm} &= \frac{1}{3!} \varepsilon\indices{_{ij}^{abc}}  R_{abcklm} \label{R0} \\
R_{0ij0kl} &= \frac{1}{3!} \varepsilon\indices{_{ij}^{abc}}  R_{abc0kl} \label{R00}.
\end{align}
The first of these equations is equivalent to $\cE = \cB$, \eqref{E=B22}, by dualizing on the $klm$ indices. Conversely,  we must show that $\cE = \cB$ implies \eqref{R00} or, equivalently, that \eqref{R0} implies \eqref{R00}. To do so, we use the Bianchi identity $\partial_{[\alpha_1} R_{\alpha_2\alpha_3\alpha_4]\beta_1 \beta_2 \beta_3} = 0$ on the curvature, which imples
\begin{equation}
\partial_0 R_{ijk \beta_1 \beta_2 \beta_3} = 3 \partial_{[i} R_{jk]0 \beta_1 \beta_2 \beta_3} .
\end{equation}
Therefore, taking the time derivative of equation \eqref{R0} gives
\begin{equation}
\partial_{[k} R_{lm]00ij} = \frac{1}{3!} \partial_{[k} R_{lm]0 abc} \varepsilon\indices{_{ij}^{abc}}, \label{curlR00}
\end{equation}
which is exactly the curl of \eqref{R00}. Now, the tensor $R_{0lm0ij}$ has the $(2,2)$ symmetry, and so does $\frac{1}{3!} R_{0lm abc} \varepsilon\indices{_{ij}^{abc}} = \mathcal{B}_{lmij}$ because of $\cE = \cB$ and the fact that $\mathcal{E}$ has the $(2,2)$ symmetry.

Using the generalized Poincaré lemma for rectangular Young tableaux, one recovers equation \eqref{R00} up to a term of the form $\partial_{[i} N_{j][k,l]}$ for $N_{jk}$ symmetric. This term can be absorbed in a redefinition of the $T_{0j0k}$ components appearing in $R_{0ij0kl}$. (In fact, the components $T_{0j0k}$ drop from equation \eqref{curlR00}, and this explains how one can get equation \eqref{R00} from \eqref{R0}, which does not contain $T_{0j0k}$ either.)

\subsubsection{Second step: $\cE = \cB \Leftrightarrow \curl(\cE - \cB)=0$ and $\bar{\cE}=0$}

Equation $\cE = \cB$ obviously implies \eqref{E2=B2}. It also implies \eqref{trace0} because the magnetic field $\mathcal{B}$ is identically traceless. To prove the converse, we introduce the tensor
\be
K_{ijk lm} = \varepsilon\indices{_{ijk}^{ab}} (\mathcal{E} - \mathcal{B})_{lmab} \, .
\ee
Equation $\bar{\cE}=0$ and the fact that $\mathcal{B}$ is traceless imply that $K$ has the $(2,2,1)$ symmetry, $K \sim \tyng{2,2,1}\,$. Equation \eqref{E2=B2} states that the curl of $K$ on its second group of indices vanishes, 
\be
K_{ijk[lm,n]} = 0 \, .
\ee
The explicit formula
\begin{equation}
K_{ijk lm} = \frac{1}{3} \left(\varepsilon_{lmpqr} \partial^{p} T\indices{^{qr}_{[ij,k]}} - \partial_{[0} T_{lm][ij,k]} \right)
\end{equation}
shows that the curl of $K$ on its first group of indices also vanishes, $\partial_{[i} K_{jkl]mn} = 0$. Using the generalized Poincaré lemma of \cite{Bekaert:2002dt} for arbitrary Young tableaux, this implies that
\be K_{ijklm} = \partial_{[i} \lambda_{jk][l,m]}\, ,\ee
where $\lambda_{jkl}$ is a tensor with the $(2,1)$ symmetry that can be absorbed in a redefinition of $T_{0ijk}$. (Similarly to the previous case, those components actually drop from the curl \eqref{E2=B2}.) One finally recovers equation $\cE = \cB$ by dualizing again $K$ on its first group of indices.

\subsection{The \texorpdfstring{$(2,1)$}{(2,1)}-tensor}
\label{app:A}

The case of the $(2,1)$-tensor is a bit more involved because it is not rectangular.

\subsubsection{First step: \texorpdfstring{$R = * R \Leftrightarrow {\mathcal E}={\mathcal B}$}{R = * R iff E=B}}
\label{app:dem1}

The self-dual equations of motion $R = * R$, see \eqref{self-duality}, can be split as
\begin{align}
R_{0 ij  kl} &= \frac{1}{3!}\varepsilon_{ij abc}R\indices{^{abc }_{kl}} \, , \label{eqb} \\
R_{0ij0k}&=\frac{1}{3!}\varepsilon_{ijabc}R\indices{^{abc}_{0k}} \, . \label{eqc}
\end{align}
Due to the Young symmetries of the Riemann tensor
the component $R_{abc0d}$ is not independent.
Explicitly, we can use the identity $R_{[\mu \nu \rho\sigma]\tau}=0$ to show $R_{\mu \nu \rho \sigma \tau}= 3 R_{\sigma[\mu \nu  \rho] \tau}$, from which
\begin{equation}
  \label{eq:Rconn}
   R_{abc 0d}=3 R_{0[abc]d}
\end{equation}
follows.
Contracting \eqref{eqb} with $\frac{1}{2!}\varepsilon^{klpqr}$
and
using the definitions of the electric and magnetic fields
we obtain
\begin{align}\label{eq:BEfoll}
  \frac{1}{2!}\varepsilon^{klpqr}R_{0 ij kl}
  =
      \frac{1}{2!3!}\varepsilon^{klpqr}\varepsilon_{ij abc}R\indices{^{abc}_{kl}}
  \qquad \Rightarrow \qquad 
  {\mathcal B}^{pqrij} ={\mathcal E}^{pqrij} \,.
\end{align}
So, ${\mathcal E}={\mathcal B}$ follows from $R = * R$.

In order to prove the converse, we must show that ${\mathcal E}={\mathcal B}$
implies \eqref{eqb} and \eqref{eqc}. 
Equation \eqref{eqb} follows from reversing the argument just given: it only remains to show that \eqref{eqc} follows from \eqref{eq:BEfoll} or, equivalently, \eqref{eqb}.
The problem is that \eqref{eqb} does not contain the $\phi_{i00}$ components, while \eqref{eqc} does. Those components appear in \eqref{eqc} with two spatial derivatives, i.e., in the form $\pd_{k}\pd_{[i}\phi_{j]00}$ (up to some factor that will not matter).

We will need to use the generalized Poincaré lemma for a tensor $T_{ijk} \sim \tyng{2,1}$.
It states that, if the curl 
on the first and second set of indices vanishes,
then the tensor $T_{ijk}$  has to be proportional to
$\pd_{k}\pd_{[i}\lambda_{j]}$ for some vector $\lambda_j$.
The role of $T_{ijk}$ will be played by \eqref{eqc}
\begin{align}
  T_{ijk} =R_{0ij0k}-\frac{1}{3!}\varepsilon_{ijabc}R\indices{^{abc}_{0k}}\, ,
\end{align}
which indeed has the right Young symmetries. It is now sufficient to show that \eqref{eqb} implies that the curl of $T_{ijk}$ on both groups of indices are zero: we then recover \eqref{eqc} up to a term of the form $\pd_{k}\pd_{[i}\lambda_{j]}$ which can be absorbed in a redefinition of $\phi_{i00}$.

Let us first show that the curl of $T_{ijk}$ on the second group of indices vanishes.
For that 
we take the time derivative of \eqref{eqb} which leads to
\begin{equation}
\pd_0 R_{0ijkl} = \frac{1}{3!}\varepsilon_{ijabc} \pd_0 R\indices{^{abc}_{kl}} \,.
\end{equation}
Next we use a consequence of the Bianchi identity~\eqref{Bianchi},
$R_{0ij kl ,0} = -2 R_{0ij0[k,l]}$,
to show that 
\begin{equation}
R_{0ij0[k,l]} = \frac{1}{3!}\varepsilon_{ijabc} R\indices{^{abc}_{0[k,l]}}\, .
\end{equation}
This is exactly the curl of \eqref{eqc} on the second group of indices.

Now, we would like to have a curl on the first group of indices.
For that, let us take again
the time derivative of \eqref{eqb}
and antisymmetrize the indices $[ijk]$
\begin{align}
 \pd_{0} R_{0 [ij k]l} =  \frac{1}{3!}\varepsilon_{abc[ij|}\pd_{0}R\indices{^{abc }_{|k]l}} \,.
\end{align}
On the left-hand side we
use \eqref{eq:Rconn} and 
the consequence $\pd_0 R_{ijk 0l} = 3 \pd_{[i}R_{jk]00l}$ of the Bianchi identity.
The right hand side can also be worked out as
\begin{equation}
 \varepsilon_{[ij | abc}\pd_{0}R\indices{^{abc }_{|k]l}} = \pd_{[i}\varepsilon_{jk]abc} R\indices{^{abc}_{0l}}- \pd_{l} \varepsilon_{[ij|abc} R\indices{^{abc}_{0|k]}} \, ,
\end{equation}
where we have used again a Bianchi identity
$R_{ij klm ,0} = -2 R_{ijk0[l,m]}$.
It is easy to check that the second term on the right-hand side vanishes on-shell.
Indeed, we take \eqref{eqb}, antisymmetrize the indices $[ijk]$, and use \eqref{eq:Rconn} to get $R\indices{^{abc}_{0k}} = \frac{1}{2}\varepsilon^{[ab|def}R\indices{_{def}^{|c]}_{k}}$. 
We now plug this into the second term to show
\begin{equation}
  \varepsilon_{[ij|abc}R\indices{^{abc}_{0|k]}}
  \sim
  \varepsilon_{[ij|abc}\varepsilon^{abdef} R\indices{_{def}^{c}_{|k]}}
  \sim
  R\indices{_{[ij|c}^{c}_{|k]}}=0 \,. 
\end{equation}
The last equality holds because the trace $R\indices{_{ijc}^{c}_{k}}$ has the $(2,1)$ Young symmetry.
So we finally arrive at
\begin{equation}
  \pd_{[i} R_{ j k]00l}
  =
      \frac{1}{3!}\pd_{[i}\varepsilon_{jk]abc} R\indices{^{abc}_{0l}}\,,
\end{equation}
which is the curl of \eqref{eqc} on the first group of indices.

As anticipated, we can now use
the generalized Poincaré lemma of \cite{Bekaert:2002dt}
to get
\begin{equation}
\label{eq:eqcprop}
R_{0ij0k}-\frac{1}{3!}\varepsilon_{ijabc}R\indices{^{abc}_{0k}} =\pd_{k}\pd_{[i}\lambda_{j]} \,.
\end{equation}
The $\pd_{k}\pd_{[i}\lambda_{j]}$ terms can be absorbed
by redefining the $\phi_{i00}$ terms of $R_{0ij0k}$.

\subsubsection{Second step: $\cE = \cB \Leftrightarrow \curl_{2}(\cE - \cB)=0$ and $\bbar{\cE}=0$}
\label{app:dem2}

Taking the curl of $\cE = \cB$ on the second pair of indices suffices to show that 
\begin{equation}\label{eq:curlEB}
\curl_{2}({\mathcal E}-{\mathcal B})=\varepsilon_{abcpq}\pd^{a}({\mathcal E}\indices{_{ijk}^{bc}}-{\mathcal B}\indices{_{ijk}^{bc}})=0 \,.
\end{equation}
The double-tracelessness
of the electric field, $\bar{\bar{{\mathcal E}}}=0$, also follows from $\cE = \cB$ because the magnetic field ${\mathcal B}$ is identically double-traceless.

To prove the converse, we introduce the tensor
\begin{equation}\label{eq:defK}
K_{ab cd} = \varepsilon\indices{^{ijk}_{cd}}({\mathcal E}-{\mathcal B})_{ijkab}\,.
\end{equation}
We have to show from \eqref{eq:curlEB} that the tensor $K$ can be written as $K_{abcd} = \pd_{[a} M_{b][c,d]}$. Indeed, this is the way in which the missing components $\phi_{0ij}$ appear in \eqref{eq:defK}: this will therefore imply $\mathcal E = \mathcal B$ up to a redefinition of the $\phi_{0ij}$. Again, the proof of this fact relies on generalized Poincaré lemmas.

The symmetries of tensor $K$ are such that 
$K_{abcd}=K_{[ab][cd]}$ and, 
because  ${\mathcal E}$ and ${\mathcal B}$ are double-traceless,
$K_{[abcd]}=0$. 
Moreover, it follows from its definition and \eqref{eq:curlEB} that its curl on the first and second group of indices vanishes,
\begin{eqnarray}
\pd_{[i}K_{ab]cd}=0 \, , \label{curlK1}\\
K_{ab[cd,i]}=0 \, . \label{curlK2}
\end{eqnarray}

Following Appendix C of \cite{Bunster:2013oaa}, we first decompose the tensor
$K \sim \ydiagram{1,1} \otimes \ydiagram{1,1}$ into three parts of irreducible Young symmetry,
\begin{equation}
K_{abcd} = R_{abcd} + (Q_{abcd} - Q_{abdc}) + A_{abcd}\, ,
\end{equation}
where
\begin{align}
R_{abcd}&=\frac{1}{2}(K_{abcd} + K_{cdab}) \sim \ydiagram{2,2} \,, \\
Q_{abcd} &= \frac{3}{2} K_{[abc]d} \sim  \ydiagram{2,1,1} \,, \\
A_{abcd}&= K_{[abcd]}=0 \sim  \ydiagram{1,1,1,1}
\end{align}
($R$ is not the Riemann tensor here: context should be clear in the following).
This permits us to use generalized Poincaré lemmas on each of the irreducible parts.
We start  with $\pd_{[i}R_{ab]cd}=0$,
which follows by using \eqref{curlK1} and \eqref{curlK2},
\begin{align}
\pd_{[i}R_{ab]cd} = \frac{1}{2}(\pd_{[i}K_{ab]cd} + K_{cd[ab,i]}) =0 \, . \label{dR=0}
\end{align}
This implies that $R_{abcd}$ can be written as 
\begin{equation}
R_{abcd} = \pd_{[a}S_{b][c,d]} \sim \ytableaushort{{}{},\pd\pd}\, , 
\end{equation}
where $S_{ab} \sim \tyng{2}$ is a symmetric tensor.
For $Q_{abcd}$, the equation $\pd_{[i}Q_{abc]d}=0$ follows directly from \eqref{curlK1}. To prove that also $Q_{abc[d,i]}=0$,
let us use the fact that the curl of $K$ in the second group of indices vanishes,
\begin{equation}
0=K_{ab[cd,i]} = \frac{1}{3}(K_{abcd,i}+ K_{abdi,c}+K_{abic,d}) = \frac{1}{3}(2 K_{abc[d,i]}+K_{abdi,c}) .
\end{equation}
This shows that $K_{abc[d,i]}=-\frac{1}{2}K_{abdi,c}$. Finally,
\begin{align}
Q_{abc[d,i]} &= \frac{3}{2} K_{[abc][d,i]} = \frac{1}{2}(K_{abc[d,i]}+K_{bca[d,i]}+K_{cab[d,i]}) \nonumber \\
            &= -\frac{1}{4}(\pd_{c} K_{abdi}+\pd_{a}K_{bcdi}+\pd_{b}K_{cadi}) = -\frac{3}{4}\pd_{[a}K_{bc]di} \nn \\
            &=0\, ,
\end{align}
which vanishes because of \eqref{curlK1}. This implies
\begin{equation}
Q_{abcd}=\frac{3}{4}\pd_{d}\pd_{[a}A_{bc]} \sim \ytableaushort{{}\pd,{},\pd}\, ,
\end{equation}
where $A_{ab}\sim \tyng{1,1}$ is an antisymmetric tensor. In the split of $K_{abcd}$ into irreducible parts, we have the combination
\begin{align}
Q_{abcd}-Q_{abdc}&= \frac{3}{4}\pd_{d}\pd_{[a}A_{bc]}-\frac{3}{4}\pd_{c}\pd_{[a}A_{bd]}\nonumber \\
                 &= \pd_{[a}A_{b][c,d]} \, .
\end{align}
Therefore, the explicit form of $K_{abcd}$ is given by
\begin{align}
K_{abcd}&= R_{abcd} + (Q_{abcd} - Q_{abde}) + 0 = \pd_{[a}S_{b][c,d]} +\pd_{[a}A_{b][c,d]} \nonumber \\
        &= \pd_{[a}M_{b][c,d]}
\end{align}
where $M_{ab} = S_{ab}+ A_{ab}$. This finishes the proof.

\section{Dimensional reduction of chiral tensors}
\label{app:dimredchiral}

In this appendix, we derive the dimensional reduction of the six-dimensional self-dual fields considered in chapter \ref{chap:selfdual} in the prepotential formalism. They directly reduce to the five-dimensional actions of chapter \ref{chap:twisted}. In this appendix, capital indices go from one to five, and lowercase indices only go from one to four.

\subsection{The exotic graviton}

\paragraph{Reduction of the field and gauge transformations.}

The prepotential $Z_{IJKL}$ and its gauge parameters split into several pieces,
\begin{align}
Z_{IJKL} \;&\longrightarrow\; Z_{ijkl},\; Z_{ij k5},\; Z_{i5j5}, \\
\xi_{IJK} \;&\longrightarrow\; \xi_{ijk},\; \xi_{i5j},\; \xi_{ij5},\; \xi_{i55}, \\
\lambda_{IJ} \;&\longrightarrow\; \lambda_{ij},\; \lambda_{i5},\; \lambda_{55}.
\end{align}
We absorb some numerical factors in the parameters to write the gauge transformations as
\begin{equation} \label{eq:6d22gauge}
\delta Z_{IJKL} = \xi_{IJ[K,L]} + \xi_{KL[I,J]} + \delta_{[I[K} \lambda_{J]L]}
\end{equation}
instead of \eqref{Weyl}.

The transformation of $Z_{i5j5}$ under Weyl rescalings is then
\begin{equation}
\delta Z_{i5j5} = \frac{1}{4} ( \lambda_{ij} + \delta_{ij} \lambda_{55} ) .
\end{equation}
Therefore, using $\lambda_{ij}$, one can gauge away $Z_{i5j5}$ by a Weyl transformation. We will therefore set $Z_{i5j5} = 0$ from now on. To preserve this condition, one must then set $\xi_{i55} = 0$ and $\lambda_{ij} = - \delta_{ij} \lambda_{55}$. 
Also, we can split $\xi_{i5j} = - \xi_{5ij}$ into its symmetric and antisymmetric parts,
\begin{equation}
\xi_{5ij} = s_{ij} + a_{ij}, \qquad s_{ij} = \xi_{5(ij)}, \qquad a_{ij} = \xi_{5[ij]} .
\end{equation}
The cyclic identity $3\,\xi_{[IJK]} = \xi_{IJK} + \xi_{JKI} + \xi_{KIJ} = 0$ implies that the $\xi_{ij5}$ component is not independent: $\xi_{ij5} = - 2 a_{ij}$.
The gauge transformations of the remaining fields are then
\begin{align}
\delta Z_{ijkl} &= \xi_{ij[k,l]} + \xi_{kl[i,j]} - \frac{1}{2} (\delta_{ik} \delta_{jl} - \delta_{il} \delta_{jk} ) \lambda_{55} \\
\delta Z_{ijk5} &= \partial_k a_{ij} + \partial_{[i} s_{j]k} - \partial_{[i} a_{j]k} + \frac{1}{2} \delta_{k[i} \lambda_{j]5}. 
\end{align}
Those are exactly the gauge transformations \eqref{eq:phigaugeapp} and \eqref{eq:Pgaugeapp} for the prepotentials of linearized gravity in five dimensions, provided we identify the fields and gauge parameters as
\begin{align}
P_{ijkl} &= Z_{ijkl}, \quad \phi_{ijk} = Z_{ijk5}, \\
\alpha_{ijk} &= \xi_{ijk},  \quad A_{ij} = a_{ij}, \quad M_{ij} = s_{ij}, \quad \xi = -\lambda_{55}, \quad B_i = \frac{1}{2}\lambda_{i5} .
\end{align}

\paragraph{Reduction of the curvature tensors.} The Einstein, Schouten and Cotton tensors of $Z_{IJKL}$ reduce as follows:
\begin{itemize}
\item Einstein:
\begin{align}
G_{ijkl}[Z] = 0, \quad G_{ijk5}[Z] = - \frac{1}{18} G_{ijk}[\phi], \quad G_{i5j5}[Z] = \frac{1}{3!^2} G_{ij}[P],
\end{align}
\item Schouten:
\begin{align}
S_{ijkl}[Z] = - \frac{1}{18} \delta_{[i[k} S_{l]j]}[P], \quad S_{ijk5}[Z] = - \frac{1}{18} S_{ijk}[\phi], \quad S_{i5j5}[Z] = \frac{1}{2.3!^2} S_{ij}[P],
\end{align}
\item Cotton:
\begin{align}
D_{ijkl}[Z] = - \frac{1}{3.18} D_{ijkl}[\phi], \quad D_{ijk5}[Z] = \frac{1}{3!^3} D_{ijk}[P], \quad D_{i5j5}[Z] = \frac{1}{3.18} D\indices{_i^k_{jk}}[\phi],
\end{align}
\end{itemize}
where the right-hand sides are given by the corresponding tensors of sections \ref{app:confphi} and \ref{app:confP} for the graviton prepotentials.

\paragraph{Reduction of the action.} We can now use those formulas to reduce the six-dimensional action \eqref{Action}. This gives
\begin{align}
S[\phi,P] = &\frac{1}{3.36} \int \dtdx[5] \left( \phi_{ijk} \dot{D}^{ijk}[P] - P_{ijkl} \dot{D}^{ijkl}[\phi] \right) \\
&-\frac{1}{9.6} \int \dtdx[5] \left( \phi_{ijk} \varepsilon^{kabc} \partial_a D\indices{_{bc}^{ij}} + \frac{1}{8}P_{ijkl} \varepsilon^{ijab}[\phi] \partial_a D\indices{^{kl}_b}[P] \right) . \nonumber
\end{align}
This is the action of section \ref{sec:actiongraviton}, up to the redefinitions $P_{ijkl} = 12 \sqrt{3} \,P'_{ijkl}$ and $\phi_{ijk} = 3 \sqrt{3} \,\phi'_{ijk}$. This result already appeared in \cite{Henneaux:2016opm}\footnote{The sign discrepancy with respect to the appendix C of that paper comes from the fact that the prepotential $\phi_{ijk}$ we use here differs by a sign from the one of reference \cite{Bunster:2013oaa} (see also footnote \ref{footnote:sign} on page \pageref{footnote:sign}).} but is made more transparent by using the appropriate Cotton tensors in five dimensions.

\subsection{The exotic gravitino}

\paragraph{Reduction of the field and gauge transformations.}

For this appendix, we write the six-dimensional prepotential as $X_{IJ}$ and also use capital letters for the gauge parameters,
\begin{equation}\label{eq:GaugeChiapp}
\delta X_{IJ} = \partial_{[I} \Lambda_{J]} + \Gamma_{[I} W_{J]} \, .
\end{equation}

For the dimensional reduction, we use the form \eqref{eq:gammasixd} of the six-dimensional gamma matrices. In particular, the block-diagonal form of $\Gamma_7$ implies that the prepotential $X_{IJ}$ only has components in the first block,
\begin{equation}
X_{IJ} = \begin{pmatrix}
\hat{\chi}_{IJ} \\ 0
\end{pmatrix}.
\end{equation}
The field $\hat{\chi}_{IJ}$ then splits into two parts, $\hat{\chi}_{ij}$ and $\hat{\chi}_{i5}$. From \eqref{eq:GaugeChiapp}, we find their gauge transformations
\begin{align}
\delta \hat{\chi}_{ij} = \partial_{[i} \hat{\eta}_{j]} + \gamma_{[i} \hat{\rho}_{j]}, \quad \delta \hat{\chi}_{i5} = \frac{1}{2} (\partial_i \hat{\eta}_5 + \gamma_i \hat{\rho}_5 + i \hat{\rho}_i ),
\end{align}
where $\hat{\eta}_I$ and $\hat{\rho}_I$ are given from the six-dimensional gauge parameters by the block form
\begin{equation}
\Lambda_{I} = \begin{pmatrix}
\hat{\eta}_{I} \\ 0
\end{pmatrix},\qquad W_{I} = \begin{pmatrix}
0\\ \hat{\rho}_{I}
\end{pmatrix}.
\end{equation}
Using $\hat{\rho}_i$, the field $\hat{\chi}_{i5}$ can be set to zero. The residual gauge transformations must satisfy $\delta \hat{\chi}_{i5} = 0$ to respect this choice; this imposes $\hat{\rho}_i = i (\partial_i \hat{\eta}_5 + \gamma_i \hat{\rho}_5)$.
The gauge transformations of $\hat{\chi}_{ij}$ are then
\begin{equation}
\delta \hat{\chi}_{ij} = 2 \partial_{[i} \eta_{j]} + \gamma_{ij} \rho, \qquad \text{with} \quad \eta_j = \frac{1}{2} \left( \hat{\eta}_j - i \gamma_j \hat{\eta}_5\right), \quad \rho = i \hat{\rho}_5 .
\end{equation}
They have exactly the form \eqref{eq:chigaugeapp} of the gauge transformations of the prepotential for the five-dimensional gravitino.

\paragraph{Reduction of the curvature tensors.}

The various curvature tensors of $X_{IJ}$ reduce as follows:
\begin{itemize}
\item Einstein:
\begin{equation}
\hat{G}_{ij} = 0, \quad \hat{G}_{i5} = - G_i[\hat{\chi}] ,
\end{equation}
\item Schouten:
\begin{equation}
\hat{S}_{ij} = i \gamma_{[i} S_{j]}[\hat{\chi}], \quad \hat{S}_{i5} = \frac{1}{2} S_i[\hat{\chi}] ,
\end{equation}
\item Cotton:
\begin{equation}
\hat{D}_{ij} = D_{ij}[\hat{\chi}], \quad \hat{D}_{i5} = i \gamma^k D_{ik}[\hat{\chi}] ,
\end{equation}
\end{itemize}
where the left-hand sides are defined by the block forms
\begin{equation}
G_{IJ}[X] = \begin{pmatrix}
\hat{G}_{IJ} \\ 0
\end{pmatrix}, \quad
S_{IJ}[X] = \begin{pmatrix}
\hat{S}_{IJ} \\ 0
\end{pmatrix}, \quad
D_{IJ}[X] = \begin{pmatrix}
\hat{D}_{IJ} \\ 0
\end{pmatrix},
\end{equation}
and the right-hand sides are given by the appropriate tensors of section \ref{app:conffermion}.

\paragraph{Reduction of the action.} Using the above formulas, the action \eqref{eq:exoticgravitino} reduces to
\begin{align}
S[\chi] = -2i \int \!dt \,d^4\!x\, \hat{\chi}^\dagger_{ij} \left( \dot{D}^{ij}[\hat{\chi}] - i \varepsilon^{ijkl} \gamma^m \partial_k D_{lm}[\hat{\chi}] \right),
\end{align}
which is exactly the action \eqref{eq:actionchi} for the gravitino in five dimensions in the prepotential formalism, up to the redefinition $\hat{\chi}_{ij} = \chi_{ij} / \sqrt{2}$.

\subsection{The exotic graviton-photon}
\label{sec:dimred}

\paragraph{Reduction of the field and gauge transformations.} The prepotential $Z\indices{^{IJK}_{AB}}$ splits into
\begin{equation}
Z\indices{^{ijk}_{ab}}\, , \quad Z\indices{^{ij5}_{ab}}\, , \quad Z\indices{^{ijk}_{a5}}\, , \quad Z\indices{^{ij5}_{a5}}\, .
\end{equation}
All the pieces have irreducible Young symmetry except for $Z\indices{^{ij5}_{ab}}$, which has components of {\tiny$\ydiagram{2,2}$} and {\tiny$\ydiagram{2,1,1}$} symmetry.
The cyclic identity $Z_{[IJKL]M} = 0$ implies
\begin{equation}
Z_{5[ijk]l} = - \frac{1}{3} Z_{ijkl5}\, ,
\end{equation}
which gives $Z_{5[ijkl]} = 0$. We can then split $Z\indices{^{ij5}_{ab}}$ in irreducible parts as
\begin{align}
Z\indices{^{ij5}_{ab}} &= R\indices{^{ij}_{ab}} + 2 Q\indices{^{ij}_{[ab]}}\, ,
\end{align}
with
\begin{align}
R_{ijab} &= \mathbb{P}_{(2,2)} \left( Z_{ij5ab} \right) = \frac{1}{2} (Z_{ij5ab} + Z_{ab5ij})\, , \\
Q_{ijab} &= \mathbb{P}_{(2,1,1)} \left( Z_{ij5ab} \right) = \frac{3}{2} Z_{5[ija]b} = - \frac{1}{2} Z_{ijab5}\, .
\end{align}
For the other two irreducible components, we make the triangular change of variables
\begin{align}
M\indices{^{ij}_a} &= Z\indices{^{ij5}_{a5}} \,, \\
N\indices{^{ijk}_{ab}} &= Z\indices{^{ijk}_{ab}} + 3 \delta^{[i}_{[a} Z\indices{^{jk]5}_{b]5}} \, .
\end{align}
The tensors $M$ and $N$ have $(2,1)$ and $(2,1,1)$ Young symmetries, respectively. The following inversion formulas are useful for explicit computations:
\begin{align}
Z\indices{^{ijk}_{ab}} &= N\indices{^{ijk}_{ab}} - 3 \delta^{[i}_{[a} M\indices{^{jk]}_{b]}}\, , \\
Z\indices{^{ij5}_{ab}} &= R\indices{^{ij}_{ab}} + 2 Q\indices{^{ij}_{[ab]}}\, , \\
Z_{ijab5} &= - 2 Q_{ijab}\, , \\
Z\indices{^{ij5}_{a5}} &= M\indices{^{ij}_a}\, .
\end{align}
With these definitions, the spatial components $\phi_{IJK}$ of the chiral tensor decompose as, according to \eqref{phiofZ},
\begin{align}
\phi_{ijk} &= \frac{1}{6} \pd^a \left( R\indices{^{bc}_{ij}} \varepsilon_{kabc} - R\indices{^{bc}_{k[i}} \varepsilon_{j]abc} \right)\, , \label{eq:phiofR}\\
\phi_{5(ij)} &= - \frac{1}{4} \partial^a M\indices{^{bc}_{(i}} \varepsilon_{j)abc}\, , \label{eq:phiofM}\\
\phi_{5[ij]} &= - \frac{1}{36} \partial^a N\indices{^{bcd}_{ij}}\varepsilon_{abcd}\, , \\
\phi_{i55} &= - \frac{1}{6} \pd^a Q\indices{^{bcd}_i} \varepsilon_{abcd}\, ,
\end{align}
up to a gauge transformation (the fact that the contribution of $Q$ to $\phi_{ijk}$ is pure gauge was proved in \cite{Bunster:2013oaa}).
Up to constant factors, equations \eqref{eq:phiofM} and \eqref{eq:phiofR} are exactly the expressions given in reference \cite{Bunster:2013oaa} for the metric and its dual in terms of the prepotentials for five-dimensional gravity. Moreover, the last two equations motivate the definitions
\begin{equation}
V_i = \pd^a Q\indices{^{bcd}_i} \varepsilon_{abcd}, \quad W_{ij} = \partial^a N\indices{^{bcd}_{ij}}\varepsilon_{abcd},
\end{equation}
for the two potentials of the vector fields in five dimensions (up to factors that will be fixed below).

The gauge and Weyl transformations of the lower-dimensional prepotentials also match: the gauge and Weyl transformations of $R_{ijab}$ and $M_{ija}$ are exactly those of the prepotentials of five-dimensional linearized gravity. The prepotentials $N_{ijkab}$ and $Q_{abci}$ have no Weyl transformation, and their gauge transformations are exactly such that $V_i$ and $W_{ij}$ transform as total derivatives.

\paragraph{Reduction of the Cotton.} The Cotton tensor of $Z$ reduces as follows:
\begin{align}
D_{abcde}[Z] &= \frac{1}{3!^{2}} \varepsilon_{abcj} \pd^j \cB_{de}[V]\, , \label{Dabcde}\\
D_{abcd5}[Z] &= \frac{1}{3 \cdot 3!^2} \varepsilon_{abcj} \partial^j \cB_d[W]\, ,\label{Dabcd5} \\
D_{ab5de}[Z] &= -\frac{1}{4 \cdot 3!} D_{abde}[M] + \frac{1}{3 \cdot 3!^2} \varepsilon_{abj[d}\pd^j \cB_{e]}[W]\, ,\label{Dab5de} \\
D_{ab5d5}[Z] &= \frac{1}{2 \cdot 4!}D_{abd}[R] - \frac{1}{2 \cdot 3!^{2}} \varepsilon_{abij}\pd^i \cB\indices{^j_d}[V]\, .\label{Dab5d5}
\end{align}
Here, the magnetic fields are \cite{Bunster:2011qp}
\begin{equation}
\cB_{ij}[V] = \varepsilon_{ijkl}\pd^k V^l, \quad \cB_{i}[W] = \frac{1}{2} \varepsilon_{ijkl} \pd^j W^{kl} .
\end{equation}
The definitions and properties of the Cotton tensors of the gravity prepotentials can be found in \cite{Lekeu:2018kul}.
\paragraph{Reduction of the action.} Using the formulas above, the reduction of the action \eqref{actionZ} is direct.

On the other hand, the action of five-dimensional gravity is given in the prepotential formalism by
\begin{align}
S[\Phi, P] = \int\!dt\, d^4\!x\, \big( &D_{ijkl}[\Phi] \, \dot{P}^{ijkl} - D_{ijk}[P] \,\dot{\Phi}^{ijk} \\
&- P_{ijkl}\, \varepsilon^{ij pq} \pd_p D\indices{^{kl}_q}[P] - \frac{1}{2} \Phi_{ijk} \, \varepsilon^{abkl} \pd_l D\indices{^{ij}_{ab}}[\Phi] \big) \, , \nonumber
\end{align}
and the action of a five-dimensional vector field is
\begin{align}
S[v,w] = \frac{1}{2} \int\!dt\, d^4\!x\, \big( &\cB_i[w] \dot{v}^i - \frac{1}{2} \cB_{ij}[v] \dot{w}^{ij} \\
&- \frac{1}{2} w_{ij} \varepsilon^{ijkl} \pd_k \cB_l[w] - \frac{1}{2} v_i \varepsilon^{ijkl} \partial_j \cB_{kl}[v] \big) \, ,\nonumber
\end{align}
where $v_i$ is a vector and $w_{ij}$ a two-form.

Comparing with the reduction of \eqref{actionZ}, we can check that the two match provided we identify the prepotentials as
\begin{equation}
R_{ijkl} = 4 \sqrt{2} \, P_{ijkl}\,,\quad M_{ijk} = 2 \sqrt{2} \, \Phi_{ijk}\,,\quad V_i = \sqrt{6}\, v_i\,,\quad W_{ij} = 3\sqrt{6}\, w_{ij}\,. 
\end{equation}

\section{Maximal supergravity in five dimensions}
\label{app:redlinsugra}

In this appendix, we write the action and supersymmetry transformations of linearized maximal supergravity in five dimensions \cite{Cremmer:1979uq} in the first-order (prepotential) formalism. It can be obtained by dimensional reduction of the $\cN = (4,0)$ or the $\cN = (3,1)$ theory. Comparison of the two enables us to fix the $\beta_1, \dots, \beta_{10}$ coefficients that were left undetermined in section \ref{sec:susy31}. Capital indices $A,B,\dots$ are $usp(8)$ indices (running from $1$ to $8$).

\subsection{Field content and linearized first-order action}
\label{sec:5daction}

\paragraph{Field content.} The field content of maximal supergravity in five dimensions is the following \cite{Cremmer:1979uq}: one metric $g_{\mu\nu}$, $8$ gravitinos $\Psi_\mu^A$, $27$ vectors $V^{AB}_\mu$, $48$ spin $1/2$ fields $\Psi^{ABC}$, and $42$ scalar fields $\Phi^{ABCD}$. They satisfy the appropriate $usp(8)$ irreducibility and reality conditions (there is no chirality in five dimensions). For fermions, those involve the five-dimensional analogue of the $\cB$-matrix, defined in this case by $\gamma_\mu^* = - \cB_{(5)} \gamma_\mu \cB_{(5)}^{-1}$.

\paragraph{Linearized action.} The first-order action for the linearized theory is the sum of the following five terms:
\begin{itemize}
\item The metric is described by two real prepotentials
\begin{equation}
\phi_{ijk} \sim \ydiagram{2,1}\,, \quad P_{ijkl} \sim \ydiagram{2,2}\,,
\end{equation}
with action
\begin{align}
S_2[\phi, P] = \int\!dt\, d^4\!x\, \big( &D_{ijkl}[\phi] \, \dot{P}^{ijkl} - D_{ijk}[P] \,\dot{\phi}^{ijk} \\
&- P_{ijkl}\, \varepsilon^{ij pq} \pd_p D\indices{^{kl}_q}[P] - \frac{1}{2} \phi_{ijk} \, \varepsilon^{abkl} \pd_l D\indices{^{ij}_{ab}}[\phi] \big) \,. \nonumber
\end{align}

\item The $8$ gravitinos are described by the prepotentials
\begin{equation}
\Theta^{A}_{ij}\, ,
\end{equation}
with action
\begin{equation}
S_{\frac{3}{2}}[\Theta] = - i \int \!dt \,d^4\!x\, \Theta^\dagger_{Aij} \left( \dot{D}^{Aij}[\Theta] - i \varepsilon^{ijkl} \gamma^m \partial_k D^A_{lm}[\Theta] \right) \,.
\end{equation}

\item The $27$ vectors are described by the two potentials
\begin{equation}
V_i^{AB}\,, \quad W_{ij}^{AB}\,,
\end{equation}
with action
\begin{align}
S_1[V,W] = \frac{1}{2} \int\!dt\, d^4\!x\, \big( &\dot{V}^{*i}_{AB} \cB^{AB}_i[W] - \frac{1}{2} \dot{W}^{*ij}_{AB}\cB^{AB}_{ij}[V] \\
&- \frac{1}{2} W^*_{ABij} \varepsilon^{ijkl} \pd_k \cB^{AB}_l[W] - \frac{1}{2} V^*_{ABi} \varepsilon^{ijkl} \partial_j \cB^{AB}_{kl}[V] \big)\,.\nonumber
\end{align}

\item The action for the 48 spin $1/2$ fields is
\begin{equation}
S_{\frac{1}{2}}[\Psi] = i \int \!dt \,d^4\!x\, \Psi^{\dagger}_{ABC} \left(\dot{\Psi}^{ABC} - \gamma^{0} \gamma^{i} \pd_{i} \Psi^{ABC} \right) \,.
\end{equation}

\item The 42 scalars are described by the usual Hamiltonian action
\begin{equation}
    S_0[\Phi,\Pi] = \frac{1}{2} \int \!dt \,d^4\!x\, \left(2\, \Pi^{*}_{ABCD} \dot{\Phi}^{ABCD} - \Pi_{ABCD}^{*} \Pi^{ABCD} - \pd_{i} \Phi_{ABCD}^{*} \pd^{i} \Phi^{ABCD} \right)\, .
\end{equation}

\end{itemize}

\subsection{Dimensional reduction of the \texorpdfstring{$\cN = (4,0)$}{N = (4,0)} theory}
\label{sec:SUSY40}

\paragraph{Reduction of the action.} Five-dimensional maximal supergravity in the prepotential formalism, as described above, can be obtained from direct dimensional reduction of the $\cN = (4,0)$ theory using the following identifications (we write the higher-dimensional quantities with a hat):
\begin{itemize}
\item The prepotential $\hat{Z}_{IJKL}$ for the exotic graviton splits as
\begin{equation}
\hat{Z}_{ijkl} = 12 \sqrt{3}\, P_{ijkl}\,, \quad \hat{Z}_{ijk5} = 3\sqrt{3}\, \phi_{ijk}\,, \quad \hat{Z}_{i5j5}=0.
\end{equation}
\item The chiral $2$-forms $\hat{A}^{AB}_{IJ}$ give
\begin{equation}
\hat{A}^{AB}_{i5} = V_i^{AB} / \sqrt{2}\,, \quad \hat{A}^{AB}_{ij} = - W_{ij}^{AB} / \sqrt{2}\,.
\end{equation}
\item The chiral fermionic prepotential $\hat{\chi}^A_{IJ}$ is
\begin{equation}
\hat{\chi}^{A}_{IJ} = \begin{pmatrix} \hat{\chi}^{A+}_{IJ} \\ 0 \end{pmatrix}\,, \quad \hat{\chi}^{A+}_{ij} = \Theta^{A}_{ij} / \sqrt{2}  \,, \quad \hat{\chi}^{A+}_{i5}=0\,.
\end{equation}
\item The scalars stay the same, and the Dirac fields are simply
\begin{equation}
\hat{\Psi}^{ABC} = \begin{pmatrix} \Psi^{ABC} \\ 0 \end{pmatrix}\, .
\end{equation}
\end{itemize}
The form of the gamma matrices we use in the reduction implies that the $\cB$ matrix appearing in the reality conditions in six dimensions takes the form
\begin{equation}\label{eq:realitymatrix65}
    \cB_{(6)} = \begin{pmatrix} \cB_{(5)} & 0 \\ 0 & -\cB_{(5)} \end{pmatrix}\, ,
\end{equation}
where $\cB_{(5)}$ is the analog matrix in five dimensions. Therefore, the reality conditions in six and five dimensions agree.

\paragraph{Reduction of the supersymmetry transformations.} We can also find the supersymmetry transformations in five dimensions from the reduction of those of the $\cN = (4,0)$ theory. In this reduction, the supersymmetry parameter reduces as\footnote{The factor of $i$ comes from the form \eqref{eq:realitymatrix65} of the reality matrix: it is such that the reality condition $\hat{\epsilon}^*_A = \Omega_{AB} \cB_{(6)} \hat{\epsilon}^B$ in six dimensions implies correctly $\epsilon^*_A = \Omega_{AB} \cB_{(5)} \epsilon^B$ in five.}
\begin{equation}\label{eq:SUSYparam40}
\hat{\epsilon}^{A} = \begin{pmatrix} 0 \\ i \epsilon^{A} \end{pmatrix} \,.
\end{equation}
With these identifications, we get the five-dimensional supersymmetry transformations\footnote{Some care must be taken in the supersymmetry transformation of the gravitino prepotentials $\Theta^{A}_{ij}$: it picks up an extra term when enforcing the gauge condition $\Theta^{A}_{i5} = 0$.}
\begin{align}
\delta P_{ijkl} &= - \frac{i \alpha_1}{12 \sqrt{6}} \,{\mathbb P}_{(2,2)} \left( \bar{\epsilon}_{A} \gamma_{ij} \Theta_{kl}^{A} \right) \\
\delta \phi_{ijk}&= \frac{\alpha_1}{6\sqrt{6}}  \, {\mathbb P}_{(2,1)}\left(\bar{\epsilon}_{A} \gamma_{k} \Theta_{ij}^{A}\right) \\
\delta \Theta^{A}_{ij} &= - \frac{\alpha_1}{12\sqrt{6}} \left( 2 \varepsilon_{qrkl}\pd^{r}P\indices{_{ij}^{kl}} \gamma^{q} + i \varepsilon_{pqrk}\pd^{r} \phi\indices{_{ij}^{k}}\gamma^{pq} - i \pd^r \phi\indices{^{kl}_{[i}} \varepsilon_{j]rkl} \right) \gamma^{0} \epsilon^{A} \nonumber \\
&\quad - \alpha_2 \left( \frac{i}{2} W^{AB}_{ij} +   \gamma_{[i}V^{AB}_{j]} \right) \Omega_{BC} \gamma^{0}\epsilon^{C} \\
\delta W^{AB}_{ij} &=-\alpha_2 \left(2  \bar{\epsilon}_{C} \gamma_{[i} S^{[A}_{j]}[\Theta]\Omega^{B]C} + \frac{1}{4} \Omega^{AB} \bar{\epsilon}_{C} \gamma_{[i}S^{C}_{j]}[\Theta]\right)+ \alpha_{3} \left(i\sqrt{2} \bar{\epsilon}_{C}\gamma_{ij} \Psi^{ABC}\right) 
\end{align}
\begin{align}
\delta V^{AB}_{i} &=-\alpha_2i \left( 2 \bar{\epsilon}_{C} S^{[A}_{i}[\Theta] \Omega^{B]C}+ \frac{1}{4} \Omega^{AB} \bar{\epsilon}_{C} S^{C}_{i}[\Theta]\right) + \alpha_3 \sqrt{2} \left( \bar{\epsilon}_{C} \gamma_{i} \Psi^{ABC} \right) \\
\delta \Psi^{ABC} &= -\alpha_3\frac{1}{2\sqrt{2}} i\gamma^{ij} \gamma^{0} \left(B^{[A B}_{ij}[V]  \epsilon^{C]} - \frac{1}{3}\Omega^{[AB}B^{C]D}_{ij}[V]\Omega_{DE} \epsilon^{E} \right)  \nonumber \\
&\quad +\alpha_3\frac{1}{\sqrt{2}}\gamma^{i}\gamma^{0} \left(B^{[AB}_{i}[W] \epsilon^{C]}-\frac{1}{3} \Omega^{[AB}B^{C]D}_{i}[W] \Omega_{DE}  \epsilon^{E} \right)\nonumber \\
&\quad+\alpha_4 \left(i \Pi^{ABCD}\Omega_{DE}\gamma^{0} \epsilon^{E} + i \pd_{i} \Phi^{ABCD} \Omega_{DE} \gamma^{i} \epsilon^{E}\right) \\
\delta \Phi^{ABCD} &= \alpha_4 \left( -2 i \bar{\epsilon}_{E} \Psi^{[ABC}\Omega^{D]E} -\frac{3}{2}i \bar{\epsilon}_{E} \Omega^{[AB} \Psi^{CD]E}\right) \\
\delta \Pi^{ABCD} &= \alpha_4 \left( -2i \bar{\epsilon}_{E} \gamma^{0} \gamma^{i} \pd_{i} \Psi^{[ABC}\Omega^{D]E} - \frac{3}{2}i \bar{\epsilon}_{E}\gamma^{0} \gamma^{i} \Omega^{[AB}\pd_{i} \Psi^{CD]E}\right) \, .
\end{align}
The constants $\alpha_i$ are determined up to an overall normalization\footnote{Requiring that the supersymmetry variations of the five-dimensional linearized metric and gravitinos have the usual normalization
\[ 
\delta h_{\mu\nu} = \bar{\epsilon}_A\,\gamma_{(\mu} \Psi^A_{\nu)}\, , \quad \delta \Psi^A_{\mu} = \frac{1}{4}\, \partial_{\rho} h_{\mu\nu}\, \gamma^{\nu\rho}\epsilon^A \, ,
\]
fixes $\alpha_1 = 3\sqrt{6}$ (i.e., $\kappa^2 = 1/2$), as can be seen by comparing with the formulas of section \ref{sec:susylingrav}.} by the relations
\begin{equation}\label{eq:relalphas}
\frac{2\alpha_1^2}{(3!)^3} = \alpha_2^2 = \frac{2\alpha_3^2}{3} = \frac{\alpha_4^2}{2} \equiv \kappa^2
\end{equation}
that follow from the supersymmetry algebra, as we saw in section \ref{sec:susy40alg}.

\subsection{Dimensional reduction of the \texorpdfstring{$\cN = (3,1)$}{N = (3,1)} theory}

\paragraph{Formulas for dimensional reduction.} The dimensional reduction of the $\cN = (3,1)$ theory in six dimensions goes as follows:
\begin{itemize}
\item The reduction of the prepotential $Z_{IJKLM}$ was done in section \ref{sec:dimred}.

\item The reduction of the chiral two-forms and exotic gravitinos are the same as in the case of the $\cN = (4,0)$ theory, but with the appropriate symplectic indices.

\item From the reduction of $\hat{\theta}^\tta_{IJK}$, we get the prepotential for the gravitinos,
\begin{equation}
\hat{\theta}^\tta_{IJK} = \begin{pmatrix} \hat{\theta}^{\tta +}_{IJK} \\ 0 \end{pmatrix}\, , \quad \theta^\tta_{ij} = 3\hat{\theta}^{\tta +}_{ij5}\, ,
\end{equation}
and also one spin $1/2$ field
\begin{equation}
\psi^\tta = \varepsilon^{ijkl} \partial_i \zeta^\tta_{jkl}\, , \quad \zeta^\tta_{jkl} =  i\sqrt{\frac{3}{4}} \left( i \hat{\theta}^{\tta +}_{jkl} + \gamma_{[j} \hat{\theta}^{\tta +}_{kl]5} \right)\, .
\end{equation}
This is the usual Kaluza-Klein reduction of a Rarita-Schwinger field. It was done explicitly in the prepotential formalism in \cite{Lekeu:2018kul} to which we refer for details. We have added an extra factor of $i$ in $\psi^\tta$ with respect to that paper in order to take the five-dimensional reality condition into account.
\item From the vector field, we have the two potentials
\begin{align}
v^{\tta\ttb}_i = \hat{V}^{\tta\ttb}_i\, , \quad w^{\tta\ttb}_{ij} = \hat{W}^{\tta\ttb}_{ij5}
\end{align}
for vector fields in five dimensions, and also the scalar fields
\begin{align}
\phi^{\tta\ttb} = \hat{V}^{\tta\ttb}_5\, , \quad \pi^{\tta\ttb} = \frac{1}{3!} \, \varepsilon^{abcd} \pd_a \hat{W}^{\tta\ttb}_{bcd}\, .
\end{align}

\item Because of the chirality conditions, the Dirac fields reduce as
\begin{equation}
\hat{\psi}^{\tta\ttb\alpha} = \begin{pmatrix} \psi^{\tta\ttb\alpha} \\ 0 \end{pmatrix}, \quad \hat{\tilde{\psi}}^{\tta\ttb\ttc} = \begin{pmatrix} 0 \\ i \psi^{\tta\ttb\ttc} \end{pmatrix}\, .
\end{equation}

\item The scalar fields stay the same.

\item For the supersymmetry parameters, we have
\begin{equation}
\hat{\epsilon}^{\tta} = \begin{pmatrix} 0 \\ i \epsilon^{\tta} \end{pmatrix}\, , \quad \hat{\tilde{\epsilon}}^{\alpha} = \begin{pmatrix} \epsilon^\alpha \\ 0 \end{pmatrix} \,.
\end{equation} 
\end{itemize}

\paragraph{Index split.} To make contact with the $usp(8)$-invariant five-dimensional theory, we split its indices as $A = (\tta\, , \alpha)$, where $\tta=1,\dots,6$ are $usp(6)$ indices and $\alpha = 1 , 2$ are $usp(2)$ indices. The $8\times 8$ matrix $\Omega_{AB}$ matrix is then block diagonal,
\begin{equation}
(\Omega_{AB}) = \begin{pmatrix} \Omega_{\tta\ttb} & 0 \\ 0 & \varepsilon_{\alpha\beta} \end{pmatrix} \,.
\end{equation}
The corresponding splitting of the fields is then
\begin{eqnarray}
\Theta^{a}= \theta^{a}\, , \quad \quad \Theta^{\alpha} = \theta^{\alpha} \, ,
\end{eqnarray}
and
\begin{align}
  W^{\tta \ttb}_{ij}&= w^{\tta \ttb}_{ij} +\frac{1}{2\sqrt{6}} \Omega^{\tta \ttb} w_{ij}\, , & W^{\tta \alpha}_{ij} &=\frac{1}{\sqrt{2}} w^{\tta \alpha}_{ij} \, ,&  W^{\alpha \beta} &= -\frac{3}{2\sqrt{6}} w_{ij} \varepsilon^{\alpha \beta}  \, , \\
  V^{\tta \ttb}_{i}&= v^{\tta \ttb}_{i} +\frac{1}{2\sqrt{6}} \Omega^{\tta \ttb} v_{i} \, , & V^{\tta \alpha}_{i} &=\frac{1}{\sqrt{2}} v^{\tta \alpha}_{i} \, , & V^{\alpha \beta} &= -\frac{3}{2\sqrt{6}} v_{i} \varepsilon^{\alpha \beta}  \, , \\
 \Psi^{\tta \ttb \ttc} &= \psi^{\tta \ttb \ttc}+\frac{1}{2} \Omega^{[\tta \ttb}\psi^{\ttc]}\, ,& \Psi^{\tta \ttb \alpha} &= \frac{1}{\sqrt{3}} \psi^{\tta \ttb \alpha} \, , & \Psi^{\tta \alpha \beta} &= -\frac{1}{3} \psi^{\tta} \varepsilon^{\alpha \beta}\, ,\\
  \Phi^{\tta \ttb \ttc \ttd} &= -\sqrt{\frac{3}{2}} \Omega^{[\tta \ttb}\phi^{\ttc \ttd]} \, ,& \Phi^{\tta \ttb \ttc \alpha} &= \frac{1}{2}\phi^{\tta \ttb \ttc \alpha} \, , & \Phi^{\tta \ttb \alpha \beta} &= \frac{1}{2\sqrt{6}} \phi^{\tta \ttb} \varepsilon^{\alpha \beta} \, ,\\
   \Pi^{\tta \ttb \ttc \ttd} &= -\sqrt{\frac{3}{2}} \Omega^{[\tta \ttb}\pi^{\ttc \ttd]} \, , & \Pi^{\tta \ttb \ttc \alpha} &= \frac{1}{2}\pi^{\tta \ttb \ttc \alpha} \, , & \Pi^{\tta \ttb \alpha \beta} &= \frac{1}{2\sqrt{6}} \pi^{\tta \ttb} \varepsilon^{\alpha \beta} \, .
\end{align}
These coefficients are determined uniquely (up to signs that can be absorbed by field redefinitions) from the following two conditions: 1) the compatibility of $usp(8)$ and $usp(6)\oplus usp(2)$ irreducibility conditions, and 2) the normalization of the action.
For example, for the scalar fields, those two conditions are
\begin{equation}
\Omega_{AB} \Phi^{ABCD} = 0 \quad \Leftrightarrow \quad \Omega_{\tta\ttb} \phi^{\tta\ttb} = 0\,, \; \Omega_{\tta\ttb}\phi^{\tta\ttb\ttc \alpha} = 0
\end{equation}
and
\begin{equation}
\Phi^*_{ABCD} \Phi^{ABCD} = \phi^*_{\tta\ttb\ttc \alpha} \phi^{\tta\ttb\ttc \alpha} + \phi^*_{\tta\ttb} \phi^{\tta\ttb}\, .
\end{equation}
Reality conditions are also compatible. Using these formulas, the reduction of the $\cN = (3,1)$ theory gives exactly the action of maximal supergravity in five dimensions presented in subsection \ref{sec:5daction}.

\paragraph{Comparison of supersymmetry transformations.} It is also straightforward to reduce the supersymmetry transformations of section \ref{sec:susy31} to five dimensions. On the other hand, we can use this splitting of indices in the supersymmetry transformations of section \ref{sec:SUSY40}. Comparing the two enables us to fix the $\beta_i$ constants of the $\cN = (3,1)$ theory as
\begin{align}
\beta_1    & = \frac{1}{\sqrt{3}} \alpha_1 = -6 \,\alpha_2 = 2\sqrt{6} \,\alpha_3 \, ,            &\beta_6    & = \frac{1}{\sqrt{6}}\alpha_3 \, ,                                                    \\
\beta_2    & = -\frac{\sqrt{2}}{3\sqrt{3}} \alpha_1 = 2\sqrt{2}  \alpha_2 \, ,                    &\beta_7    & = \frac{1}{\sqrt{6}} \alpha_3 = \frac{1}{2\sqrt{2}} \alpha_4 \, ,                    \\
\beta_3    & = -\frac{\alpha_2}{2}= \frac{1}{\sqrt{6}} \alpha_3  \, ,                             &\beta_8    & = \frac{1}{\sqrt{2}} \alpha_3 = \frac{\sqrt{3}}{2\sqrt{2}} \alpha_4 \, ,             \\
\beta_4    & = -\frac{\alpha_2}{2}= \frac{1}{\sqrt{6}} \alpha_3=\frac{1}{2\sqrt{2}} \alpha_4 \, , &\beta_9    & = -\frac{\sqrt{3}}{2} \alpha_4 \, ,                                                  \\
\beta_5    & = -\frac{1}{\sqrt{2}} \alpha_2 \, ,                                                  &\beta_{10} & = -\frac{\alpha_4}{2} \, ,                                                             
\end{align}
in terms of the $\alpha_i$ constants of the $(4,0)$ theory. This is compatible with relations \eqref{eq:relalphas}, and gives the relations \eqref{eq:relbetas} announced in section \ref{sec:susy31}.

\section{Gamma matrices and spinors}
\label{app:gammamatrices}

We follow the conventions of \cite{VanProeyen:1999ni,freedman_vanproeyen_2012}. Gamma matrices are defined by
\begin{equation} \label{eq:cliff}
\{ \gamma^\mu, \gamma^\nu \} = 2 \eta^{\mu\nu},
\end{equation}
where the flat metric $\eta_{\mu\nu}$ is of ``mostly plus" signature, $\eta = \text{diag}(-+\dots  +)$. 
Useful identities on the spatial gamma matrices are
\begin{align}
\gamma_i \gamma_j &= \gamma_{ij} + \delta_{ij} \\
\gamma_i \gamma^{i j_1 \dots j_n} &= (d-n) \gamma^{j_1\dots j_n} \\
\frac{1}{d-1} \left( \gamma_{ij} - (d-2) \delta_{ij} \right) \gamma^{jk} &= \delta^k_i \label{eq:gammadelta} \\
\gamma_{ij} \gamma^{k_1 \dots k_n} &= \gamma\indices{_{ij}^{k_1 \dots k_n}} - 2 n\, \delta_{[i}^{[k_1} \gamma\indices{_{j]}^{k_2 \dots k_n]}} - n(n-1) \,\delta^{[k_1 k_2}_{ij} \gamma^{k_3 \dots k_n]} . \label{eq:g2gd-2}
\end{align}
(They are of course also valid when the indices are space-time indices, provided $\delta$ is replaced by $\eta$ and $d$ by $D$.)
We have the hermiticity properties\begin{equation} \label{gammaherm}
(\gamma^\mu)^\dagger = \gamma^0 \gamma^\mu \gamma^0 ,
\end{equation}
i.e. $(\gamma^0)^\dagger = - \gamma^0$ and $(\gamma^i)^\dagger = \gamma^i$ .
In even dimensions $D = 2m$, we can introduce the chirality matrix
\begin{equation}
\gamma_* = (-i)^{m+1} \gamma_0 \gamma_1 \dots \gamma_{D-1} .
\end{equation}
which satisfies
\begin{equation}\label{eq:gammastarprop}
(\gamma_*)^2 = I, \quad \{ \gamma_*, \gamma_\mu \} = 0, \quad (\gamma_*)^\dagger = \gamma_* .
\end{equation}
It makes the link between rank $r$ and rank $D-r$ antisymmetric products of gamma matrices,
\begin{equation} \label{eq:evenduality}
\gamma^{\mu_1 \dots \mu_r} = - \frac{(-i)^{m+1}}{(D-r)!} \varepsilon^{\mu_r \dots \mu_1 \nu_1 \dots \nu_{D-r}} \gamma_{\nu_1 \dots \nu_{D-r}} \gamma_*
\end{equation}
(notice the index ordering).
In odd dimensions $D=2m+1$, there is no $\gamma_*$ and the analogue of this relation is
\begin{equation}
\gamma^{\mu_1 \dots \mu_r} = \frac{i^{m+1}}{(D-r)!} \varepsilon^{\mu_1 \dots \mu_D} \gamma_{\mu_D \dots \mu_{r+1}} .
\end{equation}
We use the convention $\varepsilon_{012\dots (D-1)} = +1 = - \varepsilon^{012\dots (D-1)}$ for the totally antisymmetric $\varepsilon$ tensor. Using these relations, one can prove the identities
\begin{align}
\varepsilon^{ijl_1 \dots l_{d-2}} \gamma_{l_1 \dots l_{d-2}} &= i^{m+1} (d-2)! \gamma^{ij} \gamma_0 \hat{\gamma}, \label{eq:gij} \\
\gamma^0 \gamma^{ijk} &= \frac{(-i)^{m+1}}{(d-3)!} \varepsilon^{ijk l_1 \dots l_{d-3}} \gamma_{l_1 \dots l_{d-3}} \hat{\gamma}, \label{eq:gijk}
\end{align}
where $m = \lfloor D/2 \rfloor$ and where we define $\hat{\gamma}$ to be the chirality matrix $\gamma_*$ in even space-time dimension and the identity matrix in odd space-time dimension,
\begin{equation}
m = \lfloor D/2 \rfloor = \lfloor (d+1)/2 \rfloor, \quad \, \hat{\gamma} = \left\{ \begin{array}{ll}
\gamma_* &\text{ if $D$ is even} \\
I &\text{ if $D$ is odd.}
\end{array} \right.
\end{equation}
The spatial $\varepsilon$ tensor is $\varepsilon_{12\dots d} = +1 = \varepsilon^{12\dots d}$, and spatial indices are contracted with the spatial metric $\delta_{ij}$.
With these definitions, equations \eqref{eq:gij} and \eqref{eq:gijk} are valid in all dimensions. The $\hat{\gamma}$ matrix satisfies
\begin{equation}
(\hat{\gamma})^\dagger = \hat{\gamma}, \quad \hat{\gamma}^2 = I, \quad \hat{\gamma} \gamma_\mu = (-1)^{D+1}\, \gamma_\mu \hat{\gamma} = (-1)^{d}\, \gamma_\mu \hat{\gamma}.
\end{equation}
Using relation \eqref{eq:gij} and its dual, the following identity can also be proved:
\begin{equation}\label{eq:g2epsilon}
\varepsilon_{i_1 \dots i_{d-1}j} \gamma^{jk} = \frac{1}{2}(-1)^{d-1}(d-1)\, \delta^k_{[i_1} \varepsilon_{i_2 \dots i_{d-1}] pq} \gamma^{pq} .
\end{equation}

\subsubsection{Spinors and gamma matrices in six spacetime dimensions}

We now write some useful formulas in the special case of six space-time dimensions ($D = 6$, $d = 5$).

Gamma matrices in six spacetime dimensions are denoted by a Greek capital $\Gamma$ letter. The chirality matrix is written $\Gamma_7$ and is defined as the product of all $\Gamma$-matrices,
\be
\Gamma_7 = \Gamma_0 \Gamma_1 \Gamma_2 \Gamma_3 \Gamma_4 \Gamma_5 \, .
\ee
The properties \eqref{eq:gammastarprop} imply that the chiral projectors
\be
P_L = \frac12 \left( I + \Gamma_7 \right), \; \; \; P_R = \frac12 \left( I - \Gamma_7 \right)
\ee
commute with the Lorentz generators.  One can thus impose the chirality
conditions   $ \psi = P_L  \psi$  ($\Leftrightarrow \Gamma_7 \psi = \psi$) or   $ \psi = P_R  \psi$  ($\Leftrightarrow \Gamma_7 \psi = - \psi$) on the spinors, which are then called (positive chirality or negative chirality) Weyl spinors.

The charge conjugation matrix $C$ is defined by the property
\be
- \left(\Gamma^\mu \right)^T =  C \Gamma^\mu C^{-1}
\ee
It is symmetric and unitary.  The matrices $C\Gamma^{\mu_1 \dots \mu_k}$ are symmetric for $k = 0, 3 \pmod{4}$ and antisymmetric for $k = 1, 2 \pmod{4}$. Defining also
\be
\mathcal{B} = - i C \Gamma^0
\ee
we have for the complex conjugate $\Gamma$-matrices
\be
\left(\Gamma^\mu \right)^* =  \mathcal{B} \Gamma^\mu \mathcal{B}^{-1}.
\ee
The matrix $ \mathcal{B}$ is unitary. It is antisymmetric and fulfills
\be 
 \mathcal{B}^*  \mathcal{B} = -I \, . \label{BSquare}
 \ee
 One has also
 \be
\left(\Gamma_7 \right)^* =  \mathcal{B} \Gamma_7 \mathcal{B}^{-1}. \label{RealGamma7}
\ee

The complex conjugate spinor $ \psi^*$ transforms in the same way as $\mathcal{B} \psi$ under Lorentz transformations or, what is the same,  $ \mathcal{B}^{-1} \psi^*$ transforms in the same way as $\psi$. Using \eqref{RealGamma7}, one furthermore sees that if  $\psi$ is a Weyl spinor of definite (positive or negative) chirality, then $\mathcal{B}^{-1} \psi^*$ is a Weyl spinor of same chirality.  The positive (respectively, negative) helicity representation is equivalent to its complex conjugate.
 It would be tempting to impose the reality condition $\psi^* = \mathcal{B} \psi$, but this is not possible: the consistency condition $\psi^{**} = \psi$ would impose $\mathcal{B}^*  \mathcal{B} = I$, but this contradicts \eqref{BSquare}. 
 
If we have
several spinors  $\psi^A$ ($A = 1, \cdots, 2n$), however, we can impose the condition
\be
\left( \psi^A \right)^* \equiv \psi_A^* = \Omega_{AB} \mathcal{B} \psi^B  \label{eq:SymMaj}
\ee
where the antisymmetric matrix $\Omega$ is the $2n \times 2n$ symplectic matrix
\be
\Omega = \begin{pmatrix}
0 & 1 & 0 & 0 &  \\
-1 & 0 & 0 & 0 & \cdots \\
0 & 0 & 0 & 1 & \\
0 & 0 & -1 & 0 & \\
& & \vdots &  & \ddots
\end{pmatrix} .
\ee
[For why the internal index is lowered as one takes the complex conjugate, see appendix \ref{app:usp8}.] Because the real matrix $\Omega$ squares to $-I$, $\Omega^2 = - I$,  the equation \eqref{eq:SymMaj} consistently implies $\psi^{A**} = \psi^A$.  Spinors fulfilling \eqref{eq:SymMaj} are called ``symplectic Majorana spinors''.  One can furthermore assume that the $\psi^A$'s are of definite chirality.  The Weyl and symplectic Majorana conditions define together  ``symplectic Majorana-Weyl spinors'' (of positive or negative chirality). For later purposes, we define $\Omega^{AB}$ (with indices up) through $\Omega^{AB}\Omega_{CB} = \delta^A_C$. The matrix $\Omega^{AB}$  is numerically equal to $ \Omega_{AB}$. 

The Dirac conjugate is defined as
\be
\bar{\psi} = i \psi^\dagger \Gamma^0.
\ee
 For symplectic Majorana spinors, it can be written as
 \be
 \bar{\psi}_A = \Omega_{AB} \left(\psi^B\right)^T C.
 \ee
 If $\psi^A$  and $\chi^A$ are two symplectic (anticommuting) Majorana spinors, then the product  $ \bar{\psi}_A \chi^A$ is a
real Lorentz scalar, which is symmetric for the exchange of $\psi^A$ with $\chi^A$,
\be
\bar{\psi}_A \chi^A = \bar{\chi}_A \psi^A \, .
\ee
More generally, the products $\bar{\psi}_A \Gamma^{\mu_1 \cdots \mu_k} \chi^A$ are real Lorentz tensors, which are symmetric under the exchange of $\psi^A$ with $\chi^A$ for $k = 0, 3 \pmod{4}$ and antisymmetric for $k = 1, 2 \pmod{4}$.
Moreover, if the spinors have definite chirality, some of these products vanish: if $\chi$ and $\psi$ have the same chirality, $\bar{\psi}_A \Gamma^{\mu_1 \cdots \mu_k} \chi^A$ vanishes for even $k$, while if $\chi$ and $\psi$ have opposite chiralities, it vanishes for odd $k$.

When the symplectic indices are not contracted, the rule for flipping the spinors is the following:
\begin{equation}
\bar{\psi}_A \Gamma^{\mu_1 \cdots \mu_k} \chi^B = (-1)^k\, \Omega^{BC} \Omega_{AD} \, \bar{\chi}_C \Gamma^{\mu_k \cdots \mu_1} \psi^D
\end{equation}
(notice the index reversal on the gamma matrix). This is useful for the computation of the supersymmetry commutators in section \ref{sec:susy40alg}.

\section{\texorpdfstring{$USp(2n)$}{USp(2n)} and reality conditions}
 \label{app:usp8}
 
The group $USp(2n)$ is defined as the group of $2n \times 2n$ complex matrices that are both unitary and symplectic,
\begin{equation}
USp(2n) = U(2n) \cap Sp(2n,\mathbb{C}) .
\end{equation}
We write $A,B\dots$ for indices ranging from $1$ to $2n$ in this appendix.
Quantities with indices upstairs transform in the fundamental,
\begin{equation}
v^A \rightarrow S\indices{^A_B} v^B, \qquad S \in USp(2n).
\end{equation}
Quantities with indices downstairs transform in the contragredient representation (i.e. with the inverse transpose $(S^{-1})^T$),
\begin{equation}
w_A \rightarrow (S^{-1})\indices{^B_A} w_B, \qquad S \in USp(2n).
\end{equation}
Therefore, the contraction $w_A v^A$ is $USp(2n)$ invariant. Because $USp(2n)$ matrices are unitary, $(S^{-1})^T = S^*$, the contragredient representation is actually the complex conjugate representation $w \rightarrow S^* w$. This motivates the notation
\begin{equation}
(v^A)^* \equiv v^*_A
\end{equation}
for the complex conjugates.
The matrices of $USp(2n)$ are also symplectic, $S^T \Omega S = \Omega$ (and $S \Omega S^T = \Omega$). Together with unitarity, this implies the property
\begin{equation}
\Omega S = S^* \Omega \qquad \forall S \in USp(2n)\, .
\end{equation}
Therefore, the quantity
\begin{equation}
w_A \equiv \Omega_{AB} v^B
\end{equation}
transforms indeed as its indices suggest, i.e. $w \rightarrow S^* w$ when $v \rightarrow Sv$.
Because both $v^*_A$ and $\Omega_{AB} v^B$ transform as quantities with indices down, the contractions $v^*_A v^A$ and $v^A \Omega_{AB} v^B$ are invariant.

Quantities with multiple indices transform in the corresponding tensor product of representations. In particular, we will make use of the irreducible antisymmetric representations of $USp(2n)$, described by totally antisymmetric tensors satisfying an $\Omega$-trace condition,
\begin{equation}
v^{A_1 A_2 \cdots A_k} = v^{[A_1 A_2 \cdots A_k]}\, ,\quad \Omega_{A_1 A_2} v^{A_1 A_2 \cdots A_k} = 0 \, .
\end{equation}
This is similar to the representation theory of the orthogonal groups, where one has to impose a tracelessness condition (using the invariant tensor $\delta_{AB}$ in that case).


\backmatter

\bibliographystyle{utphys}
\bibliography{bibliography}

\end{document}